%% file: main.tex
\documentclass[final,3p,times,reversenotenum]{elsarticle}

\usepackage{amssymb}
\usepackage[colorlinks]{hyperref}
\hypersetup{pdfauthor={JLab}}  % for hyperref: removes annoying warnings for titles with "math"
\usepackage{siunitx}
\usepackage{amsmath}

\RequirePackage{lineno}

\journal{Progress in Particle and Nuclear Physics}
\input{review_tool}

\begin{document}

%\linenumbers
\begin{frontmatter}

\title{
       Physics with CEBAF at 12~GeV and Future Opportunities %\tnoteref{label1}
       }

\input{AuthorList}

\input{Abstract} 




\end{frontmatter}
\let\WriteBookmarks\relax

\tableofcontents

\input{Section1} 

\input{Section2}

\input{Section3}

\input{Section4}

\input{Section5}

\input{Summary}

\input{Acknowledgments}

\appendix
\input{AppendixA}

\input{AppendixB}

\input{AppendixC}

\input{AppendixD}

\input{AppendixE}

\bibliographystyle{elsarticle-num} 
\bibliography{References}

\end{document}

%% file: review_tool.tex
\usepackage{mfirstuc} % to uppercase the name
\newcommand{\addReviewer}[2]{
  \expandafter\newcommand\csname #1\endcsname[1]{{\textbf{ \color{#2} \capitalisewords{#1}:\,##1}}}
  \expandafter\newcommand\csname #1cor\endcsname[2]{{\color{#2} \capitalisewords{#1}:\,\st{##1}{\textbf{##2}}}}
  \expandafter\newcommand\csname #1color\endcsname{#2}
  \expandafter\newcommand\csname #1todo\endcsname[1]{{\todo[inline,color=white!70!#2, caption={}]{\textbf{\capitalisewords{#1}}: ##1}}}
}

\usepackage{tikz, marginnote} % margin notes with circles
\newcommand{\checkedby}[1]{
\ifdefined\CROSSCHECKS
  \marginnote{
    \begin{tikzpicture}
      \foreach \x [count=\xi] in {#1} {
         \node[shape=circle,inner sep=0mm,
         minimum size=2mm,
         fill=\csname \x color\endcsname] at (\xi*3mm,0) {};
       }
    \end{tikzpicture}
  }
\else
\fi
}

\usepackage{soul,color}
\definecolor{chromeyellow}{rgb}{1.0, 0.65, 0.0}
\definecolor{DodgeBlue}{rgb}{0.118, 0.565,1.000}
\definecolor{asparagus}{rgb}{0.53, 0.66, 0.42}
\definecolor{cadmiumgreen}{rgb}{0.0, 0.42, 0.24}
\definecolor{blue(ryb)}{rgb}{0.01, 0.28, 1.0}
\definecolor{periwinkle}{RGB}{181, 146, 203}
\definecolor{turquoiseblue}{rgb}{0.02, 0.55, 0.55}
\definecolor{atlanticstorm}{RGB}{57,107,127}
\definecolor{amethyst}{rgb}{0.6, 0.4, 0.8}
\definecolor{blue-violet}{rgb}{0.54, 0.17, 0.89}
\definecolor{brightlavender}{rgb}{0.75, 0.58, 0.89}
\definecolor{brightmaroon}{rgb}{0.76, 0.13, 0.28}
\definecolor{brilliantrose}{rgb}{1.0, 0.33, 0.64}
\definecolor{byzantine}{rgb}{0.74, 0.2, 0.64}
\definecolor{darkmagenta}{rgb}{0.55, 0.0, 0.55}

\addReviewer{adam}{red}
\addReviewer{johna}{blue}
\addReviewer{jq}{cyan}

\addReviewer{bob}{chromeyellow}
\addReviewer{vincent}{brown}
\addReviewer{misha}{cadmiumgreen}
\addReviewer{patrizia}{magenta}
\addReviewer{miguel}{DodgeBlue}
\addReviewer{andrew}{purple}
\addReviewer{arkaitz}{asparagus}
\addReviewer{daniel}{periwinkle}
\addReviewer{astrid}{atlanticstorm}
\addReviewer{xiaochao}{brilliantrose}
\addReviewer{eugene}{byzantine}
\addReviewer{natalia}{brightmaroon}
\addReviewer{ian}{DodgeBlue}
\addReviewer{sout}{pink}

%% file: AuthorList.tex
%\author{A. B. \ Charm$,^{1,2}$ Q.\ Up,$^3$ \\ 
%L. \ Tau,$^{3,5}$  \\


\author[LBL]{J.~Arrington} 
\author[JLab,INFNGE]{M.~Battaglieri} 
\author[JLab]{A.~Boehnlein} 
\author[JLab]{S.A.~Bogacz}
\author[UTFSM]{W.K.~Brooks}
\author[JLab]{E.~Chudakov} 
\author[ANL]{I.~Clo\"et}
\author[JLab]{R.~Ent} 
\author[Duke]{H.~Gao} 
\author[JLab]{J.~Grames} 
\author[JLab]{L.~Harwood} 
\author[CNF,UMD]{X.~Ji} 
\author[JLab]{C.~Keppel} 
\author[JLab]{G.~Krafft} 
\author[JLab,CWM]{R.~D. ~McKeown\corref{cor1}}
%\ead{bmck@jlab.org}
\author[TU]{J.~Napolitano} 
\author[JLab,CWM]{J.W.~Qiu} 
\author[JLab,INFN]{P.~Rossi} 
\author[JLab]{M.~Schram}
\author[JLab]{S.~Stepanyan} 
\author[CWM]{J.~Stevens} 
\author[IU,CEEM,JLab]{A.P.~Szczepaniak} 
\author[SLAC]{N.~Toro} 
\author[UVA]{X.~Zheng} 

\cortext[cor1]{Corresponding author, {\it email address: \href{Email}{bmck@jlab.org}}}

\address[LBL]{Lawrence Berkeley National Laboratory, Berkeley, California 94720, USA} 
\address[JLab]{Thomas Jefferson National Accelerator Facility, Newport News, Virginia 23606, USA} 
\address[ANL]{Physics Division, Argonne National Laboratory, Lemont, Illinois 60439, USA}
\address[Duke]{Duke University, Durham, North Carolina 27708, USA} 
\address[CNF]{Center for Nuclear Femtography, SURA, 1201 New York Ave. NW, Washington, DC 20005, USA } 
\address[UMD]{Maryland Center for Fundamental Physics, Department of Physics, University of Maryland, College Park, Maryland 20742, USA} 
\address[TU]{Temple University, Philadelphia, Pennsylvania 19122, USA} 
\address[CWM]{William \& Mary, Williamsburg, Virginia 23185, USA} 
\address[SLAC]{SLAC National Accelerator Laboratory, 2575 Sand Hill Road, Menlo Park, California 94025, USA} 
\address[UTFSM] {Universidad Técnica Federico Santa María, Valparaiso, Chile}
\address[UVA]{University of Virginia, Charlottesville, Virginia 22904, USA} 
\address[IU]{Physics Department, Indiana University, Bloomington, Indiana 47405, USA}
\address[CEEM]{Center for Exploration of Energy and Matter, Indiana University, Bloomington, Indiana 47403, USA}
\address[INFN]{INFN, Laboratori Nazionali di Frascati, 00044 Frascati, Italy}
\address[INFNGE]{INFN, Sezione di Genova, 16146 Genova, Italy}

%% file: Abstract.tex
\begin{abstract} 
We summarize the ongoing scientific program of the 12 GeV Continuous Electron Beam Accelerator Facility (CEBAF) and give an outlook into future opportunities.  The program addresses important topics in nuclear, hadronic, and electroweak physics, including  nuclear femtography, meson and baryon spectroscopy, quarks and gluons in nuclei, precision tests of the standard model and dark sector searches.  Potential upgrades of CEBAF and their impact on scientific reach are discussed, such as higher luminosity, the addition of polarized and unpolarized positron beams, and doubling the beam energy.

\end{abstract}

%% file: Section1.tex
\section{Overview}

The ability to predict and understand the properties of nucleons and atomic nuclei from the first principles of Quantum Chromodynamics (QCD) with quarks and gluons as the underlying degrees of freedom is one of the goals of modern nuclear physics. Electron scattering at multi-GeV -- with resolutions ten times or more smaller than the size of the proton -- is a powerful microscope for probing the partonic structure and QCD dynamics of the nucleons and nuclei. With recent advances in accelerator and detector technologies, high-performing computing, new algorithms from machine learning and artificial intelligence, we are entering a new era of QCD intensity frontier in electron scattering, allowing for unprecedented precision and measurements of new observables and processes which are unimaginable previously.

The Nuclear Physics community has resourcefully exploited these advanced accelerator facilities for studies of the fundamental interactions. This research has impacted the entirety of Nuclear Physics, as well as High Energy Physics and Astrophysics. In fact, the ``Fundamental Symmetries and Neutrinos'' cohort within Nuclear Physics continues to make strides in both experiment and theory.
  
The Continuous Electron Beam Accelerator Facility (CEBAF) at Jefferson Lab has been delivering the world's highest intensity and highest precision multi-GeV electron beams for more than 25 years. The scientific justification for a project to double the CEBAF beam energy to 12~GeV was developed~\cite{Dudek:2012vr}, and construction began in 2008. This project was completed in Fall~2017, beginning a new era at the Laboratory. A total of more than 2094 publications of Nuclear Physics results, including both experiment and theory, have appeared in refereed journals.  A sizable fraction of these papers appeared in prestigious journals of high impact beyond the physics community such as Nature and Science, demonstrating the broader scientific impact of the results coming out of Jefferson Lab. 

The 12-GeV era includes a suite of new experimental equipment, including the Super Bigbite Spectrometer (SBS) in Hall~A, the CEBAF Large Acceptance Spectrometer (CLAS)-12 detector complex in Hall~B, the Super High Momentum Spectrometer (SHMS) in Hall~C, and the GlueX spectrometer and polarized photon beam in the newly completed Hall~D. Data are currently being acquired, and new results are being published, from all of these laboratory assets.

Innovations that exploited existing equipment including the latest additions have also yielded ground-breaking Nuclear Physics results. A few among these include imaging the nucleon in three dimensions via the extraction of Generalized Parton Distributions (GPDs) using Deeply Virtual Compton Scattering (DVCS) and the Transverse Momentum Dependent (TMD) and Parton Distribution Functions (PDF) from Inclusive and Semi-Inclusive Deep Inelastic Scattering (SIDIS); isospin decomposition of nucleon structure functions using a tritium gas target; observation of $J/\psi$ photoproduction from the proton near the threshold, limits on the photo-coupling of the $uudc\bar{c}$ pentaquark, %and the extraction of the matter radius of the proton,
%{\bf [Revise this after Hall C results are out]};
the resolution of the ``Proton Radius Puzzle'' through the PRad experiment with a major impact on atomic physics; and new information on neutron matter from parity violating electron scattering on $^{208}$Pb (PREX-II) and $^{48}$Ca (CREX) with implications for the structure of neutron stars.

Future additions to the experimental equipment contingent include the Solenoidal Large Intensity Device (SoLID) for measurements of SIDIS for nucleon tomography in three-dimensional momentum space, Parity-Violating Deep Inelastic Scattering (PVDIS) pushing the phase space in the search of new physics and of hadronic physics, and precision measurement of $J/\psi$ production from the proton near threshold to access the quantum anomalous energy contribution to the proton energy (mass of the proton in its rest frame ); the Measurement Of Lepton-Lepton Elastic Reaction (MOLLER) experiment, a test of the standard model and search for new physics ; new large angle tagging detectors (TDIS in Hall~A and ALERT in Hall~B); the neutral particle spectrometer (NPS) and the compact photon source (CPS) for a slate of approved experiments in Hall C; and an intense $K_L$ beamline that would serve new experiments in the GlueX spectrometer. 

Lattice QCD (LQCD), which offers a first-principle numerical
approach to QCD, predicted masses for new exotic mesons and baryons, and has matured to the point that it can reliably calculate  branching ratios of exotic decays.  Recent predictions for the full decay width of exotic $\pi_1$, as well as the first three-body Dalitz plot, from LQCD calculations
provide the much needed support and new opportunities for experimental searches of exotics at Jefferson Lab.  
%Although it is difficult, if not impossible, for LQCD to calculate the GPDs and TMDs directly to predict the internal landscape of nucleons due to its Euclidean space-time formulation, new developments in theory have made it possible to extract GPDs, TMDs and other quark/gluon correlations from LQCD calculations, complementary to the on-going phenomenological extractions of 3D hadron structure based on experimental data from Jefferson Lab and other facilities.
Due to its Euclidean space-time formulation it has been more 
difficult for LQCD to calculate the GPDs and TMDs directly and to predict 
the internal landscape of nucleons. However, new developments in theory 
have now made it possible to extract GPDs, TMDs and other quark/gluon 
correlations from LQCD calculations, testing and providing complementarity to the on-going 
phenomenological extractions of 3D hadron structure based on 
experimental data from Jefferson Lab and other facilities.

Furthermore, a new round of upgrades to CEBAF are presently under technical development. One of these is a potential energy upgrade to 24~GeV using novel magnet designs in the existing recirculation arcs. Another is a potential for intense polarized positron beams, which would allow for new measurements in nucleon tomography, and provide precision extraction of contributions from higher order electromagnetic processes. 

Clearly, CEBAF combines an illustrious history with an exciting future outlook. Our purpose in this article is to document this position, and provide a basis for future discussions as to how the Nuclear Physics community can best make use of this unique high intensity and high precision facility.

\subsection{Specific Scientific Accomplishments}

CEBAF and Jefferson Lab have produced results which impact all of Nuclear Physics, as well as High Energy Physics, Atomic Physics, and Astrophysics. These results are delineated and detailed in the remaining sections of this review. The following, however, is a short description of some specific accomplishments of the experimental program which should be of particular interest to the broader community.

The PRad Experiment (Section~\ref{sec:PRad}) has effectively settled the so-called ``Proton Charge Radius Puzzle''~\cite{CARLSON201559,gao2021proton}. Up until 2010, we believed that the proton charge radius was known with an accuracy close to 1\%, but an experiment based on the Lamb Shift in muonic hydrogen gave a number 4\% smaller than previous results  with a 0.1\% uncertainty~\cite{Pohl10,Anti13}. The PRad collaboration made a very precise measurement of the proton form factors explicitly at very low momentum transfer squared ($Q^2$), facilitating a highly accurate extrapolation to $Q^2=0$ and extraction of the proton charge radius. This experiment would not have been possible without the precision beams of CEBAF %available at Jefferson Lab 
and an innovative and new method of performing electron-proton elastic scattering measurements.

%Heavy quarkonia photoproduction close to threshold can be used to determine the quantum anomalous energy terms in QCD, which can in turn be used to determine the mass radius of the proton~\cite{Kharzeev:2021qkd}. Based on a measurement of $J/\psi$ photoproduction from GlueX the mass radius was found to be close to 0.55~fm, considerably smaller than the charge radius of 0.84~fm. More precision results are expected soon from GlueX, CLAS-12, and Hall~C, see Section~\ref{sec:PMass}. Future measurements with SoLID are expected to increase the precision drastically.

Heavy quark photoproduction can potentially be related to the quantum 
anomalous energy terms in QCD, which may in turn be used to determine 
the mass radius of the proton~\cite{Kharzeev:2021qkd}.  New precise measurements of $J/\psi$ 
photoproduction from GlueX, CLAS-12, and in Hall C (Section~\ref{sec:PMass}) are expected to 
further elucidate these issues. 

Precise measurements of the nucleon (proton and neutron) valence quark distribution functions have long been known to be important for discriminating useful models including those inspired by QCD. Determining these distributions from measurements of deep inelastic scattering from protons and deuterons, however, is problematic~\cite{Melnitchouk:1995fc, Accardi:2016qay} . The MARATHON experiment (Section~\ref{sec:duRatio}) has carried out a novel measurement by comparing scattering from $^3$He and tritium targets, which also has implications for nucleon correlations and the EMC effect in $A=3$ nuclei.

Accurate determinations of the neutron radii of spherical doubly-magic nuclei have important implications both for the neutron star equation of state~\cite{Horowitz:2001ya} and for microscopic calculations of nuclear structure~\cite{Hagen:2015yea}. Parity violating electron scattering (PVES) in the elastic regime provides a model-independent way of determining these neutron radii, because the value of the weak mixing angle $\sin^2\theta_W$ fortuitously leads to a neutral-current coupling of the neutron that is much larger than that for the proton. The high-quality polarized electron beams from CEBAF were used by the PREX-II and CREX experiments to determine the neutron radii of $^{208}$Pb and $^{48}$Ca, respectively. See Section~\ref{sec:PREX} for details.

One important motivation for the 12~GeV upgrade of CEBAF was to search for mesons with exotic quantum numbers, due to gluonic degrees of freedom, and is one of the primary goals of the GlueX experiment. One such exotic meson candidate, the $\pi_1(1400)/\pi_1(1600)$, has been reported in the $\eta\pi$ $P$-wave by several experiments~\cite{Zyla:2020zbs}. Elastic photoproduction of the $\eta\pi$ system from the proton has been studied by GlueX, including a detailed partial wave decomposition. These results are discussed in Section~\ref{sec:expmesons}.

The LHCb experiment at CERN has confirmed their observations of a heavy $c\bar{c}$ meson decaying to $J/\psi p$, a candidate for a pentaquark. Observed by them in the weak decay of $\Lambda_b^0$, the production cross section in $\gamma p\to J/\psi p$ would discriminate between a true pentaquark and possible molecular states or some dynamic effect~\cite{PhysRevD.100.054033}. The non-observation of this state in the $J/\psi$-007 experiment in Hall~C (Section~\ref{sec:CharmPentaquark}) suggests a non-pentaquark interpretation, although there is still much room for a smaller pentaquark-$J/\psi$ coupling.
%\eugene{This statement seems to be too strong. Indeed, some models can be ruled out, but in general there is a large room for the pentaquark-J/psi coupling. The article quoted points out that polarized measurements would be more sensitive. BTW at this time only the GlueX results has been published}.

The possible existence of particles with the quantum numbers of the photon but with only weak coupling to charge leptons, so called ``dark photons'', is attractive for a number of reasons~\cite{fabbrichesi2021physics}. The precise CEBAF beams have made possible searches for these objects, with results already in hand from APEX and HPS, see Section~\ref{sec:APEXHPS}. More results are forthcoming.

\subsection{Currently Planned Experimental Program}

Following the 49$^{th}$ meeting of the Jefferson Lab Program Advisory Committee (PAC) in Summer 2021, there are a total of 85 approved experiments\footnote{A list of approved experiments is available at {\sf https://www.jlab.org/physics/experiments}} in the 12 GeV program, of which 30 have received the highest scientific rating of A.  There are 56 approved experiments still waiting to run, representing at least a decade of running in the future.
Furthermore, PAC meetings are expected to continue each summer, preceded by a call for new proposals for beam time. Clearly, CEBAF is a facility in high demand.

Jefferson Lab continues to invest in facilities that make optimum use of CEBAF's capabilities, in particular those that will produce high-impact science across different areas within Nuclear Physics and beyond. This section reviews some of these new facilities, and points to the various approved experiments that will make use of them.

The SBS 
%Super Bigbite Spectrometer (SBS) -- CPS, NPS were spelled out, why not SBS?
facility was installed in Hall~A at Jefferson Lab in Fall~2021. The principal goal is to provide large acceptance at high luminosity so that small cross sections can be measured with high precision. Particle tracking is accomplished using Gas Electron Multiplier (GEM) detectors, which are able to run at high rates while providing excellent ($\sim70~\mu$m) position resolution. The high acceptance SBS is generally paired with a complementary spectrometer and detector system for tagging the scattered electrons. The first experiments to run will focus on neutron and proton magnetic and electric form factors, followed by SIDIS measurements. See Section~\ref{sec:MomTomNucleon}.

%A long standing problem in QCD is the discrepancy between experiment and theory for the pion structure function~\cite{PhysRevC.63.025213}.
Experimental determinations and theoretical interpretations of 
the pion structure function have been controversial for many years. 
Recently, the pion valence quark distribution has been extracted from 
LQCD calculations of matrix elements analyzed in terms of QCD collinear 
factorization, showing good agreements with experiments~\cite{PhysRevD.102.054508}. 
A lattice study of pion valence parton distribution within the framework of Large Momentum Effective Theory has also been reported~\cite{PhysRevD.100.034516}.
Experimental results to date are derived from the Drell-Yan process with a pion beam. CEBAF will be used to make a different type of measurement, using electron deep inelastic scattering (DIS) from the pion cloud of the nucleon. This Tagged Deep Inelastic Scattering (TDIS) measurement will be performed with the SBS apparatus, augmented with a radial time projection chamber to tag recoil protons from the reaction $n(e,e^\prime p)X$. 
Details of this TDIS measurement are discussed in Section~\ref{sec:PionKaonStructure}.

As an ambitious general purpose apparatus,  %Solenoidal Large Intensity Device (SoLID) %already defined
SoLID is being prepared for eventual use in Hall~A. SoLID makes use of the superconducting CLEO-III magnet from Cornell, which has already been delivered to the laboratory along with the yoke iron. SoLID will exploit the full capabilities of CEBAF, by combining the high intensity beam of excellent polarization quality with a large acceptance that includes a full azimuthal acceptance. %with several experiments already approved for SIDIS and for PVDIS. -- already mentioned, also J/Psi
The apparatus allows for transverse and longitudinally polarized targets for SIDIS measurements, see Section~\ref{sec:PionKaonStructure}. The PVDIS experiment will provide a precision measurement of %search for anomalous -- the coupling is not anomalous
electron-quark neutral current couplings as a test of the Standard Model, as discussed in Section~\ref{sec:BSM}. Measurements of near threshold $J/\psi$ production are also planned for SoLID allowing for the probe of quantum anomalous energy inside the proton.

The lightest nuclei, $^4$He, $^3$He, and $^2$H, offer many opportunities for understanding the fundamental partonic behavior of bound nucleons. Disecting these reactions, however, require the ability to tag spectator fragments in the midst of reactions with multi-GeV electrons. A Low Energy Recoil Tracker (ALERT) has been commissioned to join with CLAS12 to carry out these studies. Measurements on these light nuclei will include % Deeply Virtual Compton Scattering, 
DVCS, Deeply Virtual Meson Production (DVMP), and a variety of inclusive and semi-inclusive reactions. The many facets of this program including the EMC effect are discussed in Section~\ref{sec:ALERTmeasurements}.

A copious flux of kaons can be produced by the CEBAF beam through $\phi$ photoproduction. This is the basis for a new $K_L$ beam that will be developed in the Hall~D beam line, with the kaon beam incident on a target in the GlueX detector. The primary aim is spectroscopy of strange baryons and their excitations. This is discussed further in Section~\ref{sec:expbaryons}. At the same time, data will be collected on strange mesons through diffraction. 

New facilities are under development to search for physics beyond the Standard Model, once again using the unique capabilities of CEBAF. %The Measurement of Lepton-Lepton Elastic Reaction (MOLLER) %already defined
Among these, MOLLER is a large scale effort to make a percent level measurement of parity violation in electron-electron elastic scattering. This experiment will be sensitive to new electroweak interactions with mass scales on the order of 39~TeV. %see Eq.(13), should br 39 TeV
More details are given in Section~\ref{sec:MOLLER}. A new Beam Dump Experiment (BDX) is planned that would run parasitically with MOLLER (or other high intensity running), which will search for dark sector particles produced in the Hall~A beam dump. See Section~\ref{sec:BDX}.

\subsection{Future Science Opportunities}

It is becoming increasingly clear that there are exciting scientific opportunities using CEBAF beyond the currently planned decade of experiments. % with the recently upgraded 12~GeV facility. -- no need, the whole paper is about 12 GeV
In fact, one can envision that CEBAF will continue to operate with a fixed target program at the ``luminosity frontier,'' up to $10^{39}~{\rm cm}^{-2}{\rm s}^{-1}$, with large acceptance detection systems. This regime is a factor of 100,000 higher in luminosity than the Electron-Ion Collider (EIC), to be built in the next decade, and so represents very complementary capabilities even in the era of EIC operations.

Three dimensional imaging of the quark structure of the nucleon through DVCS and SIDIS 
%deeply virtual Compton scattering (DVCS) and semi-inclusive deep inelastic scattering (SIDIS) %defined many times in this section already
will be major programs for the presently planned CEBAF program as well as at the EIC. However, significant additional information can be obtained by studying double DVCS (DDVCS), a process with two virtual photons with cross sections and interaction rates a factor of 100 lower than DVCS. Therefore this process is not viable at EIC and must be studied using CEBAF at Jefferson Lab. Modest future detector upgrades will facilitate DDVCS studies in experimental Halls~A and~B.

Some very important information about nucleon imaging would require measurements of DVCS with positrons as well as electrons. A very large community of nuclear physicists has been holding workshops to develop these concepts in the last few years. Recent proposals to the Jefferson Lab PAC to use positron beams at CEBAF have received conditional approval, pending further study of the technical realization of positron beams. Operations with positron beams (polarized and unpolarized) will open a new area of study for CEBAF in the future.

Recently, the CBETA facility at Cornell has demonstrated eight pass recirculation of an electron beam with energy recovery (four accelerating beam passes and four decelerating beam passes). All eight beams are recirculated by single arcs of fixed field alternating gradient (FFA) magnets. This exciting new technology would enable a cost-effective method to double the energy of CEBAF, enabling new scientific opportunities in meson spectroscopy and extending the kinematic range of nucleon imaging studies. Technical studies of the implementation of FFA technology at CEBAF are in progress.

The individual sections of this paper include descriptions of a large number of scientific opportunities in the future of CEBAF, including higher energy beams and intense beams of polarized and unpolarized positrons.

\subsection{Complementarity with Existing and Future Experimental Facilities Worldwide}

The electron beams delivered by CEBAF will continue to be unparalleled in intensity and precision for decades to come. Energies up to 24~GeV will be available using an innovative upgrade to the accelerator arcs, and %polarized
positron beams will be delivered to the experimental halls. These facilities will keep CEBAF uniquely capable of a large number of important measurements in nuclear and hadronic physics. This section points out the various ways that CEBAF complements existing and planned facilities at other laboratories worldwide.

\subsubsection{The COMPASS Experiment at CERN and Electron-Positron Collider Experiments}

Currently the only fixed-target facility in operation that utilizes a lepton beam with the capability of both polarized beam and polarized target is the COMPASS experiment at CERN, using a muon beam at 160 GeV/$c$. The kinematic coverage of the COMPASS experiment is complementary to that of CEBAF at Jefferson Lab with significantly larger $Q^2$ values reaching 100 (GeV/c)$^2$ and a parton momentum fraction $x$ range of 0.008 to 0.21. The COMPASS collaboration plans to take the remaining SIDIS data using a polarized $^6$LiD target as an effective polarized ``neutron'' target in the near future.

The BES-III is a particle physics experiment at the Beijing Electron-Positron collider II at the Institute of High Energy Physics in China. Belle II -- an upgrade from the Belle experiment -- is a particle physics experiment at SuperKEKB, an electron-positron collider located at KEK in Japan. Both BES-III and Belle II contribute to the study of the three-dimensional imaging of the nucleon in momentum space through the extraction of the Collins fragmentation function from pion pairs produced in electron-positron collisions.

\subsubsection{Experiments at other Nuclear and Particle Physics facilities}

There are a number of experiments/programs ongoing at nuclear and particle physics facilities worldwide using hadron beams. These include the spin physics program at the Relativistic Heavy Ion Collider (RHIC) at Brookhaven National Laboratory (BNL) with polarized proton beams, and the polarized Drell-Yan experiment (SpinQuest) at Fermi National Accelerator Laboratory (FNAL) using polarized targets, both in the United States. Internationally, there are experiments using the primary proton beam of 30 GeV, and kaon and pion secondary beams at the Hadron Experimental Facility of Japan Proton Accelerator Research Complex (J-PARC). 

FAIR, an international accelerator facility in Darmstadt, Germany, is being built at GSI Helmholtzzentrum für Schwerionenforschung, where existing accelerator facilities will become part of FAIR and will serve as the first acceleration stage. The  FAIR experiment most relevant to the Jefferson Lab and the EIC science is the Antiproton Annihilation at Darmstadt (PANDA) experiment, in which anti-proton beams will be used to produce new particles in order to understand how mass is created by the strong nuclear force. 
Another future facility that is under construction is the Nuclotron-based Ion Collider fAcility (NICA) at the Joint Institute for Nuclear Research in Dubna, Russia, proposed by the Spin Physics Detector (SPD) collaboration, to study the spin structure of the proton and deuteron and other spin-related phenomena with polarized proton and deuteron beams at a collision energy up to 27 GeV and luminosity up to 10$^{32}$ cm$^{-2}$s$^{-1}$. 

\subsubsection{The Electron-Ion Collider (EIC) in the U.S.}

A principal aim of the EIC program is the study of the QCD sea inside the nucleon and nuclei, by focusing on quark, antiquark, and gluon parton distributions at small Bjorken $x$. The EIC program will accomplish this with much higher center-of-mass energies and exceptional luminosity for an environment of polarized colliding beams.

Of course, this aim is directly complementary to the extremely high luminosity, high $x$ reach of fixed target experiments that will be accomplished using CEBAF.
Contrasting these two accelerator environments, along with other current and prior facilities, it is not only evident that CEBAF and the EIC will each greatly extend  what is currently available, but they have significant overlap in the region of moderate $x$. In fact, the evolution of the structure functions from the valence into the sea quark region is an important area and motivation for study. The overlapping kinematic regions between CEBAF and the EIC will make it possible to study this evolution in detail, using both inclusive and semi-inclusive deep inelastic scattering reactions with polarized beams and targets or colliding beams, including, for CEBAF, polarized positron beams. Such studies will advance our knowledge about how QCD works which in turn allows for the extraction of the three-dimensional structural information of quarks and gluons inside the nucleon and nuclei.

\begin{figure}[t]
\begin{center}
\includegraphics[width=0.75\textwidth]{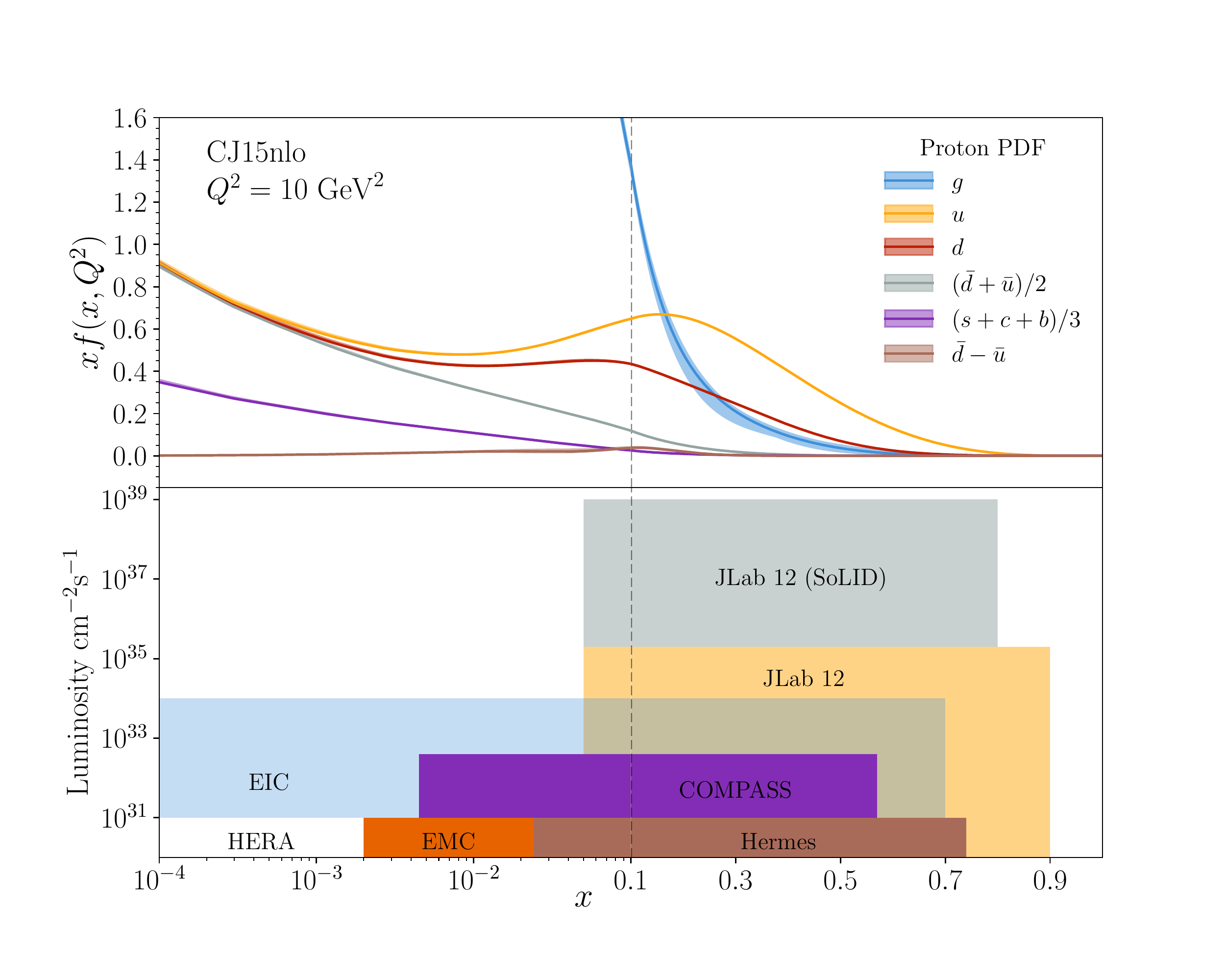}
\caption{Kinematic regions of Deep Inelastic Scattering and the comparative reach of EIC and CEBAF, as well as other facilities compared with parton distributions from CJ15~\cite{Accardi:2016qay}. CEBAF (at 12~GeV) will provide very high luminosity in the valence quark region, while providing overlap here with EIC measurements. The focus of the EIC will be on the sea at very low $x$, with orders of magnitude higher luminosity than other colliders. While not shown in the figure, a 24~GeV CEBAF provides significantly higher $Q^2$, low $x$ values in the valence region which is critical for several key measurements.}
\label{fig:CEBAFvsEIC} 
\end{center}
\end{figure}

The kinematic landscape for deep inelastic scattering at CEBAF and the EIC is shown schematically in Fig.~\ref{fig:CEBAFvsEIC}.
Precision measurements in the valence quark region requiring high luminosity are clearly the purview of CEBAF, with the 24~GeV upgrade providing important overlap into the sea quark region where the EIC is designed to probe at low $x$.

%{\bf Are there good places in the rest of the document where this physics can be referred to?}

\subsubsection{The Electron-Ion Collider (EIC) in China}

The physicist community in China, %Chinese physicists -- Chinese can be overseas
together with international collaborators, proposed a polarized electron-ion collider at the  High Intensity heavy-ion Accelerator Facility (HIAF), currently under construction in southern China.
EIC@HIAF~\cite{Anderle_2021} was proposed as an extension to HIAF in a phased approach. The ﬁrst phase of the China EIC will include 3 to 5 GeV polarized electrons on 12 to 23 GeV polarized protons (and ions at about 12 GeV/nucleon), with luminosities of 1 to 2 $\times 10^{33}$ cm$^{-2}$ s$^{-1}$. This facility with complementary kinematic reach to both Jefferson Lab and the US EIC will allow for the studies of one and three-dimensional nucleon structure, the QCD dynamics, and to advance the understanding of the strong nuclear force.

%{\bf [Maybe different groupings based on the community and the science? FRIB, Mainz, J-PARC, PANDA, BES, RHIC/RHIC-Spin, COMPASS++/Amber, Belle-II (TMD related), Fermilab SpinQuest \ldots]}

%\subsection{International Context}

%{\bf [Focus on international contributions to the laboratory program, in addition to the complementarity with international facilities. There is overlap with the section above. I believe the international context is about the science, facilities and complementarities. They should be combined or move the international facilities to this section. We can also mention the international users and contributions in the overview section briefly.]}

%% file: Section2.tex
\section{Electromagnetic Form Factors and Parton Distributions}\label{sec:section2-FF-PDF}
%The proton and neutron, collectively known as the nucleon, are fundamental particles with spin-1/2. Therefore they can have exactly two static electromagnetic moments, namely the charge and magnetic dipole moment. Although current conservation fixes the charge, the magnetic moment relies on detailed dynamics. Dirac's relativistic wave equation correctly predicted the magnetic moment of the electron, but in 1933 Otto Stern discovered that the proton's magnetic moment was anomalous. Soon afterwards, different groups showed that the neutron magnetic moment was nonzero, and therefore also anomalous. Thus began the field of hadronic dynamics, in trying to understand the underlying structure of the nucleon.

With the advent of particle accelerator technology, high energy electron scattering became an indispensable tool for understanding the internal structure of nucleons. This is because the fundamental cross section separates into an electronic part (which is accurately determined from Quantum Electrodynamics) and a hadronic part which can be formulated in terms of structure functions of various kinematic quantities, and measured in electron scattering. For elastic scattering from the nucleon, the structure functions can be written in terms of form factors $G_E(q^2)$ and $G_M(q^2)$, which represent the internal charge and magnetic moment distributions as a function of the square of the four-momentum transfer. In Deep Inelastic Scattering (DIS), where both the $q^2$  and invariant mass of the hadron fragments are large, the structure functions are decomposed into Parton Distribution Functions (PDFs) which describe the internal structure in terms of quark degrees of freedom.

These approaches have been exploited with great success. Robert Hofstadter~\cite{RevModPhys.28.214} led a program of elastic (and inelastic) electron scattering from nucleons and nuclei. The resulting distributions for charge and current elucidated some of the most fundamental properties that any dynamic theory of hadron structure must predict. Some time later, the first DIS experiments by Richard Taylor, Henry Kendall, and Jerome Friedman~\cite{RevModPhys.63.573,RevModPhys.63.597,RevModPhys.63.615} showed the partonic internal structure of the nucleon. 

However, at each new stage of experimental investigation, new challenges and inconsistencies arose. New experiments, many of which were performed at Jefferson Lab, were performed to address discrepancies, but some issues still remain.

One important example has been the ``proton radius puzzle.'' Precision measurements of the proton charge radius from muonic hydrogen disagreed with electron scattering form factor measurements extrapolated to zero momentum transfer. The PRad experiment at Jefferson Lab appears to have resolved this puzzle, with a precision measurement at very low momentum transfer $q^2$, getting a result that agrees with the muonic hydrogen experiments.

The high $q^2$ behavior of the proton elastic form factors $G_E(q^2)$ and $G_M(q^2)$ has also been a puzzle. It was tacitly assumed that the ratio of these form factors should be the proton magnetic moment over all $q^2$, based on early measurements. This appeared to be borne out by Rosenbluth separations of the elastic cross section. However, precision measurements of the ratio using recoil polarization techniques at Jefferson Lab showed that this was not the case. It appears that the discrepancy is due to higher order ``two--photon'' corrections to the cross sections, which have also been investigated at Jefferson Lab.

Experimental facilities at CEBAF will continue to tackle these questions in the years to come. A program in nucleon elastic form factors is currently underway with the Super Bigbite Spectrometer (SBS) system. Spin-- and flavor--dependent structure functions in the valence region will be probed with greater precision, with tighter control on systematic uncertainties. Higher electron beam energies and beams of polarized positrons will provide new handles to be exploited. Furthermore, elastic and inelastic scattering from pions and kaons (in tagged electron scattering from nucleons) will extend this physics into the meson sector.

%The time-honored approaches to studying the internal structure of the nucleons are elastic electron scattering to measure the electromagnetic form factors and deep-inelastic scattering (DIS) to measure the parton distribution functions (PDFs). Although much is known about these quantities, the Jefferson Lab (JLab) 12 GeV program will provide new and exciting insights into these fundamental structure measurements.
%The role of the two-photon process in elastic scattering, critical to providing precise extractions of the form factors, can be directly accessed through positron beams, which can be implemented in the future. The structure of pion and kaon could be accessed through
%processes on the nucleon target with small $t$-channel momentum transfer. 

\subsection{Elastic Form Factors at Ultra Low and High $Q^2$}
 
Probing the internal structure of the nucleon through elastic scattering began with the classical work of Hofstadter~\cite{Hofstadter56} who measured the proton radius for the first time in 1955. The precision measurement of the proton radius has again become a focus in nuclear and atomic physics experiments in recent years~\cite{gao2021proton}, motivating innovative approaches to make precise measurements of the proton form factor at ultra-low $Q^2$. Measuring the proton form factors with innovative technologies at ultra low $Q^2$ at JLab helps resolve the discrepancy among various experimental approaches. The high-precision high-$Q^2$ form factors directly probe the underlying physics mechanism of high-energy elastic scattering. 

\subsubsection*{\it Proton Radius:}
\label{sec:PRad}
\begin{figure}[ht]
%\vskip 0.5truecm
\centerline{
\includegraphics[width=0.8\textwidth]{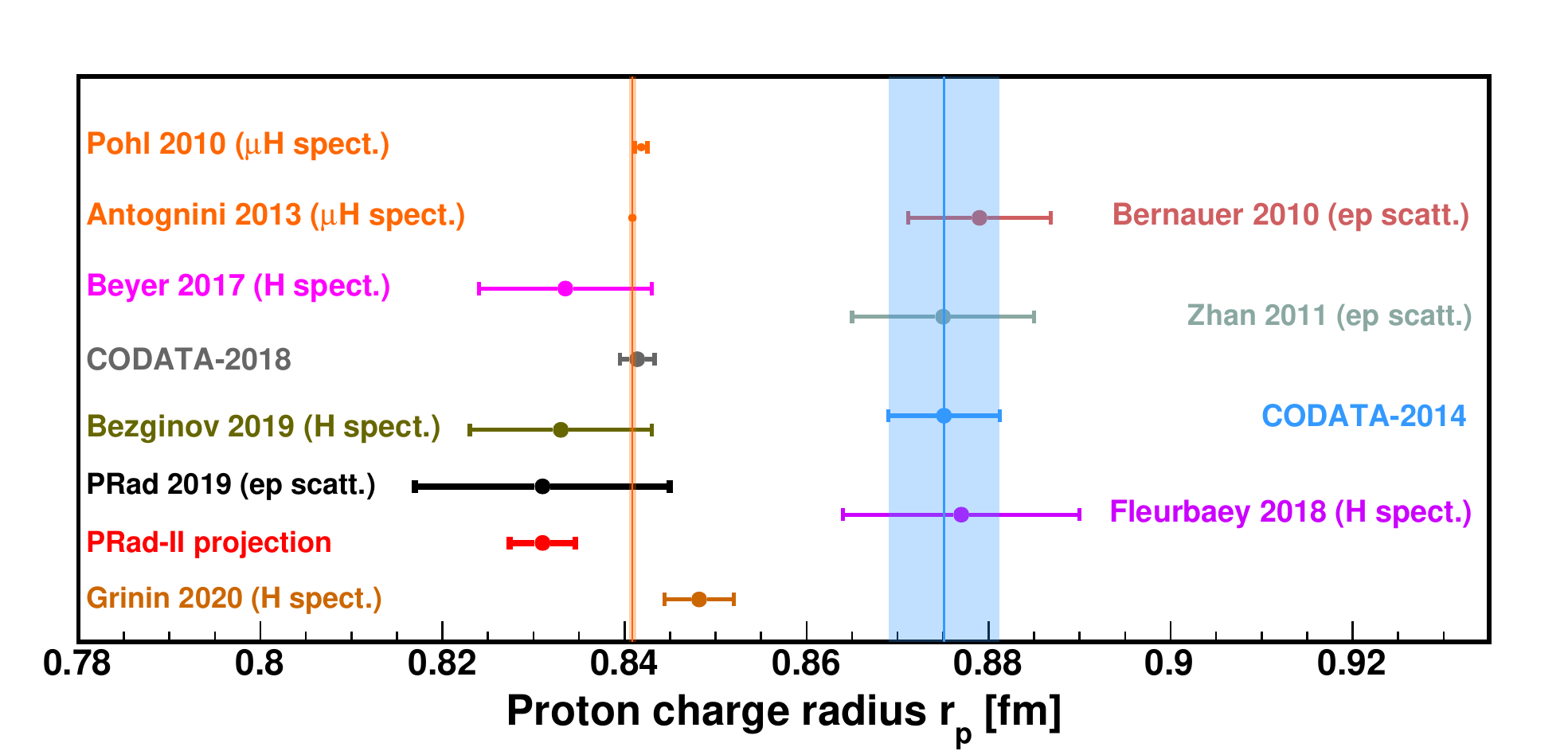}}
\caption{The projected $r_{p}$ result from PRad-II, shown along with the result from PRad and other measurements.}
\label{fig:PRad2_proj}
\end{figure}

The PRad experiment~\cite{PRad_nature:2019} performed in 2016, was the first high-precision $e-p$ experiment 
to utilize a magnetic-spectrometer-free method along with a windowless hydrogen gas target aiming at resolving the proton charge radius puzzle triggered by the muonic hydrogen spectroscopic measurements~\cite{Pohl10,Anti13}. These innovations overcame several limitations of previous $e-p$ ~experiments and reached unprecedentedly small scattering angles. The PRad result~\cite{PRad_nature:2019} is, within its experimental uncertainty, consistent with the $\mu$H results, and was one of the critical inputs in the updated CODATA-2018 value for $r_p$~\cite{CODATA_2018}. To reach the ultimate precision offered by this new method, the PRad collaboration proposed an enhanced version of PRad, the PRad-II experiment~\cite{PRad2} with several improvements including adding a second plane of tracking detectors, upgrading the hybrid electromagnetic calorimeter to an all-PbWO$_4$ crystal calorimeter, etc.. 
The approved PRad-II experiment will deliver the most precise measurement of $G^{p}_{E}$ ~reaching the lowest ever $Q^2$ (10$^{-5}$ GeV$^2$) in lepton scattering experiments, critical for the model-independent extraction of $r_p$. PRad-II will achieve a total uncertainty of 0.0036 fm on $r_p$,  which  is $\sim$ 4 times smaller than PRad and better than the most precise atomic hydrogen (H) spectroscopy result~\cite{Grinin20} with a total uncertainty of 0.0038 fm.
The projected $r_p$ from the PRad-II experiment is shown in Fig.~\ref{fig:PRad2_proj} along with recent electron scattering extractions~\cite{PRad_nature:2019,Bernauer10,Zhan11}, atomic physics measurements on ordinary hydrogen~\cite{CREMA_2017,hspec2018,Eric2019,Grinin20} and muonic hydrogen~\cite{Pohl10,Anti13}, and the CODATA values~\cite{CODATA_2014,CODATA_2018}.
The PRad-II precision will help address possible systematic differences between the most precise ordinary hydrogen and $\mu$H spectroscopy results and provide independent input for future CODATA recommendations for $r_p$ and the Rydberg constant. The precision of the PRad-II will also stimulate future high-precision lattice QCD predictions for the proton radius and contribute to new physics searches such as the violation of lepton universality.

\subsubsection*{Nucleon Form Factors at Large Momentum Transfer}

\begin{figure}[ht]
  \begin{center}
   \includegraphics[width=0.98\columnwidth]{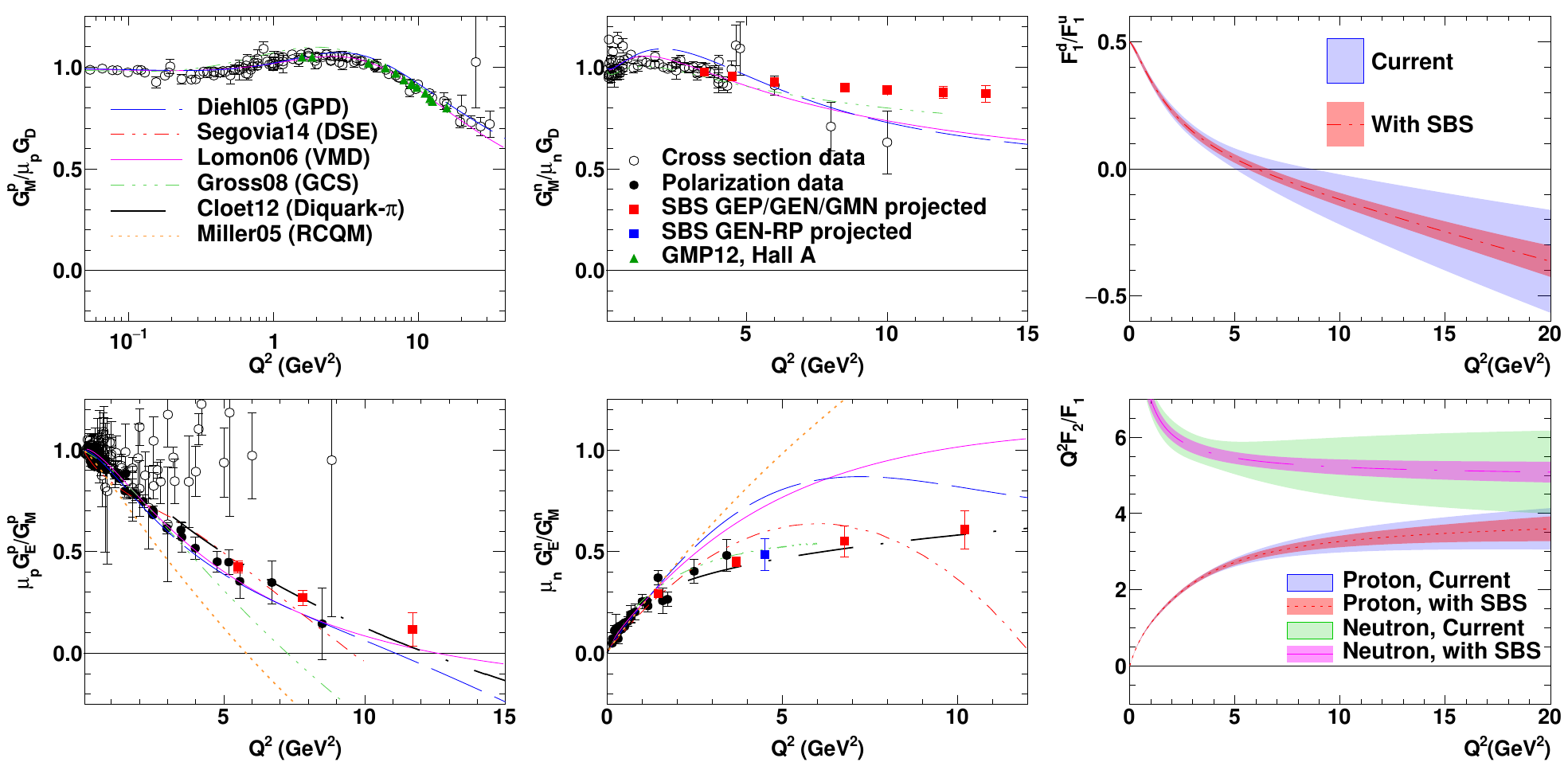}
  \end{center}
  \caption{\label{fig:SBSFF_summaryfig} Projected results from the SBS form factor program, compared to existing data, including the preliminary results for $G_M^p$ extracted from the recent high-$Q^2$ elastic $ep$ cross section measurements from Hall A (GMP12), and selected theoretical predictions. Projected SBS results are plotted at values extrapolated from the global fit described in the appendix of Ref.~\cite{Puckett:2017flj} for the proton, the Kelly fit~\cite{KellyFit} for $G_M^n$, and the Riordan fit~\cite{Riordan:2010id} for $G_E^n$. Theoretical curves are from the GPD-based model of Ref.~\cite{Diehl:2004cx} (Diehl05), the DSE-based calculation of Ref.~\cite{Segovia:2014aza} (Segovia14), the VMD model of Ref.~\cite{Lomon:2006xb,Lomon:2012pn} (Lomon06), the covariant spectator model from Ref.~\cite{Gross:2006fg} (Gross08), the quark-diquark model calculation from Ref.~\cite{Cloet:2012cy} (Cloet12), and a relativistic constituent quark model calculation by Miller~\cite{Miller:2002ig} (Miller05). Top right: projected impact of SBS program on the flavor separated form factor ratio $F_1^d/F_1^u$, with central value and uncertainty band evaluated using the Kelly fit~\cite{KellyFit}. The improvement in this ratio is mainly due to the new $G_M^n$ data. Bottom right: projected impact of SBS program on the ratio $Q^2F_2/F_1$ of Pauli and Dirac form factors for the proton and neutron, also evaluated using the Kelly parametrization~\cite{KellyFit}. The improvement in these ratios is driven by the $G_E^p$ and $G_E^n$ data. The figure is taken from Ref.~\cite{Barabanov_2021}.}
\end{figure}

At a large value of $Q^2$, the form factors should reflect a transition to the perturbatively dominated mechanism and reveal the role of orbital angular momentum of the quarks and gluons in the nucleon. One of the first completed experiments in Hall A with the upgraded JLab accelerator was a precision measurement of the proton magnetic form factor up to $Q^2 = 16$~GeV$^2$~\cite{Christy:2021snt}. This experiment nearly doubled the $Q^2$ range over which direct Rosenbluth separations of $G_E$ and $G_M$ can be performed, and confirming the discrepancy with polarization measurements (believe to be the result of two-photon exchange corrections) to larger $Q^2$ values.
The new SBS and the upgraded BigBite Spectrometer are being installed in Hall~A and will be ready for experiments starting late 2021. 
A series of SBS experiments~\cite{GEP5_PAC47-short,GEN2-short,GMN-short,GENHC-short,GENRP-short,GENTPE-short} will measure the magnetic and electric form factors of the proton and neutron and allow a determination of the flavor separated form factors
to $Q^2 = 10-12$~GeV$^2$. A complementary measurement of the neutron magnetic form factor will be performed with CLAS12 in Hall B \cite{GMN-HallB-short}.
In Fig.~\ref{fig:SBSFF_summaryfig}, plots compare the projected results of the SBS form factor experiments to various theoretical models. To visualize the impact of the SBS experiments, the uncertainty bands from a fit to the existing data is compared to a fit including the SBS projected data for the ratio, $Q^2F_2/F_1$,  of the Pauli to Dirac form factors and the ratio, $F_1^d/F_1^u$, of the flavor separated down and up quark Dirac form factors are plotted in Fig.~\ref{fig:SBSFF_summaryfig}.
The SBS form factor experiments will push into a $Q^2$ regions in which theory expects new degrees of freedom to emerge in our understanding of QCD non-pertubative phenomena in nucleon structure, e.g., log scaling 
of $F_2/F_1$ predicted in Ref.~\cite{Belitsky:2002kj}.

% - new data from LHC -> W/lepton asymmetries in pp collisions
% - SeaQuest data
% - new global QCD analysis, particularly by the JAM Collaboration,
% including simultaneous determination of PDFs and FFs, and including
% for the first time SIDIS data ... important constraints on sea quark PDFs.

\subsection{Quark Parton Distributions at High $x$}

JLab 12 GeV facility provides the unprecedented opportunities to access to high-$x$ quark distributions. New data from JLab and other facilities, including RHIC at BNL, FNAL, and the Large Hadron Collider (LHC), have provided more stringent constraints on PDFs in previously unmeasured regions at small and large values of $x$. %the parton momentum fraction $x$ -> moved to Section 1
At the same time, new analysis techniques have been developed, notably by the JLab Angular Momentum (JAM) collaboration~\cite{JAM}, using Monte Carlo methods and modern Bayesian analysis tools, which provide a more rigorous theoretical framework in which to analyze the new data.

\subsubsection*{Valence Quarks at High $x$}
\label{sec:duRatio}

One of the key properties of the nucleon is the structure of its valence quark distributions, $q_v = q(x)-\bar q(x)$.
% Valence quarks are the irreducible kernel of each hadron, responsible for its charge, baryon number and other macroscopic quantum numbers.
While many features of the valence quark PDFs have been mapped out in previous generations of experiments, the region of large %parton momentum fractions 
$x$ ($x \gtrsim 0.5$), where a single parton carries most of the nucleon's momentum, remains elusive. 
In particular, the traditional method of determining the $d$-quark PDF, from inclusive proton and deuteron structure function measurements, suffers from large uncertainties in describing the nuclear effects in the deuteron when extracting neutron structure information~\cite{Melnitchouk:1995fc, Accardi:2016qay}.% --- see the left panel in Fig.~\ref{f.donu}.

% $\bullet$ Updates from recent JAM QCD analyses

Recently, three JLab 12~GeV unpolarized DIS experiments, MARATHON~\cite{Abrams:2021xum} in Hall~A, BoNUS12~\cite{BONUS12} in Hall~B, and $F_{2d}/F_{2p}$~\cite{Niculescu:201002} in Hall-C completed data taking. These experiments aim to provide data to constrain PDFs in the high-$x$ region, especially the $d/u$ PDF ratio.
% To obtain the neutron structure function, deuterium or $^3$He target is usually used and nuclear effect needs to be corrected.
MARATHON measured the ratio of $^3$H to $^3$He structure functions, while BONUS12 tagged slow-recoiling protons in the deuteron, two different approaches to minimizing nuclear effects in extracting the neutron (and hence d) information. 
The MARATHON experiment also provided important input for the study of the nuclear EMC effect for isospin partners~\cite{Cocuzza:2021rfn}. The Hall-C experiment measured $H(e,e^\prime )$ and $D(e,e^\prime )$ inclusive cross sections in the resonance region and beyond. While there will be nuclear effects in the deuterium data, the experiment provides significant large $x$ range and reduced uncertainty to be combined with the large global data set of inclusive cross sections for PDF extraction. 

A planned experiment using PVDIS~\cite{PVDIS} on the proton with the proposed SoLID~\cite{SoLID} spectrometer will provide input on the $d/u$ ratio at high $x$ without contamination from nuclear corrections.
The experiment will measure the ratio of $\gamma Z$ interference to total structure functions, which at leading twist is sensitive to a new combination of PDFs that is not accessible to electromagnetic probes.

\iffalse
\begin{figure}[t]
\centering
\includegraphics[width=0.35\textwidth]{Figures/du_data.pdf}
\includegraphics[width=0.49\textwidth]{Figures/spinPDF.pdf}
\vspace*{-0.2cm}
\caption{{\color{blue} [UPDATE?]} (left) $d/u$ PDF ratio from the CJ15 analysis~\cite{Accardi:2016qay} (red band), illustrating the effects of nuclear corrections in deuteron DIS data (green band) and removing deuterium data altogether (hashed).
(right) Helicity PDFs for various quark flavors from the JAM15~\cite{Sato:2016tuz} Monte Carlo analysis of inclusive DIS data, using SU(3) symmetry constraints (blue bands), and the updated JAM17~\cite{Ethier:2017zbq} analysis of inclusive and semi-inclusive DIS without SU(3) constraints.}
\label{f.donu}
\end{figure}
\fi

\subsubsection*{Spin Structure Functions}
\label{sec:spinstrfunct}

As one of the high-impact flagship experiments, the $A_1^n$ experiment~\cite{A1n12GeV} in Hall~C at JLab completed data taking in 2020 with a polarized $^3$He target and a 10.4~GeV polarized electron beam. 
The polarized $^3$He target~\cite{polHe3} was upgraded to double the polarized luminosity to $2 \times 10^{36}$~cm$^{-2}$~s$^{-1}$ with in-beam polarization up to $\approx 55\%$ with a 30~$\mu$A beam, which is a new world record.
The new precision measurement on $A_1^{^3{\rm He}}$ expanded the Bjorken-$x$ of the extracted $g_1^n$ structure function to $x=0.75$.
Combined with the planned experiments to measure the proton and deuteron asymmetries $A_1^p$ and $A_1^d$ with CLAS12~\cite{CLAS12spin}, new global analyses will be able to extract the $\Delta u$ and $\Delta d$ quark helicity distributions in the high-$x$ region with much improved precision. 
%An example of a recent global analysis of inclusive and semi-inclusive polarized DIS data from the JAM collaboration is shown on the right panel in Fig.~\ref{f.donu}.

\iffalse
\begin{figure}[t]
\begin{center}
\includegraphics[width=0.49\textwidth]{Figures/spinPDF.pdf}
\caption{{\color{blue} [UPDATE?]} Helicity PDFs for various quark flavors from the JAM15~\cite{Sato:2016tuz} Monte Carlo analysis of inclusive DIS data, using SU(3) symmetry constraints (blue bands), and the updated JAM17~\cite{Ethier:2017zbq} analysis of inclusive and semi-inclusive DIS without SU(3) constraints.}
\label{f.spdf}
\end{center}
\end{figure}
\fi

Another experiment that completed data taking in Hall~C in 2020 was the d2n experiment~\cite{d2n12GeV}, which measured both $g_1$ and $g_2$ with a polarized $^3$He target. 
Extracted $g_1^n$ and $g_2^n$ structure functions allow the
    $d_2^n = \int_0^1 x^2 (2 g_1^n + 3 g_2^n)dx$ 
moment to be determined in a $Q^2$ range from 3.5 to 5.5~(GeV/c)$^2$, significantly extended from the previous range covered by JLab~6~GeV experiments~\cite{Posik:2014usi, Armstrong:2018xgk}.
A planned measurement~\cite{SoLIDd2n} of $d_2^n$ with the proposed SoLID spectrometer in Hall~A will extend the $Q^2$ range to cover $1.5 < Q^2 < 6.5$~(GeV/c)$^2$ with improved precision.
These measurements will provide a benchmark test of lattice QCD calculations of twist-3 matrix elements. 

\iffalse
\begin{figure}[ht]
\begin{center}
\includegraphics[width=0.49\textwidth]{Figures/d2.pdf}
\caption{{\color{blue} [UPDATE?]} Comparison of proton (red) and neutron (blue) $d_2$ matrix element extractions from JLab experiments and lattice QCD with the JAM15 global QCD analysis~\cite{Sato:2016tuz}.}
\label{f.d2n}
\end{center}
\end{figure}
\fi

Finally, a recently approved experiment~\cite{dalton2020measurement} in Hall~D with circularly polarized photon beam on a polarized proton target will provide the final missing piece, the high-energy part of the contributions to the GDH sum rule~\cite{Gerasimov:1965et,Drell:1966jv,HY}. 
This will complete a multi-decade-long effort to test this fundamental sum rule.

\subsubsection*{Flavor separation, SIDIS and PVDIS}
While inclusive DIS from protons and neutrons allows us to separate contributions from $u^+ = u + \bar u$ and $d^+ = d + \bar d$ PDFs at large $x$ (and similarly for the polarized $\Delta u^+ = \Delta u + \Delta \bar u$ and $\Delta d^+ = \Delta d + \Delta \bar d$ PDFs), further decomposition into $u$, $d$ and $s$ flavors, or valence and sea separation, requires additional observables sensitive to different combinations of PDFs.
Drell-Yan lepton-pair and $W$-boson production in unpolarized and polarized $pp/pd$ experiments provide evidence of asymmetries in the spin-averaged $\bar d - \bar u$~\cite{Dove:2021ejl} and spin-dependent $\Delta\bar d - \Delta\bar u$~\cite{Adam:2018bam} sea quark PDFs, respectively.
SIDIS measurements with tagged pions or kaons provide an additional powerful method for PDF flavor separation~\cite{Sato:2019yez, Ethier:2017zbq}.

Planned JLab12 SIDIS experiments with pion production~\cite{SIDIS-SeaAsymmetry} on unpolarized proton and deutron targets will provide precision data for the study of the light antiquark asymmetry in the medium-$x$ ($0.1 \lesssim x \lesssim 0.6$) region. 
SIDIS pion production with polarized proton, dueteron~\cite{CLAS12spin} and $^3$He~\cite{SIDISHe3} targets will in addition provide high-precision data for the study of the light-quark asymmetry in the polarized sea.

The strange quark sea can be constrained by SIDIS kaon production data.
However, the difficulty in the precision determination of kaon fragmentation functions at relatively low energy scales %of JLab12 and HERMES 
highlights the limitations of JLab12 in the study of the strange sea with SIDIS~\cite{Moffat:2021dji}.
On the other hand, PVDIS provides a clean alternative method for strange quark PDF extraction, without the complications of fragmentation, utilizing the interference between electromagnetic and weak interactions.
A proposed PVDIS measurement in Hall C~\cite{PVPDF} would provide precision data for the determination of the strange sea distribution.
Furthermore, PVDIS on a polarized $^3$He target with SoLID~\cite{PVDIS_polHe3} is being considered as a possible additional method for polarized strange quark extraction.

\subsection{Pion and Kaon Structure} 
\label{sec:PionKaonStructure}

In general, accessing the structure of the pion and kaon is difficult in electron scattering as these Goldstone bosons of QCD cannot easily be made into a stable target. However, chiral perturbation theory indicates that these effective degrees of freedom play an important role in nucleon structure. Through a small $t$-channel momentum transfer measurement of a tagged proton recoil (energy close to beam energy) in coincidence with a DIS event of large invariant mass $W$, HERA experiments at DESY was able to probe the pion structure of the nucleon within a theoretical assessment of the background contributions. Access to meson structure can be done at elastic scattering kinematics as well, where one can probe the meson form factors $F_{\pi, K}(Q^2)$. Here, in addition to a recoiling target baryon, the final product contains a pion or kaon. The JLab 12 GeV program has approved both this type of experiment and the DIS type to access the partonic structure of meson targets. 

The E12-06-101 experiment~\cite{CLAS12spin} will extract the pion form factor through $p(e,e'\pi^+)n$ and $d(e,e'\pi^-)pp$ with $Q^2$ extending to 6 GeV$^2$ from 2 GeV$^2$ and $-t_{\min} \sim 0.005\sim 0.2$ GeV$^2$. The proposed separation of longitudinal and transverse
structure functions is a critical check of the reaction dynamics. The charged pion electric form factor is a topic of fundamental importance to our understanding of hadronic structure. In contrast to the nucleon, the asymptotic normalization of the pion wave function is known from pion decay. There is a robust pQCD prediction in the asymptotic limit where $Q^2\rightarrow \infty$: $Q^2F_\pi(Q^2)\rightarrow 16\pi\alpha_s(Q^2)f_\pi^2$. Therefore it is an interesting question at what $Q^2$ this pQCD result will become dominant. The available data indicate that the form factor at $Q^2=2$ GeV is at least a factor of 3-4 larger. The new data will provide improved understanding of the non-perturbative contribution to this important property of the pion as well as mapping out the transition to the perturbative regime. 

The partonic structure of the pion and kaon can be accessed as at HERA through a (semi-inclusive, target-tagged) experiment leveraging the Sullivan process. The approved E12-15-006 experiment in Hall A~\cite{E12-15-006} studies the reactions $p(e,e'p)X$ and $d(e,e'pp)X$ with $M_x > 1$ GeV, using a dedicated GEM-based time projection chamber for large angular acceptance and low momentum kinematic coverage to detect the recoiling protons. To probe the soft part of the nucleon wave function, the $t$-channel momentum transfer on the nucleon is limited to 0.2 GeV$^2$ and several values of $t$ are measured for each $x$ for meson pole extrapolation. The reactions and targets allow access to both the charge and neutral pion
cloud in the nucleon. The result can be compared with parton distributions from the pion-initiated Drell-Yan process and thus offers an important check on the concept of the virtual pion target in the nucleon. A follow-up proposal
through tagging the hyperon final state studies the possibility of extracting the kaon structure function~\cite{C12-15-006A}, although theoretically the kaon pole is further away
from the accessible kinematic region and thus will have larger contamination
from background contributions.  An important prospect for these measurements is to compare with lattice QCD
calculations of the parton structure using large momentum effective theory~\cite{Ji:2020ect}. 

\subsection{Two-photon Exchange Physics with Positron Beams} 

\begin{figure}
\begin{center}
\includegraphics[width=0.49\textwidth]{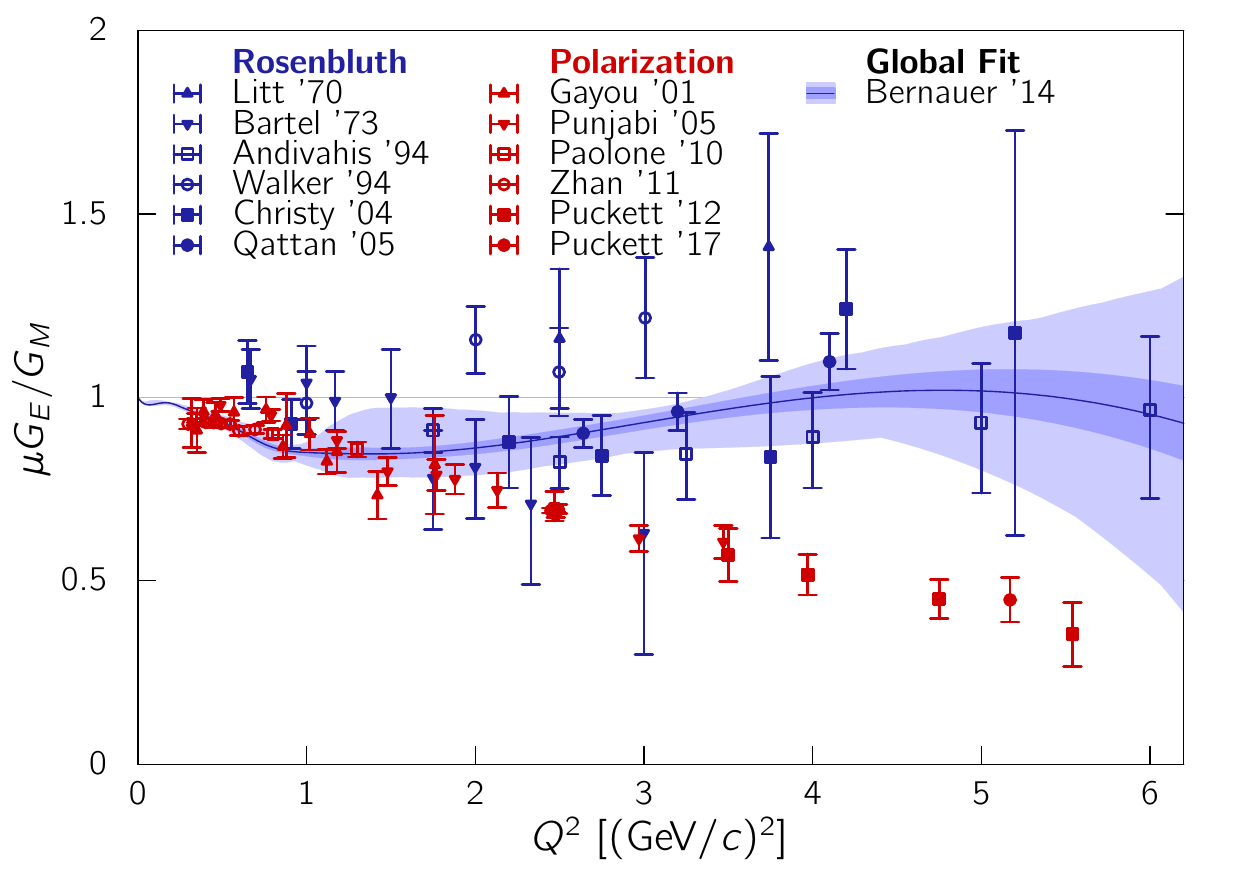}
\caption{The proton form factor ratio $\mu G_E/G_M$, as determined via Rosenbluth-type (blue points, from Litt '70\cite{Lit70}, Bartel '73\cite{Bar73}, Andivahis '94\cite{And94}, Walker '94\cite{Wal94}, Christy '04\cite{Chr04}, Qattan '05\cite{Qat05}]) and polarization-type (red points, from [Gay01\cite{Gay01}, Pun05\cite{Pun05}, Jon06\cite{Jon06}, Puc10\cite{Puc10}, Pao10\cite{Pao10}, Puc12]) experiments. Curves are from a phenomenological global fit by Bernauer \cite{Ber14} to the Rosenbluth-type world data set.}
\label{fig:tpe} 
\end{center}
\end{figure}

The discrepancy between high precision measurements of the proton's elastic form factor ($\mu_pG_E/G_M$) using  Rosenbluth separations of unpolarized cross-section data and polarization transfer in elastic electron-proton scattering exposed limitations of $1\gamma$ exchange Born approximation. The discrepancy between the two methods, see Fig.~\ref{fig:tpe}
~\cite{Accardi:2020swt}, is attributed to unaccounted hard $2\gamma$-exchange (TPE) radiative corrections. The two-photon exchange (and the large class of hadronic box diagrams) is hard to calculate without the inclusion of a significant degree of model dependence. While a few $e^+/e^-$ experiments measure a small TPE effect in the region of $Q^2<2$ GeV$^2$ \cite{Arrington:2004tpe,vepp3:2015tpe,clas:2015tpe,olympus:2017tpe,clas:2017tpe}, as precise measurements show \cite{Christy:2021snt}, it is possible that the TPE contribution is significant at higher momentum transfers. With planned measurements at $Q^2$ up to $10-16$ GeV$^2$ for different elastic form factors discussed in this section before, validation of calculations of TPE contribution to the elastic scattering at large $Q^2$ is crucial.  

Two strategies are proposed for studying TPE contribution using up to 11 GeV positron beams: the unpolarized $e^+p$/$e^-p$ cross-section ratio, and polarization transfer in polarized elastic scattering.  The next-order correction to the Born approximation contains terms corresponding to the interference of $1\gamma$- and $2\gamma$ exchange diagrams that are lepton charge sign dependent. This makes it possible to determine the size of the TPE effect in the ratio of positron to electron scattering cross-section.
The TPE contribution to the cross section ratio convolutes additional form-factors that become non-zero when moving beyond the one-photon exchange approximation. A cross-section ratio measurement is proposed for all three experimental halls with electron/positron beams (A, B, and C), with high precision spectrometers in Halls A and C \cite{AY:2021,CBS:2021} and the CLAS12 in Hall B~\cite{BBCSS:2021}.
%\cite{Accardi:2020swt}.

The polarization transfer measurement is less sensitive to hard TPE but can contribute its determination by providing additional information. As described in~\cite{Carlson:2007sp}, the ratio of transverse and longitudinal polarization transfer has different dependence to the additional form factors. The proposed polarization transfer measurement with positron beams in Hall-A will complement the cross section ratio measurements and help to further constrain TPE effects.

\section{Nuclear Femtography}

QCD is responsible for nearly all of the visible mass in the universe. However, our understanding of the nucleon in terms of its fundamental quark and gluon degrees of freedom is still miniscule when compared to our understanding the structure of atoms and molecules. The ultimate goal is to experimentally determine the quantum mechanical Wigner distribution in phase space. Measurements of PDFs integrate out all of the spatial and most of the momentum variables, so more sophisticated experiments are necessary to more completely map out the Wigner distribution.

Semi--inclusive measurements, including spin polarization observables, were provided by the pioneering measurements at HERMES, COMPASS, and the Jefferson Lab 6~GeV program, among others. Results on Generalized Parton Distributions (GPDs) and Transverse Momentum Distributions (TMDs) are now published, over limited ranges of the relevant kinematic variables. The upgraded detectors and CEBAF beam energy and intensity, as well as the potential for polarized positron beams, promise to provide a more detailed three-dimensional spatial mapping of the nucleon.

Indeed, this is a major thrust of the JLab 12 GeV facility. Mapping the (2+1)D mixed spatial-momentum images of the nucleon in terms of GPDs has been one of the important goals. GPDs expand greatly the scope of the physics in the traditional elastic form factors %FF is defined as fragmentation function later
and PDFs. On the other hand, the 3D images in the pure momentum space can be made with another generalized distributions: the TMDs. 
These femto-scale images (or femtography) will provide, among other insights, an intuitive understanding on how the fundamental properties of the nucleon, such as its mass and spin, arise from the underlying quark and gluon degrees of freedom. 

However, obtaining high-quality images from experimental data has remained a dream for more than two decades.  First, one needs to have a large amount of a specific type of experimental data -- Deep Exclusive Processes (DEP) and SIDIS -- from high-energy electron-nucleon collisions, which have not been systematically available despite the previous studies at JLab 6 GeV, HERMES, and COMPASS experiments. Second, as in any other imaging process, turning the data into images requires algorithms that can efficiently extract critical information. 
In the case of nucleon femtography, this represents a major challenge that can first be met with the JLab 12 GeV data. 

\subsection{Spatial Tomography of the Nucleon} 

\begin{figure}[htb]
\centering
\begin{minipage}{0.45\textwidth}
\centering
\includegraphics[width=0.55\textwidth]{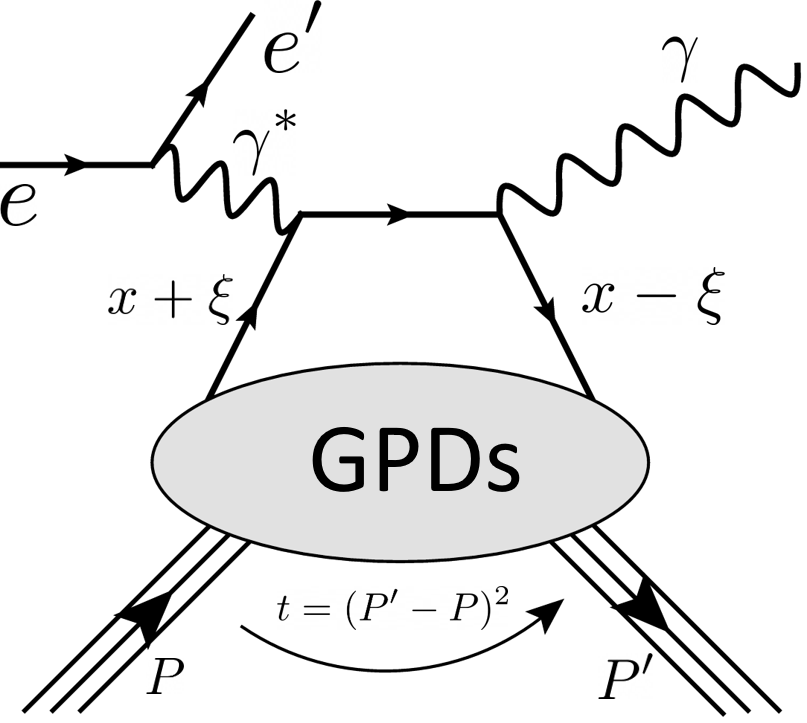}
\end{minipage}
\begin{minipage}{0.45\textwidth}
\centering
\includegraphics[width=0.55\textwidth]{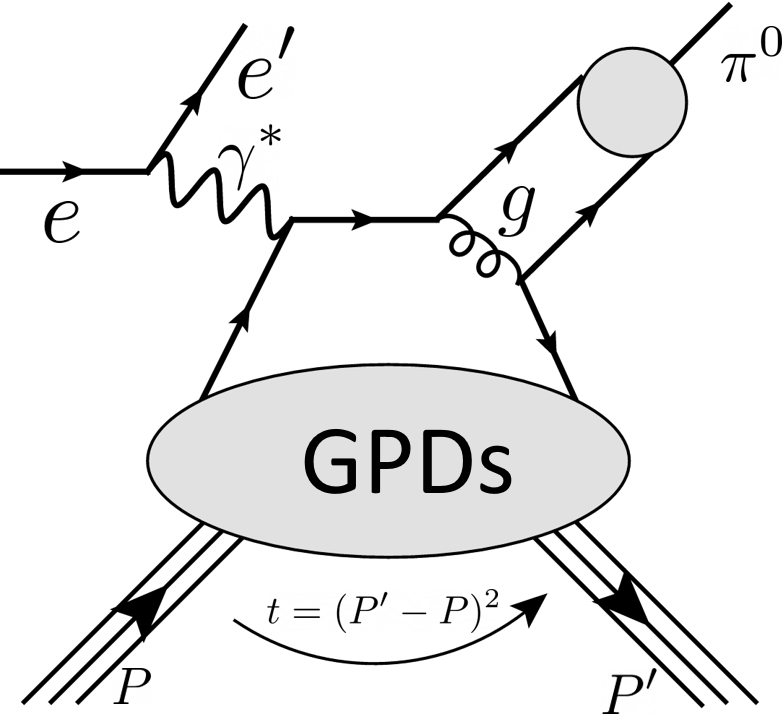}
\end{minipage}
    \caption{Deep exclusive processes in electron scattering as new type of hard scattering allowing QCD factorization and probing of the generalized structure~\cite{Ji:1996ek}.}
    \label{fig:dep}
\end{figure}

The standard approach of imaging a microscopic object is through diffractive scattering, as in optics. To obtain the phase-space quark and gluon distribution in a hadron, a new type of diffractive scattering was suggested in which a deeply-virtual photon (Bjorken limit) diffracts on a nucleon, generating a real photon or other hadrons~\cite{Ji:1996ek}. These DEPs allow probing entirely new structural information of the nucleon through QCD factorization (see Fig.~\ref{fig:dep}). The real photon production process has been called %deeply-virtual Compton scattering or 
DVCS~\cite{Ji:1996nm}, and for meson production, deeply-virtual meson production (DVMP)~\cite{Radyushkin:1996ru}.
DEPs contain the elements to learn the origin
of the nucleon mass and spin, and its gravitational properties. The information on these fundamental physical properties are encoded in the
GPDs~\cite{mueller1994wave,Ji:1996ek}.

GPDs are hybrid physical quantities reducing
to PDFs and form factors in the special kinematic limit. Thus they generally depend on Feynman $x$, momentum transfer $t= - \Delta^2= (P'-P)^2$, 
as well as the skewness parameter $\xi$. 
The GPD $E$ and $H$ defined from vector
current and its generalization to gluons 
provide the form factors $A$, $B$, $C$, and $\bar C$ of the QCD energy-momentum tensor (EMT)~\cite{Ji:1996ek}, 
\begin{equation}
\begin{split}
\label{Tmunuformfactors}
\langle{P'}|T_{q, g}^{\mu \nu}|\rangle{P}\rangle=\bar{u}\left(P^{\prime}\right)&\bigg{[}A_{q, g}(t) \gamma^{(\mu} \bar{P}^{\nu)}+B_{q, g}(t) \bar{P}^{(\mu} i \sigma^{\nu) \alpha} \Delta_{\alpha} / 2 M\\
&\;\;+C_{q, g}(t)\left(\Delta^{\mu} \Delta^{\nu}-g^{\mu \nu} \Delta^{2}\right) / M
+\bar{C}_{q, g}(t) g^{\mu \nu} M\bigg{]} u(P)\ .
\end{split}
\end{equation}
One of the combinations yields the mass form factor~\cite{Ji:2021mtz} 
\begin{equation}
    G_m(t)=\left[ MA\left(t\right) +B(t)\frac{t}{4M}
-C(t)\frac{t}{M}\right] \ . 
\end{equation}
from which one can construct the mass distribution
as well as the mass radius, $\langle r^2 \rangle_{m}
     = 6 \left|\frac{dG_{m}(t)/M}{dt}\right|_{t=0}$. 
The EMT form factors also provide the key information about 
the proton spin carried by quarks 
and gluons~\cite{Ji:1996ek}, 
\begin{equation}
     J_{q, g} = \frac{1}{2}[A_{q,g}+B_{q,g}]
\end{equation}
Finally, the form factors $C(t)$ (also called $D$-term) and $\bar C(t)$ have been related to ``pressure and shear pressure distributions''~\cite{Polyakov:2018zvc}. It was discovered that the GPDs provide phase-space images of the quarks and gluons with a fixed longitudinal momentum $x$ (momentum-dissected tomography)~\cite{Burkardt:2000za, Belitsky:2003nz}. 

Experimental observables in DVCS are parameterized by Compton Form Factors (CFFs) which as
functions of $t$, $\xi$, and $Q^2$ (which corresponds to renormalization scale of GPDs)~\cite{Belitsky:2010jw}.
At leading twist, there
are eight CFFs (four complex pairs) which are related to four relevant GPDs, $H$, $E$, $\tilde H$, $\tilde E$, which contain one additional variable $x$ integrated over in CFFs, 
\begin{equation}
     {\cal F}(\xi, t, Q^2) = \int dx F(x, \xi, t)
     \left( \frac{1}{\xi-x+i\epsilon}- \frac{1}{\xi+ x+i\epsilon}          \right)
\end{equation}
where $F$ is a generic GPD. 
From the analysis of data from HERA and HERMES at DESY, as well as the results of new dedicated experiments at JLab, and at COMPASS at CERN, the experimental constraints on CFFs have been obtained from global extraction fits~\cite{Kumericki:2016ehc,Moutarde:2019tqa}.  However, data covering a sufficiently-large kinematic range, and the many different polarization observables, have not been systematically available. Moreover, meson 
production at JLab 6 GeV has not yet shown parton dominance of scattering. The 12 GeV program at JLab will provide comprehensive information on these hard diffractive processes, entering the precision era for GPD studies.

Extracting all 8 CFFs independently at fixed kinematics require a complete set of experiments. However, given these CFFs is not sufficient to reconstruct the GPDs due to the loop integral in 
the hand-bag diagram. One either has to make some models with parameters to fit to experimental data or make combined fits with lattice QCD data.
Experimentally, one needs to explore processes that will give both $x$ and $\xi$ information, such as double DVCS or similar processes as we discussed in the next subsection. On the other hand, large-momentum effective theories developed in recent years have made possible to calculate GPDs directly on lattice~\cite{Ji:2013dva,Ji:2014gla} and some preliminary calculations can be found in Refs.\cite{Lin:2020rxa,Alexandrou:2020okk}. 
 
Experiment E12-06-114~\cite{Hyde2019} in Hall A proposed a precision measurement of the helicity dependent and helity independent {\it cross sections} for the $ep \rightarrow ep\gamma$ reaction in DVCS kinematics. The experiment considered the special kinematic range $Q^2> 2$ (GeV/c)$^2$ , $W > 2$ GeV, and $-t < 1$ GeV$^2$, with $Q^2$ extending to 9 (GeV/c)$^2$ and $x$ central region from 0.36 to 0.60.  The experiment is a follow up of the successful Hall
A DVCS run at the 5.75 GeV (E00-110). With polarized 6.6, 8.8, and 11 GeV beams incident on the liquid hydrogen target, the scattered electrons will be detected in the Hall A beam-left High Resolution Spectrometer (HRS)  %(with maximum central momentum 4.3 GeV/c) 
and the emitted photon in a slightly expanded PbF2 calorimeter. In general, the experiment will not detect the recoil proton. The $H(e, e^\prime\gamma )X$ missing mass resolution is sufficient to isolate the exclusive channel with 3\% systematic precision. The specific scientific goals are: 1) Measuring Compton cross section and comparing
with the scaling prediction of DVCS process and the dominant GDP predictions; 2) Extracting all kinematically independent observables for each $Q^2$, $x$, and $t$ point. These include five azimuthal dependencies as
$\cos(n\phi_{\gamma\gamma})$ with $n=0,1,2$, and
$\sin(n\phi_{\gamma\gamma})$ with $n=1,2$; and 3) Measuring the $t$ dependence of each angular harmonic term. 

There are two important DVCS experiments in Hall-B using CLAS12: E12-06-119\cite{E12-06-119} at 11 GeV and E12-16-010\cite{E12-16-010B} at lower energies of 6.6 and 8.8 GeV. 
These measurements will allow a large kinematical coverage and hence a more comprehensive study of GPDs. With a longitudinally polarized beam, one can extract the chiral even GPDs $H(x,\xi)$ and $E(x,\xi)$. The cross-section will be used to separate the interference of pure DVCS squared amplitude contributions to each of the Fourier moments of the cross section. The $Q^2$ dependence will allow extraction of the subtraction constant in the dispersion relation. %Finally, the $\pi^0$ cross section will also be measured to separate the longitudinal and transverse contributions. 

{\it Flavor separation:} The CFF from experimental data is a sum of contributions 
from different quark flavor weighted with charged squared. To get interesting flavor singlet information, such as in the spin sum rule, it is necessary to perform the flavor separation of GPDs. Thus it is minimally necessary to consider DVCS on the neutron. In the proposed E12-11-003\cite{E12-11-003} experiment in Hall-B, the beam spin asymmetry will be measured for incoherent DVCS scattering on the deuteron with recoiling neutron detection. In an experiment running together with E12-06-113, a similar measurement will be made with additional spectator proton measurement. It is also potentially possible to search for DVCS events on polarized neutron with a polarized deuteron target through E12-06-109 experiment. 

{\it Transversely polarized target}: 
An experiment has been proposed to measure the
target single spin asymmetry on a transversely polarized
proton~\cite{C12-12-010}. The asymmetry has
particular sensitivity on the
GPD $E(x,\xi)$ which is related to the spin flip
nucleon matrix element and hence carries important information on the up an down quark orbital angular momentum. The expected asymmetries are in the range of 20 to 40\% in the kinematics covered by C12-12-010 experiment. In addition, the double spin 
asymmetry involving longitudinally polarized electron will also be studied. 

{\it DVCS on nuclei target}: An experiment, E12-17-012\cite{E12-17-012}, has been proposed to measure coherent DVCS and DVMP with emphasis  on $\phi$ meson production. These exclusive measurements 
allow comparing the quark and gluon radii of the helium nucleus. To isolate the reaction mechanism, the low-energy
recoil nuclei are detected. Moreover, through incoherent spectator-tagged DVCS on light nuclei, the GPDs of the bound proton and neutron can be measured and compared
with the free proton and quasi-free neutron. The new apparatus used to detector low-energy nuclei recoil 
is the aforementioned ALERT, composed of a stereo drift chamber for track reconstruction and an array of scintillators for particle identification. 

\subsubsection*{Time-like Compton Scattering}

%The significant part of the physics program of JLab12 is the description of the partonic structure of hadronic matter via the GPDs. The Compton scattering is the golden reaction for mapping GPDs in the longitudinal and transverse momentum space.

\begin{figure}
\begin{center}
\includegraphics[width=0.35\textwidth]{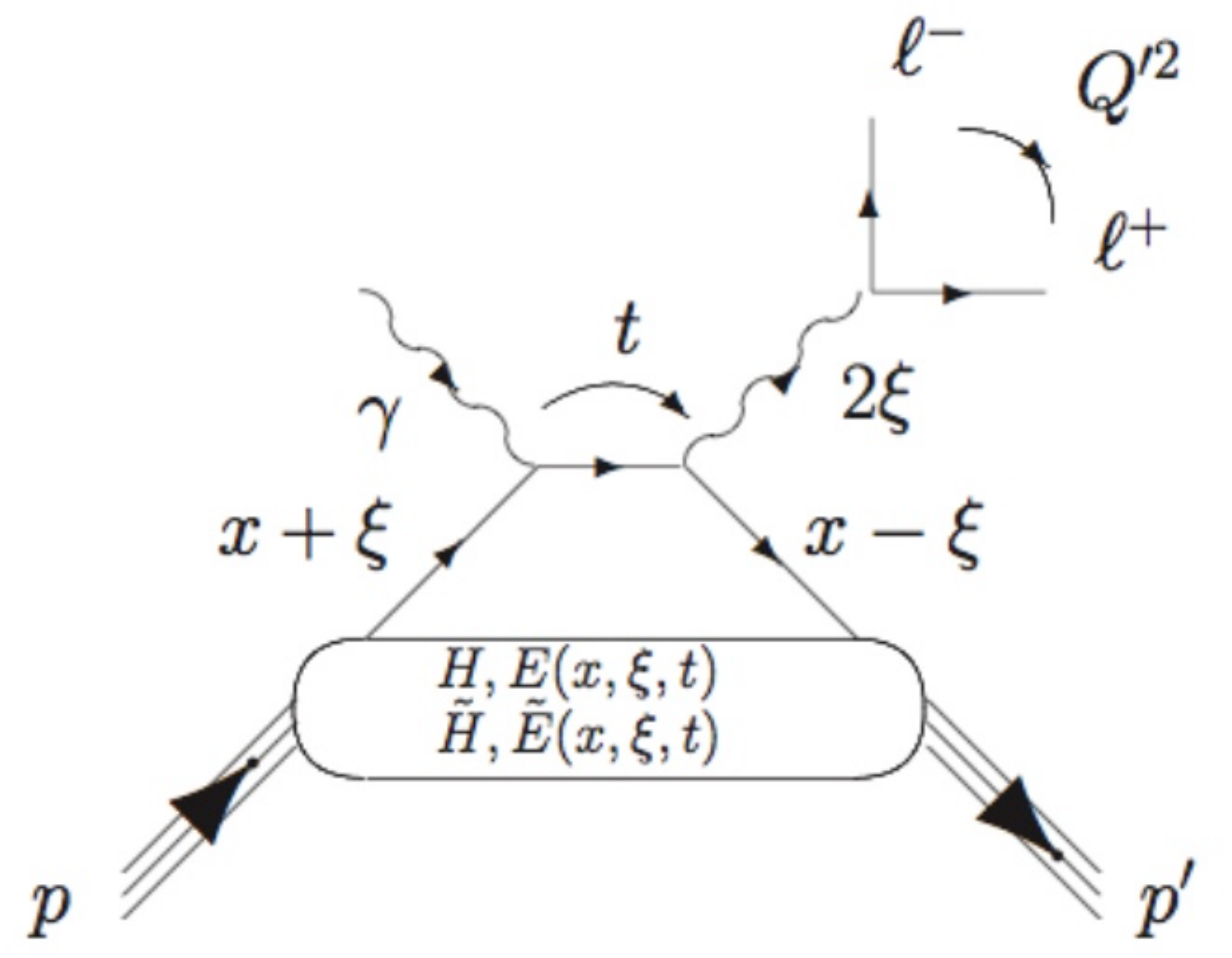}
\includegraphics[width=0.35\textwidth]{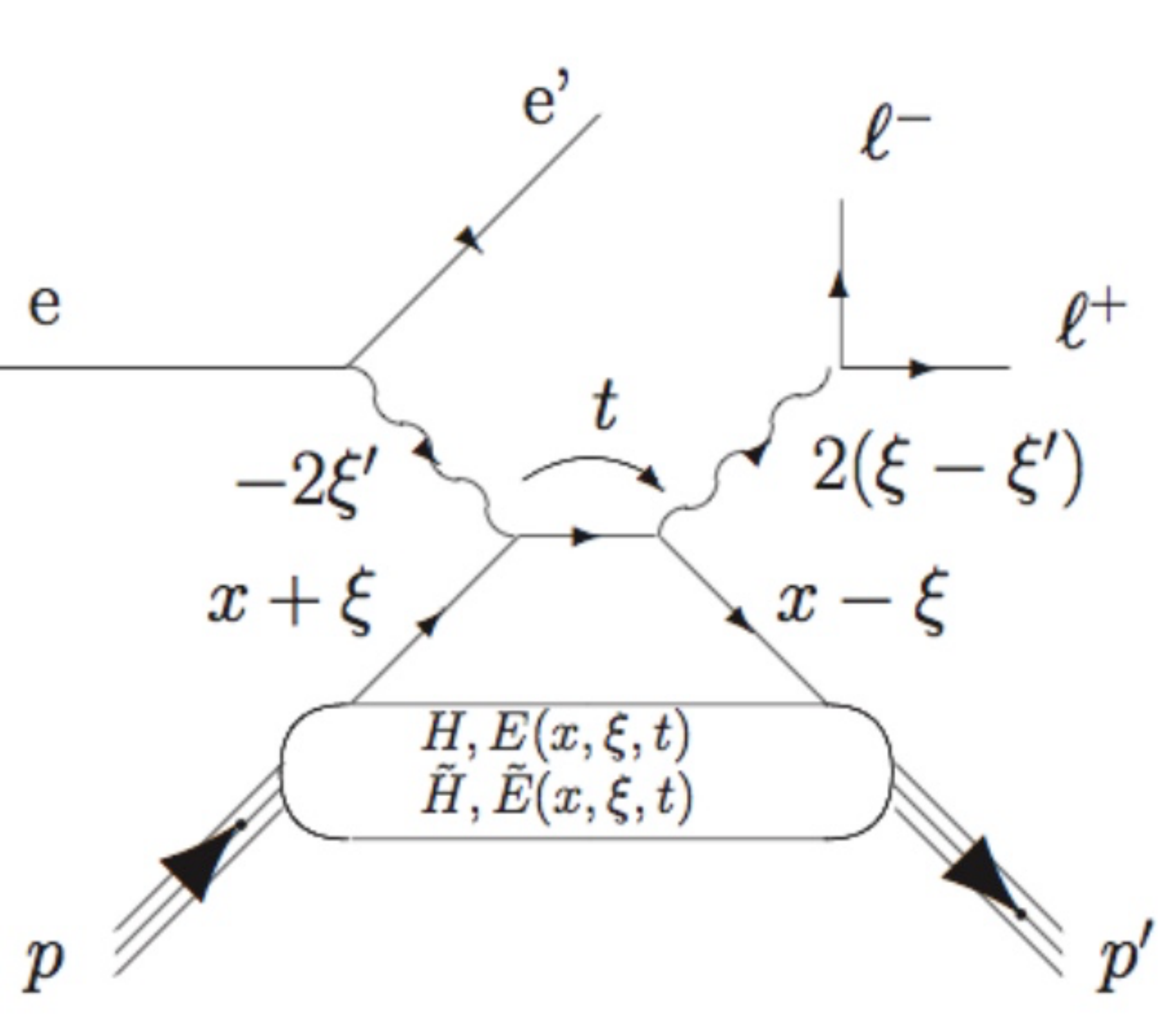}
\caption{Time-like Compton Scattering in which GPDs can also be probed, and Double DVCS (right) allowing probing $\xi\ne\xi$. In` $J/\psi$ production, the final-state time-like virtual photon is replaced by a $J/\psi$.
}
\label{fig:tcsddvcs} 
\end{center}
\end{figure}

While the most attention so far is on studies of GPD using spin (beam/target) observables and cross-sections in DVCS, the two other Compton-like processes (Fig.~\ref{fig:tcsddvcs}), Time-like Compton Scattering (TCS) and DDVCS are accessible with high energy electron beams and have much to offer. 

%A major scientific goal for future physics program with CEBAF would be to explore Double Deeply Virtual Compton Scattering (DDVCS). 

TCS is the time-reversal symmetric process of DVCS where the incoming photon is real and the outgoing photon has large time-like virtuality, see the left diagram of Fig.~\ref{fig:tcsddvcs}. The TCS hard scale is set by the virtuality $Q^{\prime 2}\equiv M^2_{l^+l^-}$ of the outgoing photon. As in the case of DVCS, the Bethe-Heitler process, $\gamma p\to p^\prime l^+l^-$, also contributes in the same final state. The formalism developed for TCS in \cite{Berger:2002tc} is based on the similar DVCS-type factorization when the TCS amplitude can be expressed as a convolution of the hard scattering kernels with GPDs appearing in CFFs.  With an unpolarized photon beam, TCS offers straightforward access to the real part of the CFFs through the interference between the Compton and Bethe-Heitler (BH) amplitudes. In the meantime, using circular photon polarization one can access the imaginary part of CCFs. 

\begin{figure}[ht]
%\vskip 0.5truecm
\begin{center}
\includegraphics[width=.490\textwidth]{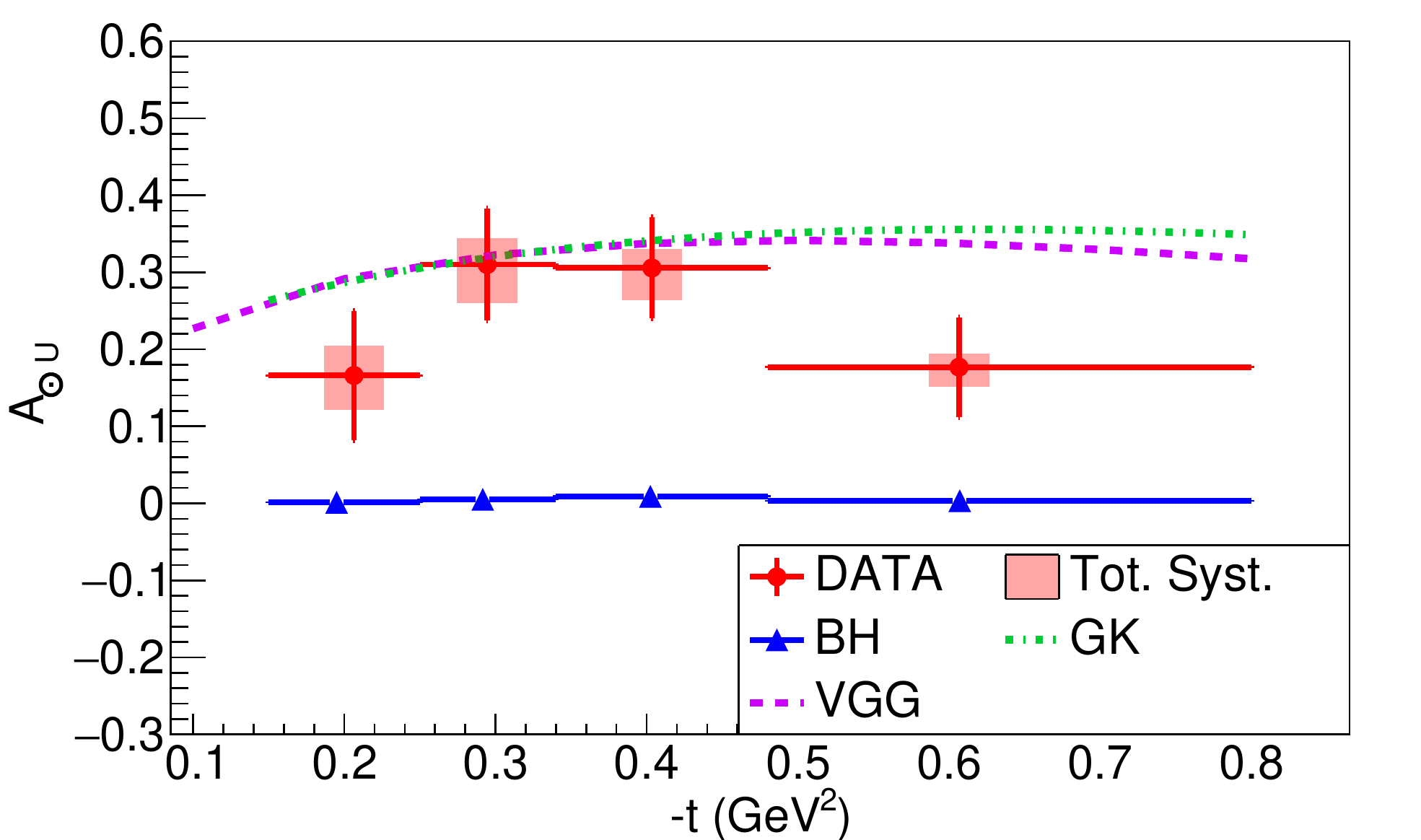}
\includegraphics[width=0.480\textwidth]{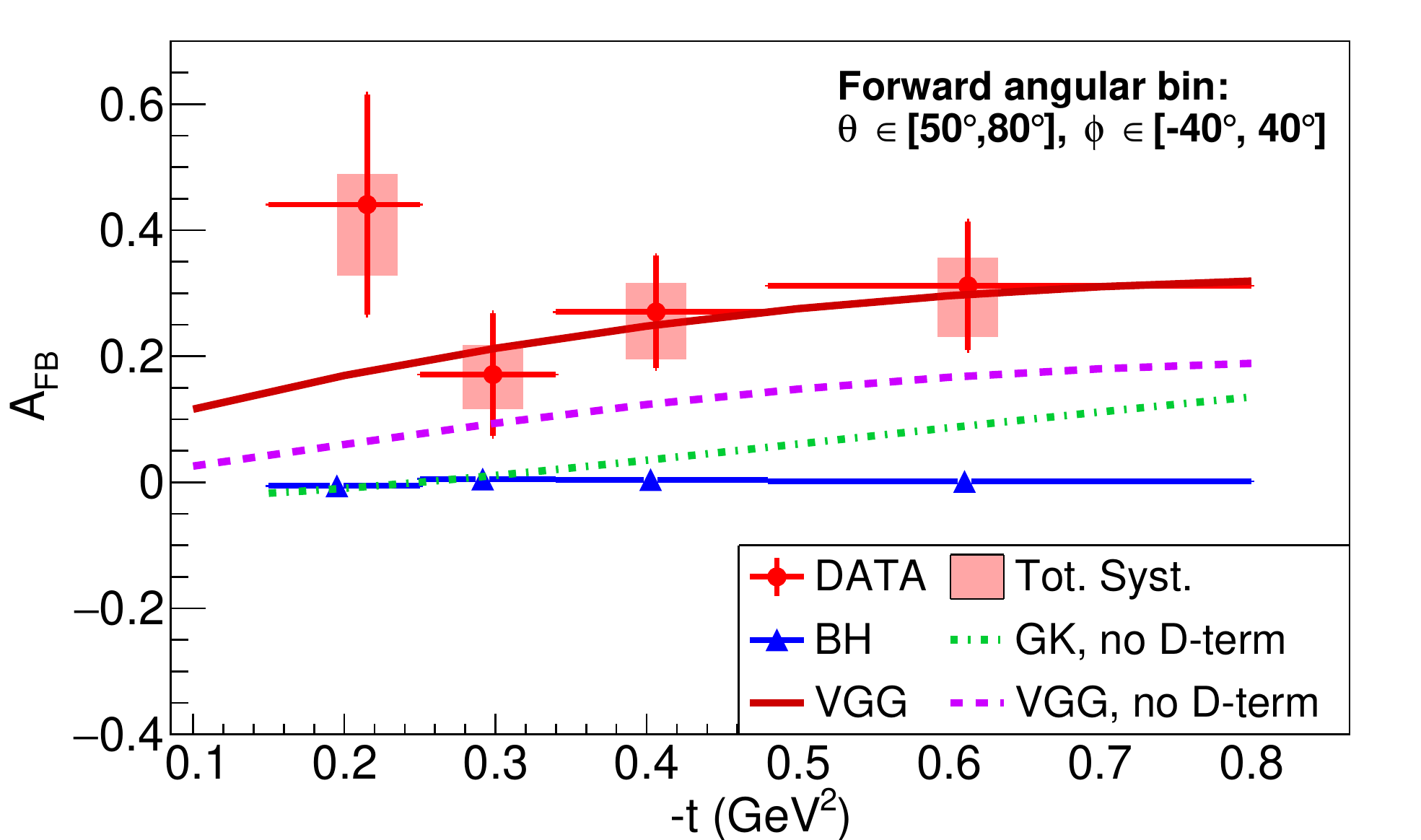}
\caption{Left: Photon polarization asymmetry $A_{\odot U}$ as a function of $-t$ at the averaged kinematic point $E_\gamma = 7.29 \pm 1.55$ GeV; $M = 1.80 \pm 0.26$ GeV. The data points are represented in red with statistical vertical error bars. The horizontal bars represent the bin widths. The shaded error bars show the total systematic uncertainty. The blue triangles show the asymmetry computed for simulated BH events. Right: FB asymmetry as a function of $-t$ in the same average kinematics. The plain line shows the model prediction for the VGG model with D-term (from \cite{PASQUINI2014133}) evaluated at the average kinematic point. The dashed and dashed-dotted lines are the predictions of, respectively, the VGG \cite{vgg1,vgg2,gprv,gmv} and the GK \cite{Goloskokov2005, Goloskokov2008, Goloskokov2009} models.}
\label{fig:clas12tcs}
\end{center}
\end{figure}

Studies of the TCS process at JLab have already begun. The experiment using CLAS12 in Hall-B, E12-12-001\cite{E12-12-001}, acquired part of the expected statistics. In Fig.~\ref{fig:clas12tcs}, the first experimental results on TCS, $\gamma p\to e^+e^- p^\prime$, are presented from a subset of obtained data \cite{clas12tcs}. The squared momentum transferred, $-t$, dependence of the $\sin{\phi}$ modulation of photon beam polarization asymmetry  (the left graph of the figure) is reported in the range of timelike photon virtualities $2.25 < Q^{\prime 2} < 9$ (GeV/c)$^2$, and average total center-of-mass energy squared $s = 14.5$ GeV$^2$. The measured values are in agreement with the predictions of GPD-based models.  The $-t$ dependence of decay lepton angular forward-backward asymmetry $A_{FB}$ is shown on the right graph of Fig.~\ref{fig:clas12tcs}. The $A_{FB}$ is proportional to the real part of the Compton amplitude. The experimental data is better described by the model with the D-term (taken from Ref. \cite{PASQUINI2014133}). This observation validates the application of the GPD formalism to describe TCS data and hints at the universality of GPDs \cite{Mueller:2012sma, Grocholski:2019pqj}. The analysis of the full data set is in progress.
%TCS studies at JLab have already started.  Both the cosine and sine moments of the weighted cross section have been measured over a wider range of momentum transfer $-t$, for outgoing time-like photon virtuality up to 9 GeV$^2$. The analysis to extract both the real and imaginary parts of Compton amplitude has been completed, preparing the publication underway. %{\it In addition to $\cos(\phi)$ moment analysis to access $Re[\tilde{M}^{--}]$ as proposed in \cite{BERG}, the forward-backward asymmetry, initially proposed for J/$\psi$ near-threshold photoproduction studies in \cite{MARCV} has been employed. The forward-backward transformation corresponds to inverting the vectors of leptons in the center of mass frame of the pair. }

Besides CLAS12, plans are in place to study TCS using SoLID in Hall-A, LOI-12-13-001. The high luminosity of SoLID will make it possible to perform a mapping of the Q$^{\prime 2}$- and $\eta$-dependence, which is essential for understanding factorization, higher-twist effects, and Next-Leading-Order (NLO) corrections \cite{Pire:2011st}. 

%{\it Time-like Compton scattering}: An experiment E12-12-001 has been proposed to measure diffractive electron-positron pair production which allows studying time-like Compton scattering and $J/\psi$ photoproduction on the proton. Both the four-fold differential cross section and the consine and sine moments of the weighted cross seciton will be measured over a wider range of momentum transfer $-t$, for outgoing time-like photon virtuality up to 9 GeV$^2$. This type of process has never been studied before except for a test run at 6 GeV. The experimental results can first be compared and contrasted with the usual DVCS processes in which the virtual photon is space-like, to test the scattering mechanism. This so-called TCS processes have extra sensitivity on the real part of the Compton form factors which are quite different in different model predictions. 
\subsubsection*{Meson Production}

At high-energy, the leading contribution to DVMP is dominated by 
the longitudinally polarized photon. However, at JLab 6 GeV, the 
experimental data does not show the dominance of the quark 
scattering process. In fact, the sub-leading contribution from
transversely polarized photon is quite significant. There has been suggestion
for the important twist-3 GPD effect with helicity-flip GPDs. 
Experimental data from 11 GeV beam will provide important
test of the deep-exclusive meson production mechanism. Experiments
have been proposed for $\pi^0$ and $\eta$ production with CLAS12 \cite{E12-06-108}
running contemporaneously with DVCS \cite{E12-06-119} experiment. Additional beam
time has also been requested at 8 GeV to make $\sigma_L$ and $\sigma_T$
separation to test the dominance of the former. 
In addition, there is an experiment E12-12-007\cite{E12-12-007} to measure $\phi$
production. Differential cross sections and beam asymmetries will be measured
as a function of the $\phi\rightarrow KK$ decay angles, $\theta$ and $\phi$, 
to extract the various structure functions. It is hoped 
that $\phi$ production can be used to extract gluon GPDs to reveal
the transverse spatial distribution of gluons in the nucleon. However, there %is also 
may be intrinsic strangeness contribution which can
only be separated if one has information through models or lattice
QCD calculations.

The Hall-A E12-06-114 experiment will also generate high-precision $\pi^0$ production data. For Hall C, the E12-09-011~\cite{E12-09-011} will perform L/T separated on Kaon Electroproduction and E12-20-007~\cite{E12-20-007} will look into backward-angle exclusive $\pi^0$ production. A summary of additional progress and opportunities for measurements in the backward-angle kinematic regime can be found in Ref.~\cite{Gayoso:2021rzj}.

\subsubsection*{Threshold $J/\psi$ production and proton mass}
\label{sec:PMass}

Near threshold $J/\psi$ production can provide unique access to the gluon GPD and the form factors of the gluon EMT, providing important information on the mass structure of the nucleon. There are approved and planned experiments for TCS in Halls A and B, and $J/\psi$ production experiments in all experimental halls, but to advance with DDVCS measurement, new high luminosity facilities will be needed.

At JLab 12 GeV, the $J/\psi$ can be produced by photon and electron beams on the proton and nuclei targets near the threshold. Besides the interest of searching for Pentaquark resonances and $J/\psi$-nucleon nuclear interactions, the experimental data can be used to study quantum anomalous energy (QAE) contribution to the proton mass and the proton mass radius~\cite{Ji:2021mtz,Kharzeev:2021qkd,Mamo:2021krl}. The most rigorous approach to analyze the threshold production is to use a heavy-quark expansion to derive a factorization theorem in terms of gluon GPDs or proton wave functions~\cite{Guo:2021ibg, Sun:2021gmi,Sun:2021pyw}. Near the threshold, the skewness parameter $\xi\to 1$, and thus the form factors of the gluon EMT could dominate over other higher-spins. Therefore data at the threshold could allow an approximate extraction of gravitational form factors $A_g(t)$, $B_g(t)$, and $C_g(t)$ at large $t$. One can also extract the QAE contribution to the 
proton mass either through Vector Dominance assumption~\cite{Kou:2021bez} or through the matrix elements of the trace part of the EMT~\cite{Ji:2021mtz}. Likewise, the mass-radius can be either extracted through vector meson or Reggeon dominance~\cite{Kharzeev:2021qkd,Mamo:2021krl,Wang:2021dis} or again through the EMT form factors~\cite{Ji:2021mtz,Guo:2021ibg}. The theoretical assumptions can be verified and the uncertainties reduced if the measurements are done closer to the threshold, and augmented by polarized photoproduction and electroproduction measurements. 

\iffalse
\begin{figure}
\begin{center}
\includegraphics[width=0.9\textwidth]{Figures/total_tdist-xsec-zm.pdf}
\caption{Projected total threshold cross section (left) and differential cross section at function of $t$ at $E_\gamma\sim$ 9.5 GeV (right) for $J/\psi$ production with SoLID detector.}
\label{fig:crosssection} 
\end{center}
\end{figure}
\fi

\begin{figure}
\begin{center}
\includegraphics[width=0.48\textwidth]{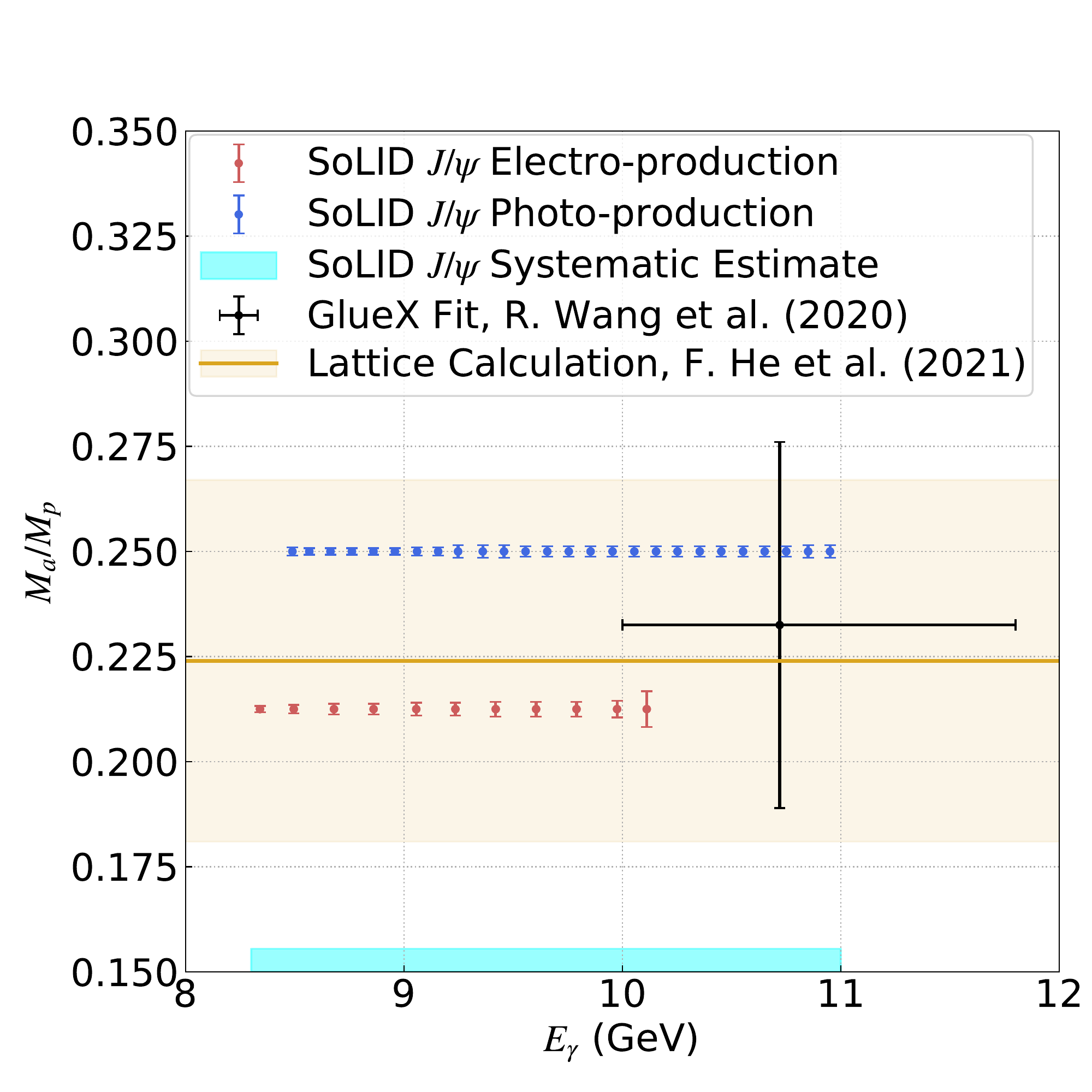}
\includegraphics[width=0.48\textwidth]{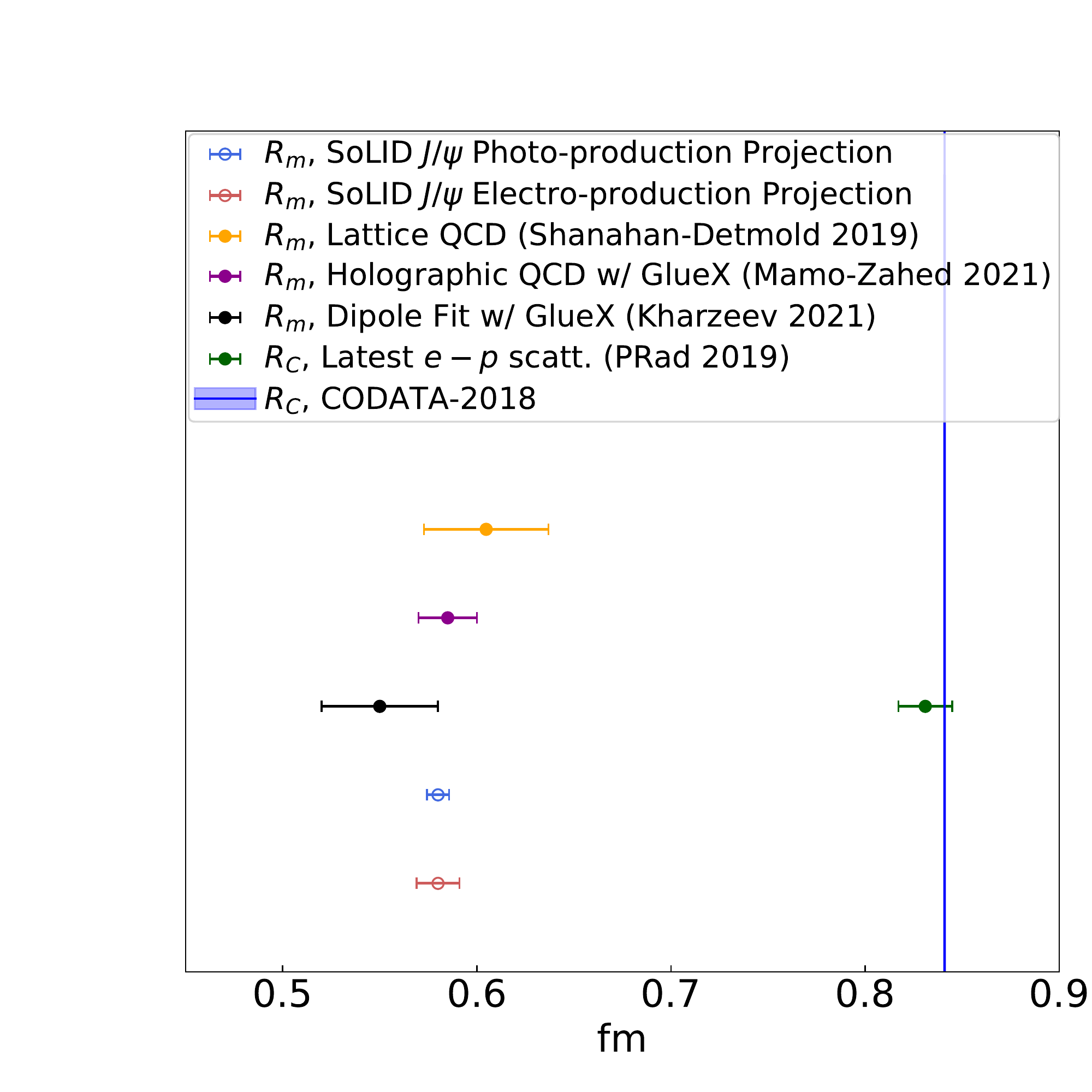}
\caption{Statistical errors on the extraction of the quantum anomalous energy contribution to the proton mass and proton mass radius from SoLID data with the vector dominance assumption. More accurate analysis is possible with QCD factorization in terms
of GPDs.}
\label{fig:mass} 
\end{center}
\end{figure}

The GlueX collaboration in Hall-D has recently published the first $J/\psi$ photoproduction data (about 470 events) ~\cite{Ali:2019lzf}, the results from a factor of 10 more data will be available soon. In Hall-C experiment E12-16-007~\cite{Meziani:2016lhg}, a similar amount of $J/\psi$ events (about 4k) were collected with a focus on the large $t$ region in search of the LHCb pentaquark, and with results to be published soon. Hall-B has ongoing experiments E12-12-001~\cite{E12-12-001} to measure TCS+$J/\psi$ in photo-production on a hydrogen target, which can have $\sim$10K events after their luminosity upgrade, and similarly E12-11-003B~\cite{E12-11-003B} utilizing a deuterium target. Hall-A has an approved experiment E12-12-006~\cite{SoLIDjpsi:proposal} using SoLID and can obtain at least another order of magnitude more events ($\sim$800k in photoproduction and $\sim$20k in electroproduction). With this large number of  threshold events, one can fit the cross section as a function of $W$ 
and $t$ %, Fig.~\ref{fig:crosssection},
to obtain all three gluon EMT form factors, and hence could shed important light on the origin of the nucleon mass. % (see Fig.~1). 
 Fig.~\ref{fig:mass} shows the possible precision one can obtain in the determination of the fractional contribution of the QAE in the proton mass following vector dominance in Ref.~\cite{Wang:2019mza} as well as the proton mechanical radius according to Ref.~\cite{Kharzeev:2021qkd} in comparison to its charge radius. A more accurate analysis is possible with QCD factorization in terms of GPDs.

\subsection{Momentum Tomography of the Nucleon } 
\label{sec:MomTomNucleon}

Experimental explorations of the spin dependent observables in SIDIS, Drell-Yan (DY), $e^+e^-$, and hadron-hadron scattering spurred extensive theoretical and phenomenological studies opening a new era of exploration of the 3D partonic structure of the nucleon. In SIDIS, involving non-collinear dynamics the nucleon structure is defined by TMD-PDF and TMD fragmentation function (TMD-FF) commonly called TMDs. Together with rapid advances of precision of lattice QCD calculations~\cite{Hagler:2009mb,Musch:2009ku,Musch:2010ka,Ji:2013dva,Ji:2020ect} and detailed predictions from theory and phenomenology, new precise experimental data on TMDs are needed. 

In the kinematic region where TMD description of SIDIS is appropriate, namely in the beam fragmentation region, $P_{hT}/z \ll Q$, the transverse momentum of the produced hadron $P_{hT}$ is generated by intrinsic momenta of the parton in the nucleon ${\bf k}_T$ and the transverse momentum of the produced hadron with respect to the fragmenting parton ${\bf p}_T$, such that
the structure functions become convolutions of TMD PDFs and TMD-FFs. The convolution integral ${\cal C}[w f D]$, for a given combination of TMD-PDF $f$ and TMD-FF $D$ is defined as~\cite{Bacchetta:2006tn}
\begin{eqnarray}
\propto \sum_q e_q^2 \int d^2{\bf k}_T \, d^2{\bf p}_T \delta^{(2)}({\bf p}_T + z{\bf k}_T - {\bf P}_{hT}) w({\bf k}_T, {\bf p}_T) f^{q}(x, ~k_T^2) D^{q}(z,~P_{hT}^2), 
\label{Eq:convo}
\end{eqnarray}
where $w$ is a kinematical factor, and the sum goes over all flavors of quarks and anti-quarks.
Well known SIDIS structure functions $F_{UU,T}$ and $F_{LL}$ will be, thus, described by convolutions
of $f_1$ and $g_1$ TMD-PDFs and $D_1$ the unpolarized TMD-FFs, with $F_{UU,T} =  {\cal{C}}\left[ f_1 D_1 \right], F_{LL}  =  {\cal{C}}\left[ g_1 D_1 \right]$. %The full list of TMD PDFs accessible in SIDIS is given in Table.\ref{tab:TMD-tables}. The TMDs depend on polarization state of the quark (rows) and polarization state of the nucleon (columns). 
The leading and higher twist non-perturbative functions
describe various spin-spin and spin-orbit correlations as corresponding operators include additional gluon and/or quark fields in the matrix element.
While the higher twist are expected to be suppressed by $P_{hT}/Q$ with respect to leading twist asymmetries, at large $P_T$ and relatively low $Q^2$, where the spin-orbit correlation are large and measurable by SIDIS experiments, they may be very significant, and play important role in development of consistent framework for analysis of SIDIS data.

One of the most important questions about the 3D structure of the nucleon is the transverse momentum dependence of the distribution and fragmentation TMDs and flavour and spin dependence of those distributions. %{\xiaochao{Is this sentence out of place?}}
At JLab, three experimental halls, A, B, and C are involved in 3D structure studies through azimuthal modulations in SIDIS for different hadron types, targets, and polarizations in a broad kinematic range, including the High Momentum Spectrometer (HMS) and Super HMS at Hall C~\cite{E12-06-104,E12-09-017,E12-13-007}, the BigBite spectrometer and SBS\cite{E12-09-018}, as well as, the SoLID detector at Hall A~\cite{SoLID-SIDIS-p,SoLID-SIDIS-He3-T,SoLID-SIDIS-He3-L}, and CLAS12 at Hall B \cite{E12-06-112,E12-07-107,E12-09-008,E12-09-009,C12-20-002}.

\begin{figure}[!h]
\begin{center}
\includegraphics[width=\textwidth]{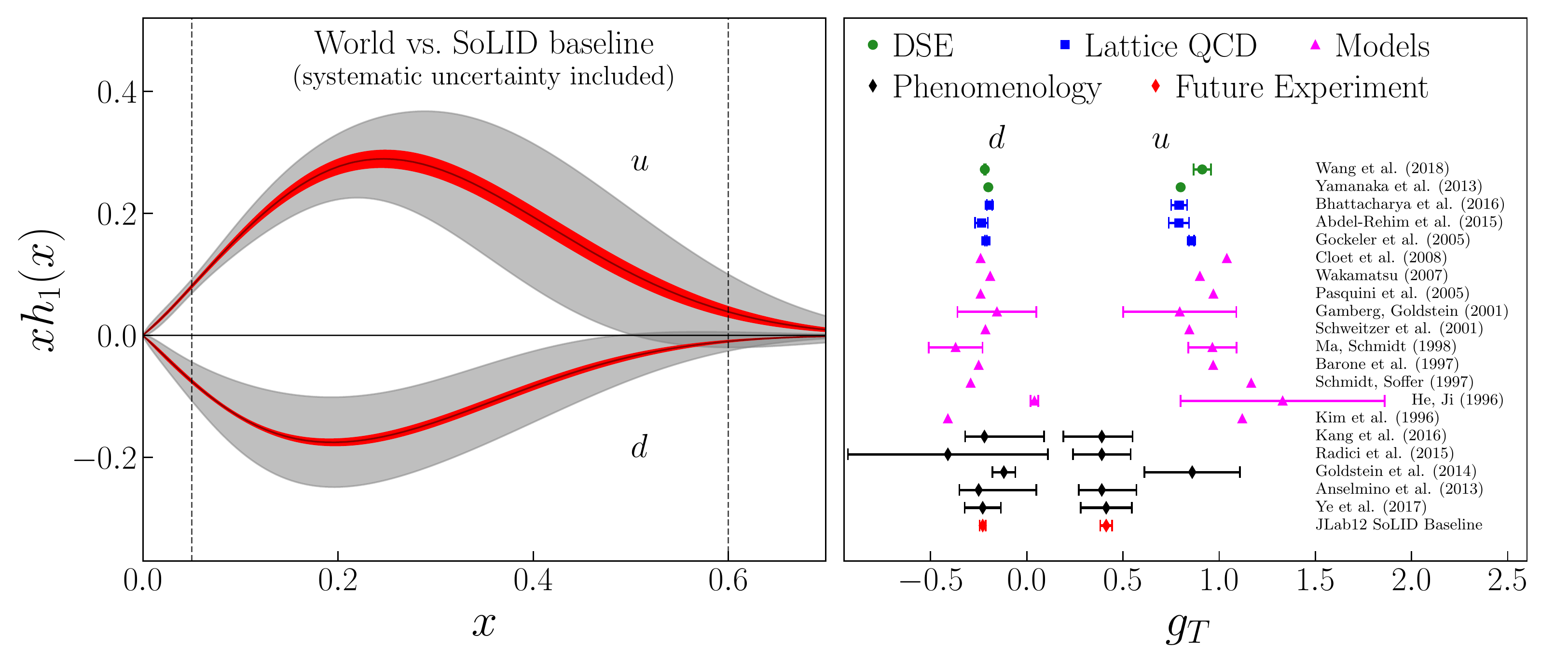}
\caption{(Left panel) The projected precision of SoLID measurements~\cite{SoLID-SIDIS-p,SoLID-SIDIS-He3-T} of transversity $h_{1}(x)$ for $u$ and $d$ quark (red bands), together with results from global analysis of the world data (grey bands). (Right panel) The extracted tensor charge for $u$ and $d$ quark together with predictions from lattice QCD, models, and phenomenological analyses of world data~\cite{YE201791}.  
\label{fig:SoLID-transversity}}
\end{center}
\end{figure}

The most celebrated SIDIS measurements on TMDs are the surprising non-zero results of the Sivers asymmetries and the Collins asymmetries~\cite{Airapetian_2005, Adolph_2014, Qian_2011}. These initial explorations established the significance of the SIDIS-TMD experiments and attracted increasingly great efforts in both experimental and theoretical studies of TMDs. The planned SoLID experiments with transversely polarized proton and neutron/$^3$He targets~\cite{SoLID-SIDIS-p, SoLID-SIDIS-He3-T} will provide the most precise measurements of Sivers and Collins asymmetries of charged pion and Kaon productions in the valance quark (large-$x$) region in 4-dimension ($x$, $Q^2$, $z$ and P$_T$).
There will be over 1000 data points in the 4-D space with high precision with a polarized $^3$He target and hundreds of data points with a polarized ${\rm NH}_{3}$ target with SoLID. These data will allow precision extractions of the Sivers functions and transversity distributions through global analyses. 
Figure~\ref{fig:SoLID-transversity} (left panel) shows the projected precision of the extracted transversity $h_{1}(x)$ from the SoLID base configuration for both the $u$ and the $d$ quark flavor compared with the current knowledge from a global analysis of the world data~\cite{YE201791}.  
In addition to providing 3-D imaging in momentum space, the Sivers functions also contain information on the quark orbital angular momentum. The transversity distributions is one of the three leading twist colinear distributions when integrated over the transverse momentum. The other two are the well-known unpolarized distributions and the helicity distributions. The integration of the transversity over $x$ is the tensor
charge. Tensor charge is a fundamental property of the nucleon which has been precisely calculated with Lattice QCD. Precision determination of the tensor charge would provide a benchmark test of Lattice QCD calculations. Figure~\ref{fig:SoLID-transversity} (right panel) shows the projection of expected precision from SoLID measurements in determining the tensor charge along with Lattice QCD calculations. Also shown are other theory/model predictions and phenomenological determinations from current world data~\cite{YE201791}.

\iffalse
\begin{figure}
\begin{center}
\includegraphics[width=.490\textwidth]{SoLID-TensorCharge.pdf}
\caption{The projected precision of extraction of the tensor charge from SoLID measurements~\cite{SoLID-SIDIS-He3-T,SoLID-SIDIS-p} along with LQCD calculations, other theory/model predictions and phenomenological determinations from current world data.}
\end{center}
\label{fig:tensorcharge}
\end{figure}
\fi 

\begin{figure}[ht]
%\vskip 0.5truecm
\begin{center}
\includegraphics[width=.490\textwidth]{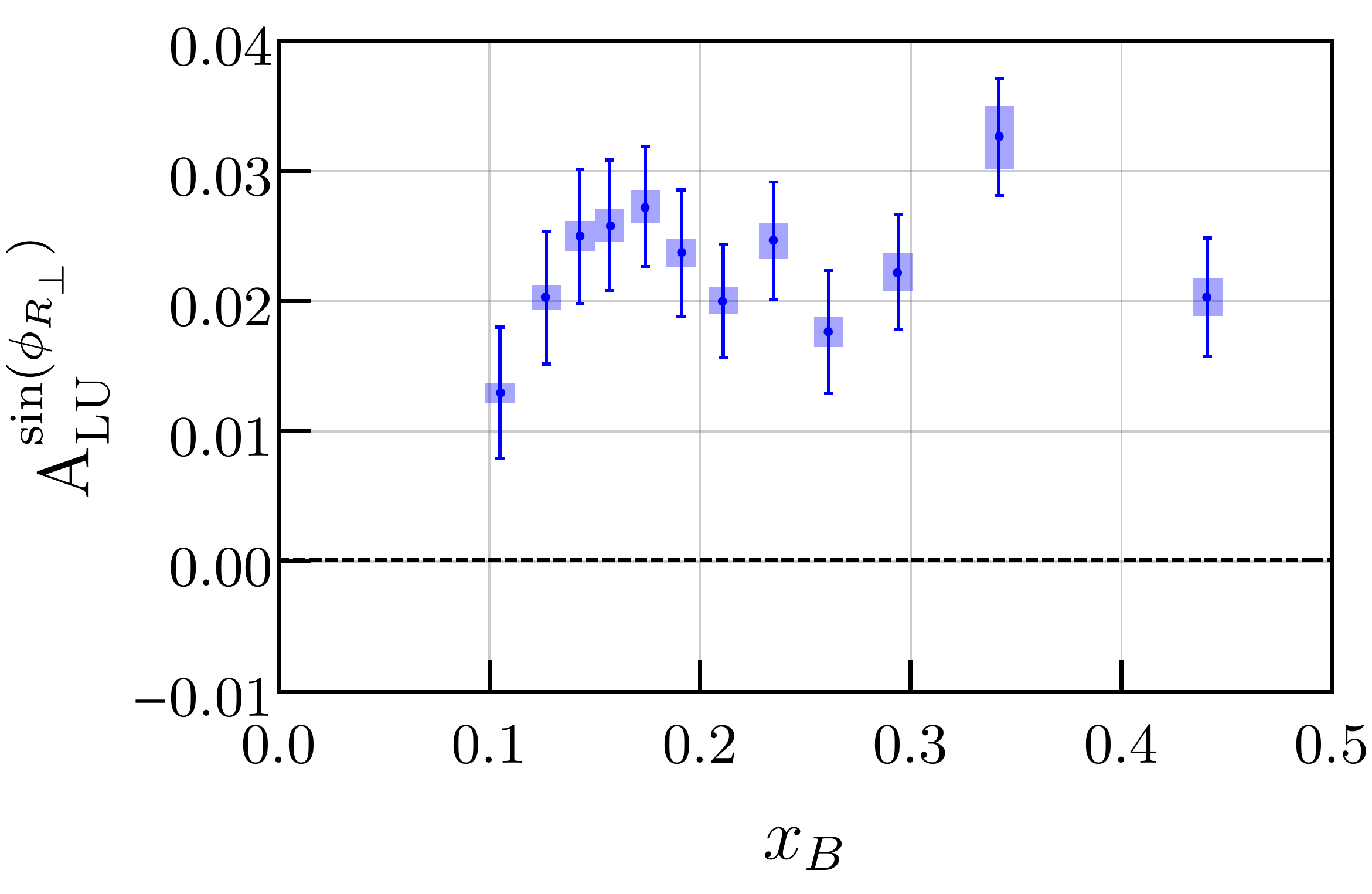}
\includegraphics[width=0.480\textwidth]{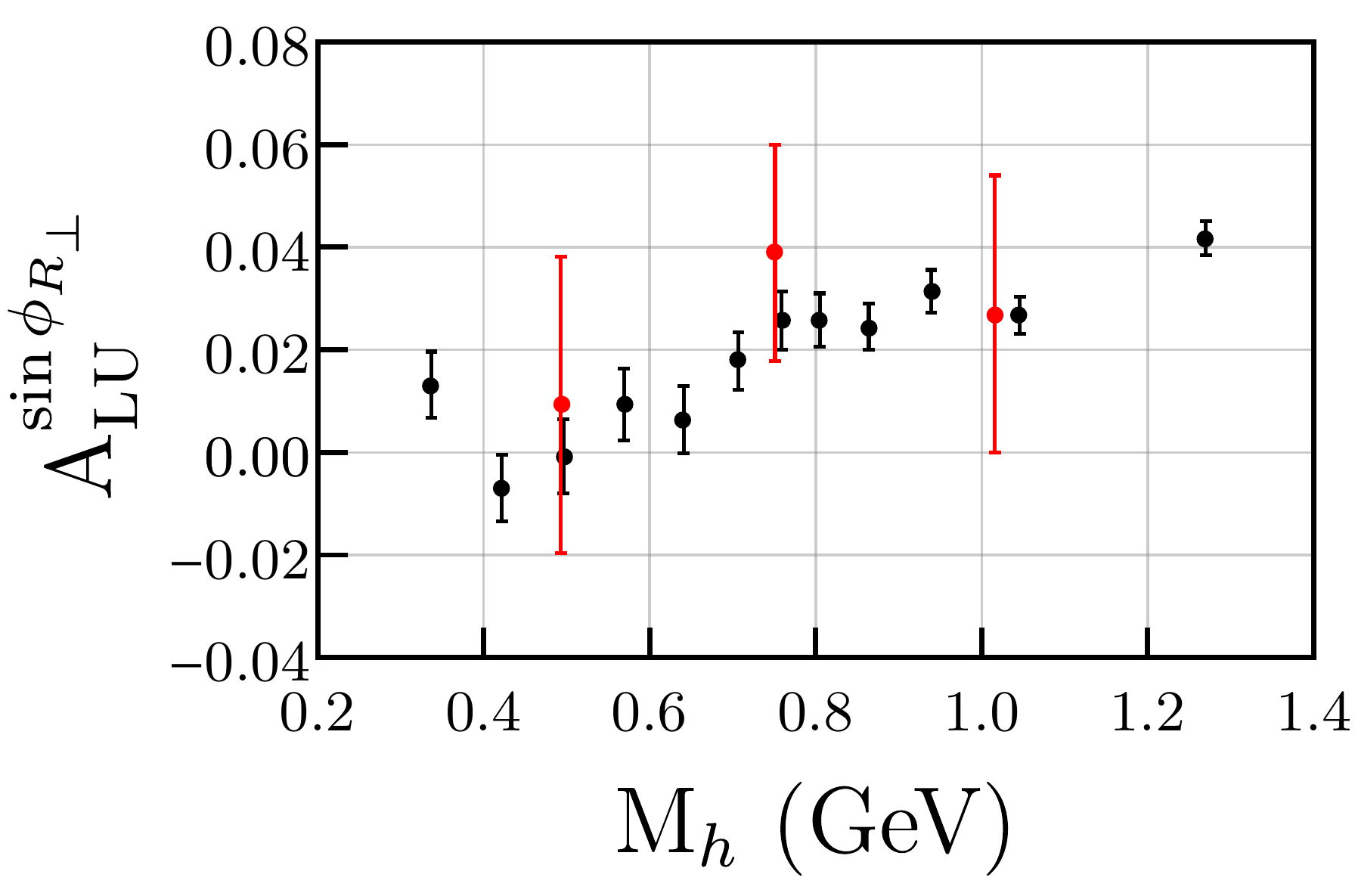}
\caption{The new CLAS12 results on beam helicity asymmetry in two-pion semi-inclusive deep inelastic electroproduction \cite{Hayward:2021psm}. The asymmetry as a function of $x$ (left) and the invariant mass of pion pairs (right). The red points on the right graph are from CLAS6 measurements \cite{Mirazita:2020lik}.}
\label{fig:clas12dih}
\end{center}
\end{figure}

Studies of correlations of final state hadrons are crucial for understanding of the hadronization process in general, and the TMD FFs, in particular. First publication of CLAS12, dedicated to correlations in two hadron production in SIDIS~\cite{Hayward:2021psm}, revealed significant correlations between hadrons produced in the current fragmentation region. Significant single-spin asymmetry has been measured, which can be related to higher twist PDF $e$, interpreted in terms of
the average transverse forces acting on a quark after
it absorbs the virtual photon~\cite{Burkardt:2008vd}. The difference of error bars of 6 GeV and 12 GeV measurements, see Fig.~\ref{fig:clas12dih}, demonstrates the impact of the beam energy on the phase space for production of multiple hadrons in the final state.

Measurements of flavor asymmetries in sea quark distributions performed in DY experiments, indicate very significant non-perturbative effects at large Bjorken-$x$, where the valence quarks are relevant~\cite{Alberg:2017ijg}.
 The measurements by E866 collaboration~\cite{Garvey:2001yq}, and more recently by SeaQuest~\cite{Nagai:2017dhp} suggest that $\bar{d}$ is significantly larger than $\bar{u}$ in the full accessible $x$-range, where non-perturbative effects are measurable.
The non-perturbative $q\bar{q}$ pairs, most likely responsible for those differences, are also correlated with spins and play a crucial role in spin orbit correlations, and in particular, single-spin asymmetries measured by various experiments in last few decades.

Collinear PDFs have flavour dependence, thus it is not unexpected that also
the transverse momentum dependence may be different for the different
flavours~\cite{Signori:2013mda}. Model calculations of transverse momentum dependence of TMDs ~\cite{Pasquini:2008ax,Lu:2004au,Anselmino:2006yc,Bourrely:2010ng} and lattice QCD results~\cite{Hagler:2009mb,Musch:2010ka} suggest that the dependence of widths of TMDs on the quark polarization and flavor may be
significant. It was found, in particular, that the average
transverse momentum of antiquarks is considerably larger than that of
quarks~\cite{Wakamatsu:2009fn,Schweitzer:2012hh}.

\subsection{Opportunities with high-intensity, positron and higher energy beams}

\subsubsection*{Double DVCS}
A major scientific goal for future physics program with CEBAF would be to explore DDVCS. 
As the subject of nuclear femtography has become a central focus of the 12 GeV science program, as well as the future EIC, DDVCS could be a major new frontier in the future. 
DVCS is primarily sensitive to the region near $x=\pm\xi$ or integrals of GPDs over internal $x$ due to the loop in the ``handbag'' diagram, and thus
cannot determine the GPDs uniquely.
The study of DDVCS enables more complete coverage of the $x-\xi$ plane giving wider access to the GPDs. 
The DDVCS process, when both the incoming and outgoing photons have large virtualities, allows for mapping the GPD's along each of the three axes ($x$, $\xi$, and $t$) as the three variables can now be varied independently. 
%The $\xi$ dependence of GPDs contains unique information about the distribution of nuclear forces and also provides a constraint on GPD models in the establishment of the total spin sum rule. 

While DDVCS has a huge advantage for mapping the GPDs, experimentally it is a challenging reaction to study.  The cross section of DDVCS is significantly smaller (more than two orders of magnitude) than that of DVCS. 
 To eliminate ambiguity and anti-symmetrization issues, the outgoing time-like photon must be reconstructed through their di-muon decays. These two conditions require a large acceptance detector capable of running at very high luminosities, $10^{37} -10^{38}$  cm$^{-2}$ s$^{-1}$, with good muon detection. %for the DDVCS studies. 

\iffalse
\begin{figure}
\begin{center}
\includegraphics[width=0.49\textwidth]{xs_dvcs_ddvcs_6gev.pdf}
\includegraphics[width=0.47\textwidth]{bsa_ddvcs.pdf}
\caption{On the left differential cross sections for DVCS+BH and DDVCS+BH processes. On the right the beam spin asymmetry for the reaction $ep\to e^\prime p\mu^+\mu^-$ (DDVCS+BH) for different virtualities of the lepton pair : Q$^2$ = 0 GeV$^2$ (thick solid line), which corresponds to a DVCS calculation, 1.5 GeV$^2$ (thick dashed line), 2 GeV$^2$ (thick dash-dotted line), 2.8 GeV$^2$ (dotted line) 3.6 GeV$^2$ (thin solid line) and 4.4 GeV$^2$ (thin dashed line). Calculations and predictions from [M. Guidal and M. Vanderhaeghen. Phys. Rev. Lett. 90 (2003) 012001]. .}
\label{fig:xsddvcs} 
\end{center}
\end{figure}
\fi

The possibility of extending studies of DVCS to DDVCS with upgraded detectors in Halls A (SoLID) and B ($\mu$-CLAS12) were discussed in letter-of-intents to PAC43 (LOI12-15-005) and PAC44 (LOI12-16-004). The SoLID DDVCS studies will take advantage of the setup proposed for the already approved J/$\psi$ experiment (E12-12-006). After adding muon detectors in forward angles, a significant set of experimental data will be collected at the luminosity of few $\times 10^{37}$ cm$^{-2}$ s$^{-1}$ to study di-muon production in deeply virtual electron scattering in the limited phase space of $Q^2>$M$_{\mu\mu}^2$. With a dedicated setup, target inside the solenoid, additional iron shielding in front of the calorimeter, GEM trackers, and muon detectors, reaction $ep\to e^\prime p^\prime \mu^+ \mu^-$ can be studied with a luminosity of $10^{38}$ cm$^{-2}$ s$^{-1}$. 
%{\it At these luminosities, studies of both space-like ($Q^2>$M$_{\mu\mu}^2\equiv Q^{\prime 2}$) and time-like ($Q^2<Q^{\prime 2}$) regimes will be possible. }

The Hall-B DDVCS setup will use the CLAS12 forward detector for muon detection and a lead tungsten calorimeter that will replace the high threshold Cherenkov counter for the detection of scattered electrons. A recoil detector inside the CLAS12 solenoid magnet for proton detection and the vertex tracker in front of the calorimeter/shield is part of the conversion of CLAS12 to $\mu$-CLAS12. It is expected that the detector will run at luminosities of $> 10^{37}$ cm$^{-2}$ s$^{-1}$. With this proposed luminosity, $\mu$-CLAS12 will be able to study DDVCS in both spacelike and timelike regions, producing high-quality results with wide kinematical coverage. 

With these upgrades to the experimental equipment in Halls A and B, SoLID and CLAS12, respectively, a DDVCS program can be realized and the GPD program at JLab can be taken to a unique new level, accessible only with large acceptance, high luminosity detectors.

\subsubsection*{DVCS with positron beam} 

An impressive amount of high-precision data start to come out and will be produced in the next several years from the experimental program developed at JLab. 
%Fine details of the inner structure of matter will be explored in measurements of cross-sections and polarization observables using unpolarized and polarized beams and targets. 
Often, however, interpretation of experimental results and extracting the information on hadron structure requires model-dependent assumption and calculations. Experimentally we measure reactions that are composed of multiple interfering elementary lepton-hadron scattering mechanisms. 
While there is no lepton charge sign difference in the obtained physics information, the comparison between electron and positron scattering, for example, allows separation of contributions of different components in the scattering process and isolates them in a model-independent way. Combining measurements with polarized electrons and polarized positrons in the elastic and DIS regime allows obtaining unique experimental observables enabling a more accurate and refined interpretation. Therefore, positron beams, both polarized and unpolarized, are essential for the experimental study of the structure of hadronic matter carried out at JLab.

The perspectives of an experimental program with positron beams at JLab are discussed in the ``$e^+$@JLab White Paper"~\cite{Accardi:2020swt} with a follow-up letter-of-intent to JLab PAC46 (LOI12-18-004~\cite{LOI12-18-004}) outlining prospects of experiments using unpolarized and polarized positron beams at CEBAF. The outlined experimental program with positron beams will contribute in a major way to the high-impact 12 GeV physics program: the physics of the two-photon exchange, the determination of GPDs, and tests of the Standard Model.

%There have been significant efforts devoted to extracting GPDs from the DVCS observables. However, with the single lepton charge measurements, separation of the reaction amplitudes cannot be done in a model-independent way. A possible approach to this separation with a single change beam is to exploit different beam energy sensitivity of DVCS and interference amplitudes [E12-13-010, E12-16-010B]. Of particular interest is the real part of the DVCS amplitude that contains the D-term [Pol99] which parameterizes the Gravitational Form Factors of the nucleon [[Pol03, Bur18, Pol18,Kum19]. Without direct measurement, the dispersion relation has been employed to determine the real part in leading order approximation [Ani07, Die07, Pol08]. However, as has been shown in recent studies, there are significant power corrections that cannot be neglected in precision DVCS phenomenology.  The only way to determine the D-term without model assumptions is to measure both the real and imaginary parts of the CFF independently. 

The combination of DVCS measurements with oppositely charged incident beams is the only unambiguous way to disentangle the contribution of different terms to the cross-section. The BH amplitude is electron charge even while the DVCS amplitude is electric charge odd. The DVCS contribution to the cross-section has a different sign for electron vs. positron scattering. Using various combinations of the difference and sum of the measured helicity ($\lambda$) and the lepton charge ($\pm$)-dependent cross-sections, one can project out the real and imaginary parts of Compton amplitudes:
\begin{eqnarray}
\sigma_\lambda^{\pm}=\sigma_{BH}+\sigma_{DVCS}+\lambda\tilde{\sigma}_{DVCS}\pm(\sigma_{INT}+\lambda\tilde{\sigma}_{INT})
\end{eqnarray}
Here $\sigma_{DVCS}$ and $\sigma_{INT}$ related to the real part of the Compton form-factor, while $\tilde{\sigma}_{DVCS}$ and $\tilde{\sigma}_{INT}$ to its imaginary part. In particular, the difference of the unpolarized cross-sections will single out $\sigma_{INT}$ that contains the real part of CFF. While the helicity-dependent charge asymmetry measures the ${\tilde \sigma}_{INT}$ and last but not least, the lepton charge integrated helicity difference measures the imaginary part of the DVCS amplitude (${\tilde \sigma}_{DVCS}$). 
%
%\begin{eqnarray}
%\Delta^C_{UU}&=&(\sigma_+^+ + %\sigma_-^+)-(\sigma_+^- + \sigma_-^-)=\sigma_{INT} \\
%\Delta^C_{LU}&=&(\sigma_+^+ - \sigma_-^+)-(\sigma_+^- - \sigma_-^-)=\lambda\tilde{\sigma}_{INT} \\
%\Delta^0_{LU}&=&(\sigma_+^++\sigma_+^-)-(\sigma_-^++\sigma_-^-)=\lambda\tilde{\sigma}_{DVCS}
%\end{eqnarray}

There were two proposals submitted to JLab PAC48 in 2020, receiving conditional approval to motivate further studies. In PR12-20-009 \cite{C12-20-009} it is proposed to use CLAS12 in Hall B to measure the unpolarized and polarized Beam Charge Asymmetries (BCAs) in DVCS on unpolarized hydrogen. The second proposal, PR12-20-012 \cite{C12-20-012}, would use the HMS of Hall C together with the Neutral Particle Spectrometer (NPS) for BCA measurement. 
%In Fig.~\ref{fig:cff}, extraction of CFFs when both positron and electron data are used compared with fits when only electron data are included. 
A significant reduction of extraction errors is expected when both positron and electron data are used compared with fits when only electron data are included. %with the combined data. 

\iffalse
\begin{figure}
\begin{center}
\includegraphics[width=0.49\textwidth]{Figures/hallc_pos_el_gpdH.pdf}
\end{center}
\caption{Figure from PR12-20-012: extraction of CFFs from fits to the experimental measurements. The first column in the left shows the
results of the helicity-conserving CFFs when both positron and electron data are used in the fit
(dark symbols), and when only the electron approved data is used (open symbols). The second and third columns
show the same information for the helicity-flip CFFs. The solid horizontal lines indicate the input
values used to generate the cross-section data.}
\label{fig:cff}
\end{figure}
\fi 

\subsubsection*{Opportunities at higher energy}

\begin{figure}
\centerline{
\includegraphics[width=0.48\textwidth]{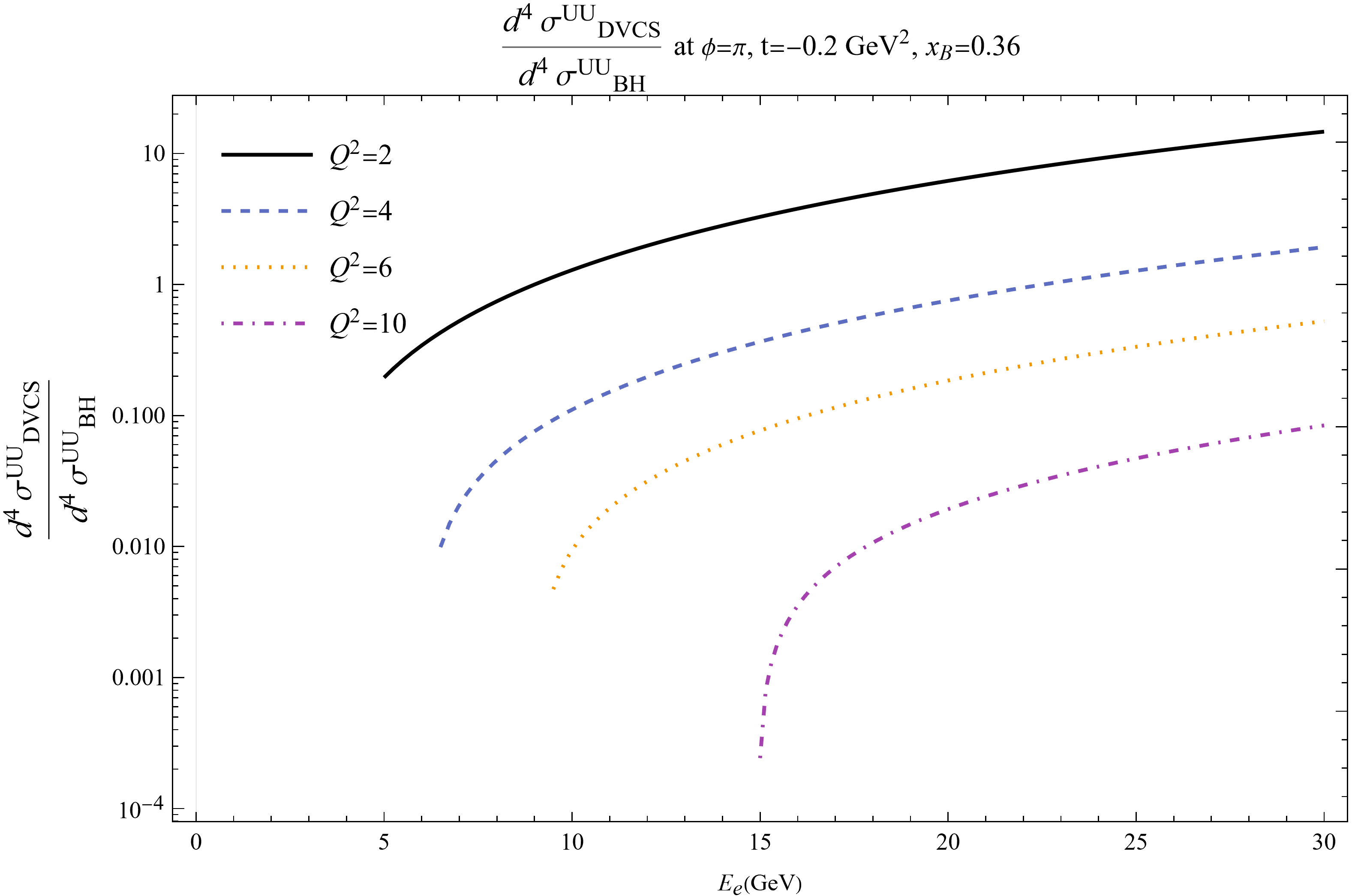}
\includegraphics[width=0.48\textwidth]{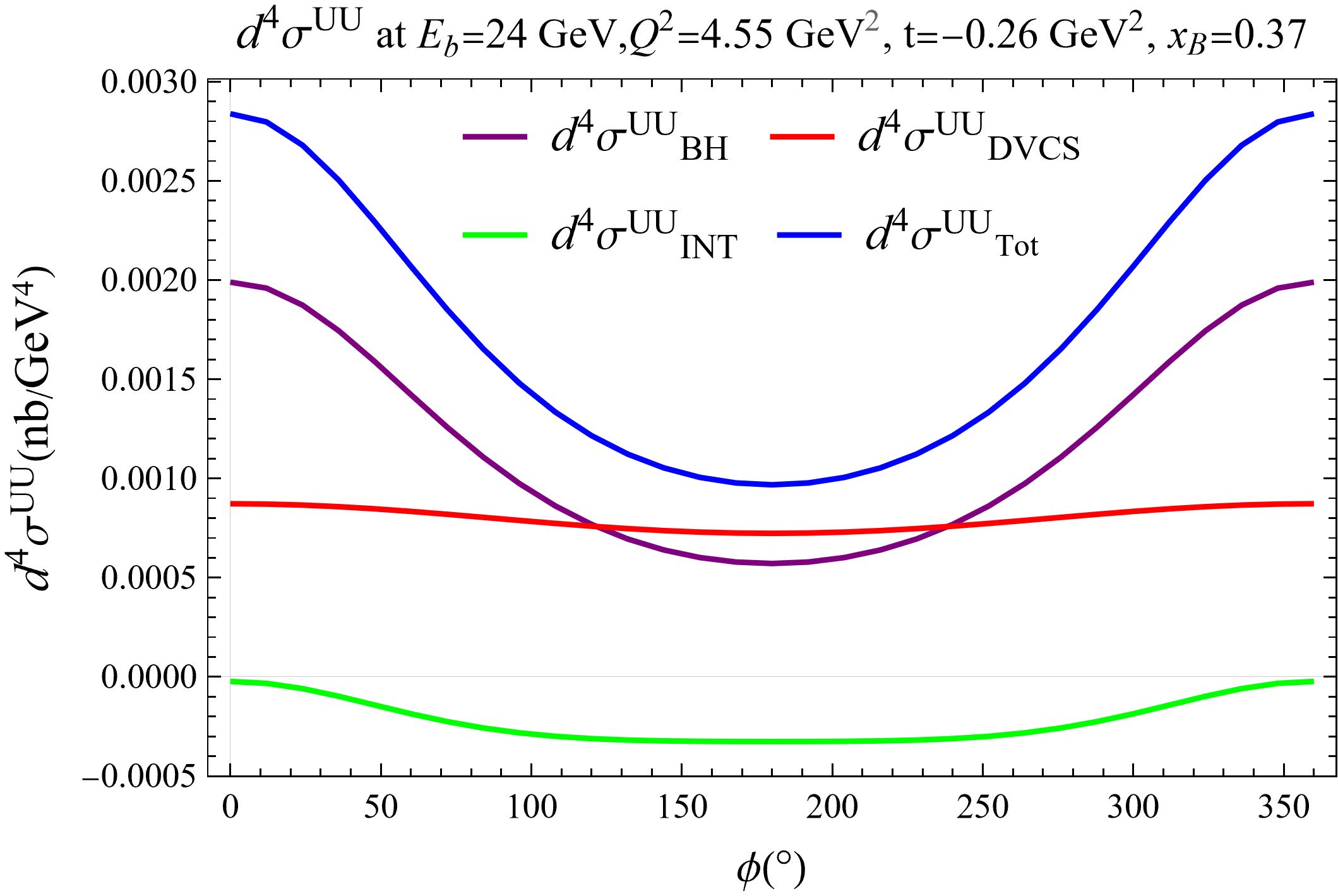}
}
\caption{(Left) Ratio of the DVCS plus DVCS-BH interference cross section over the BH cross section in unpolarized $ep \rightarrow e'p' \gamma$, plotted as a function of the electron beam energy, $E_e$. The kinematics is defined by the azimuthal angle, $\phi=180^\circ$, Bjorken $x=0.36$, four-momentum transfer square, $t=-0.2$ GeV$^2$; the four values of the initial photon four momentum squared, $Q^2 =2, 4, 6, 10$ (GeV/c)$^2$, respectively, are indicated on the right hand side of the curves. All cross sections were calculated using the formalism in Ref.~\cite{Kriesten:2019jep}, using the GPD model in Refs.~\cite{Goldstein:2010gu,Kriesten:2021sqc}. 
(Right) The unpolarized cross section for the kinematic bins from Ref.~\cite{Georges:2018kyi} with initial electron energy $E_1=$24 GeV, $Q^2= 4.5$ GeV$^2$, $t= -0.26$ GeV$^2$, $x = 0.37$; The curves correspond to contributions from the BH (purple), DVCS (red), DVCS-BH interference (green) and total (blue).}
\label{fig:DVCS/BH}
\end{figure}

JLab at 24 GeV will open up a big phase space for studying the structure of protons and neutrons, particularly for 3D nuclear femtography both
in coordinate and momentum spaces.
A 20-24 GeV beam energy presents several crucial advantages for the extraction of GPDs from data. First of all, the proposed upgrade of the current JLab settings features a guaranteed high luminosity which is strictly necessary to extract information from deeply virtual exclusive experiments. Furthermore, it will sensibly expand the accessible $Q^2$ range thus allowing thorough tests of both factorization and of the perturbative QCD behavior of GPDs in the regime $x \geq 0.05$. On the other hand, because of the sharp falloff of the BH cross section relative to the DVCS, shown in Fig.~\ref{fig:DVCS/BH},
one will be able to increase the precision  of the CFF extraction from experimental data in the $Q^2$ range already available in the 12 GeV program. A precise knowledge of the latter is, in turn, crucial for proton 3D spatial imaging. High electron energy is particularly important for meson production for which the asymptotic scaling needs larger $Q^2$. 

%DVCS \& meson production: 

%\begin{figure}[ht!]
%\begin{center}
%\vspace{0.0cm}
%\includegraphics[width=0.38\textwidth]{Figures/JLAB25-q2.png}
%\includegraphics[width=0.3\textwidth]{Figures/JLab25-xdep.png}
%\includegraphics[width=0.3\textwidth]{Figures/JLab25-pt.png}
%\end{center}
%\caption{\small Kinematical coverage of JLab for beam energies 12, 24 and 50 GeV. Events for fixed kinematic are plotted in $Q^2$ (left panel), $x_B$ (middle panel), and $P_T^2$ (right panel)}
%\label{jlab12-25}
%\end{figure}

SIDIS program at Halls A, B, and C at JLab12 \cite{E12-06-104}-- \cite{C12-20-002} with high luminosity and large acceptance detectors will provide high precision data in multi-dimensions in the valance quark (high-$x$) region to study TMDs. With the energies available, the kinematical reach of JLab12 is limited to the kinematic region of high-$x$, but low-Q$^2$ and low-P$_T$. Recent theoretical developments in TMD physics show that while this region is sensitive to the 3D nucleon structure in momentum space, non-perturbative effects could complicate the interpretation in terms of pure TMD distributions ~\cite{Scimemi:2019cmh}. TMD Studies planned for the future EIC with much higher center-of-mass (CM) energies would allow us to reach lower x, higher $Q^2$, and higher P$_T$ region, and have a cleaner theoretical interpretation for the extracted TMDs. But the much higher CM energy also means the extracted quark TMDs often being smeared by significant gluon showers/gluon radiations. The effects, such as Sivers single-spin asymmetry will also reduce at large $Q^2$. There is an energy gap between JLab12 and EIC where the “sweet region” for TMD studies lies. An energy upgrade of JLab to 24 GeV (JLab24) would fill a significant part of this energy gap to reach the “sweet region”. High luminosity and large acceptance detectors of JLab24 would allow us to access significantly higher P$_T$ and Q$^2$ range for high-$x$ where the TMD effects are still large while theory interpretation is better defined. Combining with JLab12 and EIC, JLab24 would allow for validation and extension to wider kinematics of the studies of transverse momentum dependence of various TMD distributions and fragmentation functions as well as the transition from TMD regime (P$_T$/$z\ll Q$) to the collinear perturbative regime (P$_T$/$z\sim Q$). A much higher Q$^2$ and P$_T$ range accessible at JLab24 at high-$x$ would allow for studies of $Q^2$-dependence of different spin-azimuthal asymmetries. Apart from providing important information on quark-gluon correlations, it will help validate the TMD evolution and help in building a more consistent TMD theory, extending its reach to higher P$_T$ of hadrons accessible by polarized SIDIS experiments. Extension of JLab12 kinematics to a lower-$x$ region with JLab24 will open new avenues for studies of non-perturbative sea effects.

%PDFs and FFs: Yes, this will also have significant improvement. With a possible future JLab~24~GeV energy upgrade, the kinematic reach of DIS and SIDIS measurements would increase significantly.
%It will also allow us to fully utilize SIDIS processes for flavor separation, especially for the precision study of the strange quark spin distributions.

%% file: Section3.tex
\section{ Hadron Spectroscopy}

The energies and other quantum numbers of the excited state of any physical system provide important clues to the underlying dynamics and relevant degrees of freedom. This is especially true in the case of hadrons, where the spectrum of meson and baryon excitations established the quark model, including elements of special relativity and QCD.~\cite{Godfrey:1985xj,Capstick:1986ter}

We know, however, that the full picture has not been experimentally verified. For example, we fully expect that excited gluonic degrees of freedom should manifest themselves, but these have not yet been conclusively isolated. It is generally argued that the best discovery path is through searching for so-called ``exotic'' meson states, which have quantum numbers that cannot be obtained with only quark--antiquark degrees of freedom. Previous searches, using hadron beams, have been inconclusive, but new results using polarized photon beams from CEBAF are on the horizon.

QCD and the quark model also predict a number of baryon excitations that have yet to be observed experimentally. A new program at Jefferson lab will focus on mapping the spectrum of baryons with strangeness. Excited states in this sector should be less numerous and more narrow than for the nonstrange baryons, which will ameliorate the difficulties associated with overlapping resonances.

There have been a number of narrow charmonium states discovered in recent years, which defy description in terms of the quark model. Their existence points to dynamics of multiquark states that should in principle be predicted by QCD.

Jefferson Lab is aggressively pursuing the current spectroscopic understanding of QCD dynamics. This includes photoproduction of meson and baryon states in GlueX, CLAS12, and other CEBAF facilities. It also includes new strides in Lattice QCD that not only are sorting out the hadron spectrum in concert with experimental measurements, but also pointing the way to quantifying the photoproduction cross sections.

\subsection{Role of gluonic excitations} 
\label{sec:gluerole}

Hadron spectroscopy is key to understanding the inner workings of QCD and it is particularly relevant for determining the role of gluons. On one hand gluons are responsible for the confining force that binds quarks together into color-singlet hadrons, but they also carry most of the hadron's mass and a large fraction of its spin. Thus, gluons are also expected to act as constituents affecting hadronic excitations. In the quark model, dominance of the valence constituent quarks, which explains, among other things the observed symmetry patterns of spin-orbit excitations, the anomalous magnetic moments of baryons, the OZI rule, etc., restricts the spectrum of hadron resonances to those having the quantum numbers of the quark-antiquark pair for mesons or three quarks for baryons. For example, non-strange mesons with natural parity and odd CP are forbidden in the quark model. In the past these were referred to as exotics of the second kind, to distinguish them from flavor exotics, \textit{e.g.}, baryons with positive strangeness. The  constituent  quark model has its rooting in the large-$N_c$ world of planar diagrams and strings, which is phenomenologically supported, for example, by the observed linearity of meson and baryon Regge trajectories. Thus, it is not surprising that exotic states do not appear in the conventional hadron spectrum. There are strong indications however, that such states ought to exist and have properties, \textit{e.g.}, masses and decay widths that are similar to those of the ordinary quark model  states~\cite{Cohen:1998jb}. In the language of large-$N_c$ these exotics could, for example, be associated with string junctions, which in QCD originate from gluon-gluon interactions~\cite{Montanet:1980te}. Further insights  into the nature of gluon fields that can result in hybrid hadrons have come from lattice studies of charmonia. States with a high density of the chromo-magnetic field were identified as forming a quadruplet with $J^{PC} = (0,1,2)^{-+},1^{--}$ quantum numbers~\cite{Liu:2012ze}. Phenomenologically  these can be interpreted as bound states of a color octet  $Q\bar Q$ pair and  an effective constituent gluon with $J^{PC} = 1^{+-}$. Such axial gluon states could also play a role in formation of light flavor 
hybrids~\cite{Szczepaniak:2001rg,Guo:2008yz}, 
  including those of exotic quantum numbers,  
  $J^{PC} = 1^{-+}$, referred to as the $\pi_1$ or $\eta_1$ in the isovector or isoscalar channel, respectively. 
%There exist  several  phenomenological and lattice %studies 
% of gluons configuration that 
%  may result in hybrid %hadrons~\cite{Guo:2008yz,Greensite:2014bua,Bicudo:201%5bra}.
 % and more recently were studied using lattice QCD %simulations ~\cite{Bicudo:2015bra}. 
%For example, gluon distributions that 
% can potently result in hybrid hadrons were %identified % gluon  junctions appear in the 
% distribution of the QCD action in the presence of %static quarks gluon chains connecting $Q\bar Q$ pairs %are a natural representation of QCD eigenstates in %physical gauges~\cite{Guo:2008yz,Greensite:2014bua}, %
%  This multiplet contains the exotic state with %$J^{PC} = 1^{-+}$, referred to as the $\pi_1$ or %$\eta_1$ in the isovector or isoscalar case, %respectively. 
Following the arguments 
 based on Regge-resonance duality one finds other intriguing connections. 
  The exotic meson with $J^{PC} = 0^{--}$ would lie on a daughter Regge trajectory together with the $J^{PC}=2^{--}$ state. The latter has non-exotic quantum numbers and in the quark model it corresponds to a $D$ wave excitations of the $\rho$ (isovector) or, $\omega/\phi$ (isoscalar) mesons. All these states appear in lattice simulations but, have not yet been observed experimentally. Similarly the unknown exotics $J^{PC}=0^{+-},2^{+-}$, referred to as $b_{0,2}$ for isovector or $h_{0,2}$ for isoscalar, are expected on a trajectory degenerate with that of the $b_1/h_1$ mesons. The latter are quark model states, with quarks in the $L=S=1$, spin-orbit configurations, but little is known about them. 
 Discovering quark-gluon hybrids, and exotic mesons in particular would give a clear, new direction for future research into strongly coupled QCD.

\subsection{Recent developments}
\label{sec:spectdev}
So far, the exotic $\pi_1$, which is expected to be the lightest of the hybrid mesons, has attracted most of the attention. Phenomenological models predict that it is dominated by $Q\bar Q$ pairs of spin-1 and thus should be copiously produced in photon diffractive dissociation at GlueX and CLAS12. It should be noted that at high energies, $E_\gamma > 10-20$~GeV photon interactions are mainly diffractive and dominated by the Pomeron exchange, which results mainly in production of neutral vector mesons.  In the JLab12 kinematics, however,  charge exchange reactions  have sizable cross sections and this setup ought to be ideal for searches of exotic mesons.  On the other hand, it is expected that the $\pi_1$ decays dominantly to complicated final states, like the $ b_1 \pi$ resulting in a $5\pi$ final state.  A “simpler” channel for the $\pi_1$ is the $\rho\pi$, which results in the $3\pi$ final state, and the golden channel is the $\eta^{(')}\pi$ although the branching rations for these are expected to be at the $1\%$ level.  In the past, the $\pi_1$ candidates were observed in the pion-induced reactions at VES, BNL and COMPASS, $p\bar p$ annihilation and in $\chi_c$ decays (for a review see~\cite{Meyer:2015eta}).  The main challenge in identifying the $\pi_1$ resides in complexity of amplitude analysis needed in order to isolate the resonance signal. Amplitude level analysis is necessary because the signal interferes with other nearby resonances and non-resonant processes, \textit{e.g.}, Deck production. This type of analysis can involve hundreds of parameters and to control systematic uncertainties, require sophisticated amplitude models that incorporate many constraints of the $S$-matrix theory. 
 
\begin{figure*}
\includegraphics[width=\textwidth]{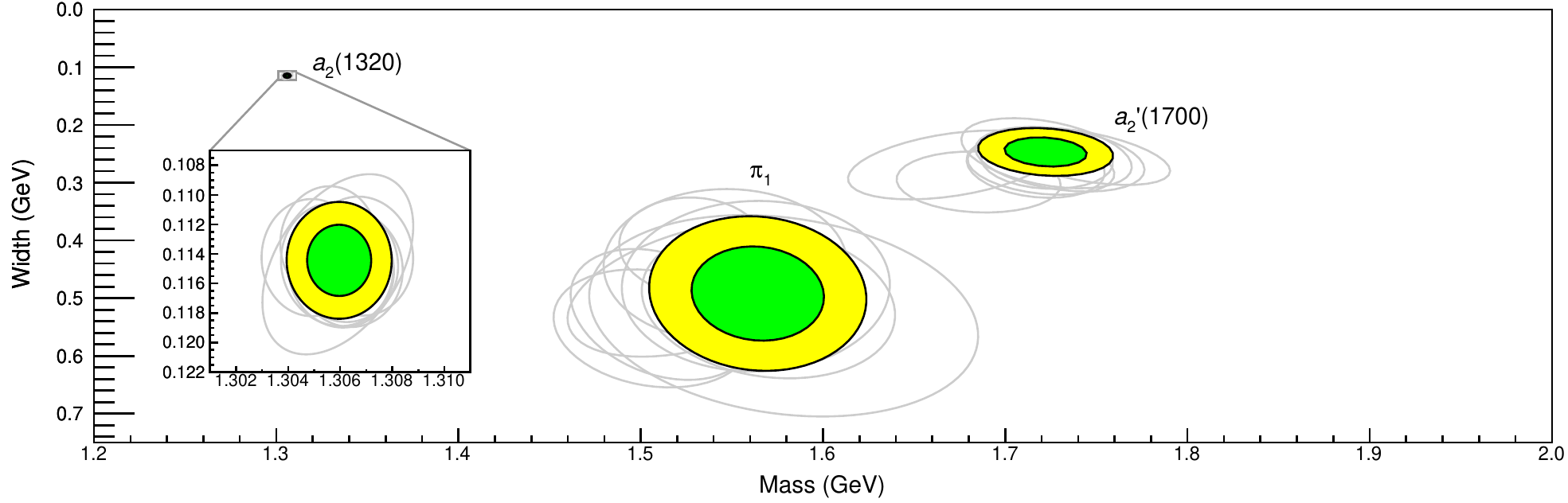}
\caption{\label{fig:poles} Positions of the poles identified as the $a_2(1320)$, $\pi_1$, and $a_2'(1700)$. The inset shows the position of the $a_2(1320)$. The green and yellow ellipses show the $1\sigma$ and $2\sigma$ confidence levels, respectively. The gray ellipses in the background show, 
 within $2\sigma$, variation of the pole position  upon changing the functional form and the parameters of the model, as discussed in the text
 }
\end{figure*}

By far the most sophisticated results on $3\pi$ partial wave analysis have been performed by COMPASS, which studied $\sim 50$~M  events from pion diffractive dissociation at $190$~ GeV~\cite{COMPASS:2018uzl,Ketzer:2019wmd}. They observed an exotic $J^{PC}=1^{-+}$ partial wave, with a mass distribution that strongly depends on the momentum transfer. At low momentum transfer $-t< 0.5$~GeV$^2$ the production appears to be consistent with the Deck process, however, for $-t> 0.5$~GeV$^2$ the Deck process diminishes and a resonant-like signal appears. When parametrized as a Breit-Wigner it yields a mass and with of $m_{\pi_1}(1600) =1600^{+110}_{-60}$~MeV and  $\Gamma_{\pi_1(1600)} = 580^{+100}_{-230}$~MeV, respectively. COMPASS has also published results on partial wave analysis of the $\eta^{(')}\pi$ final state from the same reaction~\cite{COMPASS:2014vkj}.
Their results appear to be consistent with the previous findings by E852, where a $\pi_1(1400)$ signal at a mass of approximately 1400~MeV was found decaying to $\eta\pi$ and a peak at a higher mass of $\sim 1600$~MeV was observed in the $\eta'\pi$ decay. It is difficult to reconcile with theoretical predictions existence of two exotic resonance in this mass range only a few hundred MeV apart. In Ref.~\cite{Rodas:2018owy} the JPAC collaboration performed a mass dependent analysis of the two channels using an analytical, unitary amplitude model and demonstrated that the two peaks are actually consistent with a single resonance, defined as a pole in the complex energy plane. The coupling between the two channels and the difference in thresholds for production of $\eta\pi$ and $\eta'\pi$ pairs is responsible for the resonance peaking at different positions on the real energy axis.  The resulting pole positions of the resonances determined by the JPAC analysis are shown in Fig.~\ref{fig:poles}, including the tensor mesons $a_2(1320)$ and $a_2'(1700)$ as well as the exotic $\pi_1(1600)$.

The recent analysis of coupled channel scattering lattice data, by the Hadspec collaboration~\cite{Woss:2020ayi} have confirmed existence of a single mass pole in the vicinity of the nearby decay thresholds. The calculation was done using an $SU(3)$ symmetric quark mass matrix. When extrapolated to coincide with $\pi_1$ mass obtained by JPAC, it was found that the total width, $\Gamma = 139-590$~MeV is dominated by the $\pi_1 \to b_1\pi$  channel, and branching ratios are consistent with phenomenological expectations,  $\Gamma(\pi_1 \to b_1\pi) > \Gamma(\pi_1 \to \rho\pi) > \Gamma(\pi_1 \to \eta'\pi) \sim 1 \to 12$~MeV.

\subsection{Experimental meson spectroscopy program} 
\label{sec:expmesons}

The GlueX~\cite{E12-06-102,GlueX:2014hxq,Adhikari:2020cvz} and CLAS12 MesonEx~\cite{E12-11-005} experiments provide a unique contribution to the landscape of experimental meson spectroscopy through the novel photoproduction mechanism, which has previously been relatively unexplored.  Utilizing a real, linearly-polarized photon beam in GlueX and quasi-real, low-$Q^2$ photons in CLAS12, this program covers a wide range of beam energies from $E_\gamma = 3-12$ GeV.  Unlike high-energy pion beam measurements (\textit{e.g.} the COMPASS results described above) where Pomeron exchange is dominant, for intermediate-energy photoproduction such as that explored at GlueX and CLAS12, meson production may proceed through a range of charged and neutral Reggeon exchanges.  An understanding of these exchange mechanisms is required to model the production of both conventional and exotic mesons.  

\begin{figure*}
\begin{center}
\includegraphics[width=0.45\textwidth]{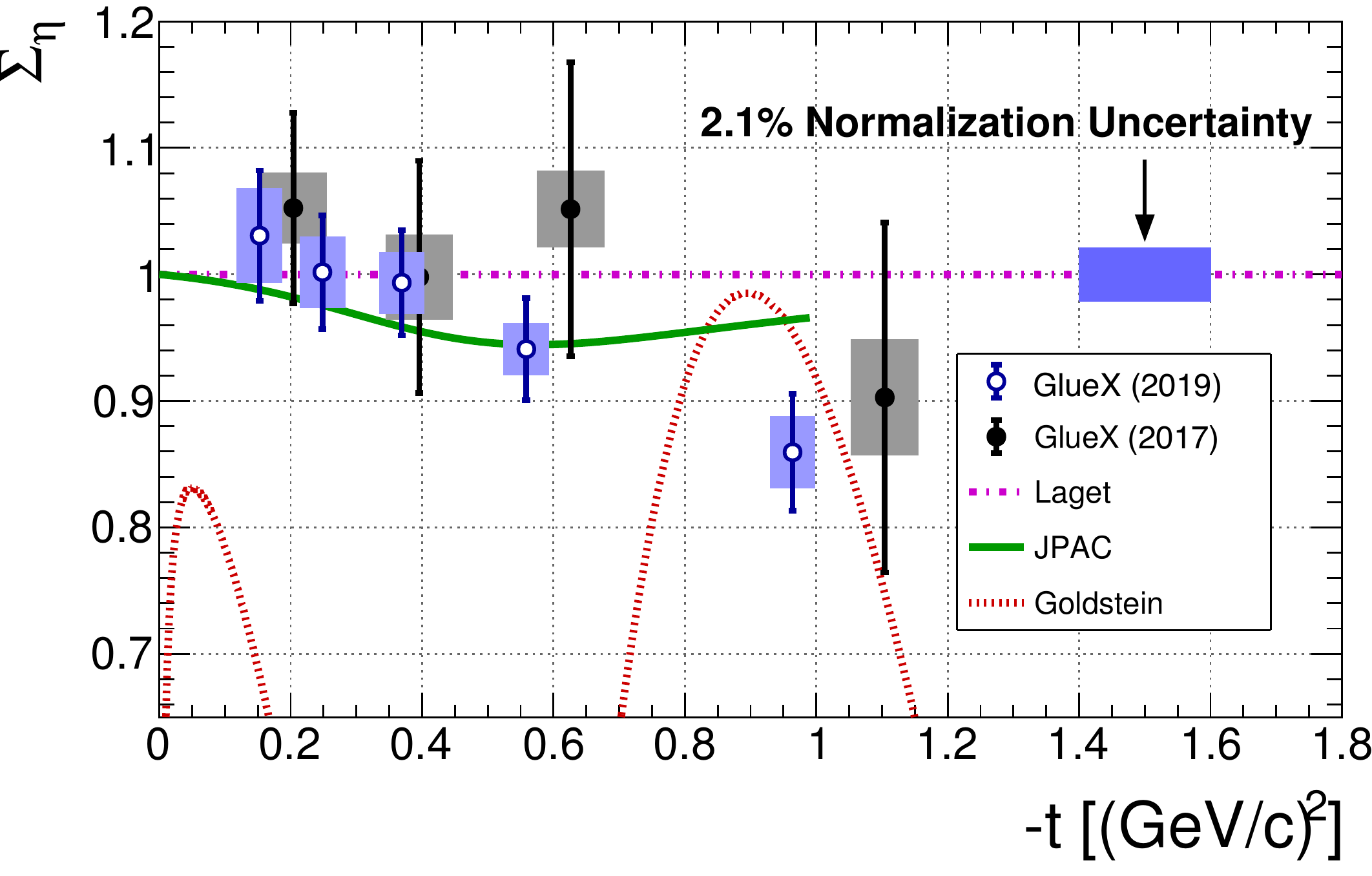}
\includegraphics[width=0.41\textwidth]{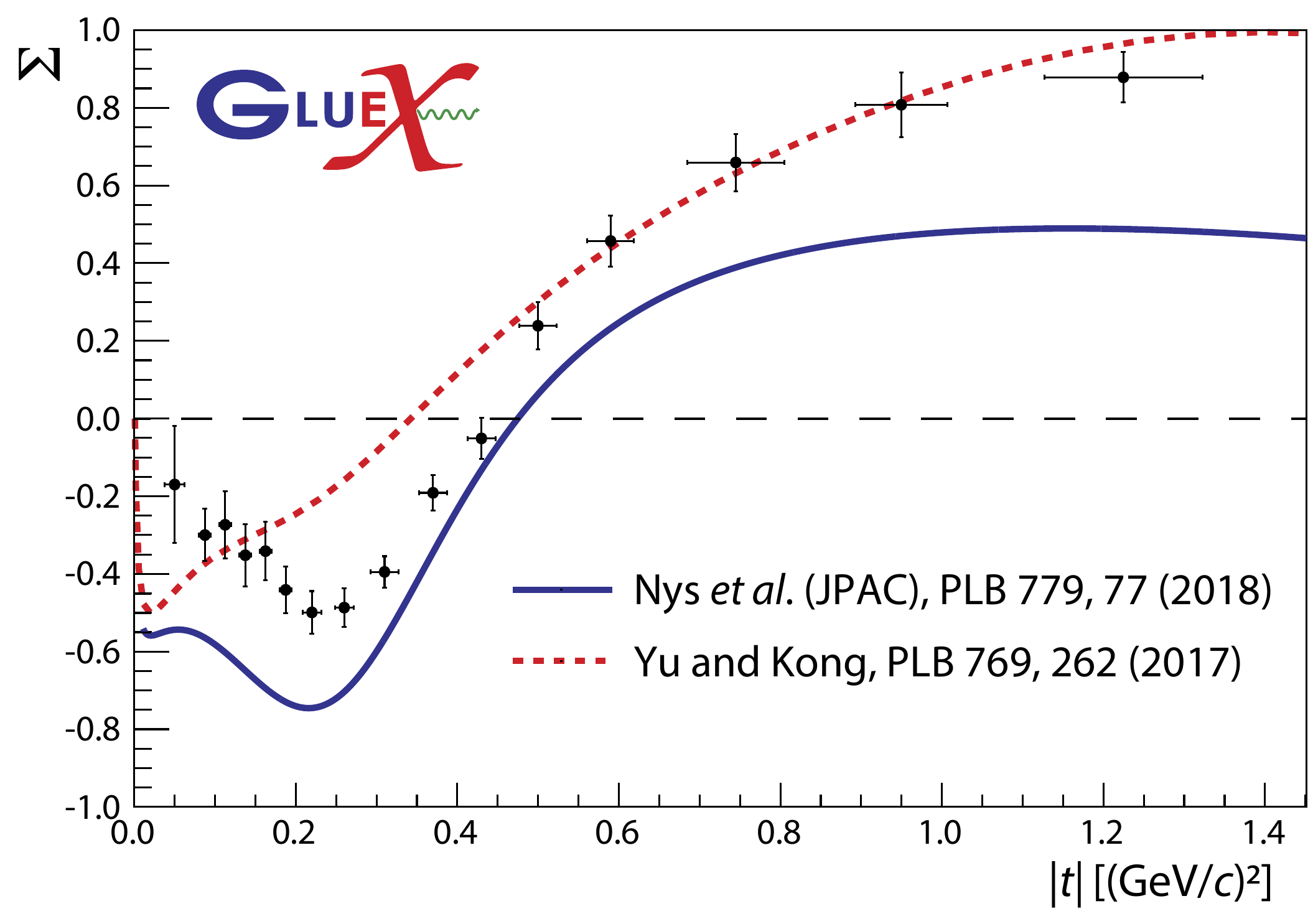}
\caption{Linearly-polarized $\Sigma$ beam asymmetries measured by GlueX for $\eta$~\cite{GlueX:2020fam} (left) and $\pi^-$~\cite{GlueX:2019adl} (right) meson production as a function of momentum transfer, $-t$.}
\label{fig:gluex_asym}
\end{center}
\end{figure*}

Thus, early measurements from GlueX have focused on a quantitative understanding of the meson photoproduction mechanism in this energy regime.  Utilizing the linearly-polarized photon beam, beam asymmetries $\Sigma$ have been measured for several single pseudoscalar mesons including the $\pi^0, \pi^-, K^+, \eta,$ and $\eta'$ over a broad range of momentum transfer, $-t$~\cite{GlueX:2017zoo,GlueX:2020fam,GlueX:2019adl,GlueX:2020qat}.  As demonstrated in Fig.~\ref{fig:gluex_asym}, these measurements indicate a dominance of natural exchange (\textit{i.e.} vector Reggeon exchange in the production of neutral pseudoscalars ($\Sigma\sim1$)), while the charge-exchange process requires both natural and unnatural exchanges with a significant $t$-dependence.  

The production of well-known vector meson and excited baryon, \textit{e.g.} $\Lambda(1520)$, resonances provide another opportunity to study these Regge exchange dynamics.  However, these studies take on the additional complication of modeling the angular distributions of the resonance decay described by the so-called Spin Density Matrix Elements (SDME).  Similar to amplitude or partial wave analyses, these SDME studies require a complete understanding of the detector response to account for acceptance and efficiencies in fits to multi-dimensional data.  First measurements of the polarized SDMEs for the reaction $\vec{\gamma}p \rightarrow \Lambda(1520)K^+$ find that natural parity exchange amplitudes are dominant~\cite{GlueX:2021pcl}.  Preliminary measurements of the polarized SDMEs for vector meson production~\cite{Austregesilo:2019tld} have statistical precision surpassing previous measurements by orders of magnitude~\cite{Ballam:1972eq} and show a strong preference for natural parity exchanges.  Detailed comparisons with theoretical models~\cite{Mathieu:2018xyc} will continue to refine our understanding of these meson production mechanisms.

\begin{figure*}
\begin{center}
\includegraphics[width=0.49\textwidth]{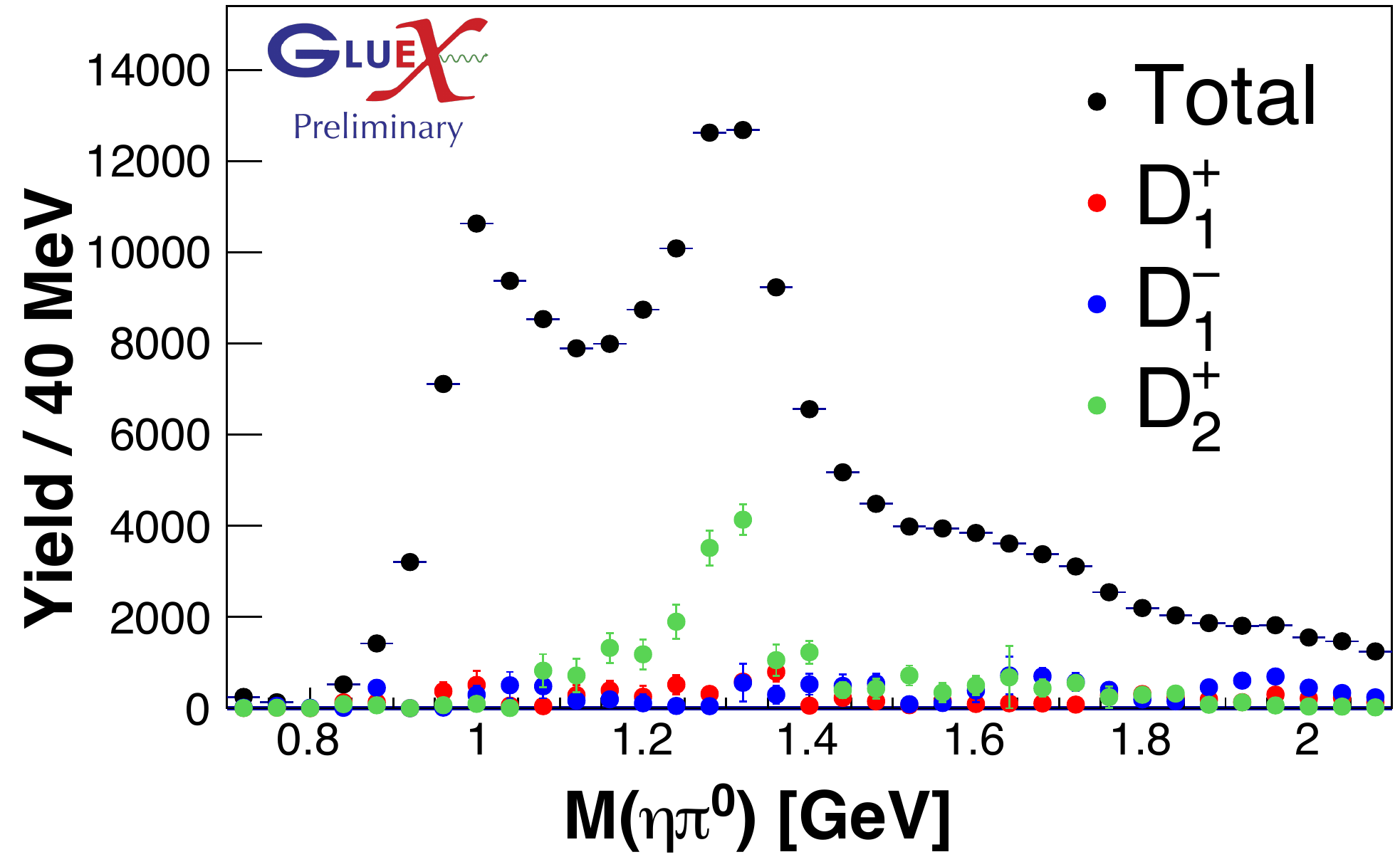}
\includegraphics[width=0.49\textwidth]{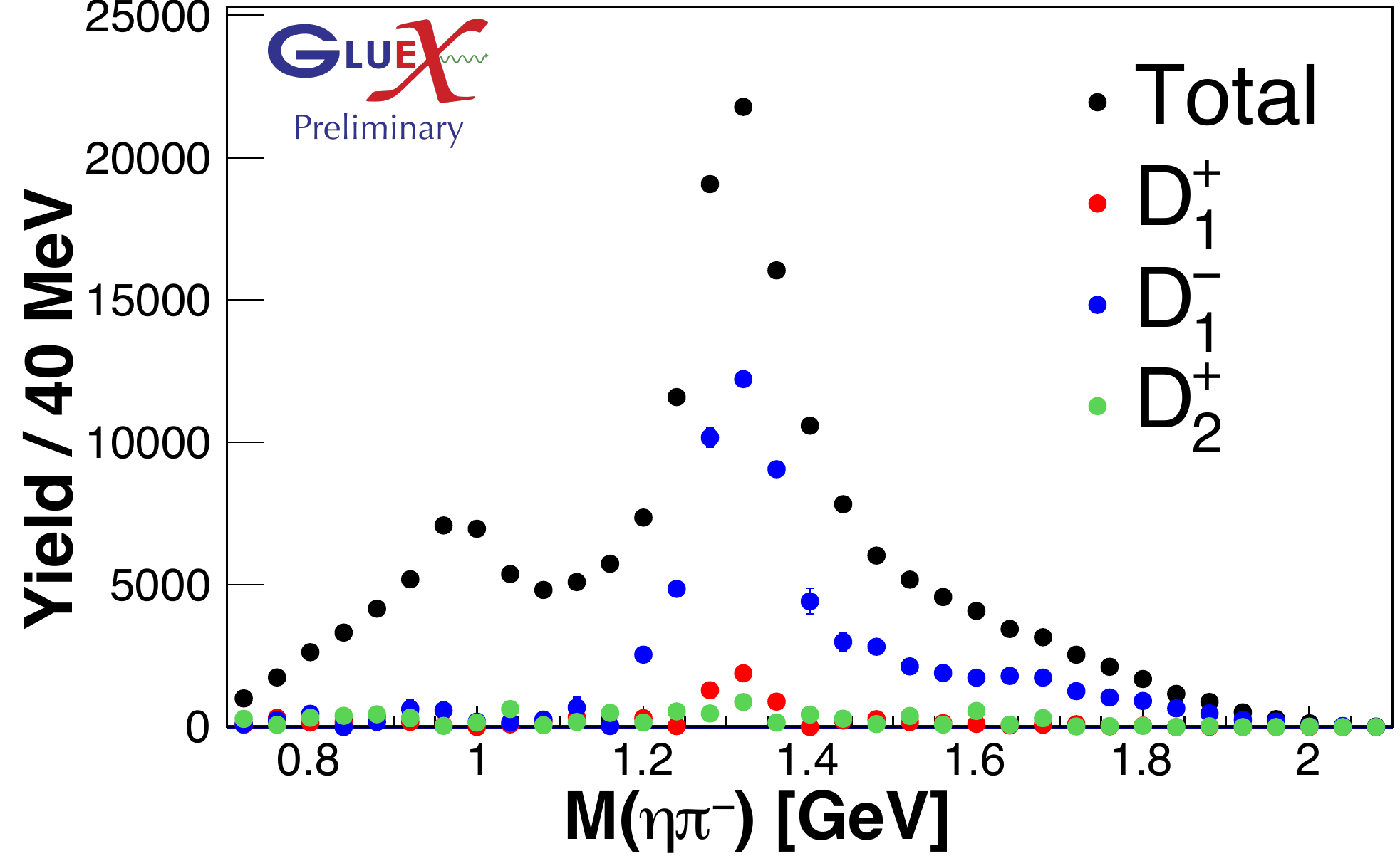}
\caption{Preliminary mass spectra and amplitude analysis results from GlueX for the reactions $\gamma p \rightarrow \eta\pi^0 p$ (left) and $\gamma p \rightarrow \eta\pi^- \Delta^{++}$ (right) with $0.1 < -t < 0.3$~GeV$^2$ and $8.2 < E_\gamma < 8.8$~GeV~\cite{gluexetapihadron}.  The total measured intensity is shown in black with colored points for the dominant tensor $a_2(1320)$ amplitudes, labeled $L_m^\epsilon$.}
\label{fig:gluex_etapi}
\end{center}
\end{figure*}

As described in Sec.~\ref{sec:spectdev}, there have been considerable recent theoretical developments and partial wave analysis of the $\eta\pi$ and $\eta'\pi$ systems from pion beam experiments.  Independent confirmation of these observations in alternative production mechanisms, such as photoproduction, are an essential component of the global spectroscopy effort.  An initial measurement of the reaction $\gamma p \rightarrow \eta\pi^0 p$ with a photon beam by CLAS observed a significant signal in the region of the $a_2(1320)$~\cite{CLAS:2020rdz}, but did not have sufficient statistics to perform a partial wave analysis.  Recent data collected by the GlueX and MesonEx experiments at higher beam energies provide the statistical precision necessary for such measurements, as shown in Fig.~\ref{fig:gluex_etapi}, where clear signals can be seen in the region of both the $a_0(980)$ and $a_2(1320)$ for both the neutral $\eta\pi^0$ (left) and charged $\eta\pi^-$ (right) systems in the preliminary analysis of the GlueX data~\cite{gluexetapihadron}.

Similar to the beam asymmetry and SDME measurements, the angle between the meson production and linear beam polarization planes is dependent on the production mechanism for the $\eta\pi$ system.  Thus, natural and unnatural parity exchange amplitudes, referred to as positive and negative ``reflectivity'' ($\epsilon$), can be determined through an amplitude analysis~\cite{Mathieu:2019fts}.  Amplitudes describing the polarization and $\eta\pi$ decay angles for $L_m^\epsilon = S_0^\pm, D_0^\pm, D_1^\pm, D_2^+$ and $D_{-1}^-$ are utilized in the fit following the model described in Ref.~\cite{Mathieu:2020zpm}.  Figure~\ref{fig:gluex_etapi} shows preliminary results for the measured intensity of the dominant waves, with evidence for tensor meson $a_2(1320)$ production in the $D_2^+$ wave in $\eta\pi^0$ produced through natural parity (vector) exchange and the $D_1^-$ wave in $\eta\pi^-$ through unnatural parity (pion) exchange.  These studies are consistent with those described above for pseudoscalar and vector meson production, indicating that for photoproduction at JLab 12 GeV energies natural exchange tends to dominate, except when unnatural (pion) exchange is permitted and the kinematics are such that the influence of the pion pole is significant (\textit{i.e.} low $-t$). Understanding these conventional meson photoproduction amplitudes lays the foundation for searches for the exotic $P$-wave in the $\eta^{(')}\pi$ system. 

% future data on flavor composition
Finally, in addition to predicting the existence of gluonic excitations in mesons, Lattice QCD calculations also predict the flavor of quarks expected to be associated with these excitations and, in fact, a spectrum of hybrid mesons containing strange quarks are expected.  A significant increase in statistics is required to study these mesons containing strange quarks as they're produced at a rate roughly an order of magnitude smaller than non-strange mesons.  Thus, higher statistics data samples are currently being collected by the GlueX and CLAS12 experiments to complete program in strange meson spectroscopy, which is required to clearly identify a pattern of gluonic excitations.  Along with additional statistical precision, the recent addition of Cherenkov detectors in GlueX~\cite{Ali:2020erv} and CLAS12~\cite{Contalbrigo:2020lnd} will provide critical separation between charged pions and kaons to separate these strange and non-strange final states.

\subsection{Experimental baryon spectroscopy program}
\label{sec:expbaryons}

Another critical component of the JLab spectroscopy program carried out over the last $\sim$15 years is the study of the spectrum and structure of excited nucleon states, referred to as the $N^*$ program.  Through measurements of exclusive electroproduction of both strange and non-strange final states, detailed electrocouplings measurements over a wide kinematic range have provided critical input to global analyses to elucidate the $N^*$ spectrum (see Ref.~\cite{Carman:2019lkk,Carman:2020qmb} for recent reviews). Studies of these $N^*$ states are currently being extended with the new CLAS12 detector in the 12 GeV era of experiments, which will significantly extend the kinematic range to $Q^2 > 5$~GeV$^2$~\cite{E12-09-003,E12-06-108A}.  

The discussion of the gluon's role in the hadron spectrum as described in Sect.~\ref{sec:gluerole} implies the search for and study of hybrid baryons with constituent gluonic excitations, and Lattice QCD calculations predict a rich spectrum of such baryons with an excitation scale comparable to that expected for hybrid mesons~\cite{Dudek:2012ag}. Hybrid baryons could be identified as supernumerary states in the $N^*$ spectrum, but they do not exhibit exotic quantum numbers, making them challenging to clearly distinguish from conventional baryons. However, measurements of the electrocoupling evolution with $Q^2$ becomes critical in the search for hybrid baryons, where a distinctively different $Q^2$ evolution of the hybrid-baryon electrocouplings is expected considering the different color-multiplet assignments for the quark-core in a conventional baryon compared to a hybrid baryon~\cite{E12-16-010}.

Finally, many hyperon spectroscopy measurements are expected from the GlueX and CLAS12 experiments, where the associated production of kaons allows one to study baryons with net strangeness including the $\Xi$ and $\Omega$~\cite{GlueX:2014hxq,E12-12-008}.  However, this program will be expanded by proposal to perform hyperon spectroscopy with a neutral kaon beam in Hall D, which was recently approved by the PAC~\cite{KLF:2020gai}.  The $K_L$ Facility (KLF) will produce a secondary beam in Hall D with a flux of $\sim10^4~K_L/s$ and utilize both hydrogen and deuterium targets inside the large-acceptance GlueX experimental setup.  Differential cross sections and hyperon recoil polarizations over a broad range of kinematics will provide significant new constrains on the partial wave analyses to search for and determine the pole positions of strange $\Lambda, \Sigma, \Xi,$ and $\Omega$ hyperon resonances, where many states are predicted by quark models and lattice QCD, which have not yet been observed.  %\textit{Include a figure here with pseudodata PWA and Cascade limits?}

\subsection{Charmonium and future opportunities at higher energy} 
\label{sec:CharmPentaquark}

The discovery of multi-quark candidates, the $XYZP$ states, observed mainly in the charmonium spectrum have revolutionized the field of hadron spectroscopy. These candidates have been observed in reactions involving complicated production and/or decay processes and some signals may be difficult to interpret due to kinematic effects which could mimic a resonance signal~\cite{Esposito:2016noz,Guo:2017jvc,Olsen:2017bmm,Brambilla:2019esw}.  Therefore, observation of these states in photo- or electro-production is needed to confirm their   existence and obtain more information on their structure.

Following the observation of the narrow pentaquark candidates $P_c^+(4312)$, $P_c^+(4440)$, and $P_c^+(4457)$ by LHCb in the $J/\psi p$ channel of the $\Lambda^0_b \rightarrow J/\psi pK^-$ decay \cite{LHCb:2015yax,LHCb:2019kea}, it was proposed to search for these states in $\gamma p \to J/\psi p$ where it can be produced directly in a much simpler $2 \to 2$ body kinematics~\cite{Wang:2015jsa,Kubarovsky:2015aaa,Karliner:2015voa,HillerBlin:2016odx}.

\begin{figure}[ht]
\begin{center}
    \includegraphics[width=0.4\textwidth]{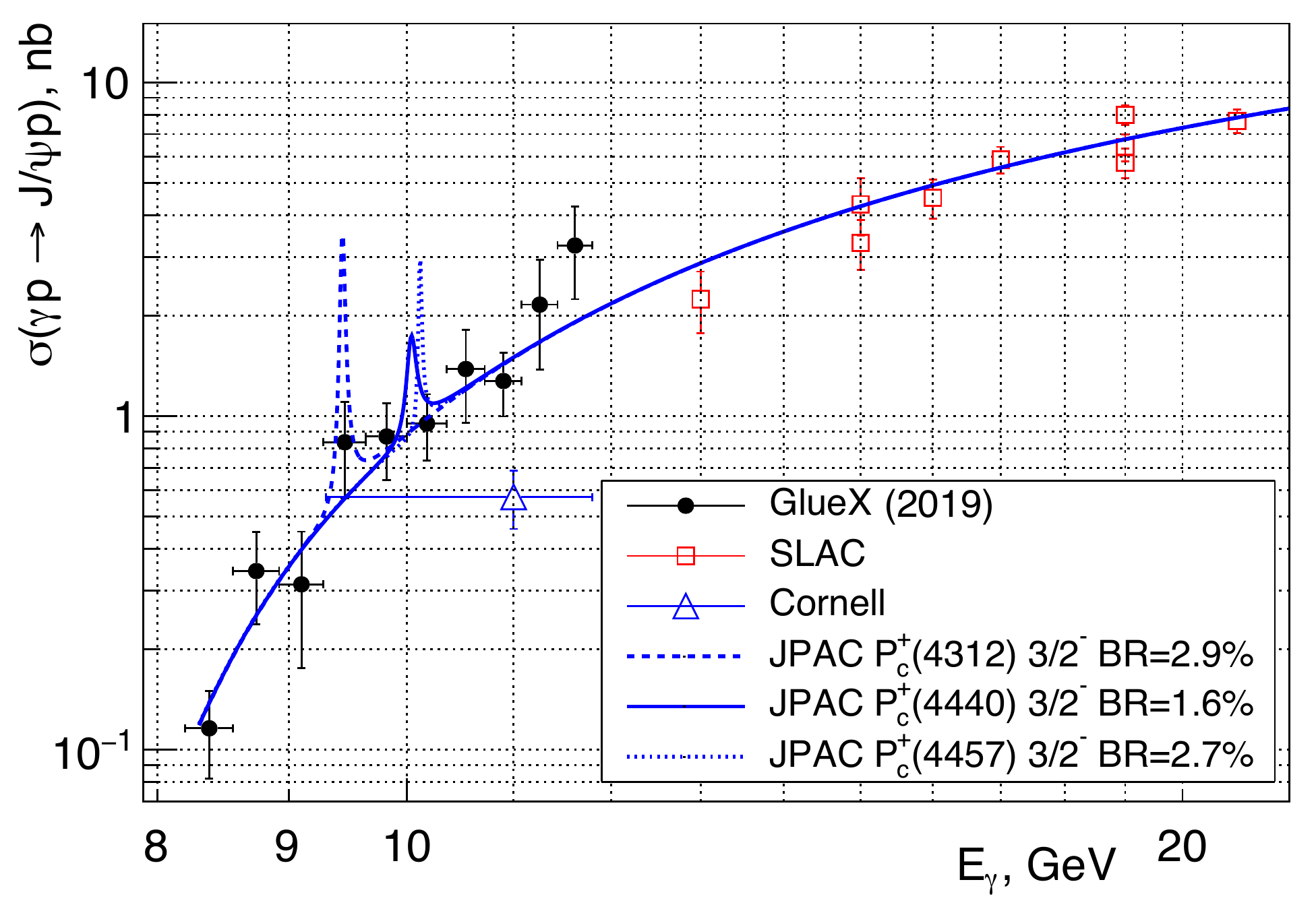}
    \includegraphics[width=0.41\textwidth]{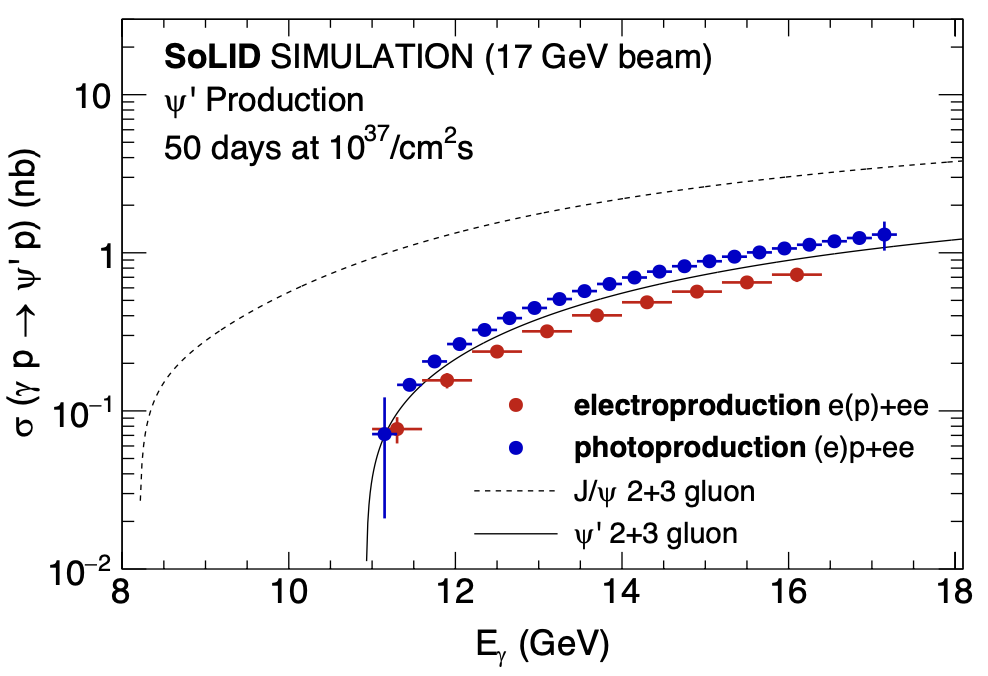}
    \caption{(left) GlueX results for the $J/\psi $ total cross section vs beam energy, compared to the JPAC model~\cite{HillerBlin:2016odx} with hypothetical branching ratios provided in the legend for $P_c^+$ with $J^P=3/2^-$ as described in Ref.~\cite{Ali:2019lzf}.  (right) Projections for SOLID $\psi(2s)$ total cross section vs beam energy for CEBAF upgrade with $E_e = 17$~GeV.
}
\label{fig:ccbar-xsec}
\end{center}
\end{figure}

The first measurements of this process at JLab were performed by GlueX~\cite{Ali:2019lzf} and are shown in Fig.~\ref{fig:ccbar-xsec}, with curves depicting the strength of hypothetical $P_c$ signals.  No structures are observed in the measured cross section, however model-dependent upper limits are set on the branching ratio of the possible $P_C \rightarrow J/\psi p$ decays.  Preliminary results from the $J\psi-007$ experiment in Hall C also observe no $P_c$ signal and will set more restrictive limits on the branching ratio~\cite{ghpJpsi007}.

At higher energies, (quasi-real) photoproduction is especially appealing since many of the $XYZP$ states could be produced directly and observed decaying to relatively simple final states, eliminating some of the kinematical effects. 
Furthermore, one can use the polarization of beam and target to achieve a precise separation of the various production mechanisms, which is not possible, for example, at hadron colliders.
Another advantage is that one can scan different center-of-mass energies by detecting the scattered electron at different angles, while keeping the beam at the nominal energy. This cannot be done by the existing $e^+e^-$ $\tau$-charm factories, where one has to tune carefully the beam energy to do so.

Three candidates stand out in particular: the   $X(3872)$, $Z_c(3900)$ and the $Y(4260)$.  The $X$ state is by far the best known. Its most unusual feature is the strength of isospin violation observed in decays,   $B(X\to J\psi\,\omega)/B(X\to J\psi\,\pi^+\pi^-) = 1.1\pm 0.4$~\cite{Zyla:2020zbs}, which is impossible for ordinary charmonium. 
Furthermore, the  mass of the $X(3872)$ is within a fraction of an MeV from the $\overline{D} {}^0 D^{0*}$ threshold, making it a good candidate for a threshold bound state.   Since the $X$ has sizeable branching fractions to $J\psi \,\rho$ and $J\psi \,\omega$, peripheral photoproduction involving light vector meson exchange can result in  sizable yields. 
The charged $Z_c(3900)^+$ is observed as a resonance in  $J\psi \, \pi^+$,  making it a good candidate for a four-quark state. Finally, there is an overpopulation of hidden-charm vector resonances. Three ordinary $\psi$ states appear in the inclusive $R_D$ measurements, leaving no room for other vectors like the $Y(4260)$. The latter can be produced diffractively.

The photoproduction cross sections for these states have recently  been estimated to be of the order of a nanobarn for photon energies $E_\gamma \sim 20$--$25$~GeV~\cite{Albaladejo:2020tzt}. 
The yields have been computed using a hypothetical detector setup based on the existing GlueX apparatus at Jefferson Lab~\cite{Adhikari:2020cvz}. Specifically, for a luminosity of $\sim 500\,\text{pb}^{-1}/\text{year}$ and even with a conservative assumption about  efficiency, one expects hundreds of events per year of data taking.
While diffractive production of $Y$ states benefits from higher energies, energies $E_\gamma \sim 20$--$25$~GeV are much more efficient in producing $X$ and $Z$ states. Simulations from the SoLID detector proposal, utilizing a 17 GeV $e-$ beam show excellent precision in determining the photo- and electro-production cross sections of the $\psi'$, as shown in Fig.~\ref{fig:ccbar-xsec} (right).

A photoproduction facility with a 24 GeV CEBAF will provide an opportunity to study these exotic states in exclusive reactions that are complementary to the ones where they have been observed so far.  A spectroscopy program at the forthcoming Electron-Ion Collider is presently under consideration~\cite{AbdulKhalek:2021gbh}. However, it is clear that a machine able to work at lower energies and with higher luminosity would be much more efficient in studying many of the $XYZP$ states.  Such a facility could provide much needed insights into the nature of these intriguing resonances.

%% file: Section4.tex
\section{QCD and Nuclei}

It has been nearly four decades since the European Muon Collaboration (EMC) published an astonishing finding on how the nucleon PDFs are strongly modified in the medium of the iron nucleus~\cite{EuropeanMuon:1983wih}. Many experiments have been done since then which both confirm this ``EMC Effect'' and have extended it to other nuclei, establishing systematic dependencies on nuclear size and density. However, although some recent studies suggest a connection to short-range correlations (SRCs) in nuclei, a full understanding of this phenomenon is still lacking.

Indeed, there are several ways in which QCD manifests itself in complex nuclei. CEBAF has contributed to this area of study, and will continue to provide new experimental data to point the way to a complete comprehension. Electron scattering gives access to a range of unique aspects of nuclear structure, providing important data relevant to nucleon modification in the nuclear medium, and quark/hadron interactions in cold QCD matter. Precision measurements of nuclear elastic, quasielastic, and inelastic scattering, especially those associated with the high-momentum part of the nucleon distributions, provide critical nuclear structure information needed in a range of other areas of nuclear and high-energy physics. Such data are needed as inputs to measurements of neutrino-nucleus scattering, nuclear astrophysics, lepton-nucleus scattering, and heavy-ion collisions, as well as providing important constraints relevant to the modeling of neutron stars. 

Studies of the partonic structure of nuclei provide insights into the impact of the dense nuclear medium on the structure of protons and neutrons and will allow, for the first time, imaging of the nuclear gluon distribution. In addition, measurements at higher energy allow for studies of hadron formation over a wide range of kinematics, as well as detailed studies of quark and hadron interaction with cold, dense nuclear matter, including color transparency studies which attempt to isolate the interaction of small-sized ``pre-hadronic" quark configurations.

The following sections summarize recent insight and highlights key future programs of measurements in nuclei, examining first measurements which probe nuclear structure at extremes of nucleon momentum, studies of the impact of the dense nuclear environment on the structure of the nucleon, and finally the use of the nucleus to study the interaction of quarks and their formation of hadrons in cold nuclear matter.

\subsection{Nuclear structure at the extremes}

Low-energy probes of nuclear structure are limited in their ability to isolate the high-momentum components of the nuclear momentum distribution. Measurements made as part of the 6 GeV program demonstrated the ability to probe these distributions with minimal corrections from final-state interactions, which limited electron scattering measurements at lower energy~\cite{Arrington:2011xs}. It also demonstrated the importance of understanding the isospin structure of SRCs which generate the high-momentum part of the momentum distribution~\cite{Arrington:2011xs,Hen:2016kwk,Fomin:2017ydn}.

At 12 GeV, our improved understanding of final-state interactions (FSI) and the expanded kinematic range available will allow for better constraints on the nucleon momentum distributions at high momentum, as well as providing reference data needed to further constrain meson-exchange and FSI corrections. In addition, targeted measurements will provide important inputs for nuclear structure relevant to neutrino oscillation experiments and nuclear astrophysics.

\subsubsection{Deuteron (and few-body nuclei) measurements at high momentum}

In the plane-wave impulse approximation, quasielastic A(e,e'p) measurements can be used to probe the spectral function of nuclei. Accessing the high-momentum part of the spectral function has proved extremely challenging, due to large FSIs that shift strength from low reconstructed momenta to large values~\cite{Sargsian:2002wc}. However, JLab measurements on the deuteron~\cite{CLAS:2007tee, HallA:2011gjn} have been used to constrain FSI  models~\cite{Sargsian:2009hf} over a wide range of kinematics, validating the models and demonstrating that for modest values of $\theta_r$ -- the recoil angle of the spectator nucleon relative to the incoming photon -- the corrections become small~\cite{CLAS:2007tee,HallA:2011gjn,Arrington:2011xs}. A new set of measurements was proposed to push deuteron measurements to extremely large nucleon momenta, exceeding 1~GeV/c~\cite{E12-10-003}. The low-momentum piece of this experiment was completed and results from the initial data set are available~\cite{HallC:2020kdm}. They show that above 700 MeV/c missing momentum, the data deviate from all existing calculations. Future data taking will emphasize improving the statistics and high-momentum coverage to provide more precise constraints for detailed calculations to have a more reliable probe of the high-momentum part of the Nucleon-Nucleon (NN) potential. Other experiments will extend this approach to study the spectral function of light nuclei~\cite{E12-20-005}
while CLAS12 measurements of D$(e,e'p)$ at 12 GeV will allow for additional studies of FSI, significantly extending the $Q^2$ and missing momentum coverage over the full range of recoil angles.

\subsubsection{Short range NN correlations}

\begin{figure}[htb]
\centerline{
\includegraphics[height=0.56\textwidth, angle=90, trim={20mm 22mm 25mm 40mm}, clip]{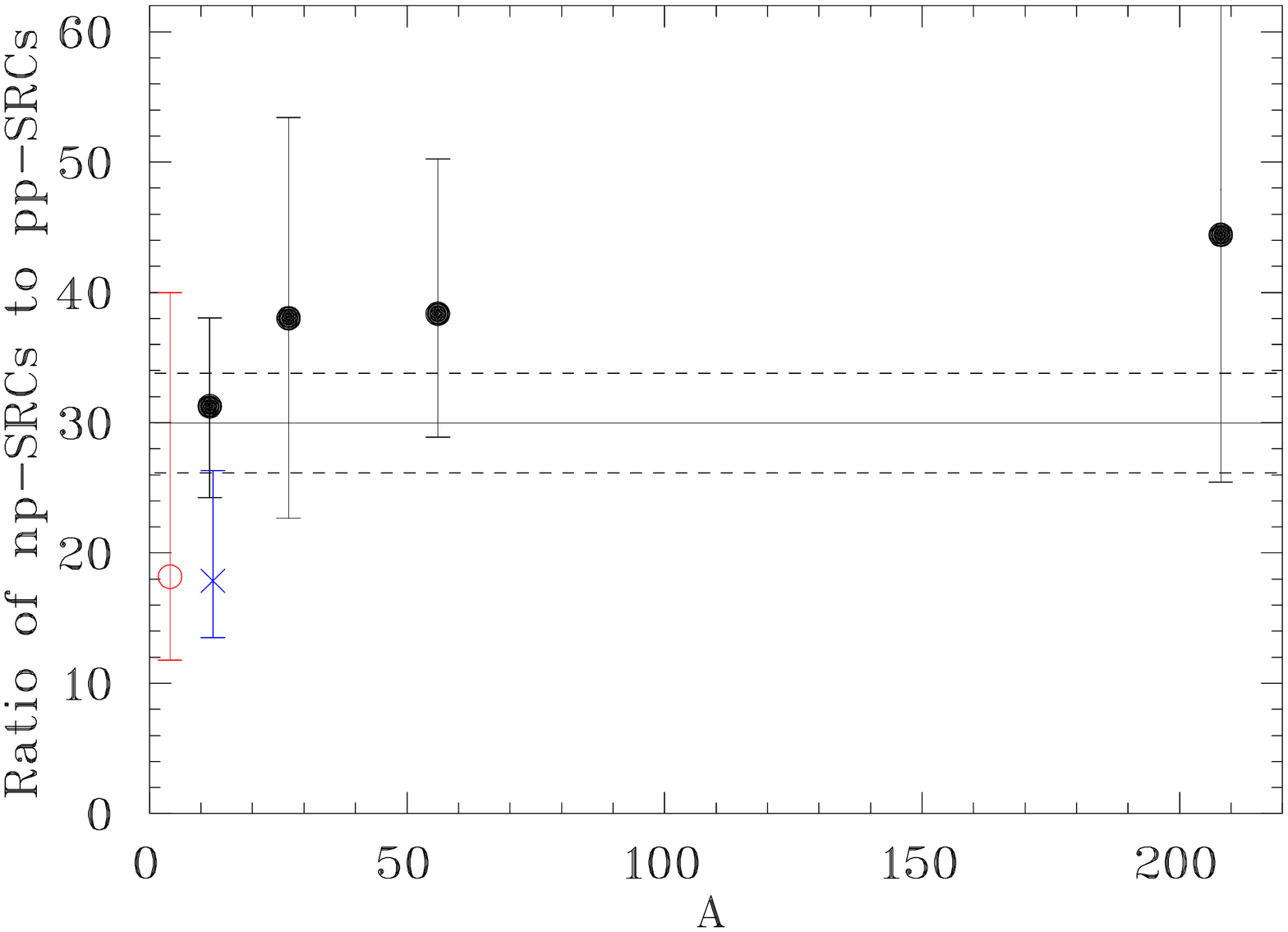}
\includegraphics[width=0.43\textwidth, trim={0mm 0mm 0mm 0mm}, clip]{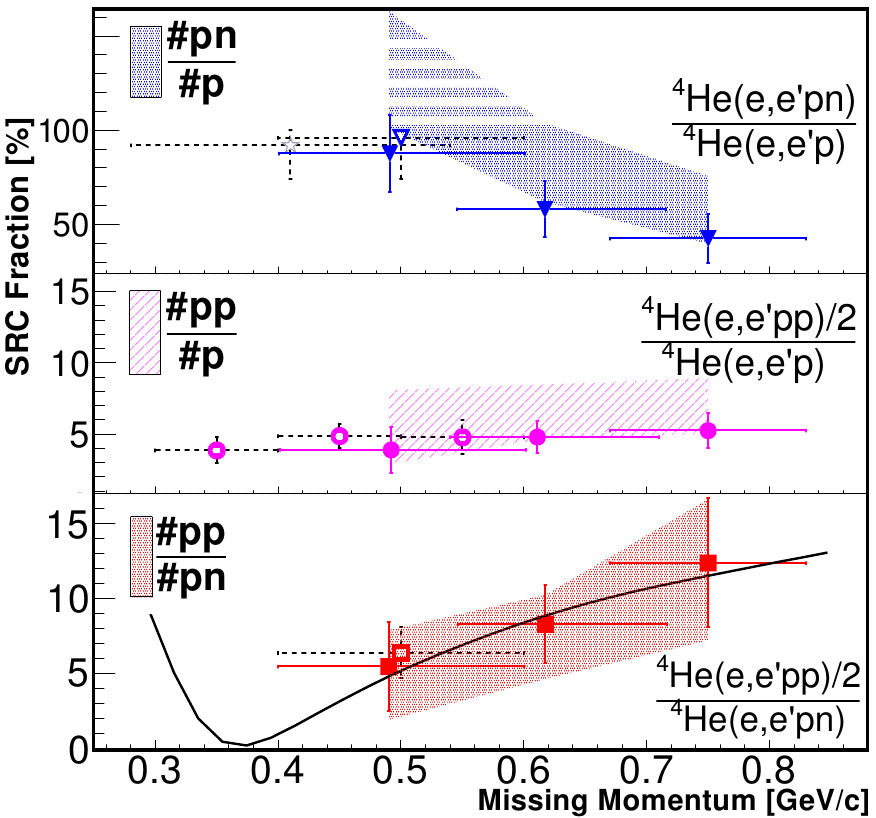}
}
\caption{(Left) Ratio of np-SRC to pp-SRCs from triple-coincidence A(e,e'pN) measurements: Blue `x'~\cite{Subedi:2008zz}, red open circle~\cite{LabHallA:2014wqo}, black solid circles~\cite{CLAS:2018xvc}; the horizontal line is the average of all measurements (np/pp = 30.0$\pm$3.8), although the value from~\cite{Subedi:2008zz} would be 50-100\% larger using the updated FSI of Ref.~\cite{CLAS:2018xvc}. (Right) Fraction of high missing momentum (e,e'p) events with a high-momentum neutron or proton spectator, and the ratio of pp to pn SRC; figure taken from Ref.~\cite{LabHallA:2014wqo}.}
\label{fig:SRC-isospin}
\end{figure}

While mapping out the high missing momentum ($P_m$) part of the spectral function presents challenges, it is still possible to study SRCs by selecting kinematics where the reaction is dominated by SRC contributions. Nucleons with momenta well above the Fermi momentum are associated with SRCs that are generated by the hard, short-range components of the NN interaction~\cite{Frankfurt:1981mk, Frankfurt:1988nt, Sargsian:2002wc, Arrington:2011xs, Fomin:2017ydn, Arrington:2022sov}. Because they are generated by two-body interactions, they have a universal structure that comes from the NN interaction, and scattering measurements in kinematics dominated by SRCs allow for studies of the nature and size of SRCs in nuclei. Inclusive scattering at modest $Q^2$ and $x>1.4$, where scattering from low-momentum nucleons is kinematically forbidden, provides sensitivity to the relative contribution of SRCs as a function of the mass number  $A$ via measurements of the inclusive $A/^2$H cross section ratios. During 6 GeV running, experiments confirmed the initial observation of SRCs~\cite{Frankfurt:1993sp} and mapped out the $A$ dependence of SRCs in light and heavy nuclei~\cite{CLAS:2003eih, Fomin:2011ng}. These data demonstrated that the contribution is sensitive to details of the nuclear structure~\cite{Seely:2009gt, Arrington:2021vuu} rather than the previously assumed average nuclear density~\cite{Gomez:1993ri}. In addition, they showed a clear correlation between the contribution of SRCs~\cite{Fomin:2011ng} and the size of the EMC effect~\cite{Seely:2009gt}, discussed further in Sec.~\ref{sec:4:emc}. Measurements of two-nucleon knockout, where both nucleons from the SRC are observed in the final state, showed dominance of np-SRC as well as a dependence of the np/pp SRC ratio as a function of the struck nucleon's momentum (shown in Figure~\ref{fig:SRC-isospin}. The dominance of np-SRCs was confirmed in additional nuclei using the  $A(e,e'p)$~\cite{CLAS:2018yvt} reaction and later through inclusive measurements taking advantage of the \textit{target} isospin structure in measurements of the $^{48}$Ca/$^{40}$Ca cross section ratio~\cite{JeffersonLabHallA:2020wrr}. Finally, measurements at $x>2$ tried to establish the presence of three-nucleon SRCs (3N-SRCs), but low-$Q^2$ measurements did not observe 3N-SRC dominance~\cite{HallA:2017ivm}, while higher-$Q^2$ data was consistent with 3N-SRC dominance but had extremely limited statistics~\cite{Fomin:2011ng}. 

An upcoming measurement~\cite{E12-06-105} will measure SRCs for additional few-body nuclei (all stable nuclei up to $^{12}$C), as well as for medium-to-heavy nuclei over a range in $N/Z$. This will allow more detailed examinations of how the EMC effect is impacted by nuclear structure in well-understood nuclei, as well as allowing for separation of mass- and isospin-dependent effects.  It will also provide the first meaningful significant test of the predicted dominance of 3N-SRCs by taking high-statistics A/$^3$He ratio data at $x>2$ and $Q^2 \approx 3$~GeV$^2$. If the presence of 3N-SRCs is established, further dedicated measurements could map out their A dependence and isospin structure through expanded A/$^3$He measurements and direct $^3$H-$^3$He comparisons~\cite{LOI12-21-001}.

\subsection{Impact on neutron stars, neutrino scattering} \label{sec:PREX}

The studies of SRCs completed so far, as well as those planned for the future, have provided significant information on details of nuclear structure that are difficult to cleanly probe in low-energy reactions, and hold the promise to provide more information on the short-range structure of the NN interaction. This information has impact on a range of other physics: neutrino-nucleus scattering measurements, including those needed to study neutrino oscillations, need reliable nuclear structure inputs including a quantitative understanding of SRCs~\cite{Kulagin:2007ju, Niewczas:2015iea, VanCuyck:2016fab}. An understanding of the high-momentum components of the nuclear wavefunction, including the isospin structure, is also important for understanding the quark distributions of nuclei~\cite{Kulagin:2004ie} and the structure of neutron stars~\cite{Frankfurt:2008zv, Higinbotham:2009zz}. In addition to the impact of SRC on these topics, there have also been dedicated measurements aimed at providing input relevant for understanding aspects of nuclear structure directly relevant to neutron stars and neutrino scattering measurements.

The PREX-II collaboration recently published a new measurement~\cite{PREX:2021umo} of the ``neutron skin thickness'' of $^{208}$Pb using PVES. Parity violation is an especially clean way to measure nuclear neutron densities~\cite{Horowitz:1999fk,Thiel:2019tkm}, owing to the fact that the $Z^0$ couples much more strongly to neutrons than to protons. This makes it particularly straightforward to relate such measurements to the neutron matter equation of state.

\begin{figure}
\centerline{
\includegraphics[width=0.85\textwidth]{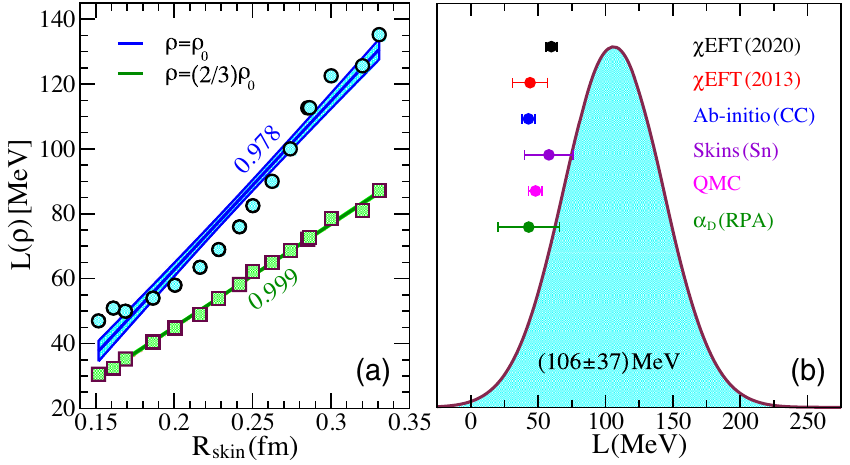}}
\caption{(a) Slope of the symmetry energy at nuclear
saturation density $\rho_0$ (blue upper line) and at $(2/3)\rho_0$ (green
lower line) as a function of the neutron skin thickness in $^{208}$Pb. The correlation coefficients are indicated. (b) Gaussian probability distribution for the slope of the symmetry energy $L=L(\rho_0)$ inferred by the PREX result. The six error bars are constraints on $L$ obtained by using different theoretical approaches.
Reprinted Figure~1 with permission from Brendan T. Reed, F.J. Fattoyev, C.J. Horowitz, and J. Piekarewicz, Physical Review Letters, 126, 172503, 2021.~\cite{Reed:2021nqk} Copyright 2021 by the American Physical Society.
\label{fig:PREXII}}
\end{figure}

The PREX-II measurement required the full integration of the CEBAF accelerator datastream into the data pipeline so that a small scattering asymmetry could be measured with high precision. The final result of the parity-violating asymmetry of $e-^{208}$Pb elastic scattering, $A_{\rm PV}=[550\pm16~({\rm stat})\pm8~({\rm syst})] \times 10^{-9}$, yields a neutron skin thickness $R_n-R_p=0.283\pm0.071$~fm. Figure~\ref{fig:PREXII} shows an interpretation~\cite{Reed:2021nqk} of the PREX-II result in terms of the ``symmetry pressure'' $L$ of pure neutron matter. One finds that $L$ is significantly larger than predictions from most available calculations, suggesting important modifications to neutron matter and the equation of state for neutron stars~\cite{Horowitz:2001ya}.
A similar measurement aiming at an extraction of the neutron skin thickness of $^{48}$Ca, called CREX, has also been carried out. The complete analysis %, including interpretation of the neutron skin thickness, 
is currently in progress. For CREX, the extracted neutron skin can be directly compared to microscopic calculations~\cite{Hagen:2015yea}.

The extraction of neutrino mixing parameters from neutrino oscillation experiments relies on the reconstruction of the incident neutrino energy and knowledge of the neutrino-nucleus interaction cross-section for various nuclei and incident neutrino energies. The energy reconstruction of the incident neutrino is often done using the yield and kinematics of particles produced from neutrino interactions in nuclei. However, none of these energy reconstruction techniques have been tested experimentally using beams of known energy. Understanding these issues is critical to the ongoing global effort to study neutrino oscillations with secondary neutrino beams and nuclear targets.

One of the early experiments of the 12 GeV eras was focused on constraining scattering from $^{40}$Ar through measurements of inclusive quasielastic scattering and single proton knockout~\cite{Dai:2018gch,JeffersonLabHallA:2020rcp}. These data will allow for tests of $\nu-^{40}$Ar scattering simulations needed for the DUNE experiment.

Because neutrinos and electrons are both leptons, they interact with nuclei in
similar ways. The ``Electrons for Neutrinos" experiment proposes to measure electron scattering from a variety of targets at a range of beam energies in CLAS12 in order to test neutrino event selection and energy reconstruction techniques and to benchmark neutrino event generators. Recently the collaboration has studied these aspects through a data-mining project using previous CLAS data from the 6 GeV era of CEBAF~\cite{e4nu}.

\subsection{Nucleon modification/Nucleons under extreme conditions}\label{sec:4:emc}

The SRCs discussed in the previous sections represent short-distance, high-density configurations of nucleon in nuclei, as well as extremely high-energy nucleons.  It is natural to wonder whether nucleons with significant overlap and with extremely large virtuality might have a modified internal structure which could contribute to the EMC effect -- the modification of nuclear PDFs relative to the sum of the individual nucleon PDFs and, beyond this, whether measurements can be made that might be directly sensitive to the internal structure of these short-distance, high-momentum nucleons.

\subsubsection{EMC effect: Nuclear parton distributions}

%\paragraph{Overview} 
\label{sec:ALERTmeasurements}

Almost 40 years after the observation by the EMC collaboration~\cite{EuropeanMuon:1983wih} that the nuclear PDF has significant modifications from the sum of proton and neutron PDFs, this observation still provides the cleanest indication that the nucleus cannot be described as a simple collection of bound, moving nucleons. Most calculations suggest that quark contributions beyond those of moving nucleons are required to explain the nuclear PDFs~\cite{Geesaman:1995yd, Arrington:2011xs, Malace:2014uea}, and there is a growing consensus that modification of the internal structure of the nucleon is required to fully explain the effect~\cite{Miller:2001tg, Sargsian:2002wc, Hen:2016kwk}. However, the most precise measurements of the EMC effect focused on medium-to-heavy nuclei~\cite{Gomez:1993ri}, and many models were able to reproduce the universal $x$ dependence and the weak $A$ dependence observed in medium-to-heavy nuclei, making it difficult to evaluate the underlying physics. The 6 GeV program at JLab added precise measurements in light nuclei~\cite{Seely:2009gt}, which showed an anomalous behavior in $^9$Be, demonstrating that the EMC effect is sensitive to nuclear structure details and does not simply scale with average nuclear density or mass~\cite{Arrington:2012ax}. The behavior in light nuclei was nearly identical to that observed in measurements of SRCs in few-body nuclei~\cite{Fomin:2011ng}, demonstrating a connection between these two phenomena~\cite{Arrington:2012ax, Hen:2016kwk} that is not yet understood~\cite{CLAS:2019vsb, Arrington:2019wky}. 
Extension of EMC effect measurements to cover additional few-body and medium-to-heavy nuclei combined with SRC measurements in the same nuclei~\cite{E12-06-105} will provide significantly better quantification of the EMC-SRC connection. It will also provide data for heavier nuclei with a range of $N/Z$ ratios, that can help separate the mass and flavor dependence of the EMC effect. Understanding the flavor dependence of the EMC effect would provide completely new inputs to test models of nuclear parton distributions. It will also help elucidate the EMC-SRC connection, as pictures where the EMC effect is caused by the presence of highly-virtual nucleons would be expected to have a flavor dependence arising from the isospin structure of SRCs, while models where they arise from a common origin, e.g. short-distance (highly-overlapping) nucleon pairs may not translate into a flavor dependence of the nuclear quark modification~\cite{Arrington:2012ax, Arrington:2019wky}.

After many decades of study it has become clear that in order to make significant further progress in understanding the origin of the EMC effect, it will require new experimental measurements that can be used to disentangle the various physical explanations. As discussed below, measurement of the spin and flavor dependence of the EMC effect would provide significant new information that could go a long way in solving the puzzles associated with the partonic structure of nuclei.

\begin{figure}[tbp]
\centerline{
\includegraphics[width=0.46\textwidth]{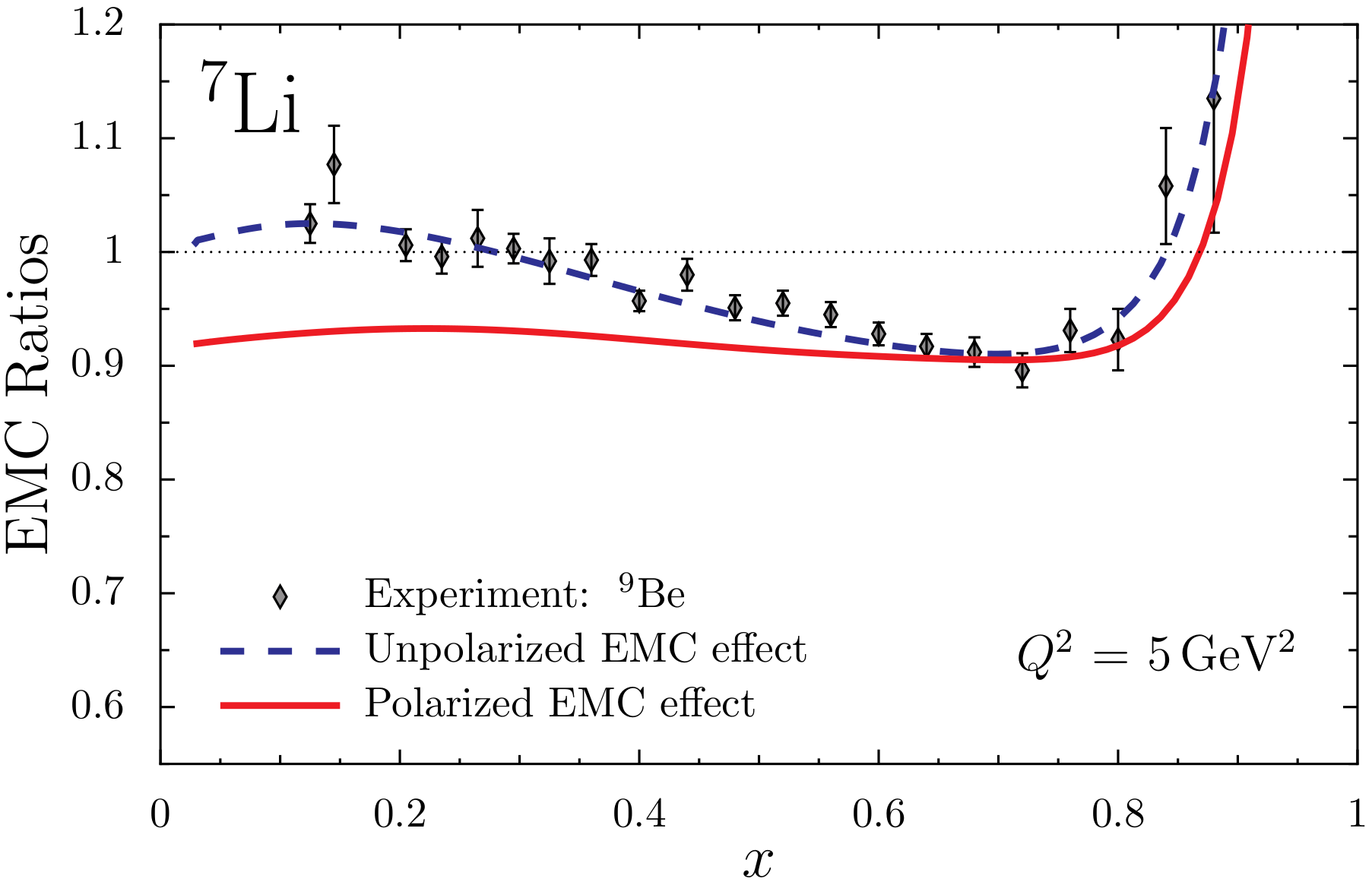} \hfill
\includegraphics[width=0.46\textwidth]{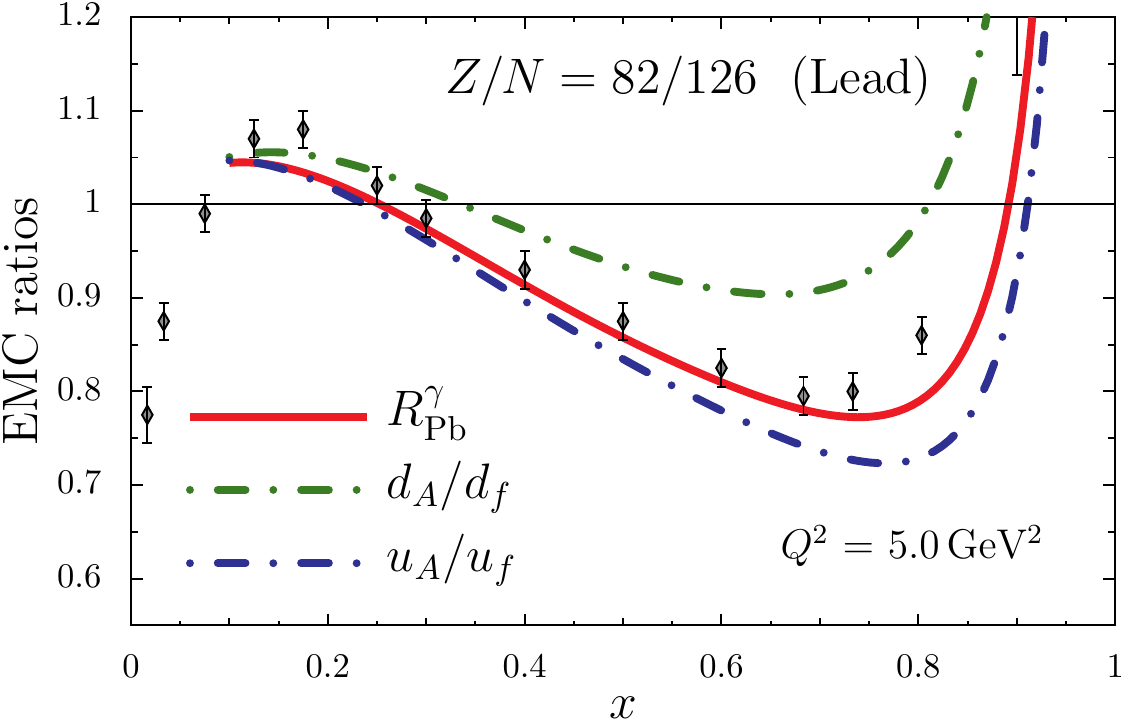}}
\caption{{\it Left panel:} Prediction for the polarized EMC effect in $^7$Li (solid line) compared to a calculation of the EMC effect (dashed line) in the same framework, figure adapted from Ref.~\cite{Cloet:2006bq}. The experimental data are from Ref.~\cite{Gomez:1993ri}. {\it Right panel:}   Predictions from Ref.~\cite{Cloet:2012td} for the flavor dependence of the EMC effect in nuclear matter with the same $Z/N$ ratio as lead. Figure adapted from Ref.~\cite{Cloet:2012td}.}
\label{fig:emc_effects}
\end{figure}

%Since t
The EMC effect and the proton spin puzzle were both discovered almost 40 years ago, and numerous experiments have been performed in the intervening decades. It is perhaps surprising that there is still no experimental data on the spin structure functions of nuclei beyond the deuteron and $^3$He. Nevertheless, there are several theoretical calculations of the polarized EMC effect in light nuclei and nuclear matter~\cite{Cloet:2005rt, Cloet:2006bq, Smith:2005ra, Tronchin:2018mvu} that predict modifications in the spin structure functions of nuclei at least as large as the usual EMC effect. The polarized EMC effect is defined by $\Delta R = g_A(x)/[P_p\,g_{1}^p(x) + P_n\,g_{1}^n(x)]$ where $g_A(x)$ is a nuclear spin structure function, $g_{1}^p(x)$ and $g_{1}^n(x)$ are %in Section 2.2 (and common literature), p and n are superscripts
the familiar nucleon results, and $P_{p/n}$ is the effective polarization of the nucleus carried by the protons/neutrons which can be obtained from detailed nuclear structure calculations. A prediction for the polarized EMC effect in $^7$Li is illustrated in the left panel of Fig.~\ref{fig:emc_effects}, and a JLab 12\,GeV experiment is planned to measure this effect~\cite{E12-14-001}. The polarized EMC effect is particularly interesting because it can distinguish between mean-field models of the EMC effect where all nucleons are modified by the nuclear medium, with those explanations based on SRCs where only those nucleons involved in these correlations are modified. The mean-field models predict a large polarized EMC effect, whereas approaches based on SRCs lead to a very small effect %EMC effect in the spin structure functions
because the nucleons in SRCs carry only a few percent of the nuclear polarization~\cite{Pudliner:1995wk}. Similarly, the nature of the SRCs is such that they depolarize the nucleons involved~\cite{Thomas:2018kcx}.

It has been known for some time that there are strong isovector forces in nuclei, which are captured for example in the asymmetry term in the semi-empirical mass formula and are largely responsible for the symmetry energy. In mean-field models of nuclear structure, such as the Quark Meson Coupling (QMC) model~\cite{Guichon:2004xg,Guichon:2006er,Tronchin:2018mvu} and those based on the Nambu--Jona-Lasinio model~\cite{Bentz:2001vc,Cloet:2005rt,Cloet:2006bq}, these isovector forces couple to the quarks in the bound nucleons, which means that for $N > Z$ nuclei, the $d$ quarks feel additional repulsion and $u$ quarks additional attraction. The opposite is the case for $N < Z$ nuclei. The sign and magnitude of these forces is constrained by the empirical symmetry energy. Because $u$ and $d$ quarks are modified differently in the nuclear medium, this produces an isovector or flavor dependent EMC effect. Predictions from Ref.~\cite{Cloet:2012td} for the size of this effect in nuclear matter with a $Z/N$ ratio similar to a lead nucleus are illustrated in the right panel of Fig.~\ref{fig:emc_effects}, where it is clear that the $u$ quarks have a much larger EMC effect than the $d$ quarks. Further suggestion for an isovector EMC effect is that calculation for iron can explain more than one-sigma~\cite{Cloet:2009qs,Bentz:2009yy} of the NuTeV anomaly~\cite{Zeller:2001hh}. In addition, evidence for a flavor dependence in the EMC effect could distinguish between mean-field explanations and those based on SRCs. Experiments have demonstrated that SRCs are predominantly $np$ pairs, which occur approximately 20 times more often that $nn$ and $pp$ pairs~\cite{Hen:2016kwk}. As such, SRCs are predominantly isoscalar and cannot produce an isovector EMC effect.

A flavor dependence in the EMC effect should manifest in a number of experiments, e.g., by contrasting structure function measurement in $^{40}$Ca and $^{48}$Ca. A novel method to measure the isovector EMC effect is via PVDIS to measure the $\gamma$-$Z$ interference structure function $F_2^{\gamma Z}(x)$ and contrast this with the usual DIS structure function to perform a flavor separation of the $u$ and $d$ quark PDFs in the same nuclear target. A proposal to perform this experiment on $^{48}$Ca is has been considered for the JLab 12\,GeV program. Interesting opportunities also exist in the comparison of SIDIS on $^3$H and $^3$He with either $\pi^+$ or $\pi^-$ detected in the final state.

\subsubsection{Nucleon structure from tagged DIS}

The discussion above around the EMC effect and its varying aspects concerns the partonic structure of the entire nucleus, and any interpretation in terms of the structure of the individual nucleons inside the nucleus is necessarily model dependent. Nevertheless, understanding how a nucleon responds to the nuclear environment is a fundamental question in nuclear physics and one pathway to this information is provided via tagged processes. These processes involve the measurement/tagging of a recoiling particle that can be used to isolate and identify a subcomponent of the original nuclear target. For example, in a experiment with a $^4$He target, tagging a recoiling $^3$He nucleus is an indication that the probe interacted with a bound neutron in the $^4$He target. Such processes were first pioneered at JLab by the BONuS Collaboration to measure the neutron $F_2(x)$ structure function by tagging the proton from a deuteron target~\cite{CLAS:2011qvj,CLAS:2014jvt}.
Recently, the BONuS12 collaboration~\cite{BONUS12,CLAS:2018unv} completed a dramatically enhanced program of tagged measurements from the deuteron to extract the structure of the free neutron by tagging low-momentum spectator nucleons.  

Two related experiments are approved that will make tagged measurements with high-momentum spectators to explore the virtuality dependence of the structure functions of protons and neutrons bound in the deuteron. These experiments will perform tagged DIS using a high-resolution spectrometer to detect the scattered electron and new large acceptance detectors (LAD and BAND) to measure recoiling protons and neutrons~\cite{Segarra:2020txy}. A similar measurement~\cite{E12-11-002} will focus on the virtuality dependence of quasielasatic e-N scattering to study the impact of the virtuality to the effective in-medium form factors. Finally, the TDIS program~\cite{C12-15-006A,E12-15-006} will make high-statistics measurements of free neutron structure using proton tagging on the deuteron~\cite{C12-15-006B}, and will also measure pion and kaon structure using spectator tagging of the proton to isolate pion structure in p(e,e'$p_s$)X and d(e,e'$p_s$)X scattering and the kaon using Lambda tagging, as discussed in Sec.~\ref{sec:section2-FF-PDF}.

Finally, ALERT~\cite{Armstrong:2017zcm,Armstrong:2017zqr,Armstrong:2017wfw} in Hall B will complement the CLAS12 spectrometer apparatus, enabling access to the low-energy nuclear remnants from electron scattering off the deuteron and $^4$He. ALERT plans a comprehensive program using tagged DIS and incoherent DVCS on these targets and can measure/tag final state protons, $^3$H, and $^3$He. Therefore, ALERT provides access to the structure of the bound protons and neutrons in the deuteron and $^4$He and will deliver data on the 3$D$ structure of these nuclei.

\subsubsection{Super-fast quarks and hidden color}

Exotic configurations have been suggested as a possible contribution to the structure in nuclei.  Overlapping baryonic states could contain exotic configurations such as 6-quark bags~\cite{Bickerstaff:1984gut}, hidden color states~\cite{Brodsky:1995rn, Brodsky:2004tq, Brodsky:2004zw}, or multi-diquark contributions~\cite{West:2020rlk}. Such states allow for more direct sharing of momentum between quarks in different nucleons, yielding predictions that they could have a significant contribution of super-fast quarks, quarks with momentum fraction well above what is easily achievable through simple smearing of the nucleon PDFs~\cite{Sargsian:2002wc,Freese:2014zda}. 

A measurement is planned~\cite{E12-06-105} that will examine DIS from the deuteron for $x>1$. Just as these kinematics isolate contributions from SRCs in quasi-elastic scattering, they can also isolate SRCs in DIS at sufficiently large $Q^2$. At this point, we are probing the PDFs of the deuteron at $x>1$, and the types of models described above predict a significant enhancement of the super-fast quark distribution compared to a simple convolution model.  In addition to looking for these exotic configurations, this also allows the possibility to test models in which the large virtualities of the high-momentum nucleons cause the EMC effect. In this case, the kinematics isolate high-momentum nucleons and probe the PDFs of the moving proton and neutron at large $x$, where several models predict a significant suppression of the PDFs~\cite{Melnitchouk:1993nk, Melnitchouk:1996vp, Hen:2016kwk}. Thus, the super-fast quark distribution could be suppressed by these off-shell effects, or could be significantly enhanced by exotic configurations of overlapping nucleons, providing important information on the microscopic structure of SRCs and a new way to understand its connection to the EMC effect.

In the presence of such non-nucleonic configurations at short-distance scales, there would be a significant enhancement of SRC-like N-$\Delta$ or $\Delta$-$\Delta$-like pairs~\cite{Ji:1985ky}. Initial estimates  suggested that such measurements would be much more sensitive at 12~GeV or even higher energies. In addition to probing the distribution of super-fast quarks in nuclei, measurements of SRC-like N--$\Delta$ and $\Delta$-$\Delta$ configurations are expected to be significantly enhanced in the presence of hidden color configuration~\cite{Ji:1985ky}, and may also be sensitive to other exotic multi-baryonic contributions. Such measurements would significantly benefit from higher energies.

%===============================================================================
%===============================================================================
\subsubsection{Femtography for Nuclei} 

The 3D imaging of quarks and gluons in nuclei presents a tremendous opportunity to study aspects of QCD that cannot occur in nucleons and to explore phenomena that may expose the explicit role of QCD in nuclei, beyond the residual van der Waals type forces studied in traditional nuclear physics methods. Just as GPDs and TMDs provide a spatial and momentum tomography for the nucleon (see Sec. II), the same quantities provide analogous information for nuclear targets. However, nuclei provide a much richer environment with which to study quark-gluon dynamics than a single nucleon, in part, because nuclei not only provide stable targets with $J=1/2$ but also those with spin $J=0$ and $J \geq 1$. For example, the deuteron with $J=1$ has 9 time-reversal-even TMDs and 9 GPDs at leading twist, each revealing different aspects of the quark-gluon structure of the deuteron.

Perhaps the simplest nucleus to study experimentally is $^4$He, which is a tightly bound system of two protons and two neutrons having zero spin, and is therefore characterized by fewer observables than targets with spin. For example, $^4$He has one electromagnetic form factor and one DIS structure function in the Bjorken limit. This simplified structure and tight binding make $^4$He an ideal nucleus with which to study QCD effects in nuclei via nuclear femtography. DVCS on $^4$He was studied at JLab, with first results for coherent scattering presented in Ref.~\cite{CLAS:2017udk} and incoherent scattering results were reported in Ref.~\cite{CLAS:2018ddh}. Further studies at 12\,GeV with DVCS and other processes such as DVMP, should provide sufficient data to extract the underlying spin-independent quark GPD for $^4$He and perhaps offer insight on the gluons. Such studies stand to reveal several interesting aspects of nuclear structure, e.g., can the color singlet nucleons be identified in the spatial distributions and do they cause interference patterns, do these interference patterns for quarks and gluons overlap and how do these correlations change when viewed at different slices in $x$ and $b_T$ in impact parameter space. 

The TMDs of nuclei provide similar opportunities. At leading twist $^4$He has two TMDs, the familiar spin-independent naive time-reversal even TMD, $f_1(x,\boldsymbol{k}_T^2)$, and the Boer-Mulders function $h_1^\perp(x,\boldsymbol{k}_T^2)$ which is naive time-reversal odd and describes the distribution of transversely polarized quarks in an unpolarized or spin-zero target. Comparison of the $(x,\boldsymbol{k}_T^2)$ dependence of these functions in nuclei with those in the nucleon will provide significant new insights into quark-gluon dynamics in the nuclear medium. 

The deuteron also provides a unique system for nuclear femtography, because it is a finely-tuned system making it sensitive to small QCD effects. As a spin-one target, it possesses a tensor polarization in addition to the familiar vector polarization of a spin-half target such as the nucleon. This tensor polarization produces three additional time-reversal even and seven additional time-reversal odd TMDs compared to a spin-half target. Deuteron TMDs and GPDs could reveal new aspects of the nucleon-nucleon interaction and how this interaction has its origins in QCD~\cite{Ninomiya:2017ggn}. Interesting studies in the 3D imaging of nuclei are also provided via comparisons between $^3$He and $^3$H, such as clean flavor separation and the study of nuclear effects in the comparison between $^3$H and the proton, which can provide guidance on the extraction of neutron results from $^3$He and deuteron targets.

A robust program in the 3D imaging of quarks and gluons in nuclei at JLab 12\,GeV stands to provide unprecedented insight into nuclear structure and QCD effects in nuclei. These studies also complement the expected program at the EIC which will be focused at higher energies and smaller $x$, and similarly smaller skewness $\xi$ in GPDs and associated Compton form factors.

\subsubsection{Impact on other physics programs}

An understanding of the EMC effect, in particular answering new questions about its spin and flavor dependence, is important for a wide range of other measurements~\cite{Cloet:2019mql}. High-energy scattering measurements from nuclei, whether $\nu$-A scattering at FNAL, A-A at RHIC or the LHC, or e-A at a future EIC, requires a good understanding of the nuclear PDFs. Understanding the flavor dependence of the EMC effect, in particular for neutron rich nuclei~\cite{Arrington:2012ax, Arrington:2015wja, CLAS:2019vsb, Arrington:2019wky} can best be addressed with PVES at JLab~\cite{Cloet:2019mql,PVEMC}. Not only does this provide the best access to flavor dependence now, but such PVES experiments would benefit significantly from higher beam energies. This is also an important issue for $^3$He, which is used as an effective neutron target for spin studies. An understanding of the polarized EMC effect, as well as an understanding of the flavor-dependence in this light but proton-rich nucleus, will be important for precision spin studies using polarized $^3$He beams and targets.

\subsection{Quark/hadron propagation in the nuclear medium}

\subsubsection{Quark/hadron propagation}
The confinement principle of QCD dictates that color charge cannot be separated from the color-neutral hadrons that contain it. This statement, however, only pertains to equilibrium conditions. Color charge can be briefly liberated from a hadron through a hard scattering. In the simplest case, a valence quark carrying color charge can undergo a high-energy scattering process that propels the quark over long distances, closely accompanied by a spray of quarks and gluons of lower energies that ultimately evolve into new hadrons. This process, called hadronization or fragmentation, can take place on distance scales of 1-100 fm or more, and its characteristics at lower energies can be studied by observing it inside an atomic nucleus. The hadronization constituents interact with the nuclear medium, modifying hadron production in ways that reveal characteristics of this fundamental process. 

Questions that can be answered by such experimental studies include: how long does the struck quark propagate before becoming bound into a forming hadron? with what mechanisms do propagating quarks interact with the nuclear medium, and at which spatial scales do they interact? how much energy do they lose in the medium? how much medium-induced transverse momentum do they acquire? is a semi-classical description of the process adequate? In SIDIS, the picture of the initial state is particularly clear at the values of $x$ accessible at JLab. A valence quark absorbs all the energy and momentum from the interaction, which can be measured directly by the scattered electron. Neglecting small contributions from intrinsic transverse momentum and Fermi momentum, the energy and momentum of the struck quark are known; thus, the initial state of the interaction between the quark and the medium is well determined. This secondary "beam" of quarks is generated throughout the nucleus with an initial position probability determined by the well-known nuclear density distribution, allowing precise modeling of the process.

In this picture, the hadron containing the struck quark will interact with the medium with a total cross section dominated by inelastic processes at the energies relevant to JLab. In the kinematics where the hadron forms earliest, which is at high relative energy $z_h=E_h/\nu$, this cross section can be determined with moderate accuracy using sufficiently precise data and simple models~\cite{BROOKS2021136171}. The experimental signature of this region is a maximal suppression of hadron production in large nuclei relative to the proton or deuteron.

An important consideration in these studies is the multi-variable dependence of observables such as multiplicity ratios and transverse momentum broadening. The nominal number of possible variables in the A(e,e'h)X reaction is up to five, typically chosen to be four-momentum transfer $\mathrm{Q^2}$, energy transfer $\mathrm{\nu}$ or Bjorken-x  $\mathrm{x_{Bj}}$, the relative energy $\mathrm{z_h}$, the momentum transverse to the direction of momentum transfer $\mathrm{p_T}$, and the azimuthal angle around that momentum transfer direction $\mathrm{\phi_{pq}}$. The HERMES Collaboration was the first to publish two-dimensional results for such observables with nuclear targets~\cite{Airapetian_2011}, finding unexpected and complex behavior that exposes the details of hadronization dynamics. This has been followed up by 3-fold differential studies by the CLAS Collaboration \cite{moran2021measurement}. Such studies provide a motivation to collect large datasets that allow binning in up to 5-fold differential bins for the hadrons that are produced most copiously, such as pions, kaons, and protons. The rarer particles, such as $\phi$, $\eta$, $\omega$, and anti-proton, can still be studied in at least one or two dimensions in the 11 GeV data.

Hadron propagation in the medium can also be studied via other reaction types, often motivated by the search for color transparency. An increased nuclear transparency to hadrons in specific kinematics is a definite prediction of QCD that is linked to important properties such as factorization of initial and final states in high-energy scattering. Experiments in the 6 GeV era of JLab saw the onset of color transparency for pions~\cite{PhysRevLett.99.242502, PhysRevC.81.055209} and $\rho$ mesons~\cite{El_Fassi_2012}. Future studies of color transparency (CT) at JLab for mesons and baryons  will strongly benefit from the high luminosity and large reach in four-momentum transfer foreseen. The A(e,e'h)X channel in diffractive and non-diffractive kinematics probes CT in meson production, while it can be used to look for proton transparency in quasi-elastic kinematics~\cite{PhysRevLett.126.082301, Bhetuwal:2022wbe}. The principal signature of CT is a reduced nuclear transparency observed as four-momentum transfer $Q^2$ increases, for fixed coherence length. While the transparency seen in the 6 GeV era was small but significant, a robust signature for light mesons is expected to be very clear as beam energy and luminosity rise.

\subsubsection{Hadron formation}
The mechanisms involved in the hadronization process that dynamically enforce color confinement are poorly known. More insight into these mechanisms can be obtained by systematic study of production of different baryon and meson types in large and small nuclear systems. Questions that can be answered by such studies include: what are the differences between formation of $q\bar{q}$ (baryon number B=0) systems and $qqq$ (B=1) systems? how do the characteristics of formation change as the number of strange quarks increases, both in mesons and baryons? how does the formation depend on hadron mass? what can we learn about the time required for complete formation of hadrons? is there evidence of diquark structure seen for baryon formation, and if so, how does it influence our understanding of proton and neutron structure?

\subsection{Future opportunities} 

There are specific measurements that would benefit significantly from an increase in the electron beam energy, and other cases where entirely new measurements are possible. Higher energies would benefit parity violating measurements, by allowing higher cross sections at fixed $Q^2$ or increased $Q^2$ values with a corresponding increase in the parity-violating asymmetries. This could improve the coverage of the parity-violating EMC measurement discussed earlier, which aims to confirm and quantify the flavor dependence of the EMC effect in $^{48}$Ca. With a sufficiently improved figure of merit, the kinematic range of the measurement could be expanded, or additional nuclei could be measured to look for flavor dependence  in isoscalar nuclei or other non-isoscalar nuclei such as $^9$Be. 

The improved $Q^2$ range provided by higher beam energies could also be used to extend the scaling studies at $x>1$, with the aim of extracting the PDFs in nuclei at $x>1$, where the cross section is dominated by scattering from SRCs. Increasing the energy would allow measurements at larger $x$ and $Q^2$, extending further into the DIS region. Such measurements on the deuteron would allow for better comparisons of models based on a simple convolution of two nucleons as opposed to those including exotic configurations (6-quark bags, hidden color, or $\Delta-\Delta$ contributions), or models with large off-shell corrections for high-momentum nucleons.

Other programs will benefit from higher energies through the increase in cross section for measurements which do not need expanded $Q^2$ coverage. Measurements of  A(e,e'p) in light nuclei, especially for polarization observables for scattering from $^3$He would benefit. Several measurements utilizing spectator tagging would also benefit, as they are frequently limited by luminosity and cross section.

\begin{figure}
    \centering
    \includegraphics[width=120mm,scale=0.35]{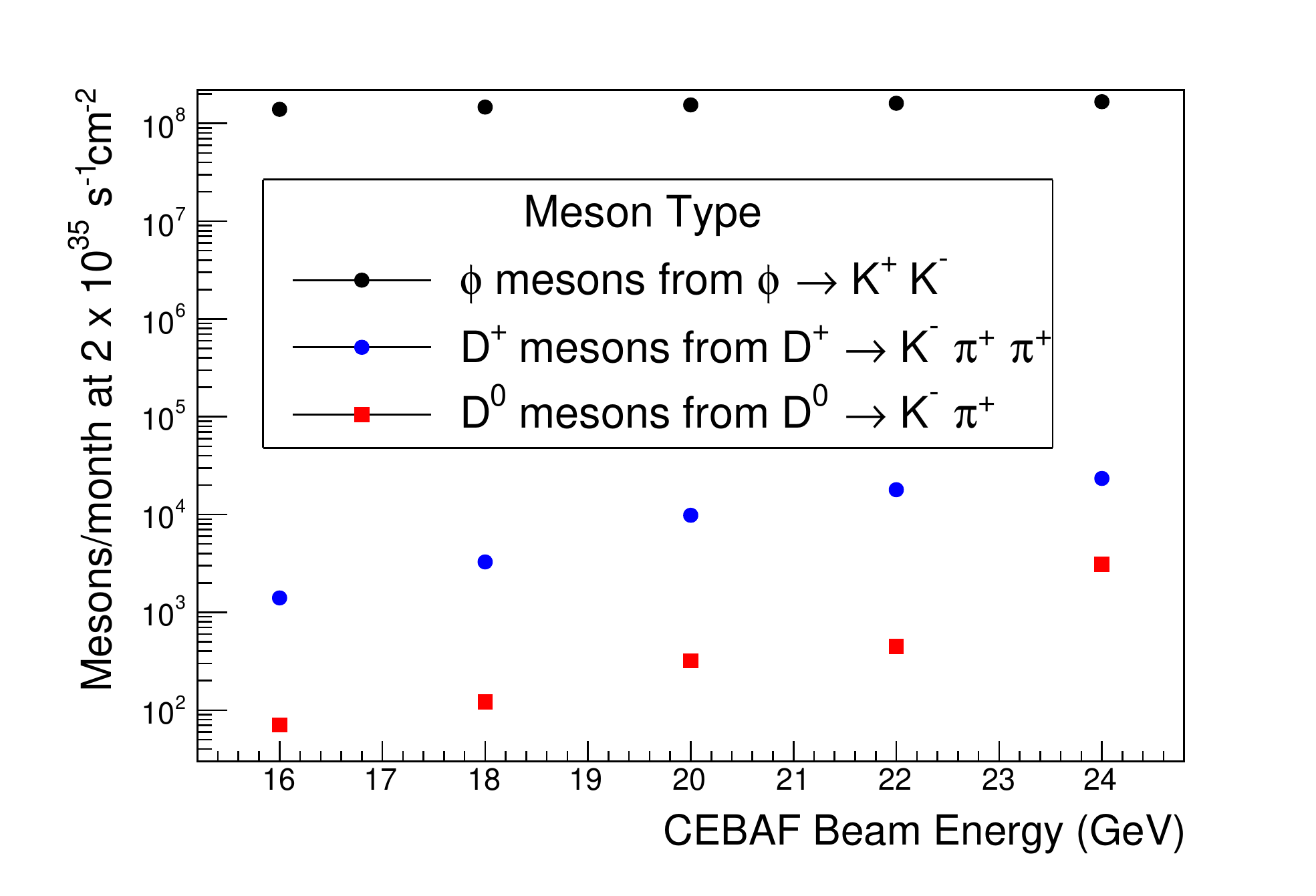}
    \caption{Production rates of heavy mesons as a function of beam energy. The calculation is based on one month of operation in CLAS12 at the already-achieved luminosity for deuterium targets of up to 2 x $\mathrm 10^{35}~cm^{-2}s^{-1}$}
    \label{fig:heavymesons}
\end{figure}

At higher energies, completely new probes of quark hadronization dynamics will be possible. For example, it will be possible to produce more massive hadrons, beyond the present-day limit of 0.78 GeV for production of the $\omega$ meson off nuclei~\cite{borquez_2021}. An example of this is illustrated in Fig.~\ref{fig:heavymesons} which shows production rates for one month of operation as a function of beam energy. The highest rates shown are for the $\phi$ meson, a 1-GeV mass two-quark system which will first be probed in these studies with an 11 GeV beam in Hall B with much lower production rates, and thus with limited statistical accuracy. With the high statistical accuracy achievable with 24 GeV beam, it will be possible to study the multi-dimensional behavior of the formation of this meson for the first time, a key factor for isolating hadronization dynamics. A further advance is shown in the figure with the D meson production rates, which is sub-threshold at current JLab energies. As seen from that figure, the phase space for charm production opens up rapidly as 24 GeV is approached. The crossing of the charm threshold enables study of these 2-GeV mass two-quark systems for the first time. This is entirely {\em{terra incognita}} with respect to current-day models of hadron formation.   

Beyond mass reach, the increase in {\em{kinematic}} reach provided by approximately doubling the beam energy will have important benefits for studies of color propagation and color transparency. For a particle of mass $m$ and lab energy $E$, the relativistic boost in the lab frame will be $E/m$. This factor will approximately double relative to the 11~GeV case. This increase will permit studies of the expected color lifetime time dilation which is expected to occur for light mesons, and will improve the sensitivity to color transparency as well, not only from the boost but also from the increased coverage in four momentum transfer $Q^2$, the primary kinematic variable used in those studies. Since color transparency is expected to increase with $Q^2$, the substantially expanded range in that variable should reveal a larger magnitude of the transparency, thus increasing sensitivity of the measurement.
Another impact of the wider kinematic range is that slightly lower $x$ will be accessible, e.g. down to 0.07-0.08. At sufficiently low $x_{Bj}$, $q\bar{q}$ pairs are produced diffractively from the virtual photon. For nuclear targets this would result in dijets or dipions passing through the medium, sharing the energy of the virtual photon. This process has been studied theoretically~\cite{PhysRevD.46.931} and the onset of this distinct mechanism could be searched for at 24 GeV.

\newpage

%% file: Section5.tex
\section{ The Standard Model and Beyond} 
\label{sec:BSM}

The Standard Model (SM) of Particle Physics has reached a concluding milestone in 2012, brought by the experimental observation of the Higgs boson by both the ATLAS and the CMS collaborations at the LHC~\cite{Aaboud:2018zhk,Sirunyan:2018kst}. 
There are still many opening questions. But overall, the SM has withheld nearly half a century of experimental examination and has so far been a quite successful framework to describe three of the four interactions of Nature. 
On the other hand, it's probably a consensus within the physics community that the SM is not the ultimate theory: At what energy scale and by what symmetry can we unify strong with electroweak? Under what framework can we add gravity? And how do we explain dark matter and dark energy?  We must therefore continue high-precision measurements of SM parameters and look hard for beyond the SM (BSM) phenomenon -- hints of new particles and new interactions -- by experimenting across the whole energy scale. One such example is the recent muon $g-2$ measurement~\cite{PhysRevLett.126.141801} that raised challenges to lepton universality, adding exciting fresh information to the search for BSM physics.

CEBAF at JLab has provided an essential tool in our pursuit of understanding the strong interaction and the nucleon and nuclei since the late 1990's. In the recent decade, two new directions have emerged in JLab's research program: studies of electroweak (EW) physics and ``dark sector'' searches for direct production of low-mass, weakly coupled new physics. Being at an energy scale between atomic physics and particle colliders, CEBAF holds a unique position in the landscape of SM and BSM study. 

Among EW physics measurements, JLab has a long history of carrying out parity-violating electron scattering (PVES) experiments. During the 6 GeV era, the Qweak experiment provided the best knowledge on the proton weak charge and improved our knowledge on the EW neutral-current (NC) axial-vector ($AV$) $C_{1q}$ couplings~\cite{Androic:2013rhu,Androic:2018kni}. The 6~GeV PVDIS experiment similarly improved our knowledge on the vector-axial ($VA$) $C_{2q}$ couplings~\cite{Wang:2014bba,Wang:2014guo}. In the coming years, the planned MOLLER experiment~\cite{Benesch:2014bas} will provide one of the most precise data on the weak mixing angle $\sin^2\theta_W$. 
With the addition of a high intensity device -- the SoLID spectrometer~\cite{Chen:2014psa} -- the PVDIS measurement will be extended over a wide $(x,Q^2)$ range and to higher precision, making it possible to simultaneously probe the $C_{2q}$ couplings and hadronic effects such as charge symmetry breaking and higher twists~\cite{PVDIS}. PVDIS on the proton will provide $d/u$ ratio at high $x$ without the need of using a nuclear model. Despite technical and theoretical challenges, the future positron beam at CEBAF opens up another direction -- possible measurements of the axial-axial ($AA$) couplings $C_{3q}$ -- that cannot be overlooked.

In the dark sector program, JLab's APEX and HPS experiments have pioneered searches for new bosons that are produced via a weak coupling to electrons and decay through the same interaction, with the massive, kinetically mixed ``dark photon'' as a canonical benchmark model.  HPS will continue to explore this parameter space in the coming years, exploring both prompt and displaced decays of dark photons and other weakly coupled bosons. In addition, the BDX proposal for a detector downstream of the Hall A beam dump is poised to explore light dark matter production. % at the dump.  
The future positron beam at JLab opens up new possibilities to search for production of weakly-coupled new physics via annihilation of beam positrons on atomic electrons, using the complementary missing-mass and missing-energy approaches. 

\subsection{ Neutral-current electroweak physics at low energies}
\label{sec:MOLLER}

At energies much below the mass of the $Z^0$ boson (the ``$Z$-pole''), $Q^2\ll M_Z^2$, the Lagrangian of the EW NC interaction relevant to electron-electron (M\o ller) scattering or electron deep inelastic scattering (DIS) off quarks inside the nucleon is given by~\cite{Zyla:2020zbs}:
\begin{eqnarray}
L_{NC} &=& \frac{G_F}{\sqrt{2}} \left[ g_{AV}^{ee}\bar e\gamma_\mu\gamma^5 e\bar e\gamma^\mu e+ 
%g_{VV}^{eq} \, \bar e\gamma^\mu e \bar q\gamma_\mu q + 
g_{AV}^{eq} \, \bar e\gamma^\mu\gamma_5 e \bar q\gamma_\mu q 
%\right. \nonumber \\ &+& \left. 
+ g_{VA}^{eq} \, \bar e\gamma^\mu e \bar q\gamma_\mu \gamma_5 q + 
g_{AA}^{eq} \, \bar e\gamma^\mu \gamma_5 e \bar q\gamma_\mu \gamma_5 q \right], \label{eq:L}
\end{eqnarray}
where $G_F=1.166\times 10^{-5}$~GeV$^{-2}$ is the Fermi constant and the $g^{ee}_{AV}$, $g^{eq}_{AV}$, $g^{eq}_{VA}$, $g^{eq}_{AA}$ are effective four-fermion couplings. 
We have omitted both $VV$ and $AA$ coupling of electron-electron interaction on the RHS of Eq.~(\ref{eq:L}) because they can only be measured at high energies~\cite{ALEPH:2005ab,Schael:2013ita}. Similarly, the $g_{VV}^{eq}$ term, being chirally identical to electromagnetic interactions of QED, is too difficult to measure at low energies and is omitted here. 
In contrary, the $AV$ and $VA$ couplings are both parity-violating and thus can be separated from QED effects by measuring parity violation observables, for example the cross section asymmetry between a right-handed and left-handed electron beam scattering, 
\begin{eqnarray}
  A_{RL} &=& \frac{\sigma_R-\sigma_L}{\sigma_R+\sigma_L}~. 
\end{eqnarray}
The coupling $g_{AV}^{ee}$ has been measured by SLAC E158~\cite{Anthony:2005pm} and is the primary operator for the planned MOLLER experiment~\cite{Benesch:2014bas} at JLab. 
The $g_{AV}^{eq}$ is best determined by combining atomic parity violation~\cite{Wood:1997zq,Guena:2005uj,Toh:2019iro} and elastic electron scattering experiments such as Qweak~\cite{Androic:2013rhu,Androic:2018kni}.
The $g_{VA}^{eq}$ requires spin-flip of quarks and can only be measured in deep inelastic scattering. 
The last term on the RHS of Eq.~(\ref{eq:L}) does not violate parity but can be measured by comparing DIS cross sections of lepton with anti-lepton DIS, which can be done once a positron beam becomes available at JLab~\cite{Zheng:2021hcf}. 

%Note: At HERA energies (at least using the high $Q^2$ bins) they could resolve the vector and axial-vector couplings, even though with large error bars.  New amplitudes are suppressed there, because they don’t resonate. At lower energies you only see the products but new amplitudes are relatively less suppressed. So there is complementarity.  The EIC may have a little bit of both.

\begin{figure}[!ht]
\begin{center}
\includegraphics[width=0.8\textwidth]{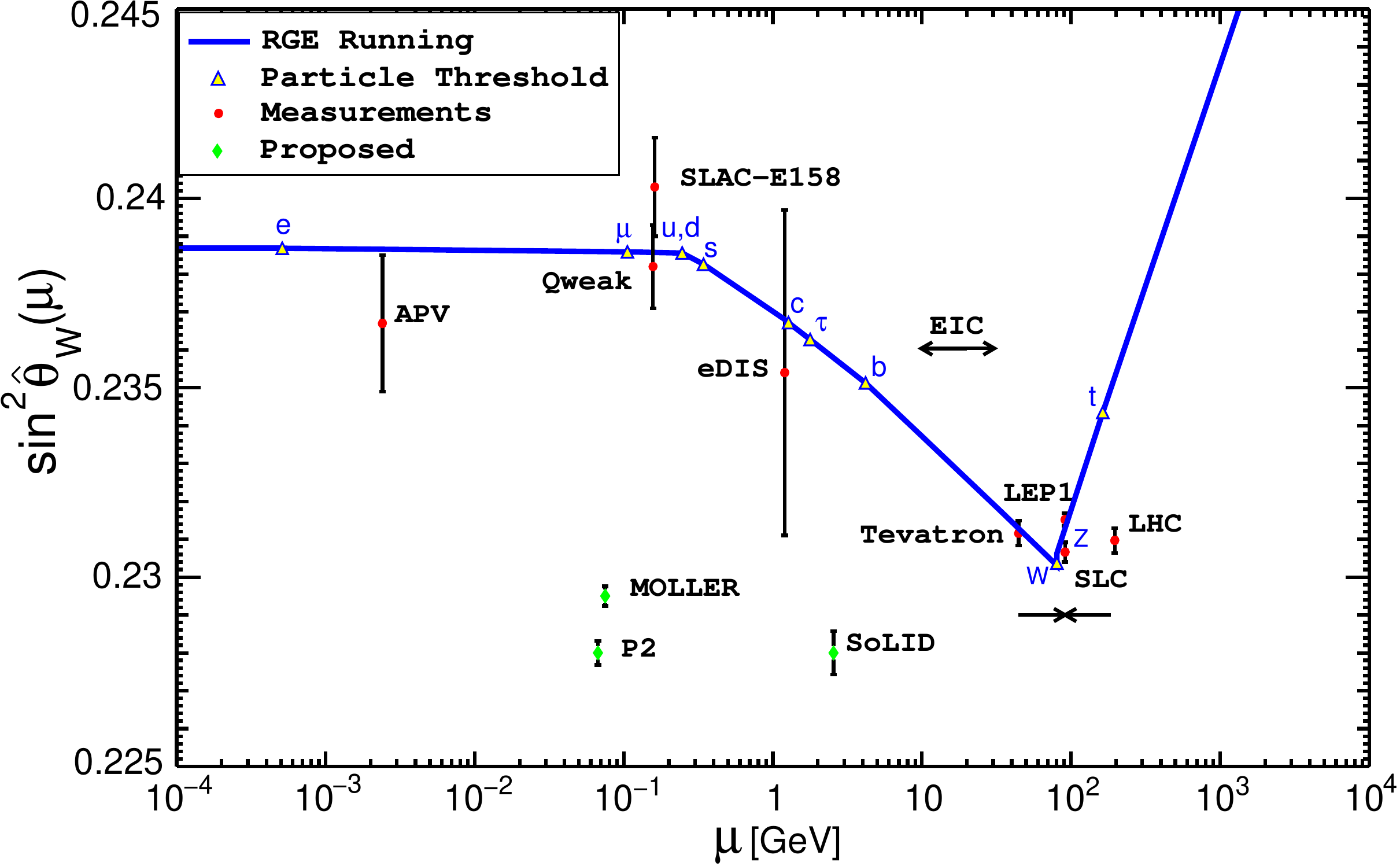}
\end{center}
\caption{Running of the weak mixing angle $\sin^2\theta_W$, updated from Ref.~\cite{Erler:2017knj} (by courtesy of R. Ferro-Hernandez). Data points for Tevatron and LHC are shifted horizontally for clarity. The expected $Q^2$ coverage of the future electron-ion coillder (EIC) where its expected data will have an impact is shown by the arrows. The achieved precision from Qweak is $\pm 0.0011$~\cite{Androic:2018kni}, and the expected precision from P2, MOLLER and SoLID PVDIS are $\pm 0.00033$~\cite{Becker:2018ggl}, $\pm 0.00028$~\cite{MOLLERCDR} and $\pm 0.00057$~\cite{Erler:2014fqa}, respectively.
}\label{fig:thetaW}
\end{figure}

All $AV$ and $VA$ effective couplings are related to the weak mixing angle $\sin^2\theta_W$, the single parameter that intertwines electromagnetic with weak interactions within the SM framework.  The present world knowledge on the weak mixing angle is shown in Fig.~\ref{fig:thetaW}. There are measurements across the energy scale from atomic to LHC experiments, but the precision of measurements much below the $Z$ pole can be improved. In the next decade, three high-precision experiments will come online: the MOLLER and the SoLID PVDIS experiments at JLab and the P2 experiment at the upcoming MESA accelerator at Mainz, with the precision of MOLLER and P2 comparable to the most precise data to date from LEP, SLC and LHC above the $Z$-pole. MOLLER and P2 experiments will potentially anchor the running of $\sin^2\theta_W$ from the low energy side, with SoLID PVDIS sitting in the transition $Q^2$ range. Low energy measurements also hold unique discovery potential if BSM physics (particles and amplitudes) exist in this energy regime and not higher. 

On the other hand, the weak mixing angle is one parameter of the EW theory that works across energy scale and link all observables, but only in the framework of SM itself.  %Should a particular BSM physics affects only a certain type of interactions (leptonic vs. semi-leptonic, or a certain chiral structure), one must dive into precision measurements of different couplings. 
By combining observables from different types of experiments, our ultimate goal is to map out all chiral combinations of the couplings for the purpose of both understanding and testing the SM. In M{\o}ller scattering, the parity-violating asymmetry for longitudinally polarized incoming electron is 
\begin{eqnarray}
 A_{PV}^{ee} &=& m E \frac{G_F}{\sqrt{2}\pi\alpha}\frac{2y(1-y)}{1+y^4+(1-y)^4}Q_W^e
\end{eqnarray}
where $Q_W^e=-2(1-4\sin^2\theta_W)$ is the weak charge of the electron, $\alpha$ is the fine structure constant, $E$ is the incident beam energy, $m$ is the electron mass, $\theta$ is the scattering angle in the center-of-mass frame, $y=1-E'/E$ with $E'$ the energy of either one of the scattered electrons. The SLAC E158 experiment measured %$Q_W^e$ to 
$g^{ee}_{AV} = 0.0190 \pm 0.0027$ and $\sin^2\theta_W^\mathrm{eff}=0.2397\pm 0.0010$(stat.)$\pm 0.0018$(syst.) at $Q^2=0.026$~(GeV)$^2$. The MOLLER experiment will improve the uncertainty of $g_{AV}^{ee}$ ($Q_W^e$) to %$\pm 0.0010$ or
relative $\pm 2.4\%$ and $\sin^2\theta_W$ to $\pm 0.00028$~\cite{MOLLERCDR}, comparable to 
%the $\pm 0.00027$ 
the results from SLC~\cite{SLD:2000leq} and 
%the $\pm 0.00021$ from 
LEP~\cite{ALEPH:2005ab}.
%Note: LEP publication had several results for sin2thetaW, from 0.00021 to 0.00019 and combined (on their first page) 0.00016
%0.0010 Tevatron PRD84(2011)012007
%0.00053 CMS 1806.00863 hep-ex

\begin{figure}[t]
\includegraphics[width=0.47\textwidth]{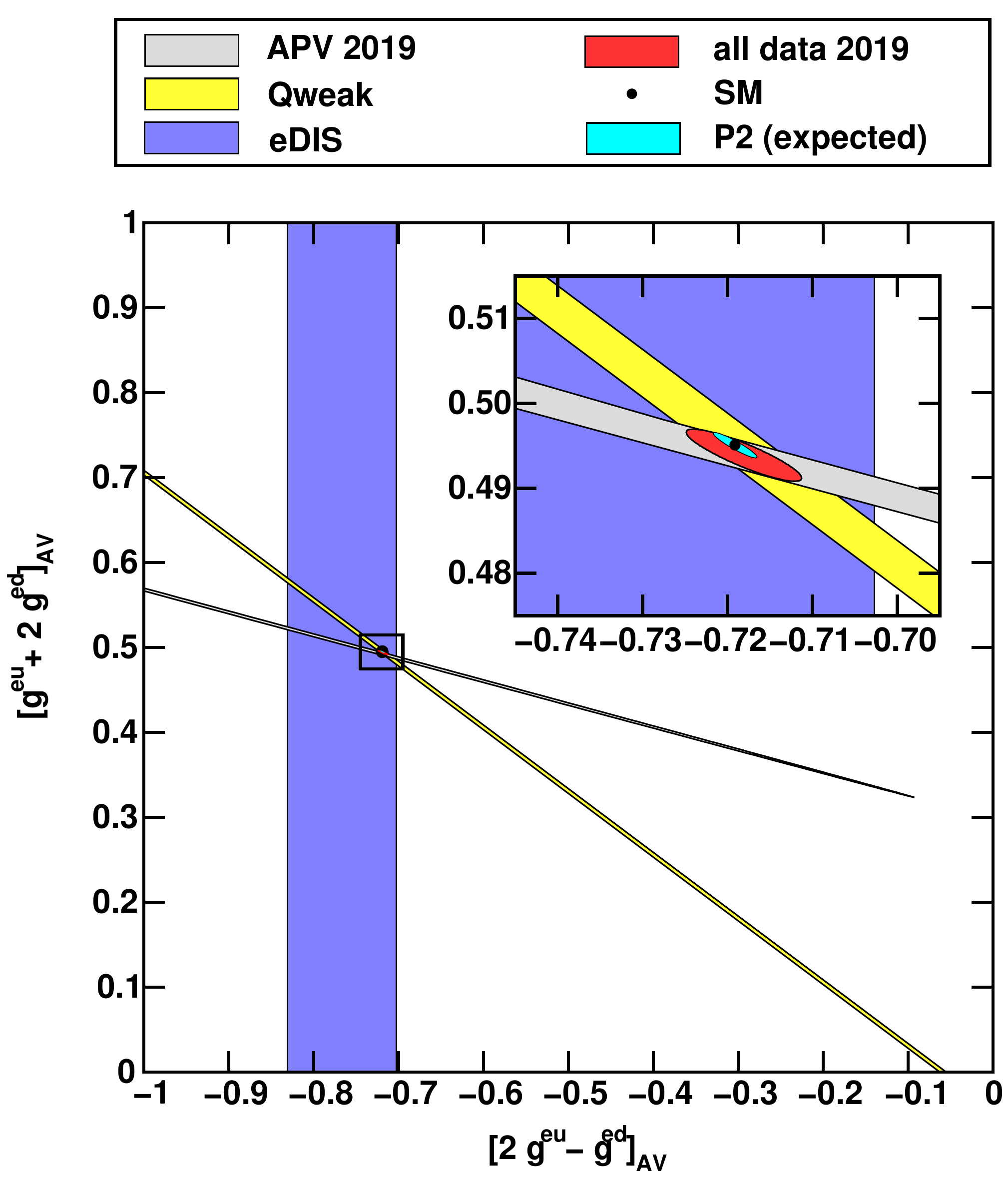}
\includegraphics[width=0.47\textwidth]{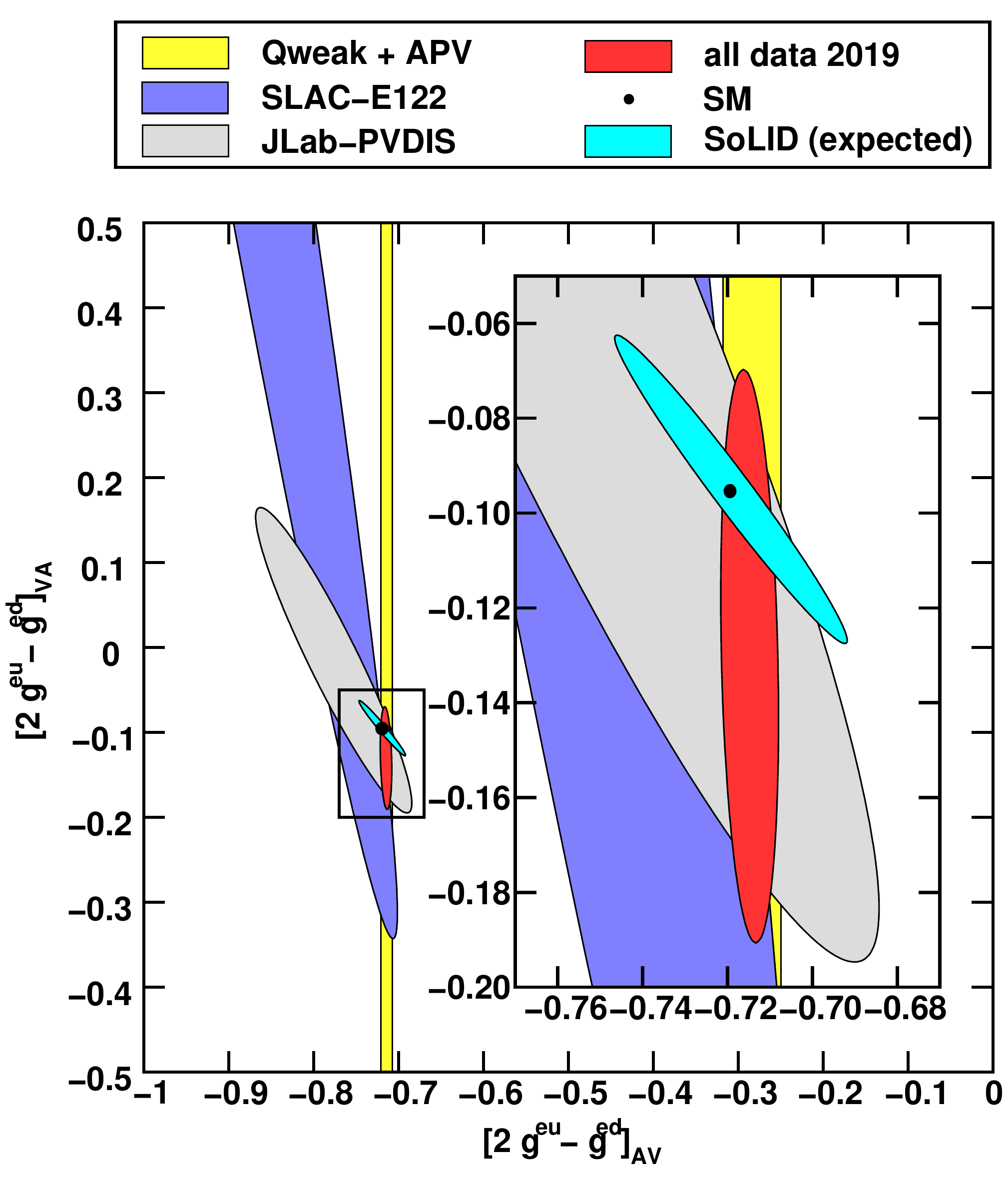}
\caption{Adapted from Ref.~\cite{Zheng:2021hcf}: Current experimental knowledge of the couplings $g_{AV}^{eq}$ and $g_{VA}^{eq}$. 
For $g_{AV}^{eq}$ (left): The latest measurement is from the 6 GeV Qweak experiment~\cite{Androic:2018kni} at JLab. The Atomic Parity Violation ("APV 2019") results shown utilized the theory calculations of Ref.~\cite{Toh:2019iro}. The "eDIS" band is a combination of the SLAC E122~\cite{Prescott:1978tm,Prescott:1979dh} and the JLab PVDIS~\cite{Wang:2014bba,Wang:2014guo} experiments. 
For $g_{VA}^{eq}$ (right), 
The latest measurement is from the PVDIS experiment~\cite{Wang:2014bba,Wang:2014guo} at JLab. 
Also indicated are the expected uncertainties from the planned P2 experiment~\cite{Becker:2018ggl} at Mainz (left) and from the planned SoLID project at JLab (right), both centered at the SM value. }
\label{fig:c1c2}
\end{figure}

The current knowledge on the $AV,VA$ electron-quark couplings are shown in Fig.~\ref{fig:c1c2}. 
The latest data on $g_{AV}^{eq}$ was from the parity-violating asymmetry in $ep$ elastic scattering during the Qweak experiment~\cite{Androic:2013rhu,Androic:2018kni}: 
\begin{eqnarray}
 A_{RL, elastic}^{ep} &=& - \frac{G_F Q^2}{4\pi\alpha\sqrt{2}} \left(Q_w^p+Q^2 B(Q^2,\theta)\right),
\end{eqnarray}
where $Q_w^p=-2(2g_{AV}^{eu}+g_{AV}^{ed})$ is the proton weak charge, $-Q^2$ is the four-momentum transfer squared, $\theta$ is the scattering angle, and $B(Q^2,\theta)$ represents the proton's internal structure. By measuring the asymmetry at $Q^2=0.025$~GeV$^2$ and extrapolating to the $Q^2=0$ point, the weak charge of the proton was determined to be $Q_w^p=0.0719\pm 0.0045$ and the weak mixing angle $\sin^2\theta_W=0.2383\pm 0.0011$. When combined with atomic parity violation experiment~\cite{Wood:1997zq,Guena:2005uj,Toh:2019iro}, the Qweak experiment provides the best constraint on the $g_{AV}^{eq}$ couplings to date. The P2 experiment planned at Mainz will improve the uncertainty by a factor 3 over Qweak, determine $Q_w^p$ to $\pm 1.83\%$ and $\sin^2\theta_W$ to $\pm 0.00033$~\cite{Becker:2018ggl}.

One may notice the stark contrast in Fig.~\ref{fig:c1c2} in the precision between $AV$ and $VA$ couplings. This is because access to the electron's axial coupling is directly provided by the spin flip of electron beam and thus is an observable in all PVES experiments. On the contrary, access to the quark's axial coupling requires quark spin flip and can only be achieved in the DIS regime, and is suppressed due to angular momentum conservation by the kinematic factor 
$Y=[1-(1-y)^2]/[1+(1-y)^2]$. For electron DIS off an isoscalar target such as the deuteron and if one considers only the light quark flavors $u$ and $d$: 
\begin{eqnarray}
  A^{e^-,\mathrm{PVDIS}}_{RL,d}
    &\approx& \frac{3G_F Q^2}{10\sqrt{2}\pi\alpha}\left[(2g_{AV}^{eu}-g_{AV}^{ed})+R_V Y(2g_{VA}^{eu}-g_{VA}^{ed})\right]~,
\end{eqnarray}
where $R_V(x) \equiv ({u_V+d_V})/({u^+ + d^+})$ 
with $q^+, q_V$ defined with parton distribution functions $q(x)$: $q^+\equiv q(x)+\bar q(x)$ and $q_V\equiv q(x)-\bar q(x)$.
Access to $g_{VA}^{eq}$ is thus best provided by ``high $y$'' settings such as the fixed-target configuration at JLab. As shown in the right panel of Fig.~\ref{fig:c1c2}, the SoLID PVDIS experiment~\cite{PVDIS} will improve over the previous 6 GeV measurement~\cite{Wang:2014bba,Wang:2014guo} by an order of magnitude.

One general and model-independent way of characterizing BSM physics search potential of an experiment is to express BSM physics in terms of contact interactions that perturb the SM Lagrangian~(\ref{eq:L}), {\it i.e.}, by replacements of the form~\cite{Erler:2014fqa},
\begin{eqnarray}
%\frac{G_F}{\sqrt{2}} g_{ij} \rightarrow \frac{G_F}{\sqrt{2}} g_{ij} + \eta_{ij}\frac{4\pi}{(\Lambda_{ij})^2}\ ,
\frac{G_F}{\sqrt{2}} g_{ij} \rightarrow \frac{G_F}{\sqrt{2}} g_{ij} + \eta_{ij}\frac{g^2}{(\Lambda_{ij})^2}\ ,
\label{eq:ciqmodified}
\end{eqnarray}
where $ij=AV,VA,AA$ and can be for either $ee$ or $eq$ interaction, $g$ is the coupling and $\Lambda$ is the mass scale of BSM physics, {\it i.e.} the coupling and the mass of the hypothetical BSM particle being exchanged. 
If the new physics is strongly coupled, $g^2 = 4\pi$, then the 90\% C.L. mass limits reached or to be reached by MOLLER, Qweak, and SoLID PVDIS on $g_{VA}^{ee}$, $g_{AV}^{eq}$ and $g_{VA}^{eq}$ are, respectively:
\begin{eqnarray}
 \Lambda_{VA, \mathrm{MOLLER}}^{ee}&=& g \sqrt{\frac{\sqrt{2}}{G_F 1.96\Delta g_{VA}^{ee}}}=39~\mathrm{TeV}, \\
 \Lambda_{AV, \mathrm{Qweak}}^{eq}&=& g \sqrt{\frac{2\sqrt{2}\sqrt{5}}{G_F 1.96\Delta Q_w^p}}=30~\mathrm{TeV},\\
 \Lambda_{VA, \mathrm{SoLID+world}}^{eq} &=& g \sqrt{\frac{\sqrt{2}\sqrt{5}}{G_F 1.96\Delta \left(2g_{VA}^{eu}-g_{VA}^{ed}\right)}}=16~\mathrm{TeV}; 
\end{eqnarray}
where the $\sqrt{5}$ for $Q_w^p$ and PVDIS cases are to represent the ``best case scenario'' where BSM physics affects maximally the quark flavor combination being measured. Similar to Qweak, the P2 experiment will pose a mass limit on $g_{AV}^{eq}$ at  49~TeV~\cite{Erler:2014fqa}. The expected uncertainty $\Delta\left(2g_{VA}^{eu}-g_{VA}^{ed}\right)=\pm 0.007$ is obtained by combining SoLID PVDIS with existing world data. If one instead look for the mass limit expected from PVDIS in any combination of $g_{AV}^{eq}$ and $g_{VA}^{eq}$ then it would be 22~TeV from SoLID alone. 
%{\color{red}{(SoLID PVDIS is 22 TeV in  https://arxiv.org/pdf/1401.6199.pdf but not purely VA)}}

Another approach to characterize BSM physics reach is to use SM effective field theory (SMEFT)~\cite{Boughezal:2021kla}. In SMEFT approach, BSM physics is described as contact interactions similar to that EW NC interactions can be treated as  4-fermion contact interactions at energies much below the $Z$-pole. Furthermore, operators in SMEFT can be generalized to beyond 4-fermion (dimension-6) terms, for example, to include general dimension-8 operators. Low energy PVES experiments such as P2 and SoLID PVDIS in fact will help to disentangle dimension-6 from dimension-8 SMEFT couplings, as these cannot be separated by data from high energy colliders alone~\cite{Boughezal:2021kla}. On the other hand, in SMEFT analysis one must be careful to limit the use of assumptions, as over-simplification will minimize the apparent value of low energy data that provide direct access to a single combination of couplings.

Looking forward, in addition to MESA and the possible energy upgrade of JLab, a series of upgrades for the LHC are being discussed~\cite{LHeCStudyGroup:2012zhm,LHeC:2020van,FCC:2018byv,FCC:2018evy} that will venture into the unexplored energy range much beyond the $Z$-pole~\cite{Britzger:2020kgg}. %Electroweak observables at high energy colliders are combinations of all couplings and separation is done through global fitting (see {\it e.g.}~\cite{Spiesberger:2018vki}).
The EIC, coming online within the next 1-2 decades, likely will have decent sensitivity to EW couplings in between JLab and high-energy colliders as well. %, though the concept of effective low-energy couplings no longer apply due to its relatively high $Q^2$.  
 Among all existing and planned facilities, CEBAF is one of the few that can provide direct access and high precision measurements of the SM effective couplings owing to both its high luminosity fixed-target settings and the relatively low beam energies, and thus holds a unique place in the test of the SM across all energy scales.  

\subsection{ Searches for Dark Sectors}
\label{sec:APEXHPS}
\label{sec:BDX}
The last decade has seen rapidly growing interest in searches for physics beyond the Standard Model at low masses and much weaker couplings to familiar matter than the SM interactions.  These are motivated by the broad framework of ``dark sectors'', i.e. any new particles neutral under SM interactions, which form a natural framework for low-mass dark matter and are potentially connected to a wide range of astrophysical and experimental anomalies.  The dark sector framework and these motivations are summarized in a recent review article \cite{Lanfranchi:2020crw} and many community reports \cite{Alexander:2016aln,Battaglieri:2017aum,DarkMatterBRNReport,Beacham:2019nyx,Agrawal:2021dbo}.  

Standard Model symmetries considerably restrict the interactions of ordinary matter with SM-neutral matter. In the absence of high-energy modifications to the Standard Model, the allowed low-energy interactions or ``portals'' correspond to mixing of a new vector $A'$ (``dark photon'' or ''U boson'') with the Standard Model photon, of a new scalar $\phi$ with the SM Higgs boson, or of a new neutral fermion $\psi$ with neutrinos. Because each of these is associated with a marginal operator ($ \frac{1}{2}\partial_{[\mu} A'_{\nu]} F^{\mu\nu}$,  $|\phi|^2 |H|^2$, and $\psi HL$), loops of heavy particles can generate these interactions with parametric strength of order one or two loop factors, $\sim 10^{-2} - 10^{-6}$.   These interaction strengths are sufficiently weak that they would not be seen in generic collider searches for high-mass new physics.  Instead, these possibilities are best explored through either moderate-energy, high-intensity experiments or unconventional low-$p_T$ LHC searches.  The intensity and bunch structure of the CEBAF beam is especially well-suited to direct searches for the production of dark-sector particles. 

Of the portals above, the dark photon (a.k.a. vector or kinetic mixing portal) mentioned above has emerged as a canonical example. It allows for simple and predictive dark matter models and is also the portal for which electron-beam searches, such as those at CEBAF, are most powerful so we will focus on this case.  For massive dark photons, kinetic mixing of strength $\epsilon$ induces couplings $\epsilon e$ to SM matter of electric charge $e$ \cite{Holdom:1985ag, Okun:1982xi}. Dark photon exchange effects are parametrically smaller than their electromagnetic counterparts and respect the same symmetries, but can be constrained either by the \emph{difference} of QED-like effects at different energy scales (\emph{e.g.}~different values of $\alpha$ inferred from electron \emph{vs.}~muon anomalous magnetic moments) or from direct production of dark photons (or dark-sector particles that couple to them) in laboratory experiments \cite{Pospelov:2008zw}. Most such searches can be divided into three categories:
\begin{itemize}
\item\textbf{Visible dark photon} searches assume that a dark photon decays back to SM particles through the same kinetic mixing interaction that mediates its production.   This assumption implies a precise lifetime and branching ratios for a given dark photon mass and production cross-section; at masses below the muon threshold, the leading decay is to $e^+e^-$ pairs with a lab-frame lifetime that can be prompt or displaced at the mm to meter scale in light of existing constraints (see e.g. \cite{Bjorken:2009mm,Lanfranchi:2020crw}).   This signal can be distinguished from QED backgrounds in the same final state by searching for a small but narrow resonance peak in the $e^+e^-$ mass distribution and/or reconstructing a displaced decay vertex. The parametric estimate of interaction strength $\epsilon \sim 10^{-2} - 10^{-6}$ arising from one- or two-loop effects also motivates exploring this full range of allowed lifetimes.
\item\textbf{Dark matter production searches} test the possibility of new invisible species being produced through dark-photon interactions. Some \emph{beam dump searches} aim to detect a small fraction of these invisible products through their subsequent scattering in a detector downstream of a high-intensity beam dump; other \emph{missing energy} or \emph{missing momentum} searches instead aim to detect the production event via low total energy deposition and/or measurement of the recoiling electron's kinematics.  A particular focus of these searches is the \emph{thermal relic target}, a band of interaction strengths $\epsilon \gtrsim 10^{-6} (m_\chi/\rm{MeV})$ for which a dark-sector particle $\chi$ annihilating to ordinary matter through the dark photon could explain the observed abundance of dark matter through thermal freeze-out (see e.g.~\cite{Battaglieri:2017aum,DarkMatterBRNReport,Berlin:2020uwy} for discussion of this target as well as the complementary sensitivity of non-JLab fixed-target experiments using electron and proton beams). 
\item\textbf{Missing mass} searches generally search for the reaction $e^+e^- \rightarrow \gamma A^\prime$, in either high-intensity colliders or positron-beam fixed-target experiments.  These experiments use the detected photon kinematics to reconstruct the mass of an undetected dark-photon candidate, and search for a peak in this distribution.  In some cases, such experiments can probe dark photons inclusively, irrespective of the decay mode (including not only the decays to SM or dark matter particles discussed above, but also more complex cascade decays into multiple SM and dark-sector particles).  
\end{itemize}
It should be noted that, while these searches are conventionally interpreted in the context of dark-photon models, they apply more generally to any boson that couples to electrons --- including $B-L$ gauge bosons or scalar bosons --- with only O(1) changes in the CEBAF constraints.

%%{\it This part needs to be finished. Include 
%(1) overview of why low-mass dark sectors are interesting. 
%(2) general phenomenological framework of vector/scalar with small coupling to electrons and/or other SM particles and kinetic mixing portal as a canonical example, 
%(3) possibilities for dark photon/mediator decay and light dark-matter production via virtual mediator, and 
%(4) exciting milestones in parameter space e.g.~for thermal dark matter and loop-level mixing.}

So far, three dark-sector search experiments have run at JLab, focused on different aspects of this physics. 
The A-Prime EXperiment (APEX) \cite{Essig:2010xa} searches for dark photons decaying promptly to $e^+e^-$ in Hall A.  Such a signal would show up as a narrow resonance in $e^+e^-$ mass over a smooth QED background, and is peaked when the combined energy of the pair is near the incident beam energy.  The APEX search uses this kinematics to optimize the signal acceptance relative to background rate, using the two High Resolution Spectrometers (HRSs) in coincidece to measure events with an electron in one arm and positron in the other.  A septa magnet is used to deflect pairs with central angle $\sim 6^\circ$ into the spectrometer acceptances.   The 2010 APEX test-run \cite{Abrahamyan:2011gv} demonstrated this technique in a narrow mass range, achieving new sensitivity in dark photon parameter space tough it has since been surpassed by resonance searches at MAMI-A1 \cite{Merkel:2014avp} and in BaBar data  \cite{Lees:2014xha}.  The full APEX run in 2019 accumulated over 100$\times$ greater true $e^+e^-$ pair statistics than the test run in a similar geometry; its ongoing analysis will explore new dark photon parameter space in the mass range 160--230 MeV.   

\begin{figure}[htbp]
\begin{center}
  \includegraphics[angle=0,width=0.6\linewidth]{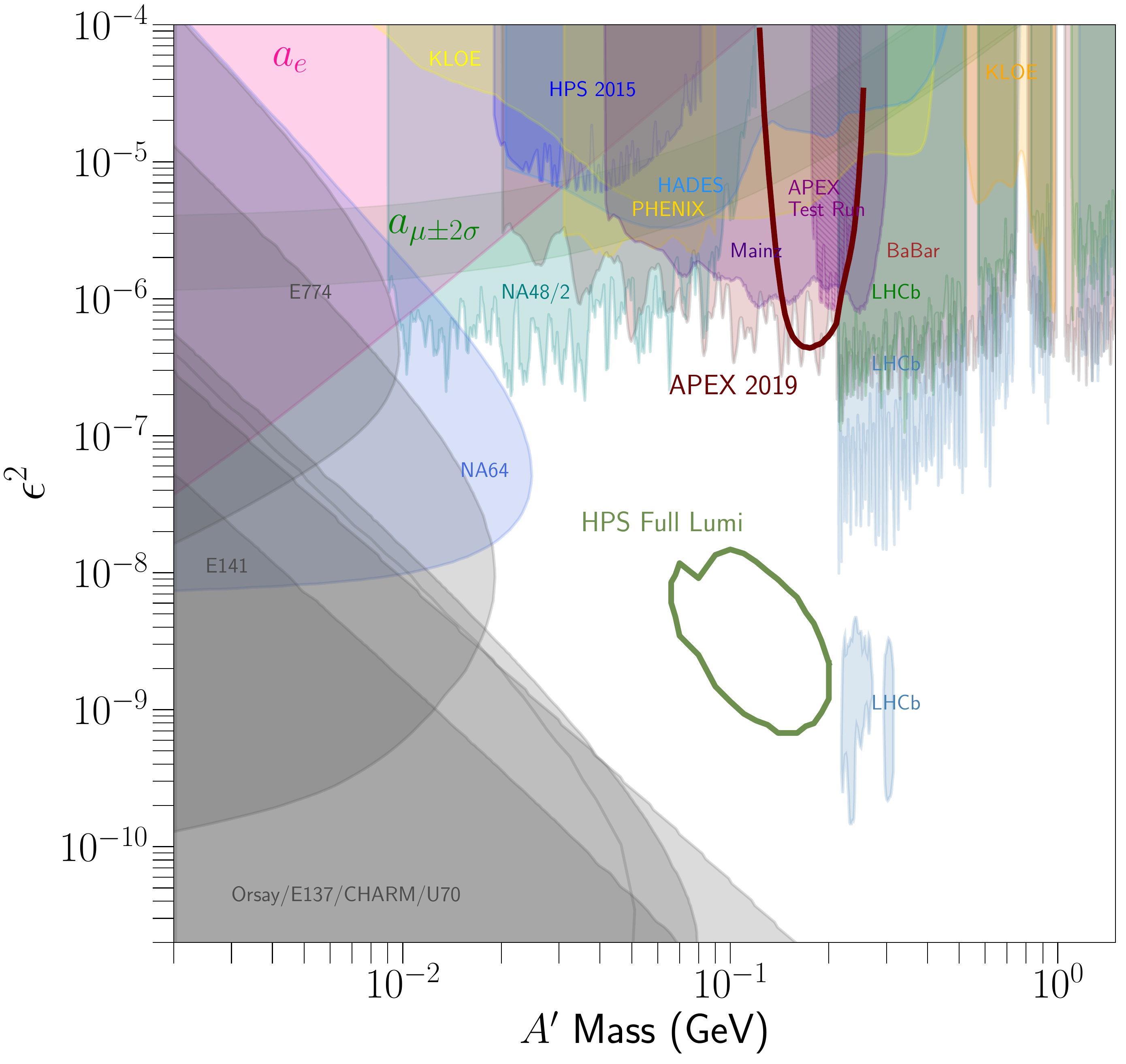}
  \caption{Current exclusion regions for a dark photon with the projected sensitivity of two fully approved and operating CEBAF experiments, APEX (2019 statistics --- magenta) and HPS (full run plan including 2019, 2021, and future running---green).  The exclusion regions come from \cite{Konaka:1986cb,Riordan:1987aw,Bjorken:1988as,
     Bross:1989mp,Davier:1989wz, Bjorken:2009mm, Abrahamyan:2011gv, Archilli:2011zc, Andreas:2012mt, Babusci:2012cr, Agakishiev:2013fwl, Adare:2014mgk, KLOE-2:2014qxg, Lees:2014xha, Merkel:2014avp, Adrian:2018scb,  Aaij:2017rft, LHCb:2019vmc}. Plots and projections courtesy of HPS collaboration and Vardan Khachatryan.}
  \label{fig:darkPhotonProj}
\end{center}
\end{figure}

 The Heavy Photon Search (HPS) in Hall B also searches for  dark photons decaying promptly to $e^+e^-$, but with a detector consisting of a 7-layer Silicon Vertex Tracker (SVT), a scintillation hodoscope, and downstream ECal \cite{Battaglieri:2014hga,Balossino:2016nly}. This detector configuration allows for precision tracking and vertexing of $e^+e^-$ pairs produced at forward angles, enabling a displaced-vertex search that opens up sensitivity to weaker dark-photon couplings, in addition to a resonace search.  HPS engineering runs at 1.1 and 2.3 GeV beam energies took place in 2015 and 2016, with the first physics results from these runs appearing in \cite{Adrian:2018scb,Solt:2020zbi}.  Following a detector upgrade, a first physics run took place in 2019 at 4.56 GeV beam energy, for which analysis is ongoing. The second physics run completed in November 2021, with more beam time available to run approved by the JLab PAC for outyears (105 PAC days).  These runs will use beam energies between $\approx 2$ and $\approx 4$ GeV, with each beam energy allowing sensitivity to a different dark photon mass range.  Projected sensitivities for the full runs of APEX and HPS, as well as current constraints, are shown in Figure \ref{fig:darkPhotonProj}. 

 A third bump-hunt search at JLab has recently been proposed using the PRad apparatus in Hall B.  Instead of a magnetic spectrometer, this proposal uses GEM tracking and calorimetry $\sim 8$ meters downstream of a thin target to reconstruct the kinematics and invariant mass of $A^\prime$ candidate $e^+e^-$ pairs.  This approach is well-suited to reconstruction of small-angle pairs, enabling a search for $3-60$ MeV $A^\prime$'s with a $2-3$ GeV beam.  The proposal calls for 60 PAC days of running to cover most of the allowed parameter space in this mass range with $\epsilon$ large enough to allow prompt $A^\prime$ decays, and is conditionally approved.
 
A complementary ongoing effort at JLab is the Beam Dump eXperiment (BDX) proposal to search for light dark matter with a parasitic detector downstream of the Hall A dump \cite{Battaglieri:2014qoa,Battaglieri:2016ggd,Battaglieri:2017qen,Battaglieri:2019nok}.  This detection scheme relies on the fact that any light dark matter (LDM) particles that interacts with electrons (as is the case in benchmark dark photon models as well as other leptophilic scenarios) will be produced as beam electrons traverse a beam dump and initiate an electromagnetic shower.  The resulting diffuse secondary beam of LDM particles emanates from the dump, and LDM traversing the detector volume can, with low probability, deposit energy in the detector via a single scattering reaction off electrons or nuclei \cite{Izaguirre:2013uxa}.  In inelastic dark matter models, excited states of DM may also decay in the detector into a lighter DM state plus an $e^+e^-$ pair, whose energy deposition offers another modality for LDM detection \cite{Izaguirre:2014dua}.  BDX proposes to use a meter-scale segmented CsI(Tl) scintillation detector surrounded by layers of an active veto. The BDX technique has been tested at JLab exposing a small prototype detector (BDX-MINI ~\cite{Battaglieri:2020lds}) to a 2 GeV electron beam and accumulating in a $\sim$6-months-time $\sim 15\%$ of the Electron-On-Target expected in BDX (fully parasitic). The stability of the system and the results obtained provided a solid ground for the full  experiment. 
BDX has been approved by JLab PAC-46, and is seeking funding for the detector and associated cavern.  Once constructed, operation of BDX will be completely parasitic as data can be taken during \emph{any} high-current operations in Hall A.  A complementary detection strategy, using a DRIFT directional light-WIMP TPC \cite{Snowden-Ifft:2018bde} in the BDX cavern, has also been proposed.  

Finally, we note that JLab experiments are also sensitive to axion-like particles (ALPs).  Unlike the portals discussed above, ALPs have an approximate shift symmetry, as a result of which their leading couplings to the Standard Model are dimension-5 operators (for example, a coupling $\frac{4\pi \alpha_s c_g}{\Lambda}a G^{\mu\nu} \tilde G_{\mu\nu}$ to gluons, motivated by the strong CP problem, as well as analogous couplings to photons).  The coupling discussed above allows ALP photoproduction off protons \cite{Aloni:2019ruo}.  This process is also the basis for a recent ALP search at GlueX \cite{GlueX:2021myx}, which explores new ALP parameter space in the 180--480 MeV and 600--720 MeV regions.  

\subsection{ Rare Processes } 
%\section{The JEF experiment in Hall D}
\label{sec:jefexperiment}

The JLab Eta Factory (JEF) experiment~\cite{pro-JEF17, pro-JEF14} will
perform precision measurements of various $\eta^{(\prime)}$ decays
with emphasis on rare decay modes, using the GlueX
apparatus~\cite{Adhikari:2020cvz} and an upgraded Forward
electromagnetic Calorimeter (FCAL-II).  Significantly boosted
$\eta^{(\prime)}$ mesons will be produced through $\gamma
+p\rightarrow\eta^{(\prime)} +p$ with an 8-12 GeV tagged photon beam.
Non-coplanar backgrounds will be suppressed by tagging
$\eta^{(\prime)}$ with recoil proton detection. The $\eta^{(\prime)}$
decay photons (and leptons) will be measured by FCAL-II with a
high-granularity, high-resolution PbWO$_4$ crystal core in the central
region that minimizes shower overlaps and optimizes the resolutions of
energy and position. A capability of tagging highly-boosted
$\eta^{(\prime)}$ in combination with a state-of-the-art FCAL-II
offers two orders of magnitude improvement in background suppression
and better control of systematics compared to the other
$\eta^{(\prime)}$ experiments, such as A2-MAMI~\cite{Adlarson:2019nwa,
  Kashevarov:2017kqb}, WASA-at-COSY~\cite{Husken:2019dou},
KLOE-II~\cite{Berlowski:2019hst}, BESIII~\cite{Shuang-shiFang:2015vva}
and the proposed future REDTOP~\cite{Gatto:2019dhj}.  JEF is a unique
eta-factory, producing $\eta$ and $\eta^{\prime}$ simultaneously at
similar rates ($\sim 6\times 10^7$ tagged $\eta$ and $\sim 5\times
10^7$ tagged $\eta^{\prime}$ per 100 days), with no competition in
rare neutral decay modes.

\begin{figure}[!htb]
\begin{center}
  \includegraphics[viewport=21 49 504 524,clip,angle=0,width=0.6\linewidth]{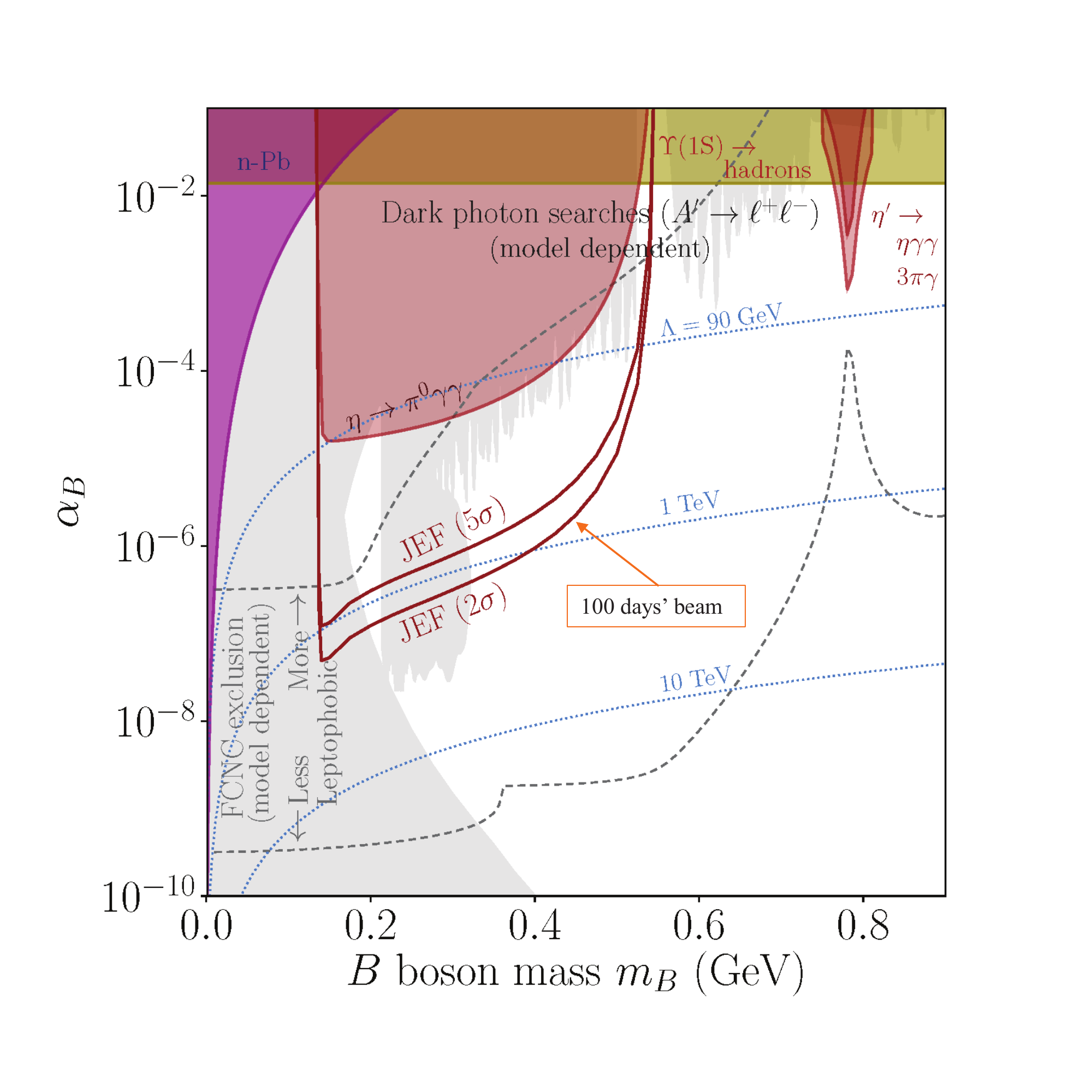}
  \caption{Current exclusion regions for a leptophobic gauge boson
    $B^{\prime}$~\cite{Gan:2020aco, Tulin:2014}, with the projected
    $2\sigma/5\sigma$ sensitivity reach for the JEF experiment via $\eta
    \to \gamma + B^\prime (\to \gamma + \pi^0)$. Color-shaded regions
    and curves are model-independent. These include constraints from
    rare $\eta$, $\eta^\prime$ decays (red), hadronic $\Upsilon (1S)$
    decays~\cite{Aranda:1998fr} (yellow), and low-energy $n$-Pb
    scattering~\cite{Barbieri:1975xy} (purple). The gray shaded regions
    and dashed contours are model-dependent and involve leptonic
    couplings via kinetic mixing $\varepsilon_x = x \frac{e g_B}{(4\pi
      )^2}$: these regions are excluded by dark photon searches for dilepton
    resonances~\cite{Babusci:2012cr,Konaka:1986cb,Riordan:1987aw,Bjorken:1988as,
     Bross:1989mp,Davier:1989wz,Banerjee:2019hmi,Tsai:2019mtm,Astier:2001ck,
     Bernardi:1985ny,Batley:2015lha,LHCb:2019vmc},
    $A^\prime \to \ell^+ \ell^-$, for $\varepsilon_{0.1}$; and the gray
    dashed contours are upper limits on $\alpha_B$ from FCNC $b \to s
    \ell^+ \ell^-$ and $s \to d
    \ell^+\ell^-$~\cite{Dror:2017nsg,Dror:2017ehi}, for
    $\varepsilon_{0.001}$ (upper line) and $\varepsilon_1$ (lower
    line). The blue dashed contours denote the upper bound on the mass
    scale $\Lambda$ for new electroweak fermions needed for anomaly cancellation.}
  \label{b-reach-map}
\end{center}
\end{figure}

The $\eta^{(\prime)}$ meson has the quantum numbers of the vacuum
(except the parity). Its strong and electromagnetic decays are either
anomalous or forbidden at the lowest order due to symmetries and
angular momentum conservation. This enhances the relative importance
of higher order contributions or new weakly-coupled interactions. A
study of the $\eta^{(\prime)}$ decays provides a rich
flavor-conserving laboratory to test the isospin-violating sector of
low-energy QCD and to search for new physics beyond the Standard Model
(SM)~\cite{Gan:2020aco}.  The JEF experiment will primarily focus on
the following areas:

\begin{itemize}
\item {\em Precision tests of low-energy QCD}. The $\eta \rightarrow
  3\pi$ decay promises an accurate determination of the quark mass
  ratio, $ {\cal Q} =(m_s^2-\hat{m}^2)/(m_d^2-m_u^2)$ with $\hat{m} =
  (m_u+m_d)/2$. This decay is caused almost exclusively by the isospin
  symmetry breaking part of the Hamiltonian $\sim (m_u-m_d)(u{\bar
    u}-d{\bar d})/2$.  Moreover, Sutherland's
  theorem~\cite{Bell:1968,Sutherland:1966} forbids electromagnetic
  contributions in the chiral limit; and contributions of order
  $\alpha$ are also suppressed by $(m_u+m_d)/\Lambda_\mathrm{QCD}$.
  These single out $\eta\rightarrow 3\pi$ to be the best path for an
  accurate determination of ${\cal Q}$~\cite{Leutwyler:1996,
    Bijnens:2001, Colangelo:2018jxw}.  A low-background measurement of
  the rare decay $\eta \rightarrow \pi^0 \gamma\gamma$ provides a
  clean, rare window into ${\cal O}(p^6)$ in chiral perturbation
  theory~\cite{Bijnens:1992}. This is the only known meson decay that
  proceeds via a polarizability-type mechanism. The Dalitz
  distribution measured by JEF will offer sufficient precision for the
  first time to explore the role of scalar meson dynamics and its
  interplay with the vector meson dominance. The measurements of the
  transition form factor of $\eta$ and $\eta^{\prime}$ via the
  $\eta^{(\prime)}\rightarrow e^+e^-\gamma$ decays will reveal the
  dynamic properties of those mesons, providing important input to
  calculate hadronic light-by-light corrections to the anomalous
  magnetic moment of the muon~\cite{Aoyama:2020ynm}.

\item {\em A search for various gauge boson candidates in the MeV--GeV
  mass range, probing three out of four highly motivated portals
  coupling the SM sector to the dark sector}. A leptophobic vector
  boson ($B^\prime$)~\cite{Tulin:2014} coupling to baryon number can
  be searched for via $\eta, \eta^\prime \rightarrow B^\prime
  \gamma\rightarrow \pi^0\gamma\gamma$ for $0.14< m_{B^\prime} < 0.62$
  GeV, and $\eta^\prime \rightarrow B^\prime \gamma\rightarrow
  \pi^+\pi^-\pi^0\gamma$ for $0.62 < m_{B^\prime}< 1$ GeV.  The
  leptophilic vector
  bosons~\cite{Fayet:2007ua,Reece:2009un,Bjorken:2009mm,Batell:2009yf}
  can be searched for in the decays of $\eta, \eta^\prime \rightarrow
  A^\prime \gamma\rightarrow e^+e^-\gamma$.  A
  hadrophilic~\cite{Batell:2018fqo,Liu:2019} scalar can be probed in
  $\eta\to \pi^0 S \to \pi^0 \gamma \gamma , \: \pi^0 e^+ e^-$ for
  $m_S < 2m_\pi$, and in $\eta, \eta^\prime \to \pi^0 S \to 3\pi , \:
  \eta^\prime \to \eta S \to \eta\pi\pi$ for $m_S >
  2m_\pi$. Axion-Like Particles (ALP)~\cite{
    Dobrescu:2000jt,Aloni:2018vki,Nomura:2008ru,Freytsis:2010ne} can
  be explored via~$ \eta, \eta^\prime \to \pi\pi a \to \pi\pi
  \gamma\gamma , \: \pi\pi e^+ e^-$.  Figure~\ref{b-reach-map} gives an
  example for the sensitivity of the JEF experiment. With 100 days of
  beam time, a study of $\eta \to \gamma + B^\prime (\to \gamma +
  \pi^0)$ will improve the existing model-independent bounds by two
  orders of magnitude, with sensitivity to the baryonic fine structure
  constant $\alpha_B$ as small as $10^{-7}$, indirectly constraining
  the existence of anomaly cancelling fermions at the TeV-scale.

\item {\em Fundamental symmetry tests}.  The $\eta^{(\prime)}$ meson
  is the eigenstates of C, P, CP, and G ($I^GJ^{PC} = 0^+0^{-+}$),
  representing a natural candidate to test discrete symmetries.
  Particularly, the $\eta^{(\prime)}$ meson is among very few
  self-conjugate particles existing in the nature for testing
  charge-conjugation symmetry. A search for C-violating
  $\eta^{(\prime)}$ decays (such as $\eta^{(\prime)} \to 3\gamma$,
  $\eta^{(\prime)} \to 2\pi^0 \gamma$, and $\eta^{(\prime)} \to \pi^0
  e^+ e^-$) and a mirror asymmetry in the Dalitz distribution of
  $\eta^{(\prime)} \to \pi^+\pi^-\pi^0$ will offer the best
  direct-constraints for new C-violating, P-conserving reactions.

\end{itemize}

\subsection{Future Opportunities and Positron Beam Prospects}

\subsubsection{EW NC couplings with a positron beam and at higher energies}
The addition of a positron beam to CEBAF will open up new landscape of EW physics study. On the topic of neutral-current EW physics, one possibility is to measure the axial-axial component of electron-quark coupling $g_{AA}^{eq}$. Historically, there has been only one measurement of the muonic $g_{AA}^{\mu q}$, experiment NA41 at CERN~\cite{Argento:1982tq}, that provided $(2g_{AA}^{\mu u}-g_{AA}^{\mu d})=1.58\pm 0.36$, analyzed with the latest knowledge on the $g_{AV,VA}$ of electrons. For the electron-quark $g_{AA}^{eq}$, it can be accessed by comparing the electron to positron scattering cross section on an isoscalar target. The yield ratio of $e^+p$ and $e^-p$ scattering was measured at SLAC and was found to be consistent with zero at the 0.3\% level~\cite{Fancher:1976ea}. 
The DIS cross sections of $e^+p$ and $e^-p$ scattering were measured at HERA and a global fit of the $g_{A,V}$ couplings was performed~\cite{ZEUS:2016vyd,H1:2018mkk}, but their sensitivity to BSM physics is different from JLab and the concept of low-energy effective coupling does not apply to HERA kinematics. The possibility of measuring the ratio of $e^+D$ to $e^-D$ DIS cross sections using the SoLID spectrometer is being explored~\cite{PR12-21-006}. In addition to possibly extracting $2g_{VV}^{eu}-g_{AA}^{ed}$, the measurement will provide data on the nucleon interference structure function $F_3^{\gamma Z}$, which are linked to the valence PDF in the parton model. 

The PVDIS measurement can be extended using a 22 GeV beam and SoLID. Under the same run conditions, the higher energy beam will provide a tighter constraint on the $g_{AV,VA}^{eq}$ coupling and the measurement is less susceptible to hadronic effects, which may allow us to utilize PV asymmetry measured in the full kinematic region for SM study and further reduce the uncertainty. 

\subsubsection{Dark sector search with a positron beam}
Positron beams also open up new opportunities for dark-sector searches that exploit the reaction $e^+ e^- \rightarrow A^\prime \gamma$ or the resonant reaction $e^+ e^- \rightarrow A' $ in the interaction of beam positrons with atomic electrons.  The latter reaction can occur for degraded positrons in a thick target so long as $m_{A^\prime} < \sqrt{2 m_e E_+}$, i.e. $m_{A^\prime} < 110$ (150) MeV for a 12 (24) GeV positron beam, and yields a huge enhancement to the production rate compared to bremsstrahlung. The availability of high energy, continuous, and high intensity positron beams at JLab allows exploration of substantial new regions in dark photon parameter space. Two complementary experimental techniques have been considered to exploit these production modes: 

In the first case, the dark photon $A'$ is produced by the interaction of the positron beam on a thin target via the process $e^+ e^- \rightarrow A' \gamma$. By detecting the associated photon, the $A'$ will be identified and its (missing) mass measured. This technique has been used by the PADME experiment at LNF-Italy. The proposed experiment will take advantage of the JLab high-energy and high intensity positron beam extending significantly the $A'$ mass range by a factor of four witth two order of magnitude higher sensitivity to the DM-SM coupling constant. 

The second set up  will use an active thick target and a total absorption calorimeter to detect remnants of the light dark matter production in a missing energy experiment.  $A'$ resonant production by positron annihilation on atomic electrons, with subsequent invisible decay, produces a distinctive peak in the missing energy distribution, providing a clear experimental signature for the signal. This experiment has the potential to  cover a wide parameter space including the thermal relic target milestone.  The use of polarization observables in the annihilation process could provide an additional tool to suppress backgrounds.

%% file: Summary.tex
\section{Summary}
The 12 GeV era of CEBAF is well underway and, as discussed in section 1.1, many important experimental results are being reported. A diverse and exciting program of approved experiments is planned for the next 8-10 years. In addition, there is great potential to upgrade CEBAF to realize higher luminosities and energies as well as the production of polarized and unpolarized positron beams. With the realization of these upgrades there is a rich and unique program of experimental nuclear and particle physics that will extend the life of the facility well into the 2030’s and beyond. CEBAF will thus provide the capability to address the high luminosity frontier of hadronic nuclear physics, complementary to the Electron Ion Collider and other international facilities. 

%% file: Acknowledgments.tex
\section*{Acknowledgments}
We acknowledge the assistance of many colleagues in the preparation of this work, including
Alberto Accardi,
Harut Avagyan,
Ryan Bodenstein,
Cameron Bravo,
Jian-Ping Chen,
Pavel Degtiarenko,
Latifa Elouadrhiri, 
Charles Hyde, 
Liping Gan,
Yuxun Guo, 
Sami Habet,
Mark Jones, 
Vladimir Khachatryan,
Tianbo Liu,
Simonneta Liuti, 
Zein-Eddine Meziani, 
Kent Paschke,
Sarin Philip,
Jorge Piekarewicz,
Yves Roblin,
Mark Stefani,
Riad Suleiman,
Michael Tiefenback,
Eric Voutier,
Weizhi Xiong.

This work was supported by the U.S. Department of Energy, Office of Science, Office of Nuclear Physics, including contracts DE-AC05-06OR23177, DE-AC02-05CH11231, the U.S. National Science Foundation, the Chilean ANID PIA/APOYO AFB180002 and ANID-Millennium Science Initiative Program - ICN2019 044.

%% file: AppendixA.tex
\section{Experimental Equipment}
\label{sec:appendixa}
%\subsection { Overview } 

\input{AppA_HallA}

\input{AppA_HallB}

\input{AppA_HallC}

\input{AppA_HallD}

%\subsection{ Other Existing Equipment and Upgrades to Planned Base Equipment} 

%% file: AppA_HallA.tex
\subsection{Hall A}
\label{sec:app-halla}
Hall A has transitioned from being a focusing spectrometer Hall to a long-term experimental installation Hall, most notably facilitating the Super BigBite Spectrometer (SBS), the Measurement of a Lepton-Lepton Electroweak Reaction (MOLLER), and the Solenoidal Large Intensity Device (SoLID) equipment and programs. One existing High Resolution Spectrometer (HRS), constructed in the 1990's to access
a momentum range of 0.3 to 4 GeV/c with high momentum resolution, is also still available. To facilitate in particular parity-violating experiments, the beamline -- including a precision Unser current monitor, several beam position monitors, and Compton and Møller polarimeters -- was and continues to be upgraded to operate with unprecedented precision up to 11 GeV.

The SBS program involves two open, single bend, resistive spectrometers, and two large standalone calorimeters (ECal for electromagnetic and HCal for hadron calorimetry). This apparatus will enable a complete study of the unpolarized nucleon form factors G$_E$ and G$_M$ for both the proton and neutron at high Q$^2$, as well as approved semi-inclusive and tagged DIS studies. 

MOLLER will measure the PV asymmetry in electron-electron scattering by rapidly flipping the longitudinal polarization of electrons that have been accelerated to 11 GeV and observing the resulting fractional difference in the probability of these electrons scattering off atomic electrons in a high power liquid hydrogen target. This asymmetry is proportional to the weak charge of the electron, which in turn is a function of the electroweak mixing angle, a fundamental parameter of the electroweak theory. 

The planned SoLID apparatus will enable a broad program of experiments at the high luminosity frontier also requiring large acceptance, made possible by developments in both detector technology and simulation accuracy that were not available in the early stages of the 12 GeV program planning. The spectrometer is, moreover, designed with a unique reconfiguration capability to optimize for either PVDIS or SIDIS/threshold production of the J/$\Psi$ meson.

%% file: AppA_HallB.tex
\subsection{Hall B		}
\label{sec:app-hallb}

Hall-B is the smallest of the current four experimental halls at Jefferson Lab. The Hall-B physics program covers all area of the lab science: nucleon structure (SIDIS, GPDs, TMDs), hadron spectroscopy (meson and baryon), nuclear structure (SRCs and quark hadronization in nuclear medium), and BSM physics (Dark Sector). The approved experimental program started in 2018 and will continue for the next $\sim$10 years.

The Hall is equipped with a permanent 4$\pi$ acceptance detector (CLAS12) but different experimental setups for dedicated measurements (PRAD, HPS) are also in place\footnote{CLAS12, PRad and HPS can not run in parallel.}. This paragraph describes the experimental equipment available in the Hall and plans for a future upgrade to pursue the physics program described in the first section of this document.\\ 

The CEBAF Large Acceptance Spectrometer for operations at 12 GeV beam energy (CLAS12) is used to study electro-induced nuclear and hadronic reactions. This spectrometer provides efficient detection of charged and neutral particles over a large fraction of the full solid angle. CLAS12
is based on a dual-magnet system with a superconducting torus magnet that provides a largely azimuthal
field distribution that covers the forward polar angle range up to 35$^o$, and a solenoid magnet and detector covering the polar angles from 35$^\circ$ to 125$^\circ$with full azimuthal coverage. Trajectory reconstruction in the
forward direction using drift chambers and in the central direction using a vertex tracker results in momentum
resolutions of $<1\%$ and $<3\%$, respectively. Cherenkov counters, time-of-flight scintillators, and
electromagnetic calorimeters provide identification of the scattered electron and produced hadrons. Fast triggering and high data-acquisition
rates allow operation at a luminosity of 10$^{35}$ cm$^{-2}$s$^{-1}$. These capabilities are being used in a broad program
to study the structure and interactions of nucleons, nuclei, and mesons, using polarized and unpolarized
electron beams and targets for beam energies up to 11 GeV.

The PRad (Proton Radius) experiment measured the charge radius of the proton with the highest precision ever reached in electron-scattering. The experiment required a dedicated setup consisting of an electromagnetic calorimeter and a GEM-based tracker, as well as a partial modification of the beam line. An upgraded version of the experiment (PRad-II) has been recently approved. The PRad Collaboration, in coordination with the Hall-B staff, is currently preparing the new set up that is expected to reduce by a factor of four the error bars on the proton radius.

The HPS (Heavy Photon Search) experiment uses the electron beam on a nuclear target to produce the postulated new boson A' (heavy photon) expected to mediate the interaction between the Standard Model particles and the Dark Sector. The experiment requires different beam energies (1.1. GeV to 6.0 GeV) to cover unexplored areas in the parameter space (A' coupling vs. mass). The HPS detector, installed in the Hall-B alcove downstream of the CLAS12 detector, is made by an electromagnetic calorimeter, a silicon-vertex tracker and a hodoscope.
\\

%% file: AppA_HallC.tex
\subsection{Hall C}
\label{sec:app-hallc}

In the upcoming era, Hall C at Jefferson Lab will
be the only high luminosity, flexible, large-scale installation
electron scattering facility in the world. This will enable critical benchmark measurements for nucleon and nuclear experiments at the EIC and elsewhere. The existing focusing magnet spectrometers HMS and SHMS (High Momentum and Super High Momentum Spectrometers, respectively) with well-shielded detector huts, high power cryotargets, and a 1 MWatt beam dump, allow routine operation at luminosities of $10^{38}$ to $10^{39}$.
Other equipment, such as the Neutral Particle Spectrometer (NPS), Compact Photon Source (CPS), high power cryogenic targets, and multiple
polarized targets are available to use in various combinations.  The hall itself provides floor space for a variety of
experimental configurations utilizing existing or new detectors or
equipment.

The base detectors for Hall C are the HMS and SHMS. These several msr acceptance devices can reach momenta of 11 GeV/c and 6+ GeV/c respectively  with momentum bites exceeding 10\%. 
Momentum reconstruction resolution and accuracy at the 0.1\% level are routinely achieved.
An electron beam energy measurement facility is provided by a string of upstream dipoles. These spectrometers can be placed at scattering angles as small as 5.5
and 10.5 degrees with sub-mrad pointing accuracy. They may be used singly or in
coincidence with each other or with other detectors, and have well-shielded
detector stacks.  This provides a unique ability to measure small
cross sections which demand demand high luminosity and facilitate the
careful study of systematic uncertainties.

When the magnetic spectrometers are rotated to large angles, the hall,
with its large beam height provides the flexibility to install a
large variety of additional detectors.  Historically, this flexibility
has allowed for large acceptance devices for parity violating
electron scattering, spectrometers optimized for hypernuclear physics, 
and other equipment. In the future, it can accommodate devices such as
the Super BigBite and BigBite (SBS and BB) open detector
spectrometers, the large hadronic calorimeter HCal, or new detectors such as NPS. The new NPS is a $62\times74~\textrm{cm}^2$
$\textrm{PbWO}_4$ calorimeter for photon and $\pi^0$ detection.

A variety of targets and beamline instruments are available in Hall C.
The standard target assembly allows for liquid
hydrogen/deuterium, dense ${}^3\textrm{He}$/${}^4\textrm{He}$, and other nuclear targets - all of
which may be operated at high luminosities.  Other targets have
included dynamically polarized hydrogen and deuterium targets
($\textrm{NH}_3$ and $\textrm{ND}_3$ as well as polarized
${}^3\textrm{He}$. Beam polarization can be measured to the
sub-percent level with both Moller and Compton polarimetry. As well as producing mixed photon/electron beam with a Bremsstrahlung
radiator, pure photon beams can be produced with the new CPS.  With electrons removed from the photon beam, the photon
flux on power-limited polarized targets can be maximized.

%% file: AppA_HallD.tex
\subsection{Hall D}
\label{sec:app-halld}

\begin{figure}[!htb]
\centering
  \includegraphics[angle=0,width=0.8\linewidth]{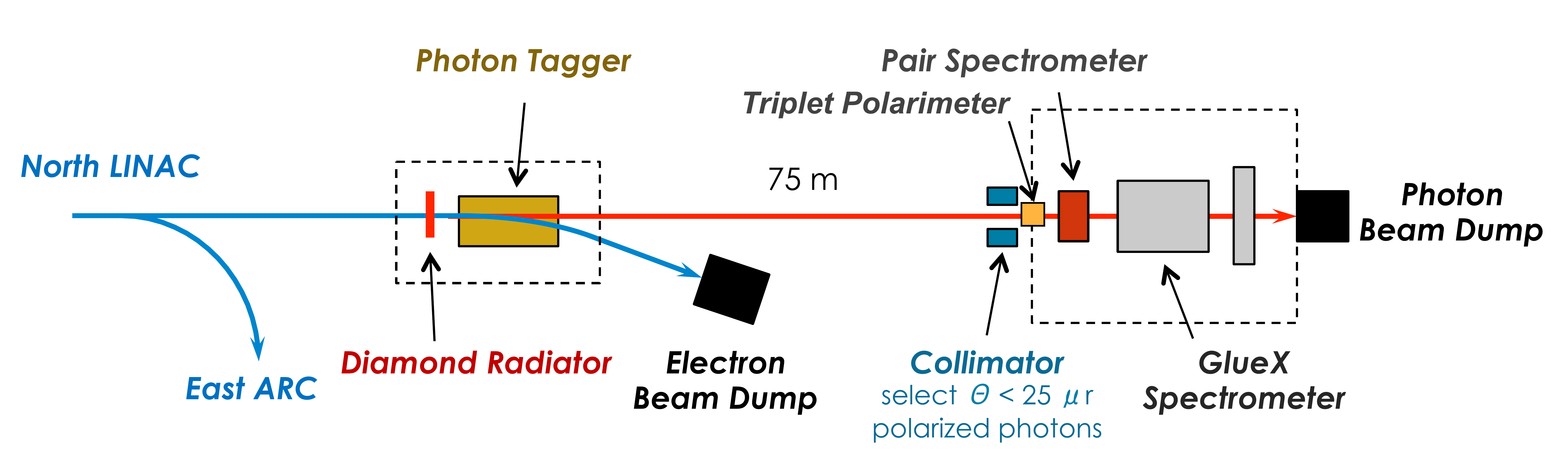}%
  \caption[Hall D beamline]{Hall D beamline layout~\cite{Adhikari:2020cvz}. 
    \label{fig:halld_beamline}
  }
\end{figure}

\begin{figure}[!htb]
%\centering
  \begin{minipage}[]{0.36\linewidth}
    \includegraphics[angle=0,width=0.99\linewidth]{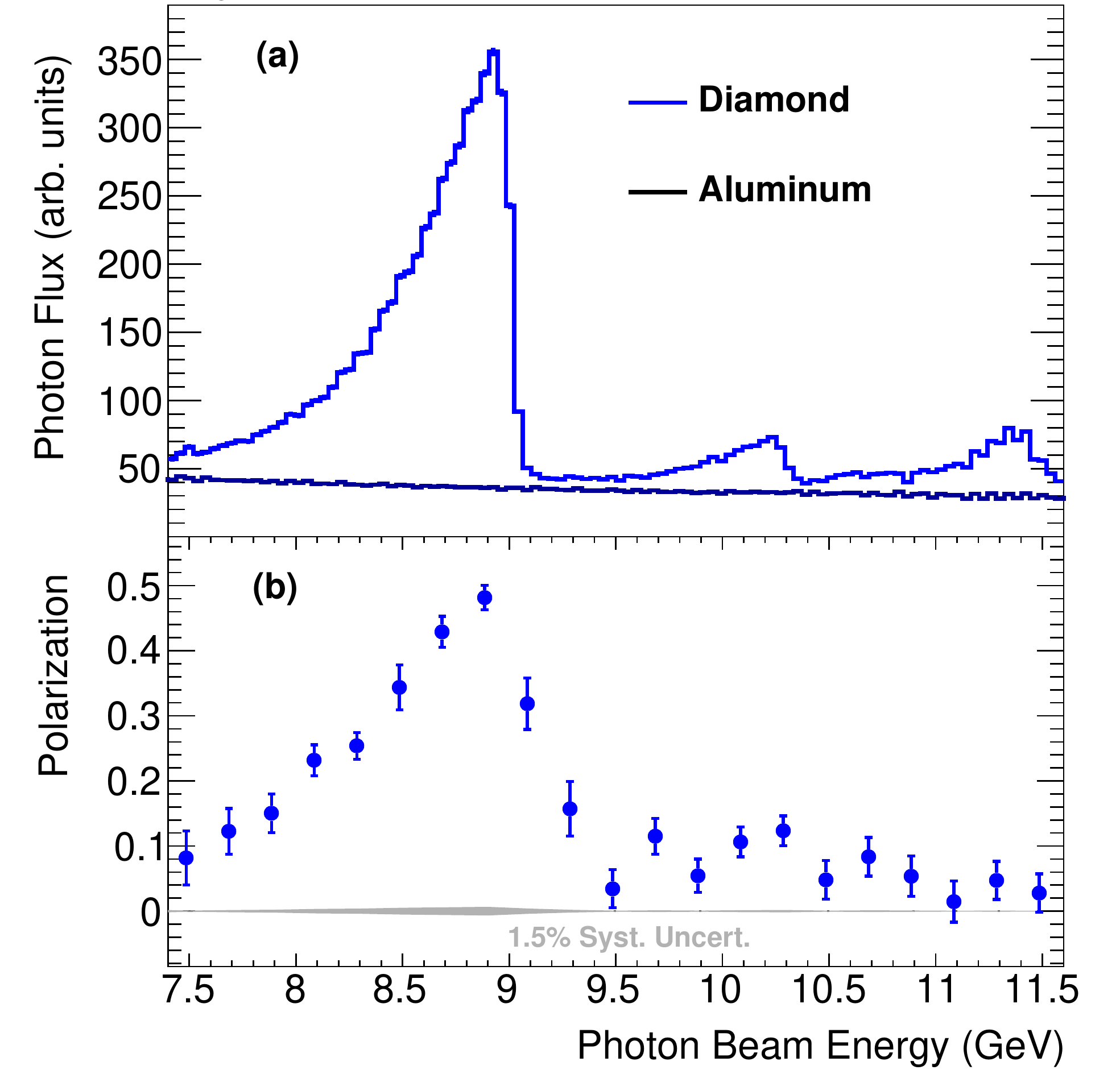}
  \end{minipage}
  \begin{minipage}[]{0.63\linewidth}
    \includegraphics[angle=0,width=0.99\linewidth]{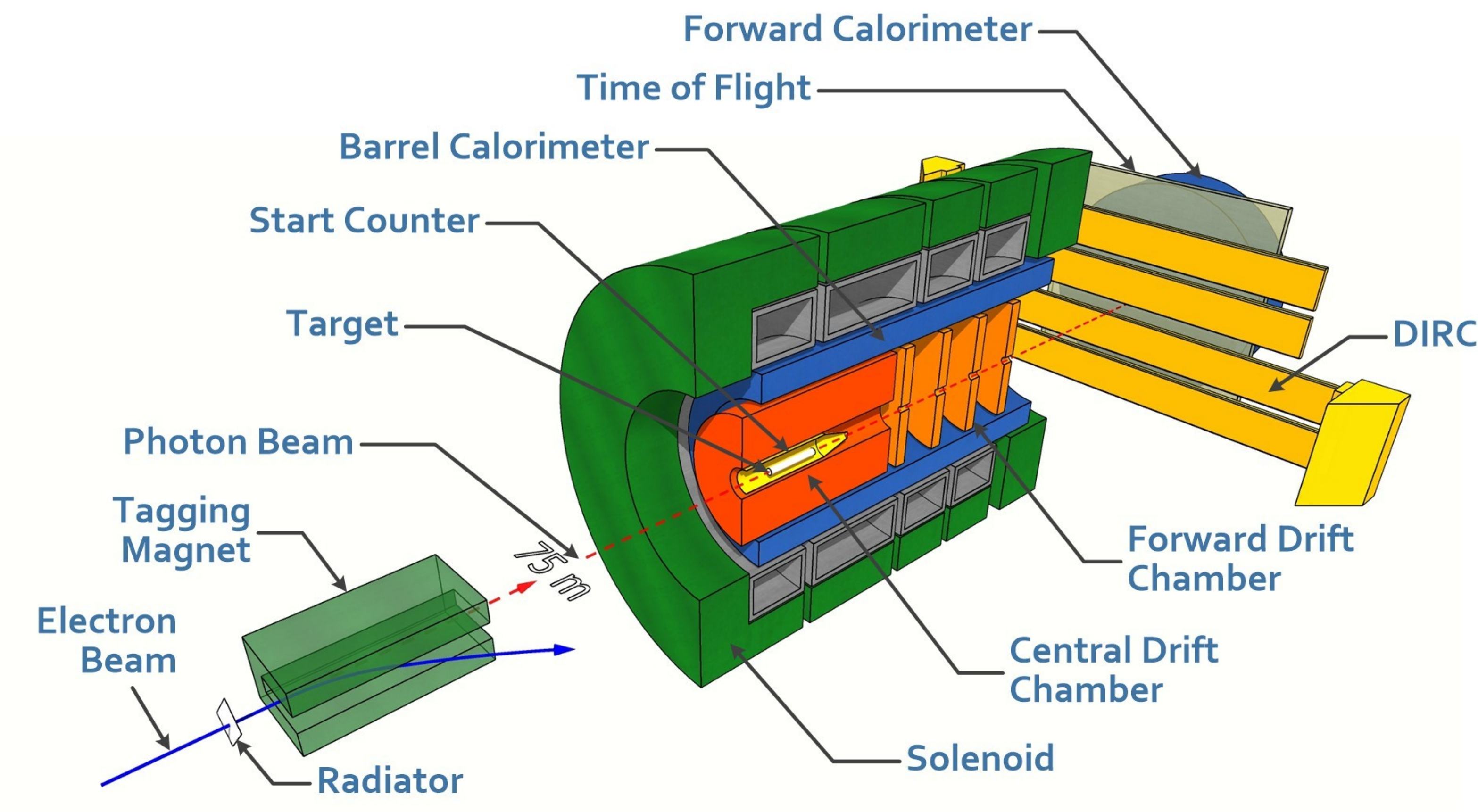}%
  \end{minipage}
  \caption{{\it Left, (a):} Spectrum of the photon beam produced with a 0.05~mm diamond radiator and the 5~mm collimator and measured with the Pair Spectrometer. compared with the spectrum obtained with an Aluminum radiator. 
  {\it Left, (b):} Linear polarization of the beam produced with a 0.05~mm diamond radiator and 5~mm collimator.
  {\it Right:} GlueX spectrometer layout~\cite{Adhikari:2020cvz}. The DIRC PID detector was installed for the GlueX-II experiment that started in 2020. GlueX-I had finished data taking in 2018.
    \label{fig:halld_spectrometer}
  }
\end{figure}

Hall D uses a beam of polarized real photons and a nearly hermetic magnetic spectrometer~\cite{Adhikari:2020cvz}, optimized for the GlueX experiment (see Section~\ref{sec:expmesons})- for the meson spectroscopy studies. The beamline layout is schematically shown in Fig.~\ref{fig:halld_beamline}.
The electron beam is extracted from CEBAF at 5.5 passes (up to 12~GeV) to the Tagger Hall, where it passes through a 0.02~-~0.05~mm (2-5$\cdot{}10^{-4}~X_0$) thick diamond radiator and is deflected by a dipole magnet to a beam dump. The electrons that radiated 25-96\% of the initial energy are deflected by the dipole magnet into the tagger scintillator detectors. Tagging provides a ${\approx}0.1\%$ energy resolution for the beam photons. 
The radiator is aligned in order to provide the coherent radiation peaking at about 75\% of the beam energy. A 3.4 or 5.0~mm collimator 75~m downstream of the radiator increases the fraction of coherently produced photons in the photon beam. The collimated beam is characterized using the Pair Spectrometer consisting of a thin converter, a dipole magnet, and scintillating hodoscopes. It also includes a Triplet Polarimeter which allows to measure the linear polarization of the beam. Downstream, the beam arrives at the GlueX spectrometer (shown in Fig.~\ref{fig:halld_spectrometer}).

The GlueX experiment uses a 30~cm long liquid Hydrogen target. Liquid Helium and solid targets have been used for other experiments. The trajectories of charged tracks are detected with the help of the Central and Forward Drift Chambers (CDC and FDC), while photons are measured in the Barrel and Forward Calorimeters (BCAL and FCAL). The Start Counter positioned around the target, the Time-of-Flight (TOF) counter, and BCAL provide the timing measurements, used for event selection and Particle Identification (PID). At the end of 2019 the spectrometer was augmented with a Cherenkov detector DIRC~\cite{Ali:2020erv} for using in the GlueX-II experiment. The pipeline front-end electronics provide both the event selection (trigger) and Data Acquisition (DAQ). The luminosity is limited by the accidental rate in the tagger counters, as well as the DAQ performance. The GlueX experiment uses a relatively open trigger based on the calorimeter signals, which provides a high efficiency for most of photoproduction processes at the photon beam energies $>4$~GeV. GlueX-II runs at a post-collimator photon flux in the coherent peak of ${\approx}$50~MHz, and the DAQ event rate of 80~kHz. The data are recorded at a rate of about 1~GB/s. The momentum resolution of the spectrometer is 1-5\% for charged particle trajectories, depending on the momentum and the polar angle. The calorimeters' resolution is ${\approx}6\%/\sqrt{E}\oplus{}3\%$. For exclusive reactions the kinematic fit uses the measured energy of the beam photon, which allows to improve the resolution considerably.

The approved JEF experiment (see Section~\ref{sec:jefexperiment}) requires an upgrade of the lead glass based FCAL . The central area of about 80$\times$80~cm$^2$ will be equipped with crystals, providing a better energy and spatial resolutions. Preparations for the upgrade are in progress. The approved GDH experiment (see Section~\ref{sec:spinstrfunct}) will require a polarized target. The KLF experiment (see Section~\ref{sec:expbaryons}) will require considerable changes in the beam line. The changes will be reversible and it is assumed that after the KLF experiment the photon beam will be restored.

%% file: AppendixB.tex
\section{Positron beams at CEBAF} 
\label{sec:appendixb}
To fully explore the potential of CEBAF, beams of polarized and unpolarized positrons with quality and modes of operation similar to the electron beam are highly desired. For many types of electron scattering experiments, conducting the measurement with a positron beam will expand the physics reach to include the lepton-charge difference, a domain that was not explored before at JLab. 
In 2018, the JLab Positron Working Group (https://wiki.jlab.org/pwg), with over 250 members from 75 institutions submitted a Letter of Intent titled {\it Physics with Positron Beams at Jefferson Lab 12 GeV} \cite{afanasev2019physics}, promoting a series of experiments using positron beams that could occupy CEBAF operations for more than 3 years. Recently, two proposals focusing on DVCS were conditionally approved based upon the availability of $>$100 nA polarized or $>$1 $\mu$A unpolarized beam intensities, respectively. 

In contrast to using a real-photon polarized gamma-ray source or exploiting self-polarization in a storage ring, a new technique for generating the positron polarization is considered. Referred to as PEPPo (Polarized Electrons for Polarized Positrons)~\cite{PhysRevLett.116.214801}, the polarization of an electron beam is transferred through the electromagnetic shower effect within a high-$Z$ target, initially by polarized bremsstrahlung and then by polarized $e^+$/$e^-$ pair creation~\cite{POTYLITSIN1997395}. This technique was experimentally tested at the CEBAF injector and demonstrated a very high transfer of polarization from an 8.2~MeV/$c$ electron beam to positrons with polarization approaching the electron beam polarization ($\sim$85\%). Most importantly, this technique is essentially independent of the initial electron beam energy, providing great flexibility in the choice of the electron injector to be used.

\begin{figure}[tbh]
 \centering
 \includegraphics[width=0.8\textwidth]{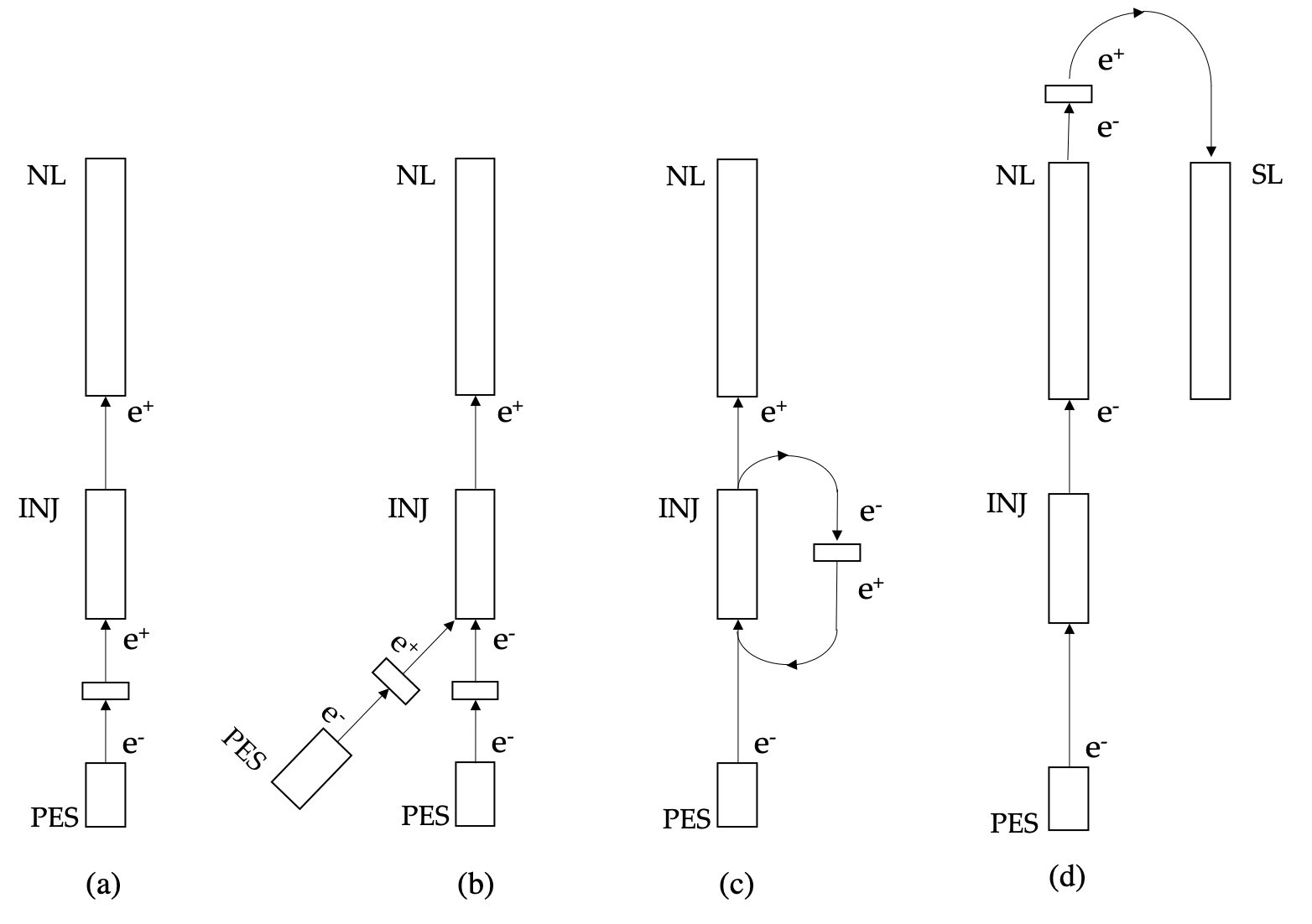}
 \caption{Four $e^+$ production schemes utilizing beams of different energies produced from a Polarized Electron Source (PES) are considered: (a) 10~MeV, (b) 123~MeV, (c) 123~MeV recycler, and (d) 1090~MeV options.}
 \label{AppB_Figure1}
\end{figure}

A conceptual design study is underway to develop continuous-wave positron beams for CEBAF. Four potential schemes are depicted in Fig.~\ref{AppB_Figure1}, each having a different “footprint” within the existing CEBAF configuration. Scheme “a” is attractive because 10~MeV is below the photo-neutron production threshold of most materials, however, suffers from the lowest yield and highest demand on the polarized electron source.  On the other hand, scheme “d” would provide a significantly higher $e^+$ yield, but likely has the largest radiological and construction footprint of the four. Rather, both schemes “b” or “c” are a better compromise between yield and footprint, each nominally generating 123 MeV positrons for acceleration to high energy, similar in function to how the electron beam at CEBAF is accelerated today.

\subsection{Polarized Positron Production Target}

\begin{figure}[tbh]
 \centering
 \includegraphics[width=0.8\textwidth]{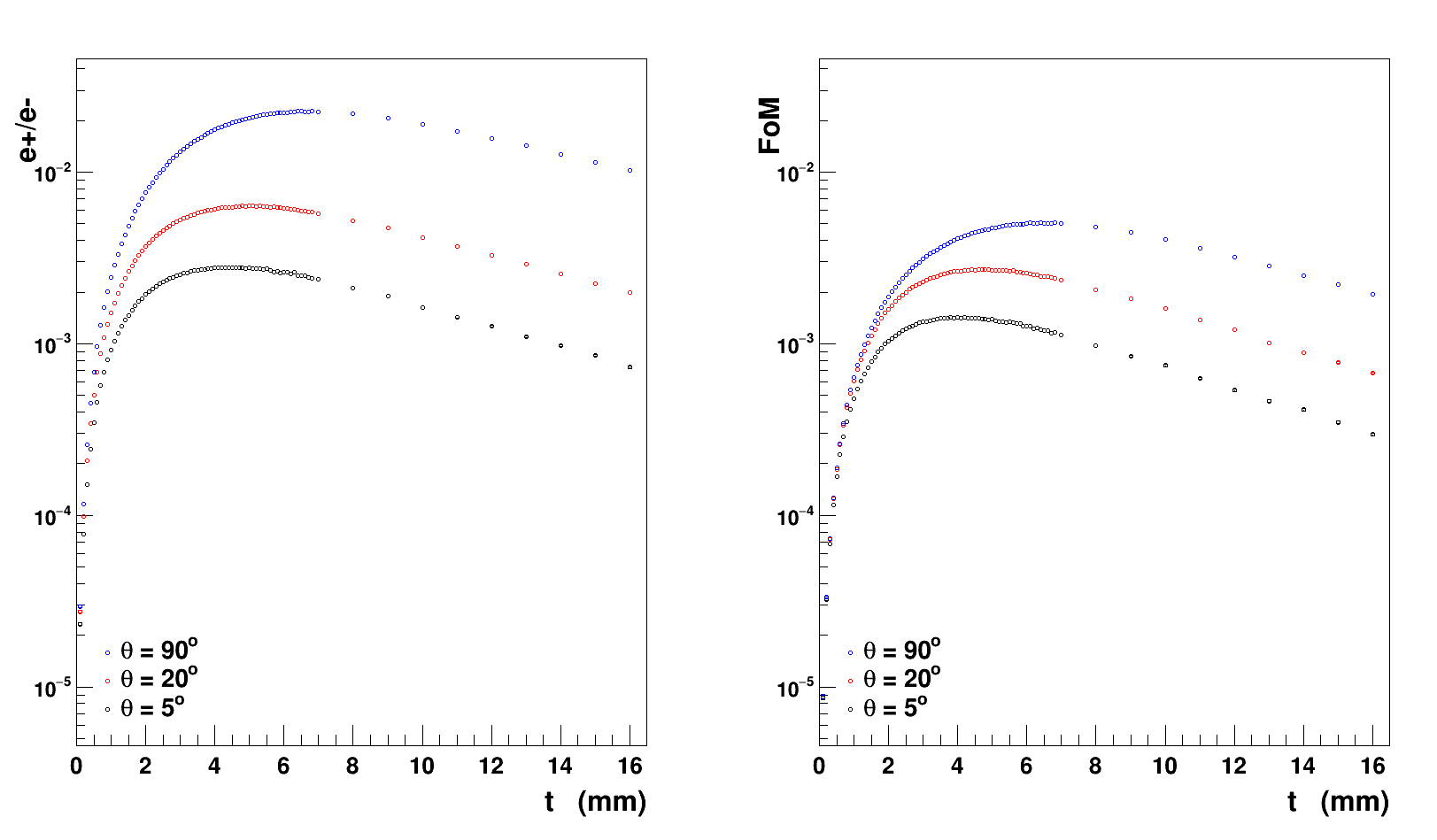}
 \caption{Geant4 simulation of $e^+$/$e^-$ yield (left) and FoM (right) versus target thickness.}
 \label{AppB_Figure2}
\end{figure}

Three important metrics of a polarized positron conversion target are the total Yield (number of useful positrons per incident beam electron), the average longitudinal polarization of the beam $\bar{P_{z}}$ and the statistical polarized Figure of Merit (FoM=Yield$\times \bar{P_{z}}^{2}$). The total yield of $e^+$/$e^-$ pairs depends on both the electron beam energy and the thickness of the target; the target should be thick enough to generate a healthy electromagnetic shower, yet not behave as an absorber.  Figure~\ref{AppB_Figure2} shows that a tungsten converter of 3-5 mm is optimal when the electron beam energy is 123 MeV, yielding approximately 1-10 useful $e^+$ for every 1000 incident $e^-$. Assuming approximately 1\% of the positrons are collected into the positron bunch for acceleration, an electron beam current of 1~mA is needed (i.e. 123 kW) for the production of a 10-100~nA polarized $e^+$ beam.

If a single conversion target is used as both the $e^-$ radiator and $e^+$ generator, it must have the ability to dissipate tens of kW of beam power and be closely integrated with the positron collection. This requires a sophisticated vacuum volume which includes the target, a vacuum beam line,  some form of a powerful solenoid magnetic matching device, and shielding of the surrounding area.  Both solid targets composed of titanium alloys which spin and liquid metal jet targets are presently being considered for the target material.  An initial engineering study for a 2-3 Tesla dc solenoid matching device is also planned.

An alternative, although less efficient option, is a two-target split design, which may benefit from the development of the so-called “Compact Photon Source” (CPS) presently underway at JLab \cite{ali2017workshop}. A CPS-type photon generator and beam power absorber could serve as the first stage, generating an intense partially polarized photon beam from a high-$Z$ radiator, which is then directed at a separate positron production target. The current JLab CPS is designed for GeV-range electron beams with beam power up to 30-60 kW, producing the photon beam and absorbing the bulk of the electron beam power inside the heavily shielded enclosure. Lower electron beam energies would allow for a more compact design, while also providing adequate radiation shielding.

\subsection{Polarized Electron Source for Positron Beams}

Highly spin-polarized electron beams are generated at CEBAF within a high voltage dc photogun via photo-emission from a superlattice GaAs/GaAsP semiconductor device, yielding an average longitudinal polarization $>$85\%.  A 123~MeV PEPPo driven positron source will require polarized electron beams of $\sim$1 mA (86 C/day) for a week or more, implying that photocathode charge lifetimes $>$1000~C are needed. Measurements at CEBAF have shown the operating lifetime of GaAs/GaAsP superlattice could be extended to 450~C or more~\cite{Proceedings:GramesPSTP2017} by increasing the beam emission area to dilute the effects of ion back-bombardment. A new photo-gun with a larger emission area $>$20 mm$^{2}$ may reasonably achieve the additional factor of two in lifetime enhancement needed.  Notably, the polarized electron source must also provide a high degree of spin polarization while providing high average currents.  Using a precision Mott scattering polarimeter, the spin polarization from a GaAs/GaAsP photocathode has been demonstrated to remain high ($>$85\%) and independent of beam intensity up to over 1~mA. 

\subsection{Positron Collection}
Because of the large emission angles and varied energies of pair-created positrons, a positron collection system is needed to reduce the emission angle and transform the energy profile into an acceptable range that is suitable for the CEBAF accelerator. The useful “slice” of the transverse-longitudinal 6-D $(x,x',y,y',E,t)$ distribution is selected by using a combination of transverse focusing magnets, charge separating dipoles, acceptance limiting collimators and RF cavities for bunching and acceleration.
\begin{figure}[tbh]
 \centering
 \includegraphics[width=1.0\textwidth]{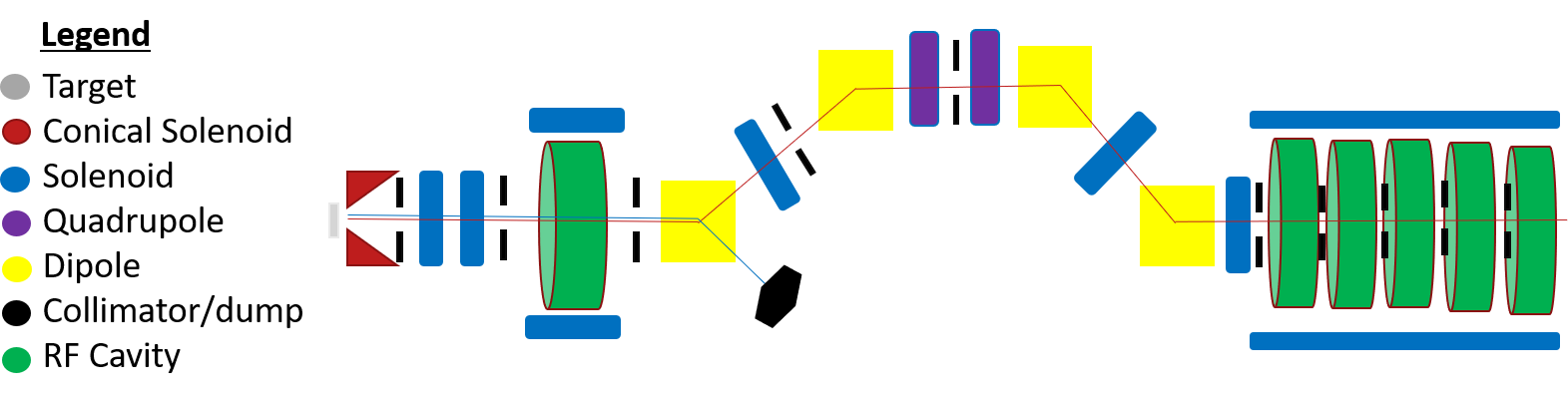}
 \caption{A positron collection scheme.}
 \label{AppB_Figure3}
\end{figure}

One such collection scheme is shown in Fig.~\ref{AppB_Figure3}.  A large-aperture high-field tapered solenoid is used first to collect the large energy-angle positron distribution, then solenoid magnets and collimators are used to define the transverse momentum acceptance, and an energy chirping RF cavity is used to later improve bunch compression. The first dipole magnet separates positrons from electrons which are dumped, while the dispersion defines the positron momentum acceptance further at an aperture. A dipole and quadrupole chicane compresses long bunches and accelerates positrons to 123 MeV, which are then injected into the first linac of CEBAF. Using the particle tracking code General Particle Tracer~\cite{Proceedings:Stefani2021}, augmented for spin tracking, the settings of the beam line components were optimized using nominal values of acceptance for the first CEBAF recirculating linac.   Beginning with the positron distribution produced by a 123~MeV electron beam interacting with a 4~mm thick tungsten target  Table~\ref{AppB_Table1} lists pamrameters of the  polarized and unpolarized positron beams optimized and approaching the CEBAF acceptance values.   This study suggests that a 1~mA polarized electron beam would produce a 66 nA positron beam with a polarization larger than 66\%, which is acceptable for acceleration in CEBAF to an energy as high as 12~GeV.   The unpolarized positron distribution is unacceptably large in this initial study, however, in on-going optics design activities the re-injection chicane of CEBAF is now being optimized to have a larger momentum acceptance to the north linac.

An alternative acceleration scheme employs a compact Fixed Field Alternating gradient (FFA) accelerator. An FFA-type beamline uses fixed field, combined function magnets (which can also be permanent magnets) with large energy acceptances to control the trajectories of the beam. FFAs have been demonstrated at several facilities, including EMMA~\cite{EMMA-Pos07} and CBETA~\cite{PhysRevLett.125.044803}. Initial studies collecting unpolarized positrons are considered in the energy range of 25-35 MeV (see Table~\ref{AppB_Table1}). This suggests an energy boost ratio of about 6:1 is needed to achieve 123 MeV, which is well within the abilities of even more demanding FFA schemes (e.g. 30:1 for FFA@CEBAF). A notable benefit of an FFA is that fewer accelerating cavities are required to achieve the required beam energy, mitigating unnecessary beam loss within SRF cavities.

\begin{table}
\centering
\begin{tabular}{l|l|l|l}
\hline
 Parameter       & Unpolarized $e^+$ &  Polarized $e^+$ &  CEBAF Acceptance\\   
\hline
Efficiency  & $1.98\times10^{-4}$ & $0.66\times10^{-4}$ &  \\
Mean Energy  & 123 MeV & 123 MeV  & 123 MeV \\
$\frac{\Delta P}{\bar{P}}$ & 10\% & 4\%  & 2\% \\
$\epsilon_n$  & 105 mm-mrad & 38 mm-mrad  & $<$40 mm-mrad \\
Bunch Length  & 12 ps & 2.4 ps & $<$4 ps \\
Transverse rms  & 1.8 mm & 1.2 mm & $<$3 mm \\
Polarization & $\sim8$\% & $\sim66$\% &\\
\hline
\end{tabular}
\caption{Polarized and unpolarized positron beam properties after the scheme shown in Fig.~\ref{AppB_Figure3} prior to injection into the CEBAF recirculating linacs.}
\label{AppB_Table1}
\end{table}

\subsection{12 GeV CEBAF Acceptance of $e^-$ and $e^+$ Beams}

The injection chicane properties (aperture and dispersion) control the CEBAF beam acceptance. These are normally configured for low emittance and low momentum spread beams, but the configuration has considerable flexibility. For instance, the chicane dispersion can be configured to accept up to 2\% momentum dispersion and for the low currents anticipated for positron operation, the acceptable RMS beam radius may be as high as several millimeters matching a normalized emittance acceptance of 40 mm-mrad for a beam energy of 120 MeV. Because this principal limiting aperture is very localized, it can be readily modified to increase its acceptance. After injection, the beam momentum is increased by a factor of 9 in the first linac (the north linac). The result of this strong adiabatic damping is that the momentum acceptance of the accelerator is dominated by the injection chicane. The transverse emittance is similarly strongly damped, and the injection chicane again provides a principal limitation.

\begin{figure}[tbh]
 \centering
 \includegraphics[width=1.0\textwidth]{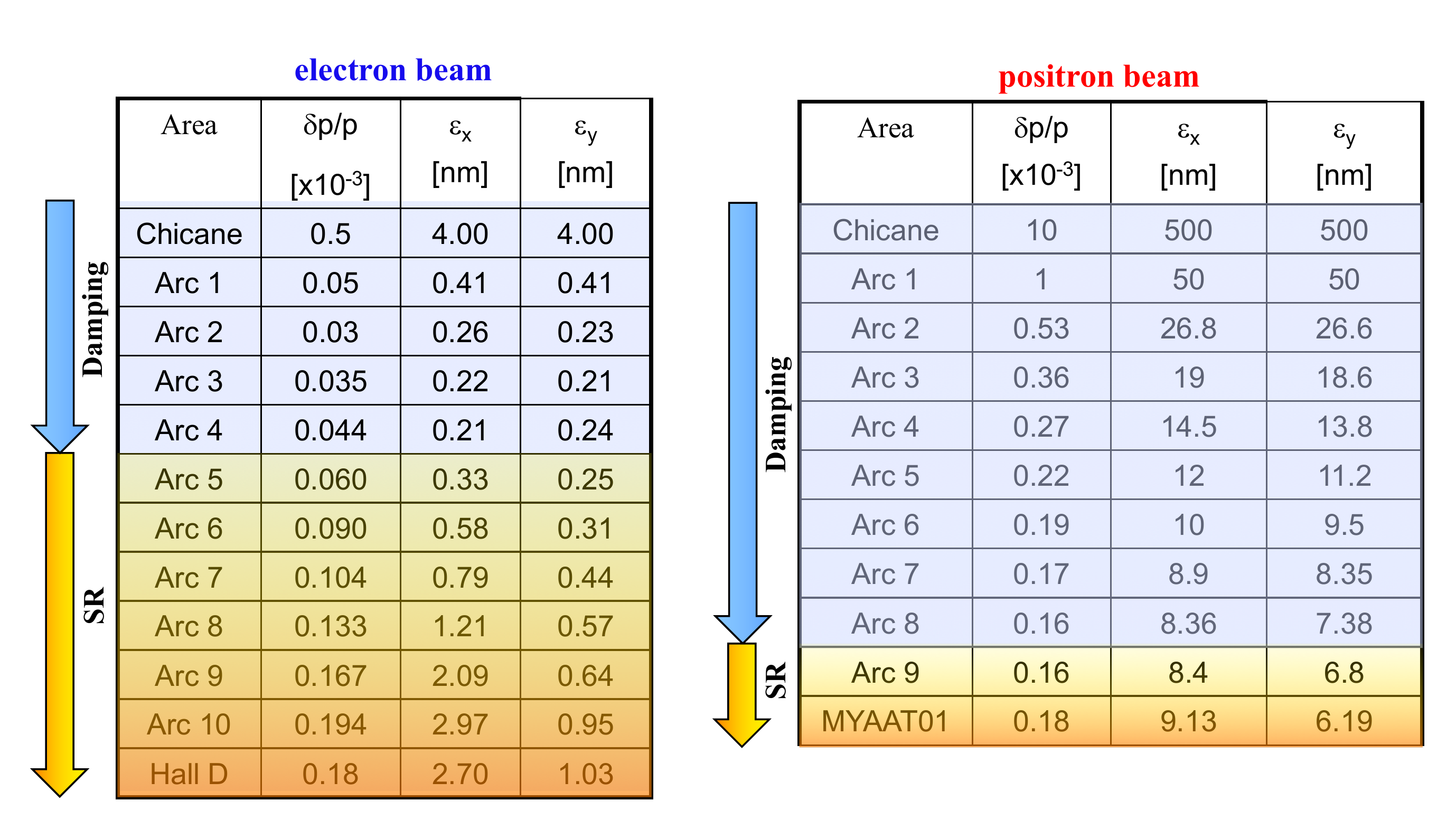}
 \caption{Comparison of simulated electron (left) and positron (right) beam properties \cite{Proceedings:RoblinJPOS2017}. The light blue arrow indicates the prominence of acceleration damping effects, and the light orange ones corresponds to the dominance of the effects of synchrotron radiations. The emittances are geometric and the momentum spread is RMS.}
 \label{AppB_Figure4}
\end{figure}

Estimated beam parameters are shown in Fig.~\ref{AppB_Figure4} comparing electron and positron beam momentum spread and geometric transverse emittance from the north linac entrance through to the experimental halls at 12~GeV. Two main regimes are affecting the beam properties: the acceleration damping within the CEBAF accelerating sections, and the synchrotron radiation in the recirculating arcs. As can be seen from Fig.~\ref{AppB_Figure4}, the dynamics of the momentum spread of electron beams is dominated by synchrotron radiation. On the other hand, positron beams essentially benefit from acceleration damping which results in the same momentum spread than electron beams, despite a much larger initial momentum spread. The large positron beam emittance at the injector entrance is also strongly reduced by acceleration effects which result in a final emittance 4-5 times larger than the electron beams, resulting in beam sizes only about twice as large.

\subsection{Reversing CEBAF Magnets for Positron Beams}

The CEBAF magnetic transport system contains over 2100 magnets whose polarity must be configured separately for either electron or positron operation.  Fortunately, more than 90\% ($>$1900) are correctors and quadrupole magnets with bipolar power supplies able to drive positive or negative amperage. These can be used for electron or positron beams without any changes in hardware. Some trim system magnets are used for 30~Hz modulation, position modulation and fast feedback. It is expected that these will remain the same as well.

All of the recirculation dipole magnets in CEBAF are powered by unipolar power supplies without polarity reversal capacity, namely 21 recirculation and dogleg units, 12 extraction units and 6 units in the end-station transport and beam dump lines.  All 39 units and their respective shunt modules to distribute current to magnet strings require an engineered solution for reversing the current and a firmware upgrade of the controls; the cost of such effort has been estimated and is not unreasonable. The four experimental halls have a small number of mixed bipolar and unipolar power supplies for their beamline magnets that would similarly need to be addressed. The only permanent dipole is in Hall D to prevent beam into the hall in case of the tagger magnet failure. This magnet may need to be rotated to accommodate positron beams. 
%achieve the same outcome. 

All CEBAF magnets would also need to provide reliable magnetic fields for either polarity.  The existing bipolar magnet field maps have been examined for polarity invariance as an internal self-consistency validation. In some low-field magnets the earth’s field is apparent, but only as a measurement offset. Generally, observations are consistent with no change in magnetic field performance with bipolar current operation.

The unipolar meter-scale recirculation dipoles have not been tested for a calibration change after polarity inversion. The relative size of effects associated with remnant fields (10-15 Gauss), sudden power supply trips which introduce uncontrolled flux as the dipole field collapses, and occasional adjustment or correction of hardware during the life span of the magnets suggest the magnet iron is magnetically “soft” enough such that field strengths in use in the CEBAF accelerator result in no persistent calibration shifts after field reversal and restoration.  The calibration curves are expected to be polarity-invariant, so that inverting the current will invert the field as long as the hysteresis cycle is respected.  In order to demonstrate this, typical magnets operated at CEBAF would require testing in a magnet measurement lab, for restoration of field after polarity inversion to quantify any small systematic effects.  Local nuclear magnetic resonance core field measurements would suffice for this purpose.

\subsection{Conclusions and Future Work}

The Jefferson Lab Positron Working Group has developed a world-class selection of experiments which may be accomplished using polarized (and unpolarized) positron beams with energies up to 12~GeV.  In response, a Laboratory Directed Research and Development (LDRD) project investigated the parameter space of a continuous-wave linac-driven source of polarized positrons for CEBAF.  That effort produced a layout, based upon the PEPPo approach and integrated with the existing accelerator, which when optimized delivers polarized positrons with an energy of 123 MeV and within the transverse and longitudinal acceptance injection of the two 12~GeV recirculating linacs.  The combination of momentum damping in the accelerating sections and emittance damping due to synchrotron radiation in the recirculation arcs suggests positron beams with energy spreads and geometric emittances within a factor of a few of those achieved with 12 GeV electron beams are possible.  A  review of the transport system suggests all of the magnets at CEBAF could operate in reversed polarity for positrons, once magnet tests are completed and engineered solutions are developed.  The state-of-the art in polarized electron sources and high power targets suggest there are no "show stopping" limitations for this scheme to be realized at Jefferson Lab, although significant design and engineering research and development remains. 

In conclusion, a new Positron Beams Research and Design working group has been established in the Accelerator Division at Jefferson Lab which includes colleagues from US and international national laboratories, universities, and industry.   The group is presently exploring the following topics:

\begin{itemize}
    \item Design of a compact milliAmpere polarized electron injector
    \item Technical description of a 123 MeV positron injection beam line
    \item Reconfiguration of the CEBAF re-injection chicane for increased momentum and emittance acceptance
    \item Design, development and testing of high power 10-100~kW targets for positron production
    \item Design of a high-field adiabatic matching solenoid magnet
    \item Spin rotator designs of 100 MeV to 12 GeV polarized positron beams
    \item Start to end CEBAF positron beam optics model and simulation
    \item Evaluating positron production schemes compatible with simultaneous electron beam operations
\end{itemize}
The working group is planning to publish a technical report by the end of 2022, and would like to begin the engineering of a prototype target and positron collection beam line by 2023.

%% file: AppendixC.tex
\section{CEBAF Energy Upgrade}
\label{sec:appendixc}

Extending the energy reach of the CEBAF accelerator up to 24 GeV should be possible within the existing tunnel layout~\cite{harwood:2001, leemann1}. Such an increase can be achieved either by increasing the overall energy per pass, increasing the number of recirculations in the accelerator, or some combination of both. Other considerations in the overall upgrade path are related to reuse of existing 12 GeV accelerator components. Preliminary studies have produced several options that could be considered for further development~\cite{mckeown:2020}. Here, the option of adding two, high energy acceptance Fixed Field Alternating Gradient (FFA) arcs to 12 GeV CEBAF is discussed~\cite {IPAC21:2018bde}.

Applying this idea to achieve significantly more beam recirculations in CEBAF to reach higher beam energy is attractive for several reasons~\cite{mckeown:2020}. First is that multiple beam energies can be confined and recirculated in the same, albeit more complicated, FFA beam line. A second reason this approach is attractive is that the existing CEBAF linacs, perhaps marginally enhanced in energy, will be all that is needed to achieve the higher energy. A third reason this approach may be attractive is that the existing ARC beamlines for the first four beam passes could remain part of the upgraded accelerator; the FFA lines would occupy roughly the area where the existing ARC 9 lies in the east region of CEBAF, and the other FFA arc would reside near the present ARC A.

\subsection{Modern Fixed Field Alternating Gradient (FFA) Technology }
 
 Initially deployed in cyclotrons in Japan for accelerating heavy particles, two small-scale electron accelerators have been built and operated successfully using the FFA idea. In the UK, the multipass EMMA~\cite{machida:2012} accelerator at Daresbury Laboratory successfully contained and accelerated electron beam through a factor of 1.7 in energy in a single beamline with non-varying magnetic field. More pertinent to a CEBAF-like accelerator, the Cornell-BNL ERL Test Accelerator (CBETA)~\cite{CBETA-DR:2018bde} recently demonstrated eight-pass beam recirculation with energy recovery (four accelerating beam passes and four decelerating beam passes) through a complete TESLA-style SRF
cryomodule~\cite{CBETA-PRL:2018bde}. The beam energies on the different beam passes were 42, 78, 114, and 150 MeV (energies spanning nearly a factor of 4) with a single beamline configured consisting of fixed-field permanent magnets. CBETA demonstrates experimentally simultaneous transport of multiple  beams and the wide dynamic energy range possible in the FFA transport system beyond the requirements for an upgrade to CEBAF.

\begin{figure}[tbh]
  \centering
  \includegraphics[width=0.6\textwidth]{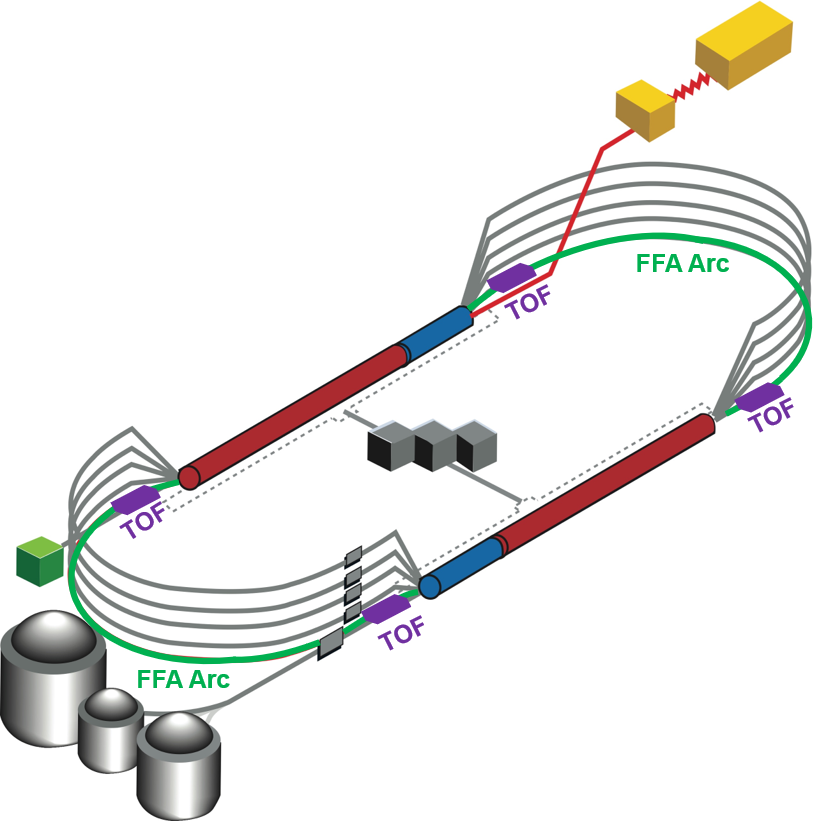}
  \caption{Schematic layout of FFA CEBAF.}
  \label{fig:FFA_CEBAF}
\end{figure}
\subsection{Energy 'Doubling' Scheme}

Encouraged by this recent success, a proposal was formulated to increase CEBAF energy from the present 12~GeV to 20-24~GeV by replacing the highest-energy arcs, ARC 9 and Arc A, with a pair of Fixed Field Alternating-Gradient (FFA) arcs, with very large momentum acceptance to recirculate the beam for 6-7 additional passes. Such a scheme would nearly double the energy using the existing CEBAF SRF cavity system. The new pair of arcs configured with FFA lattice would support simultaneous transport of 6-7 passes with energies spanning a factor of two. This wide energy bandwidth was achieved using the non-scaling FFA principle~\cite{Trbojevic-NSFFA:2018bde} implemented with Halbach-derived permanent magnets~\cite{CBETAmagnets:2018bde}.  CBETA's maximum energy was 150~MeV, whereas CEBAF upgrades plan to extend this technology to higher beam energies. 
The design schematic is shown in Fig.~\ref{fig:FFA_CEBAF} In the accelerator passes 1-4 would be accomplished through the current 12~GeV CEBAF. Passes 5-10 (six passes) would be facilitated by constructing two 'CBETA-like' beam-lines indicated schematically as 'green' arcs with 'purple' Time of Flight (TOF) chicanes in Fig.~\ref{fig:FFA_CEBAF}. 
\subsection{Multi-pass Linac Optics}
\begin{figure}[ht]
  \centering
  \includegraphics[width=0.6\columnwidth]{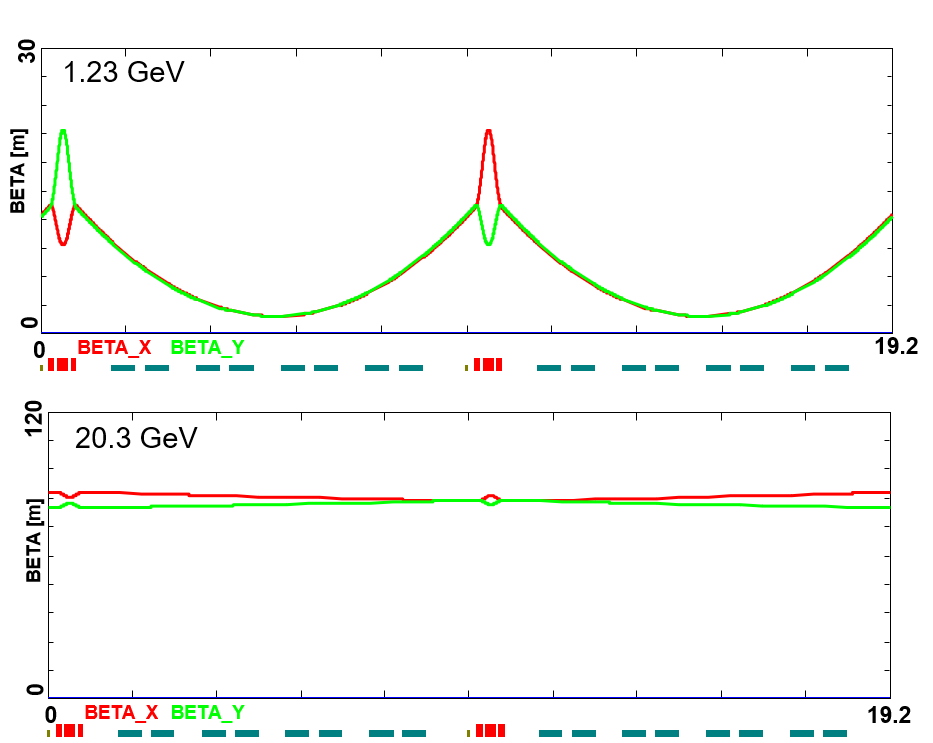}
  \caption{'Twin-Cell' periodic triplet lattice at the initial and final linac passes : 1.23 GeV and 20.3 GeV. Initial triplets, configured with 45 Tesla/m quads, are scaled with increasing momentum along the linac.}
  \label{fig:Twin_Cell}
\end{figure}

One of the challenges of  the multi-pass (10+) linac optics is to provide uniform focusing in a vast range of energies, using fixed field lattice. Here, we configured a building block of of linac optics as a sequence of two triplet cells with reversed quad polarities flanking two cryomodules, as illustrated in Fig.~\ref{fig:Twin_Cell}, with a stable periodic solution covering energy ratio 1:18.
\subsection{Proof-of-principle FFA Arc Cell}

\begin{figure}[htb]
  \centering
  \includegraphics[width=0.7\columnwidth]{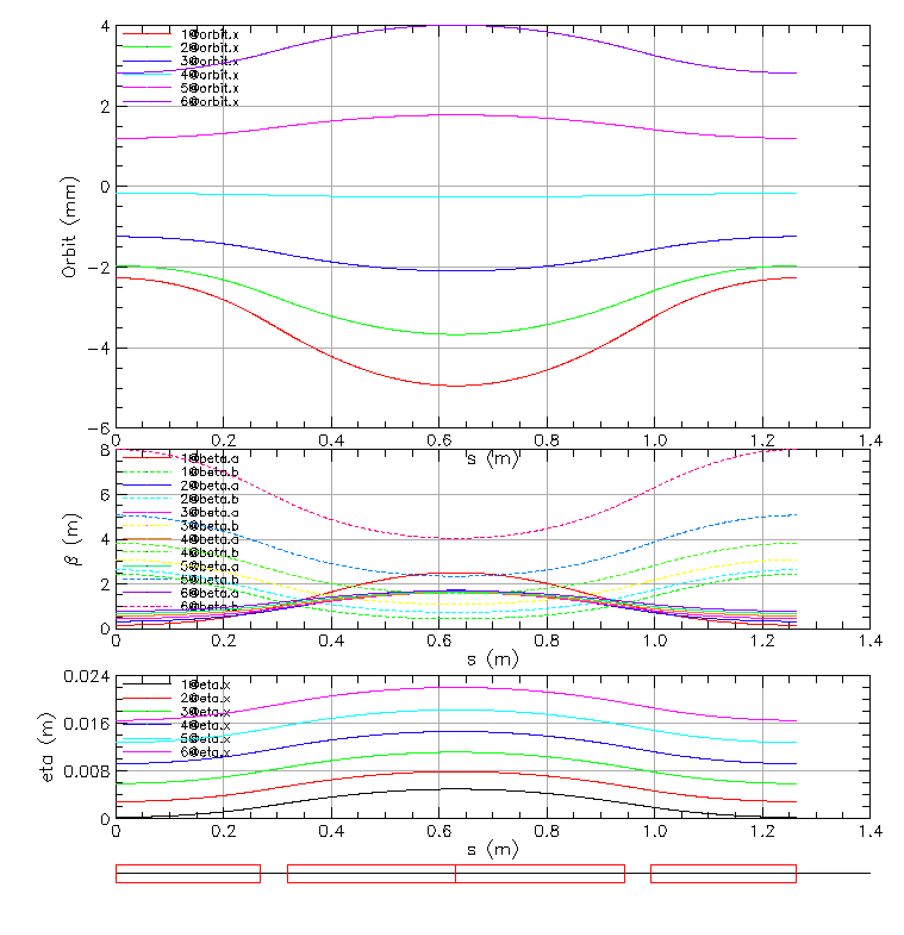}
  \caption{MAD-X optical functions for the 10-22~GeV energy doubler FFA arc cell.}
  \label{fig:22GeV-arc-cell}
\end{figure}

The parameters of the main arc cell for the energy doubler are given in Table~\ref{tab:22GeV-arc-cell}.  This lattice uses very high gradients and the 10-22~GeV beams are all confined to a region $-5~\mbox{mm}<x<4~\mbox{mm}$. The orbits and optics of the unit cell for the different energies are shown in Fig.~\ref{fig:22GeV-arc-cell}.
\begin{table}[!hbt]
  \centering
  \small
  \begin{tabular}{lcccc} 
  \hline
  %\toprule
  Element & Length & Angle & Dipole & Gradient \\
   & [m] & [$^\circ$] & [T] & [T/m] \\
  \hline\hline
  %\midrule
  BF & 0.625 & 0.5 & 0.681 & 250.91 \\
  O & 0.05 & 0 & & \\
  BD & 0.5382 & 0.5 & 0.941 & -233.13 \\
  O & 0.05 & 0 & & \\
  \hline
 \\
  %\bottomrule
  \end{tabular}
  \caption{Energy doubler FFA arc cell.}
  \label{tab:22GeV-arc-cell}
\end{table}
\subsection{Energy Doubler FFA Racetrack}
\begin{figure}[htb]
  \centering
  \includegraphics[width=0.8\columnwidth]{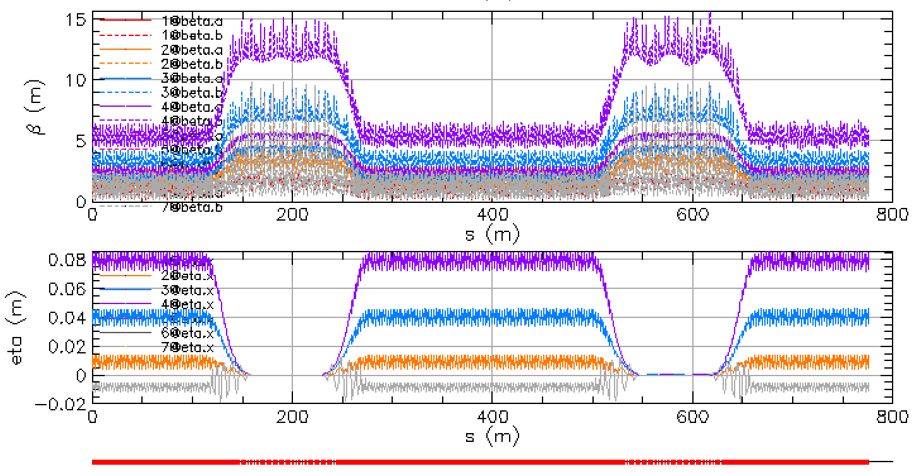}
  \caption{MAD-X optical functions for the entire 10-22~GeV energy doubler FFA racetrack lattice.}
  \label{fig:22GeV-racetrack}
\end{figure}

Figure~\ref{fig:22GeV-racetrack} shows how the beta functions are expanded and the dispersion matched adiabatically to zero in the straight sections of the racetrack-shaped lattice.  The larger beta functions allow longer cells compatible with the higher-beta optics in the main CEBAF linacs.
\subsection{1.2 GeV FFA Booster Injector} %\NoCaseChange{1.2 GeV} FFA Booster Injector} 
The current CEBAF facility is configured with a 123~MeV injector feeding into a racetrack recirculating linear accelerator (RLA) with a 1090~MeV linac on each side.  The 123~MeV minimum makes optical matching in the first linac virtually impossible due to extremely high energy span ratio (1:175). Thus, it is proposed to replace the first pass by a new FFA-based booster, outputting an energy of $123+1090=1213$~MeV into the South Linac.

\begin{figure}[htb]
  \centering
  \includegraphics[width=0.8\columnwidth]{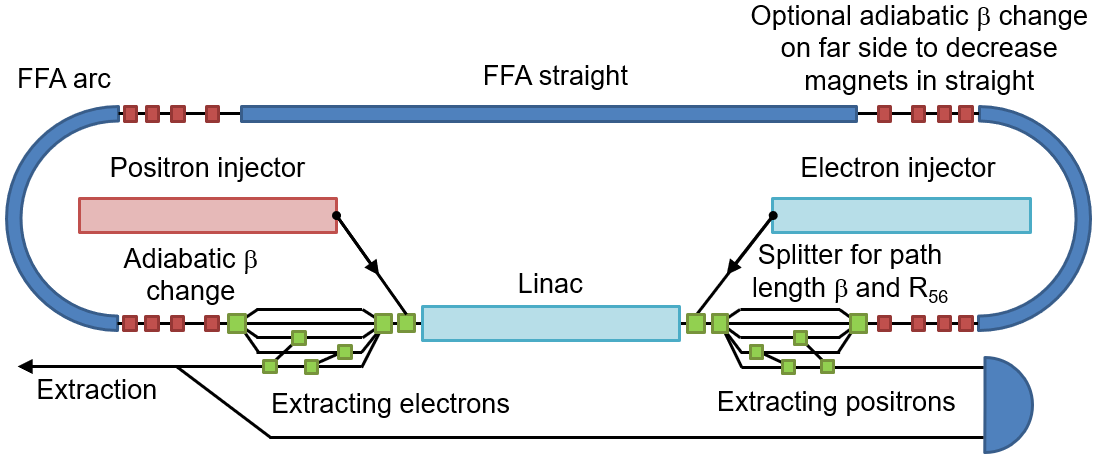}
  \caption{Proposed 1.2~GeV FFA booster for CEBAF.}
  \label{fig:booster-schem}
\end{figure}

This booster will have injector linacs of energy up to $1213/6=202$~MeV, delivering either electrons or positrons (for circulation in the opposite direction).  The booster resembles CBETA, with a linac on one side surrounded by splitter lines and an FFA return loop (Fig.~\ref{fig:booster-schem}).  The booster linac operates at the same energy as the injector, meaning five passes in the booster linac (four passes in the FFA return loop) produces 1213~MeV.  Energy tunability from 50 to 100\% is produced by reducing the number of booster linac passes to three and the energies of both linacs to $1213/8=152$~MeV.  

\subsection{Synchrotron Radiation Effects on Beam Quality}
 Staying within CEBAF footprint, while transporting high energy beams 10-24 GeV) calls for increase of the bend radius at the arc dipoles (packing factor of  the FFA arcs increased to about 87.6\%), to suppress adverse effects of the synchrotron radiation on beam quality. Arc optics was designed to ease individual adjustment of momentum compaction and the horizontal emittance  dispersion, $H$, in each arc.
Tab.~\ref{tab:SREmittance} lists arc-by-arc dilution of the transverse, $\Delta \epsilon$, and longitudinal, $\Delta \sigma_{\frac{\Delta E}{E}}$, emittance due to quantum excitations calculated using analytic formulas:  
\begin{equation}
  \Delta E = \frac{2 \pi}{3} r_0 ~mc^2~ \frac{\gamma^4}{\rho}\,
  \label{eq:Emit_dil_1}
\end{equation}
\begin{equation}
  \Delta \epsilon_N = \frac{2 \pi}{3} C_q r_0 <H> \frac{\gamma^6}{\rho^2}\,,
  \label{eq:Emit_dil_2}
\end{equation}
\begin{equation}
  \frac{\Delta \epsilon_E^2}{E^2} = \frac{2 \pi}{3} C_q r_0~ \frac{\gamma^5}{\rho^2}\,,
  \label{eq:Emit_dil_3}
\end{equation}

Here, $\Delta \epsilon^2_E$ is an increment of energy square variance, $r_0$ is the classical electron radius, $\gamma$ is the Lorentz boost and $C_q = \frac{55}{32 \sqrt{3}} \frac{\hbar}{m c} \approx 3.832 \cdot 10^{-13}$~m for electrons (or positrons).
The horizontal emittance dispersion in Eq.~\ref{eq:Emit_dil_2},  is given by the following formula: $H = (1+\alpha^2)/\beta \cdot D^2 + 2 \alpha \ D D' + \beta \cdot D'^2$ where $D, D'$ are the bending plane dispersion and its derivative, with averaging over bends defined as: $<...>~=~\frac{1}{\pi}\int_{bends}...~d\theta$.
\begin{table}[!hbt]
  \centering
  \small
  \begin{tabular}{lccccc} 
  \hline
  %\toprule
  Beamline & Beam Energy & $\rho$ & $\Delta E$  & $\Delta \epsilon^x_N$ & $\Delta \sigma_{\frac{\Delta E}{E}}$  \\
    & [GeV] & [m] & [MeV]&  [mm~mrad] &  [\%]\\
  \hline\hline
  %\midrule
  Arc 1 & 1.2 & 5.1 & 0.02 & 0.003 & 0.0003\\
  ... & ... & ... & ... & ... & ...\\
  Arc 8 & 8.8 & 30.6 & 9 & 12 & 0.022\\
  FFA 9 & 9.91 & 70.6 & 6 & 13 & 0.026\\
  ... & ... & ... & ... & ... & ...\\
  FFA 19 & 20.44 & 70.6 & 109 & 37 & 0.15\\
  FFA 20 & 21.42 & 70.6 & 132 & 47 & 0.17\\
  FFA 21 & 22.38 & 70.6 & 157 & 60 & 0.020\\
  FFA 22 & 23.31 & 70.6 & 185 & 76 & 0.023\\
  \hline
  %\bottomrule
  \end{tabular}
  \caption{Energy loss and cumulative emittance dilution (horizontal and longitudinal) due to synchroton radiation at the end of selected  180$^{\circ}$ arcs (not including Spreaders, Recombiners and Doglegs). Here, $\Delta \sigma_{\frac{\Delta E}{E}} = \sqrt{\frac{\Delta \epsilon_E^2}{E^2}}$}.
  \label{tab:SREmittance}
\end{table}
An 11.5 pass, 24~GeV design would deliver a normalized emittance of  \SI{76}{mm~mrad} with a relative energy spread of $2.3 \cdot 10^{-3}$ and each electron looses 976 MeV traversing the entire accelerator. Additional recirculations become less effective due to increasingly large energy loss from synchrotron radiation.
\subsection{Conclusions and Future Work}
It appears possible to roughly double the energy of CEBAF by implementing an FFA in place of the highest pass of the present configuration~\cite{IPAC21:2018bde}. Our work expands upon CBETA efforts, and shows a promising possible way forward for CEBAF after the 12 GeV era. Initial studies into the beam dynamics, possible machine layouts, and magnet designs have been made and paint a positive picture. However, considerable work remains to validate this concept, including, but not limited to, full start-to-end beam dynamics simulations, detailed magnet designs, diagnostics, controls, and detailed engineering. Furthermore, positron acceleration in the CEBAF machine must be investigated in detail.     

A Working Group has been established including Jefferson Lab, Cornell, and BNL scientists to further develop this concept. At present the group is pursuing:
\begin{itemize}
  \item Development of  a prototype arc cell based on permanent magnet technology spearheaded by CBETA
  \item Adiabatic arc architecture with gradually increased bending to alleviate complexity of a horizontal time-of-flight switchyard to be configured between current CEBAF Spr/Rec and the new FFA arcs
  \item New design of multi-pass linac optics based on triplet focusing (140 deg. phase advance per cell, at the lowest energy pass) implemented with permanent quads.
  \item New Injector design (1.2 GeV) including three C-75 cryomodules in a 5-pass FFA re-circulator
\end{itemize}
The group is targeting the end of 2022 for the completion of this stage of the design’s development work.

%% file: AppendixD.tex
\section{Computation for NP} 
\label{sec:appendixd}
Similar to other scientific disciplines, Nuclear Physics is on the threshold of a new paradigm of discovery with the establishment of new experimental facilities, advances in theory, and expanded access to computational resources, and application of data science techniques.  

In this appendix, we cover the importance of data curation for analysis, particularly for AI/ML techniques, Streaming Readout and Data Science.  Improvements needed in computing infrastructure and services are covered, as we complete the transition of the JLab Scientific Computing Model to seamlessly utilize distributed computing including HPC centers and a possible national data infrastructure.  This will be accomplished by the use of common and community tools for scaling and interoperability. We also note the necessity of developing workflows to enable theory-experiment integration to support some of the scientific activities outlined in this document.

%Technical advances:\\ \\
\subsection{Data Curation}
The JLab experimental and operational data sets are invaluable and must be curated to enable new research in data science techniques. To produce robust and reproducible data analytics for production systems (detector, accelerator, etc.), curated data set and a model repository is a requirement.  Preparing a collection of benchmark data sets with detailed Datasheets would allow JLab researchers to study new techniques to improve scientific discovery. Similarly, developing a model repository with detailed provenance would allow us to ensure reproducibility, models reuse, and define ``golden model''.  All production or published research must be captured in a online repository to track changes in the source code during software development on all essential efforts. The repositories should have unit tests and continuous integration with the ability to easily rollback to a previous build. In addition, the computing infrastructure should be linked to automate the builds of new containers

%%FAIR data??

\subsection{Streaming Readout}

Within the experimental computing program, recent efforts have been focused on the need to adapt the computing model to better accommodate to changes in data acquisition (DAQ)/readout schemes and advanced analytics techniques. A streaming architecture will simplify the readout schemes from the standard “event building” as well as provide more flexibility both to accommodate detector systems that have large disparities in readout times, and to accommodate more complicated geometries for detectors. Streaming readout architectures are in advanced prototypes for 12 GeV experiments and are integral to the design of SOLID. The streaming architecture for DAQ will have implications for offline processing with the possibility to move some types of processing closer to the detectors by running, for example, AI/ML algorithms in FPGAs.

\subsection{Infrastructure}
The full scientific process is a meta-workflow that connects individual data processing workflows from theory and experiment.
In order to support the growing computing needs for the JLab science mission, infrastructures that support integrated workflows and data production will need to be developed to augment or replace existing functionality to provide maximal flexibility and scalability. In order to process data for the different Halls in a timely manner, ideally, the data would be automatically transferred to onsite resources, HTC/HPC cluster(s) and/or to remote resources for processing and analysis. The computing systems must be designed to handle data volumes at scale and the ability to dynamically integrate a variety of geographically distributed computing resources. This infrastructure based on common tools has been developed by the high energy physics community and is in use for multiple experiments.
The infrastructure could be design to seamlessly integrate with EIC workflows and future workflows on High Performance Data Facility (HPDF).

Virtualization can address the different software and OS requirements from the experimental Hall.  For example, Docker containers can be used to build new instances of the development and production software and environment stack. The production containers would then be converted to Singularity images and available to be used across all computing platforms (laptops, HPC  and HTC cluster, GPU farms, etc.).

\subsection{Computing and services}
The heart of this distributed computing model will be the virtualized service cluster which provides all the central services for local and geographically distributed computing infrastructure. This computing infrastructure would be developed using containerized components and orchestration. We propose using an open-source container orchestration system, such as  Kubernetes, to automate application deployment, scaling, and management. %We choose to back our cluster with the containerd container runtime, which forms the core runtime engine of the popular Docker tool. Some of the key motivations for using containers and container orchestration was the ability to quickly reinstall machines and workloads. 
Additionally, it enables high availability by easily and dynamically moving containers off problematic nodes to healthy ones. A tool like Kubernetes allows for compute resource management by dynamically adding and/or removing services as required by the current computing load. The use of containers allows to decouple the build, test, and deploy workflows enabling more reliable deployment. They also provide a clean solution for conflicting dependencies when running numerous services on the same node. 
%Kubernetes enables a light weight layer to maintain multiple nodes and many containers.
In additional to traditional computing workflows, the data science effort would also benefit from a virtual/modular workflow that can expand as the compute requirements grows. As such, the data science effort can also leverage the aforementioned tools and expand by using Kubeflow.
%\subsubsection{Workflows and automation}
%\subsubsection{Reproducibility and re-usability}
%\subsubsection{Hardware}
%CPU, GPU, TPU, FPGAs, and Neuromorphic Computing

\subsection{Data Science - Methods, Algorithms, and Applications}
%New advancements in machine learning and artificial intelligence could provide a better accelerator and detector operational efficient. The use of machine learning techniques to identify anomalies in an accelerator and detector systems has show promising results [expand and provide references]. Augmenting these techniques with uncertainty quantification would provide a robust and actionable techniques with a quantifiable level of confidence.      
%The use of Bayesian optimization, evolutionary algorithms, and reinforcement learning has shown promising results for design and control optimization problem [expand and \cite{}].
%Physics informed machine learning can provide enhance techniques to leverage the power of deep learning while still incorporating know physics equations [expand and \cite{}]. 
%Online ML model on FPGA can be use to improve data efficiency and targeted "trigger" [expand and \cite{}].\\ \\
The use of machine learning (ML) and artificial intelligence (AI) has significantly grown in the past 5 years and new techniques are continuously being develop to address scientific challenges. In order to provide solutions for the NP community, JLab will work on addressing the basic research needs defined in the DOE workshops and town-halls. 
Specifically, JLab will focus on developing and integrating new capabilities in ML-based uncertainty quantification techniques. The inclusion of UQ into the models is critical for study in anomaly detection, particle identification, etc. Augmenting these studies with UQ would provide robust and actionable techniques with a quantifiable level of confidence. In addition, an area of growth and potential large operational efficiency improvements lie in advancements in safe and robust optimization research for design and control. The use of Bayesian optimization, evolutionary algorithms, and reinforcement learning has shown promising results for the optimization problem. Another unique area of research that JLab can lead is in the use of domain aware ML for NP. Developing a multi-disciplinary team of experimentalists, theorists, and data scientists can yield new solutions in which the solution would leverage known physics equations and fold them into the ML model to provide a reliable solution requiring fewer data points.

\begin{figure}[tbh]
  \centering
  \includegraphics[width=1.0\textwidth]{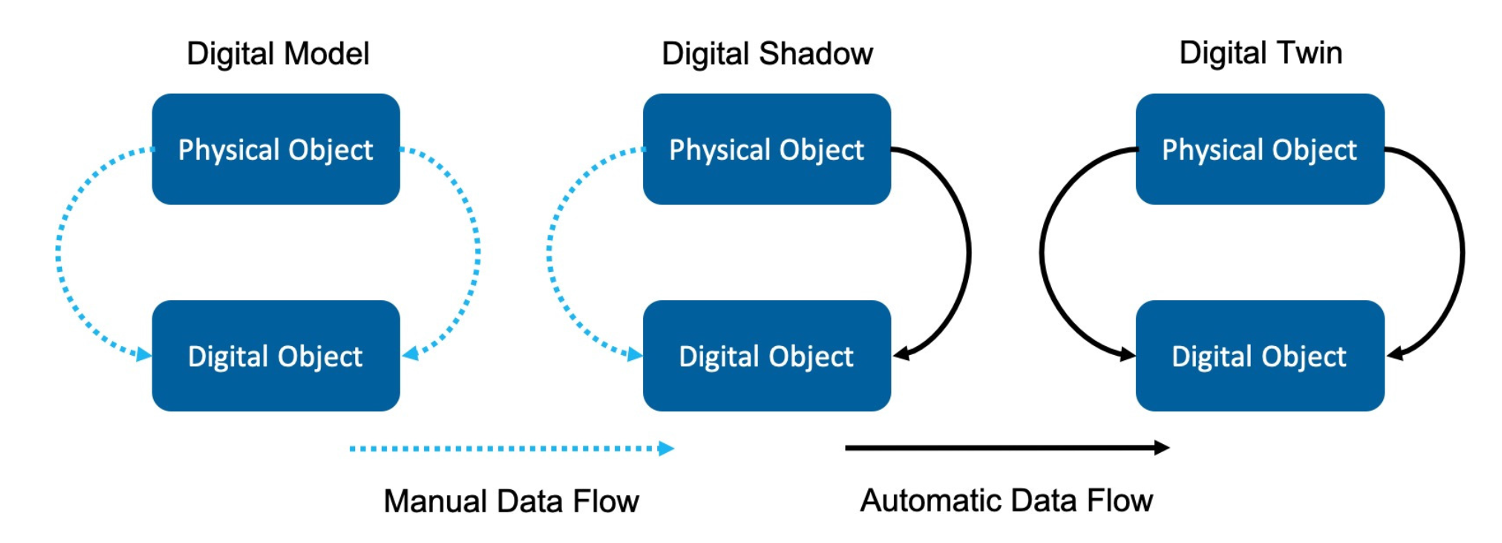}
  \caption{Schematic diagram of digital twin~\cite{DT9103025}.}
  \label{DigitalTwin}
\end{figure}

The recent work in the experimental Halls, specifically Halls B and D, have shown promising results in areas such as anomaly detection and reconstruction. 
Hall B has made several advancements in AI/ML to improve the performance of the tracking system and triggering systems. 
Additionally, there are some ML efforts with strong overlap between the experimentalist and the theorists. 
The majority of these machine learning tools are well motivated and developed using manual and sequential techniques, however, are limited to a select few Halls. 
In order to improve on these results, the use of leading edge techniques in ML should be applied along with the inclusion of uncertainty quantification and physics informed models (when applicable). 
Unfortunately, these offline analyses are typically slow and include a lot of manual intervention.
In fact, the majority of DOE efforts in AI/ML are restricted to one-directional manual data flow from the physical object to an offline analysis, as shown in Figure \ref{DigitalTwin}.
Although these studies can yield great results, they lack, in their implementation, the ability to perform these tasks continuously in near real-time.
Continuous evaluation of the analytical results and model predictions is required in order to determine if the model is still valid as the system evolves.
In order to elevate these efforts, a common infrastructure is needed that should include at its core a fully realized digital twin model. 
A digital twin is a bi-directional data exchange between a physical object and its digital representation, commonly known as a digital object. 
The physical and digital objects share a common data repository, ensuring that the data used by either is identical. With a fully realized digital twin, in-situ analysis can be conducted using the digital object without impacting the physical object, for example:
\begin{enumerate}
    \item Statistical and causal analysis
    \item Anomaly detection and classification
    \item Forecasting component fatigue and failures
    \item Continuous calibration and optimization
\end{enumerate}
%Figure \ref{DigitalTwinAndAna}, illustrate a high level workflow that include the ingest of new data from the physical object to the digital object. Various %traditional and machine learning based analytic are them performed to provide continuous information.
%\begin{figure}[tbh]
%  \centering
%  \includegraphics[width=1.0\textwidth]{Figures/DTandAna.png}
%  \caption{Schematic diagram of digital twin with various analytical tools.}
%  \label{DigitalTwinAndAna}
%\end{figure}
Several of the recently funded Scientific User Facility projects are developing analytical tools used to monitor, diagnose, and optimize these facilities.
However, they are currently not addressing the issue regarding long term usability of these tools.
Jefferson Lab is in a good position to incorporate these techniques into the digital twin framework, given the upcoming experiments and computing facilities.
Needless to say, the digital twin framework can be extended to include other key systems within the complex (accelerator, facilities, etc.)
As such, the digital twin framework could eventually accommodate all critical systems that require real time monitoring and analysis/optimization.
The framework would need to be modular in order to easily include new analytical tools as new techniques are developed.
In order to support a digital twin framework, a computing infrastructure will be required to efficiently move the data from the experimental Halls to the digital twin framework for near real-time analytic. 
In fact, this framework could provide an ideal use case for the anticipated High Performance Data Facility hosted at Jefferson lab.

%% file: AppendixE.tex
\clearpage
\section{Approved 12 GeV Era Experiments to Date (August 2021)}
\label{sec:appendixe}

\centerline{\bf Approved Experiments Proposed During 2006--2009}
\begin{tabular}{ll}
Experiment & Title\\
\hline
E12-06-101 & Measurement of the Charged Pion Form Factor to High $Q^2$\\
E12-06-102 & Mapping the Spectrum of Light Quark Mesons and Gluonic Excitations\\
& with Linearly Polarized Photons\\
E12-06-104 & Measurement of the Ratio R=sigmaL/sigmaT in Semi-Inclusive Deep-Inelastic Scattering\\
E12-06-105 & Inclusive Scattering from Nuclei at $x >1$ in the quasielastic and deeply inelastic regimes\\
E12-06-106 & Study of Color Transparency in Exclusive Vector Meson Electroproduction off Nuclei\\
E12-06-107 & The Search for Color Transparency at 12 GeV\\
E12-06-108 & Hard Exclusive Electroproduction of pi0 and eta with CLAS12\\
E12-06-109 & The Longitudinal Spin Structure of the Nucleon\\
E12-06-110 & Measurement of Neutron Spin Asymmetry $A_{1_n}$ in the Valence Quark Region\\
& Using an 11 GeV Beam and a Polarized $^3$He Target in Hall C\\
E12-06-112 & Probing the Proton's Quark Dynamics in Semi-Inclusive Pion Production at 12 GeV\\
E12-06-113 & The Structure of the Free Neutron at Large x-Bjorken\\
E12-06-114 & Measurements of Electron-Helicity Dependent Cross Sections of\\
& Deeply Virtual Compton Scattering with CEBAF at 12 GeV\\
E12-06-117 & Quark Propagation and Hadron Formation\\
E12-06-119 & Deeply Virtual Compton Scattering with CLAS at 11 GeV\\
E12-07-104 & Measurement of the Neutron Magnetic Form Factor at High $Q^2$\\
& Using the Ratio Method on Deuterium\\
E12-07-105 & Scaling Study of the L-T Separated Pion Electroproduction Cross Section at 11 GeV\\
E12-07-107 & Studies of Spin-Orbit Correlations with Longitudinally Polarized Target\\
E12-07-108 & Precision Measurement of the Proton Elastic Cross Section at High $Q^2$\\
E12-07-109 & Large Acceptance Proton Form Factor Ratio Measurements at 13 and 15 (GeV$/c$)$^2$\\
& Using Recoil Polarization Method\\
E12-09-002 & Precise Measurement of $\pi^+/\pi^-$ Ratios in Semi-inclusive Deep Inelastic Scattering\\
& Part~I: Charge Symmetry violating Quark Distributions\\
E12-09-003 & Nucleon Resonance Studies with CLAS12\\
E12-09-005 & An Ultra-precise Measurement of the Weak Mixing Angle using Moller Scattering\\
E12-09-007 & Studies of partonic distributions using semi-inclusive production of kaons\\
E12-09-008 & Studies of the Boer-Mulders Asymmetry in Kaon Electroproduction\\
& with Hydrogen and Deuterium Targets\\
E12-09-009 & Studies of Spin-Orbit Correlations in Kaon Electroproduction\\
& in DIS with polarized hydrogen and deuterium targets\\
E12-09-011 & Studies of the L-T Separated Kaon Electroproduction Cross Section from 5--11 GeV\\
E12-09-016 & Measurement of the Neutron Electromagnetic Form Factor Ratio Gen/GMn at High $Q^2$\\
E12-09-017 & Transverse Momentum Dependence of Semi-Inclusive Pion Production\\
E12-09-018 & Measurement of the Semi-Inclusive pion and kaon electro-production in DIS regime\\
& from transversely polarized 3He target with the SBS\&BB spectrometers in Hall A\\
E12-09-019 & Precision Measurement of the Neutron Magnetic Form Factor up to $Q^2$=18.0 (GeV$/c$)$^2$\\
& by the Ratio Method
\end{tabular}

\clearpage
\centerline{\bf Approved Experiments Proposed During 2010--2014}
\begin{tabular}{ll}
Experiment & Title\\
\hline
E12-10-002 & Precision measurements of the $F_2$ structure function at large $x$\\
& in the resonance region and beyond\\
E12-10-003 & Deuteron Electro-Disintegration at Very High Missing Momentum\\
E12-10-006 & An update to PR12-09-014: Target Single Spin Asymmetry in Semi-Inclusive\\
& Deep-Inelastic Electro Pion Production on a Transversely Polarized $^3$He Target at 8.8 and 11 GeV\\
E12-10-007 & Precision Measurement of Parity-Violation in Deep Inelastic Scattering\\
& Over a Broad Kinematic Range\\
E12-10-008 & Detailed studies of the nuclear dependence of F2 in light nuclei\\
E12-10-011 & A Precision Measurement of the eta Radiative Decay Width via the Primakoff Effect\\
E12-10-103 & Measurement of the $F_{2_n}/F_{2_p}$, $d/u$ Ratios and $A=3$ EMC Effect\\
& in Deep Inelastic Scattering off the Tritium and Helium Mirror Nuclei\\
E12-11-002 & Proton Recoil Polarization in the 4He(e,e'p)3H, 2H(e,e'p)n, and 1H(e,e'p) Reactions\\
E12-11-003 & Deeply Virtual Compton Scattering on the Neutron with CLAS12 at 11 GeV\\
E12-11-005 & Meson spectroscopy with low $Q^2$ electron scattering in CLAS12\\
E12-11-006 & Heavy Photon Search at Jefferson Laboratory\\
E12-11-007 & Asymmetries in Semi-Inclusive Deep-Inelastic Electro-Production of Charged Pions\\
& on a Longitudinally Polarized $^3$He Target at 8.8 and 11 GeV\\
E12-11-008 & A Proposal for the DarkLight Experiment at the Jefferson Laboratory Free Electron Laser\\
E12-11-009 & The Neutron Electric Form Factor at $Q^2$ up to 7 (GeV/c)$^2$\\
& from the Reaction d(e,e'n)p via Recoil Polarimetry\\
E12-11-101 & PREX-II: Precision Parity-Violating Measurement of the Neutron Skin of Lead\\
E12-11-106 & High Precision Measurement of the Proton Charge Radius\\
E12-11-107 & In Medium Nucleon Structure Functions, SRC, and the EMC effect\\
E12-11-108 & Target Single Spin Asymmetry inSemi-Inclusive Deep-Inelastic ($e,e^\prime\pi^\pm$) Reaction\\
& on a Transversely Polarized Proton Target\\
E12-12-001 & Timelike Compton Scattering and J/psi photoproduction on the proton\\
& in $e^+e^-$ pair production with CLAS12 at 11 GeV\\
E12-12-002 & A study of meson and baryon decays to strange final states with GlueX in Hall D (C12-12-002)\\
E12-12-004 & CREX: Parity-Violating Measurement of the Weak Charge Distribution of $^{48}$Ca to 0.02 fm Accuracy\\
E12-12-006 & Near Threshold Electroproduction of J/$\psi$ at 11 GeV\\
E12-12-007 & Exclusive Phi Meson Electroproduction with CLAS12\\
E12-13-003 & An initial study of hadron decays to strange final states with GlueX in Hall D\\
E12-13-005 & Measurement of $^{16}$O($\gamma,\alpha$)$^{12}$C with a bubblechamber and a bremsstrahlung beam\\
E12-13-007 & Measurement of Semi-Inclusive $\pi^0$ Production as Validation of Factorization\\
E12-13-008 & Measuring the Charged Pion Polarizability in the $\gamma \gamma \rightarrow \pi^+ \pi^-$ Reaction\\
E12-13-010 & Exclusive Deeply Virtual Compton and Neutral Pion Cross-Section Measurements in Hall C\\
E12-14-001 & The EMC Effect in Spin Structure Functions\\
E12-14-002 & Precision Measurements and Studies of a Possible Nuclear Dependence of R\\
E12-14-003 & Wide-angle Compton Scattering at 8 and 10 GeV Photon Energies\\
E12-14-005 & Wide Angle Exclusive Photoproduction of $\pi^0$ Mesons\\
E12-14-009 & Ratio of the electric form factor in the mirror nuclei $^3$He and $^3$H\\
E12-14-011 & Proton and Neutron Momentum Distributions in $A=3$ Asymmetric Nuclei\\
E12-14-012 & Measurement of the Spectral Function of $^{40}$Ar through the ($e,e^\prime p$) reaction
\end{tabular}

\clearpage
\centerline{\bf Approved Experiments Proposed During 2015--2021}
\begin{tabular}{ll}
Experiment & Title\\
\hline
E12-15-001 & Measurement of the Generalized Polarizabilities of the Proton in Virtual Compton Scattering\\
E12-15-008 & An isospin dependence study of the Lambda-N interaction through the\\
& high precision spectroscopy of $\Lambda$ hypernuclei with electron beam\\
E12-16-001 & Dark matter search in a Beam-Dump eXperiment (BDX) at Jefferson Lab -- 2018\\
& update to PR12-16-001\\
E12-16-007 & A Search for the LHCb Charmed ÒPentaquarkÓ using\\
& Photoproduction of J/$\psi$ at Threshold in Hall C at Jefferson Lab\\
E12-16-010 & A Search for Hybrid Baryons in Hall B with CLAS12\\
E12-17-003 & Determining the Unknown $\Lambda-n$ Interaction by Investigating the $\Lambda-nn$ Resonance\\
E12-17-004 & Measurement of the Ratio $G_{E_n}/G_{M_n}$ by the Double-polarized $^2H(e,e^\prime n)$ Reaction\\
E12-17-005 & The CaFe Experiment: Short-Range Pairing Mechanisms in Heavy Nuclei\\
E12-17-006 & Electrons for Neutrinos: Addressing Critical Neutrino-Nucleus Issues\\
E12-17-008 & Polarization Observables in Wide-Angle Compton Scattering at large $s$, $t$ and $u$\\
E12-17-012 & Partonic Structure of Light Nuclei\\
E12-19-001 & Strange Hadron Spectroscopy with a Secondary $K_L$ Beam in Hall D\\
E12-19-002 & High accuracy measurement of nuclear masses of $\Lambda$ hyperhydrogens\\
E12-19-003 & Studying Short-Range Correlations with Real Photon Beams at GlueX\\
E12-19-006 & Study of the $L-T$ Separated Pion Electroproduction Cross Section at 11 GeV\\
& and Measurement of the Charged Pion Form Factor to High $Q^2$\\
E12-20-004 & PRad-II: A New Upgraded High Precision Measurement of the Proton Charge Radius\\
E12-20-005 & Precision measurements of A=3 nuclei in Hall B\\
E12-20-007 & Backward-angle Exclusive pi0 Production above the Resonance Region\\
E12-20-008 & Polarization Transfer in Wide-Angle Charged Pion Photoproduction\\
E12-20-010 & Measurement of the Two-Photon Exchange Contribution\\
& to the Electron-Neutron Elastic Scattering Cross Section\\
E12-20-011 & Measurement of the high-energy contribution to the Gerasimov-Drell-Hearn sum rule\\
E12-20-013 & Studying Lambda interactions in nuclear matter with the $^{208}$Pb($e,e^\prime K^+$)$^{208}_\Lambda$Tl\\
E12-21-005 & Double Spin Asymmetry in Wide-Angle Charged Pion Photoproduction
\end{tabular}
\clearpage